\documentclass[aps,prb,superscriptaddress,notitlepage,nopacs,amsmath,amstex,amssymb,citeautoscript,longbibliography,floatfix,letter,twocolumn]{revtex4-2}
\usepackage{graphicx, amssymb, color, amsmath}
\usepackage{nicefrac}
\usepackage{multirow, array, booktabs}
\usepackage{longtable}
\usepackage{float}
\usepackage{romannum}
\usepackage{mathtools}
\usepackage{bbm}
\usepackage{mathrsfs}
\usepackage{IEEEtrantools}
\usepackage{relsize}
\usepackage{mathrsfs}
\usepackage{mathtools}
\usepackage{xparse}
\usepackage{subfigure}
\usepackage{adjustbox}
\usepackage{longtable}
\usepackage{bm}
\usepackage{color}
\usepackage{braket}
\usepackage{standalone}
\usepackage{multirow}
\usepackage{tikz}
\usepackage{mathrsfs}
\usepackage{dsfont}
\usepackage{comment}
\usepackage{amsmath}
\usepackage[colorlinks,bookmarks=true,citecolor=blue,linkcolor=red,urlcolor=blue]{hyperref}
\usepackage{cleveref}
\usepackage{orcidlink}
\renewcommand{\vec}[1]{\mbox{\boldmath$#1$}}

\graphicspath{{./figures/}}
\begin{document}
\pagenumbering{arabic}

\title{Displacement field stabilizes even-denominator and partonic fractional quantum Hall states in the $\mathcal{N}{=}2$ Landau levels of Bernal-stacked bilayer graphene}

\author{Rakesh K. Dora\orcidlink{0009-0009-0043-2982}}
\email{prakeshdora1729@gmail.com}
\affiliation{Institute of Mathematical Sciences, CIT Campus, Chennai, 600113, India}
\affiliation{Homi Bhabha National Institute, Training School Complex, Anushakti Nagar, Mumbai 400094, India}
\affiliation{Department of Physics, Indian Institute of Technology Bombay, Mumbai, MH 400076, India}

\author{Udit Khanna\orcidlink{0000-0002-3664-4305}}
\email{udit.khanna.10@gmail.com}
\affiliation{Theoretical Physics Division, Physical Research Laboratory, Navrangpura, Ahmedabad-380009, India}

\author{Ajit C. Balram\orcidlink{0000-0002-8087-6015}}
\email{cb.ajit@gmail.com}
\affiliation{Institute of Mathematical Sciences, CIT Campus, Chennai, 600113, India}
\affiliation{Homi Bhabha National Institute, Training School Complex, Anushakti Nagar, Mumbai 400094, India}

\date{\today}

\begin{abstract}
Bernal-stacked bilayer graphene (BLG), in which a graphene layer is stacked atop another and laterally shifted by a lattice constant, offers remarkable tunability in its single-particle states under applied magnetic and displacement fields. Owing to this tunability, recent transport and scanning tunneling microscopy experiments in the presence of a perpendicular magnetic field and finite interlayer displacement fields have observed even-denominator fractional quantum Hall (FQH) states at half-filling in the first excited, namely, $\mathcal{N}{=}2$, Landau level (LL) of BLG. In contrast, at zero displacement field, a gapless composite fermion Fermi liquid (CFFL) is realized at half-filling of the $\mathcal{N}{=}2$ LL. Motivated by these experiments, we compute the phase diagram as a function of the displacement field in the half-filled $\mathcal{N}{=}2$ LL of BLG by studying the competition between the CFFL and the Moore-Read state—a candidate even-denominator FQH state—by calculating their thermodynamic energies in this setting. We find that the modified effective Coulomb interaction, induced by changes in the single-particle states with increasing displacement field, softens the inter-electronic repulsion at short distances, thereby stabilizing the Moore-Read state over the CFFL in the $\mathcal{N}{=}2$ LL of BLG. We also study the nature of FQH states at fillings $2/5$, $3/7$, $4/9$, and $6/13$ in the $\mathcal{N}{=}2$ LL of BLG. Our results suggest that, with increasing displacement field, the Jain composite-fermion states at $3/7$, $4/9$, and $6/13$ transition into states with distinct topological order that are well-captured by parton wave functions.
\end{abstract}

\maketitle

\section{Introduction}
\label{sec: introduction}
Fractional quantum Hall (FQH) states~\cite{Tsui82} are strongly interacting states carrying topological order and long-range entanglement that emerge at ultra-low temperatures in two-dimensional electron systems (2DESs) subjected to strong perpendicular magnetic fields~\cite{Tsui82}. The magnetic field quenches the kinetic energy of the electrons into a discrete set of energy levels, known as Landau levels (LLs), thereby enhancing the role of electron–electron interactions. The Coulomb interaction between the electrons can stabilize incompressible FQH states at certain fractional fillings of a LL. FQH states have been observed in many 2DESs, most prominently in semiconductor devices based on GaAs/AlGaAs-quantum wells/heterostructures~\cite{Tsui82} and graphene~\cite{Novoselov04, Zhang05}. The latter include monolayer graphene (MLG), which is a single atomic layer of carbon atoms that hosts a pristine 2DES, as well as bilayer and trilayer graphene, which also realize 2DESs~\cite{Neto09, Aoki13}.

The observed sequence and nature of FQH states in a given physical system are governed by the single-particle states within the LLs. For example, the lowest LL (LLL), indexed by $n{=}0$, in narrow GaAs quantum wells stabilizes FQH states that are distinct from those in its second LL (SLL), indexed by $n{=}1$~\cite{DasSarma07, Halperin20}. The strong short-range Coulomb repulsion inherited from the nodeless analytic LLL wave functions stabilizes odd-denominator FQH states along the Jain sequence of LL fillings, $\nu{=}s/(2ps\pm1)$, where $s$ and $p$ are integers. These states are understood as integer quantum Hall (IQH) states of non-interacting composite fermions (CFs), which are bound states of electrons and an even number of vortices~\cite{Jain89, Jain07}. Moreover, at even-denominator fillings in the LLL of the type $\nu{=}1/(2p)$, the states are compressible and are interpreted as composite fermion Fermi liquids (CFFLs)~\cite{Halperin93}. By contrast, the weaker short-range repulsion in the SLL, arising from a node in its single-particle wave function, stabilizes even-denominator FQH states~\cite{Willett87}, as well as odd-denominator FQH states whose nature is distinct from their LLL counterparts~\cite{Pan08, Choi08, Kumar10, Zhang12}. These $\nu{=}1/(2p)$ FQH states can be understood as paired states of CFs~\cite{Greiter91, Greiter92a, Read00, Morf98, Scarola02b, Sharma23}, in which there is an effective attraction between CFs that arises from the suppression of the short-range repulsion between the electrons in the SLL, which destabilizes the CFFL to turn it into a superconductor.

Similarly, in wide GaAs quantum wells, the Coulomb interaction is modified by the out-of-plane component of the single-particle wave functions, which can be tuned by varying the carrier density and the well-width. In wide quantum wells, this can result in stabilizing even-denominator states in the LLL itself, such as those at $\nu{=}1/2$, $1/4$~\cite{Suen92, Suen94, Luhman08, Shabani09a, Shabani09b, Singh23, Singh25}. Moreover, tuning the carrier density can drive a sequence of phase transitions at a given filling factor~\cite{Shabani13}. For example, at $\nu{=}1/2$, a CFFL at low density can evolve into a Moore-Read (MR) state~\cite{Moore91}, which can be viewed as a $p$-wave paired state of CFs~\cite{Greiter91, Greiter92a}, at intermediate density~\cite{Singh25}.

A variety of FQH states have also been observed in experiments on graphene-based devices. At low energies, in contrast to non-relativistic electrons in GaAs systems, the single-particle electronic spectrum in graphene can be Dirac-like, which is described by a multi-component spinor. Moreover, these systems host additional isospin or pseudospin quantum numbers—the valley and orbital degrees of freedom—alongside the usual spin~\cite{Min08, Barlas12}. Furthermore, the relative weights of the single-particle spinor components in graphene and its multilayer avatars, such as bilayer and trilayer graphene, can be tuned by external parameters, including a potential difference between inequivalent sublattice sites, an interlayer displacement field, or a perpendicular magnetic field~\cite{Aoki13}. Consequently, these parameters can modify the effective Coulomb interaction within an LL, thereby controlling the nature of the resulting FQH states.

Below, we summarize the FQH states observed by tuning the aforementioned parameters in graphene-based systems. The zeroth LL (ZLL), labeled $\mathcal{N}{=}0$, in MLG primarily hosts odd-denominator FQH states~\cite{Du09, Dean11, Feldman12, Feldman13, Amet15}. However, in the presence of a potential difference between the inequivalent sublattice sites, even-denominator FQH states emerge in the ZLL, in the vicinity of parameter regimes where LLs from opposite valleys cross, thereby modifying the effective interaction through LL mixing~\cite{Zibrov17}. A qualitatively similar phenomenon arises in the ZLL ($\mathcal{N}{=}0$) of the monolayer-like band in the Bernal-stacked trilayer graphene, where an interlayer displacement field stabilizes FQH states at half-fillings~\cite{Chanda25}. Even-denominator FQH states have been observed in the $\mathcal{N}{=}3$ LL of MLG~\cite{Kim19, Sharma22}.

In recent years, remarkable control over the isospin degrees of freedom in bilayer graphene (BLG) has been achieved through an interlayer displacement field. This has led to the discovery of many valley-correlated odd-denominator~\cite{Zibrov16, Huang21, Kumar24, Huang25} and even-denominator states~\cite{Zibrov16, Huang21, Kumar24, Hu24} in the ZLL of BLG, which is spanned by spin, valley, and $\mathcal{N}{=}0,1$ orbital degrees of freedom. The ZLL states with $\mathcal{N}{=}1$ orbital character host even-denominator FQH states at half-fillings, whereas no such states have been observed in $\mathcal{N}{=}0$ orbital LLs~\cite{Kumar24}. The ability to tune the orbital character of the ZLL states with displacement fields allows inducing phase transitions of even-denominator FQH states into compressible CFFLs. Moreover, the $\mathcal{N}{=}1$ LL has orbital character that is intermediate between $n{=}0$ and $n{=}1$ LLs, whose relative amplitudes can be tuned with the magnetic field~\cite{Apalkov11}. This tuning can also drive a transition from a paired CF state, like the MR, that occurs at low magnetic fields to a CFFL at high magnetic fields, as observed experimentally~\cite{Zhu20a} and in numerics~\cite{Zhu20a, Balram21b}.  

In this article, we focus our attention on the first excited LL of Bernal-stacked BLG, namely, the spin- and valley-polarized $\mathcal{N}{=}2$ LL, in which a recent transport experiment observed a phase transition from a CFFL to an incompressible state at $\nu{=}1/2$ with increasing displacement field~\cite{Kumar25}. Moreover, an FQH state at $\nu{=}1/2$ has also been observed in the $\mathcal{N}{=}2$ LL in a scanning tunneling microscopy (STM) experiment with an intrinsic displacement field between the layers~\cite{Hu24}. However, previous experiments without a displacement field did not observe the $\nu{=}1/2$ state in the $\mathcal{N}{=}2$ LL~\cite{Diankov16}, suggesting that a finite displacement field is required to stabilize the even-denominator state. The experimentally observed even-denominator FQH state at half-filling of the $\mathcal{N}{=}2$ LL of BLG is likely to lie in the MR universality class as evinced by the FQH effect observed at fillings that flank it, which can be attributed to the daughter states of the parent MR state~\cite{Levin09a}.

Encouragingly, in our theoretically/numerically-computed phase diagram we find that increasing the displacement field does drive the CFFL into the MR state. We employ the four-band model of BLG, in which the $\mathcal{N}{=}2$ LL spinor states contain components of the $n{=}0$, $n{=}1$, and $n{=}2$ LLs of non-relativistic electrons. The weights of these components can be tuned with the displacement field. Beyond a certain displacement-field strength, we find that the effective interaction in the $\mathcal{N}{=}2$ LL favors pairing of CFs over their Fermi sea. A previous study obtained a similar phase diagram by calculating the overlap of the MR state with the exact ground states obtained through exact diagonalization, which is limited to small system sizes~\cite{Papic11}. However, we have obtained the phase diagram by computing the energies of the CFFL and MR states in the thermodynamic limit. 

In addition to the state at $\nu{=}1/2$, we have investigated the nature of FQH states at $\nu{=}2/5, 3/7$, $4/9$, and $6/13$ in the $\mathcal{N}{=}2$ LL of BLG. The FQH states along this sequence can be described by Jain CF wave functions, which provide an accurate description in the $n{=}0$ LL but do not provide a good description in the $n{=}1$ LL~\cite{Jain07, Balram21}. Instead, FQH states in the $n{=}1$ LL can be described in a unified manner by a class of wave functions known as parton wave functions~\cite{Jain89b, Balram18, Balram18a, Balram19, Balram21, Bose23, Balram21}. Thus, we also study the competition between Jain CF and parton states in the $\mathcal{N}{=}2$ LL of BLG by tuning the parameters of the four-band model, which are controlled by the displacement and magnetic fields. 

Finally, we have also considered stripe states of electrons~\cite{Fogler96, Koulakov96}, which compete with FQH states near the half-filling of a LL. However, we note that we have not considered the possibility of other charge-density wave states, such as Wigner crystals and bubble phases of electrons or CFs~\cite{Lam84, Fogler96, Goerbig04a, Archer13, Zuo20}, which can potentially be competitive in the regime of our interest~\cite{Yutushui26}. Our main results are summarized in the phase diagrams shown in Fig.~\ref{fig: schematic_phase_diagram}, which are obtained by combining our theoretical results with the experimental observations. 

The rest of the article is organized as follows.  In Sec.~\ref{ssec: single_particle_Hamiltonian}, we discuss the single-particle Hamiltonian of Bernal-stacked BLG and the corresponding LL spectrum. In Sec.~\ref{ssec: interaction_Hamiltonian}, we present the interaction Hamiltonian projected into the $\mathcal{N}{=}2$ LL of BLG, and in Sec.~\ref{ssec: candidate_states}, we list the candidate FQH states considered in this study. We compute the energy of FQH and stripe states in Sec.~\ref{sec: Energies_of_candidate_states}. The phase diagram of the competing candidate states at the fillings $2/5$, $3/7$, $4/9$, $6/13$, and $1/2$ in the $\mathcal{N}{=}2$ LL of BLG is discussed in Sec.~\ref{sec: results}. We close this article by summarizing the main results and discussing the connection to other works in the literature, as well as possible future directions in Sec.~\ref{sec: discussion}. Some additional results are presented in Appendices~\ref{app: 6_13_parton_WQW}-\ref{app: thermodynamic_energies_candidates_different_LLs}.  

\section{Model and methods}
\label{sec: model_methods}

\subsection{Single-particle Hamiltonian and Landau levels}
\label{ssec: single_particle_Hamiltonian}
\begin{figure}
    \centering
    \includegraphics[width=0.99\columnwidth]{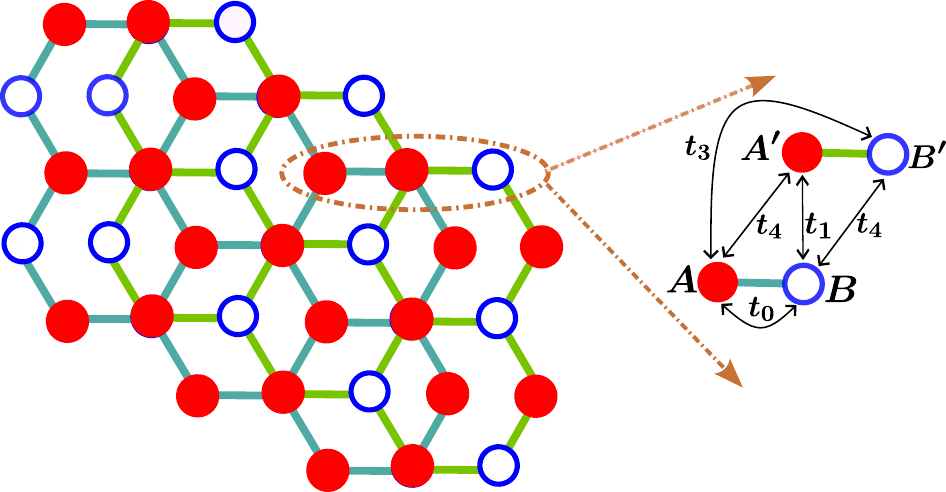}
    \caption{Schematic lattice structure of Bernal-stacked bilayer graphene. The left-hand side shows the top view of the bilayer graphene, with the unit cell highlighted by a brown dashed ellipse. The side view of the unit cell is shown on the right-hand side. The inequivalent sublattice sites in each layer are represented by filled red circles for $A$ (bottom layer) and $A^{'}$ (top layer) sites, and empty blue circles for $B$ (bottom layer) and $B^{'}$ (top layer) sites. The intralayer hopping between inequivalent sites within each layer is represented by $t_0$. The hopping between nearest interlayer sites, $A^{\prime}$ and $B$, is denoted by $t_1$. Similarly, the hoppings between the interlayer equivalent sublattice sites---$A{\leftrightarrow}A^{\prime}$ and $B{\leftrightarrow}B^{\prime}$---within the unit cell are denoted by $t_4$. The hopping between the farthest sites, $A$ and $B^{\prime}$, in the unit cell is denoted by $t_3$, whose strength we set to zero in this study.}
    \label{fig: schematic_BLG_lattice}
\end{figure}

We model the single-electron problem in the Bernal-stacked BLG using the Slonczewski–Weiss–McClure tight-binding parametrization~\cite{McClure57}. The unit cell of Bernal BLG contains four lattice sites: the bottom-layer $A$ and $B$ sublattices, and the top-layer $A^{\prime}$ and $B^{\prime}$ sublattices, arranged such that $A^{\prime}$ lies directly above $B$ (see Fig.~\ref{fig: schematic_BLG_lattice}). Within each layer, the electronic motion is described by the nearest-neighbor hopping between the inequivalent $A$ and $B$ sites with an amplitude $t_{0}{=}2.61$ eV, while between layers, electrons can hop vertically between aligned sites $A^{\prime}$ and $B$ sites with an amplitude $t_{1}{=}0.361$ eV, as well as between the next-nearest interlayer equivalent sites---i.e., $A {\leftrightarrow}A^{\prime}$ and $B {\leftrightarrow }B^{\prime}$---with amplitude $t_4{=}0.138$ eV~\cite{Hunt17}. A longer-range hopping between $A$ and $B^{\prime}$ sites, which gives rise to the trigonal warping of the Fermi surface, is neglected here since its effect is small under the strong magnetic fields that we consider~\cite{Khanna23}. We also account for a local electrostatic potential offset $\Delta{=}0.015$ eV~\cite{Hunt17} between the Bernal $(B, A^{\prime})$ and non-Bernal $(A, B^{\prime})$ sites, reflecting the difference in their stacking environments. 

\subsubsection{Bilayer graphene Landau levels on the plane}
As in MLG, the low-energy physics of BLG near the charge neutrality point is governed by states in the vicinity of the two inequivalent corners of the Brillouin zone, $\boldsymbol{K}$ and $\boldsymbol{K}^{\prime}$, which serve as valley quantum numbers. The low-energy effective Hamiltonian in the vicinity of the $\boldsymbol{K}$ and $\boldsymbol{K}^{\prime}$ points in the continuum limit can be written in the basis $\left(A, B, A^{\prime}, B^{\prime}\right)$ as~\cite{McCann06} 
\begin{align}
\label{eq: intrinsic_hopping_Hamiltonian}
    H_{\alpha} &=\begin{pmatrix}
    0 & v_0 \pi^{\dagger} & -v_{4}\pi^{\dagger}& 0\\
    v_0\pi & \Delta & t_1 & -v_4\pi^{\dagger}\\
    -v_4\pi & t_1 & \Delta& v_0 \pi^{\dagger}\\
    0 & -v_4\pi & v_0 \pi & 0
    \end{pmatrix}.
\end{align}
Here, $\pi{=}\hbar(\alpha q_x{+}iq_{y})$ is the canonical momentum operator with $\vec{q}{=}(q_x, q_y)$ being the momentum defined relative to $\boldsymbol{K},\boldsymbol{K}^{\prime}$ points with $\alpha{=}{+}1$ for $\boldsymbol{K}$ valley, and $\alpha{=}{-}1$ for $\boldsymbol{K}^{\prime}$ valley. The velocity parameters, $v_{0}{=}\sqrt{3}a_0t_{0}/(2\hbar){=}8.44{\times}10^{5}$ m/s and $v_{4}{=}\sqrt{3}a_0 t_{4}/(2\hbar){=}4.47{\times}10^{4}$ m/s~\cite{Hunt17}, are related to the corresponding hopping parameters $t_0$ and $t_4$, respectively, via the graphene lattice constant $a_0{=}2.46$\AA. The nature of the single-particle wave functions can be tuned via an interlayer displacement field and a magnetic field. In the presence of an electrostatic potential $U$ induced by the interlayer displacement field, the Hamiltonian of Eq.~\eqref{eq: intrinsic_hopping_Hamiltonian} gets an additional term given by
\begin{align}
    \label{eq: displacement_field_HAmiltonian}
    H_{U}&=\begin{pmatrix}
        -U/2 & 0& 0&0\\
        0 & -U/2 & 0& 0\\
        0&0&U/2&0\\
        0&0&0&U/2
    \end{pmatrix}.
\end{align}
In the STM experiment under consideration, there is an intrinsic offset of $U{=}30$ meV~\cite{Hu24}. The presence of a uniform out-of-plane magnetic field in the quantum Hall regime is incorporated by changing the canonical momentum $\pi$ to the gauge-invariant mechanical momentum $\Pi{=}\pi{+}eA/c$, where $A{=}A_{x}{+}iA_{y}$ with $\boldsymbol{A}{=}(A_{x}, A_{y})$ being the vector potential. The operators $\Pi,~\Pi^{\dagger}$ satisfy the canonical commutation relation $\left[\Pi, \Pi^{\dagger}\right]{=}{-}2\hbar e B_{\perp} \alpha$, where $B_{\perp}$ is the strength of the perpendicular magnetic field. Thus, the operators $\Pi,~\Pi^{\dagger}$ can be identified with the harmonic oscillator ladder operators $a$, $a^{\dagger}$ that satisfy $\left[a, a^{\dagger}\right]{=}1$ as: $\Pi{=}\left(\sqrt{2}\hbar /\ell\right)a^{\dagger}$ for $\alpha{=}{+}1$, and $\Pi{=}\left(\sqrt{2}\hbar /\ell\right)a$ for $\alpha{=}{-}1$, where $\ell{=}\sqrt{\hbar c/(eB_{\perp})}$ is the magnetic length. To this end, in the vicinity of the $\boldsymbol{K}$ valley, the single-particle Hamiltonian in a perpendicular magnetic field is
\begin{align}
\label{eq: single_particle_magnetic_Hmiltonian_K_valley}
 H_{+}&=  
 \hbar\omega_0\begin{pmatrix}
        -\gamma_{U} & a & -\gamma_{4}a & 0\\
        a^{\dagger} & \gamma_\Delta\gamma_1-\gamma_{U} & \gamma_1 & - \gamma_{4}a \\
        -\gamma_{4}a^{\dagger} & \gamma_1 & \gamma_\Delta \gamma_1+\gamma_U & a\\
        0 & - \gamma_{4}a^{\dagger} & a^{\dagger}& \gamma_U
    \end{pmatrix}.
\end{align}
Here, we have defined $\hbar\omega_0{=}\sqrt{2}\hbar v_0 /\ell{\approx}30.68\sqrt{B_{\perp}[T]}$ meV, $\gamma_1{=}t_1/(\hbar\omega_0){\approx}11.7/\sqrt{B_{\perp}[T]}$, $\gamma_4{=}v_4/(v_0){\approx}0.053$, $\gamma_{\Delta}{=}\Delta/t_1{\approx}0.042$, and $\gamma_U{=}U/(2\hbar\omega_0){\approx}0.49/\sqrt{B_{\perp}[T]}$, for $U{=}30$ meV. The Hamiltonian $H_{-}$ near $\boldsymbol{K}^{\prime}$ valley can be obtained by replacing $a{\rightarrow}{-}a^{\dagger}$ in Eq.~\eqref{eq: single_particle_magnetic_Hmiltonian_K_valley}. In the following, we present results for the $\boldsymbol{K}$ valley; similar results follow at the $\boldsymbol{K}^{\prime}$ valley. 

\begin{figure}
    \centering
    \begin{tabular}{cc}
     \includegraphics[width=0.49\columnwidth]{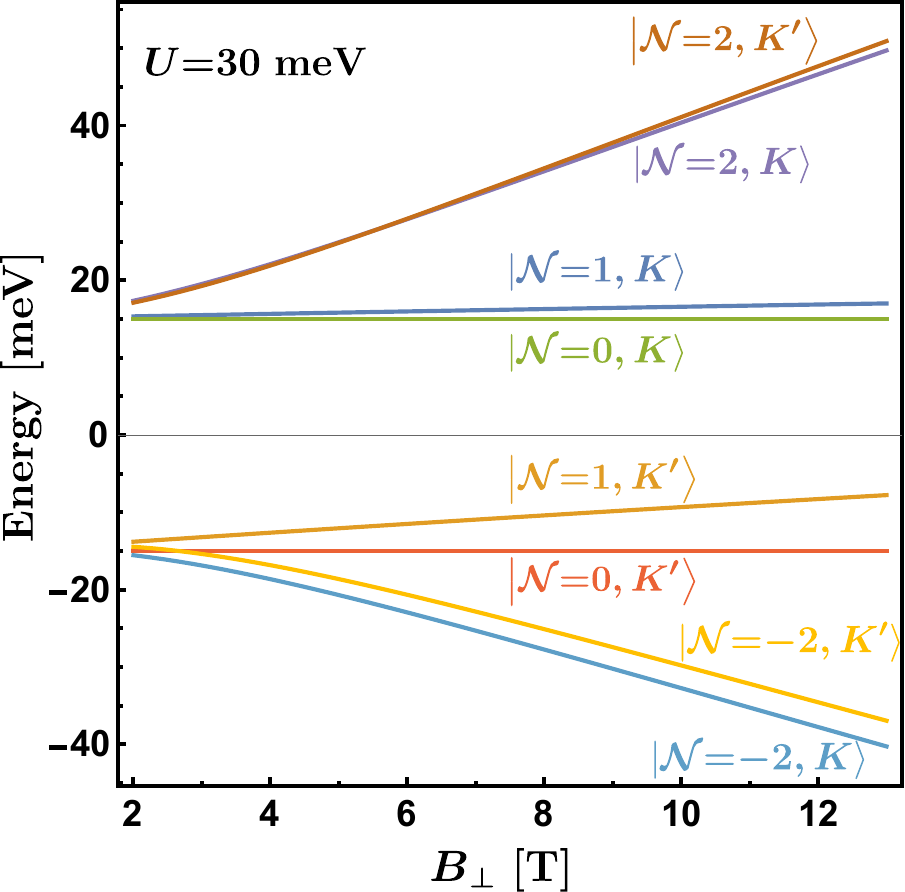}
    \includegraphics[width=0.49\columnwidth]{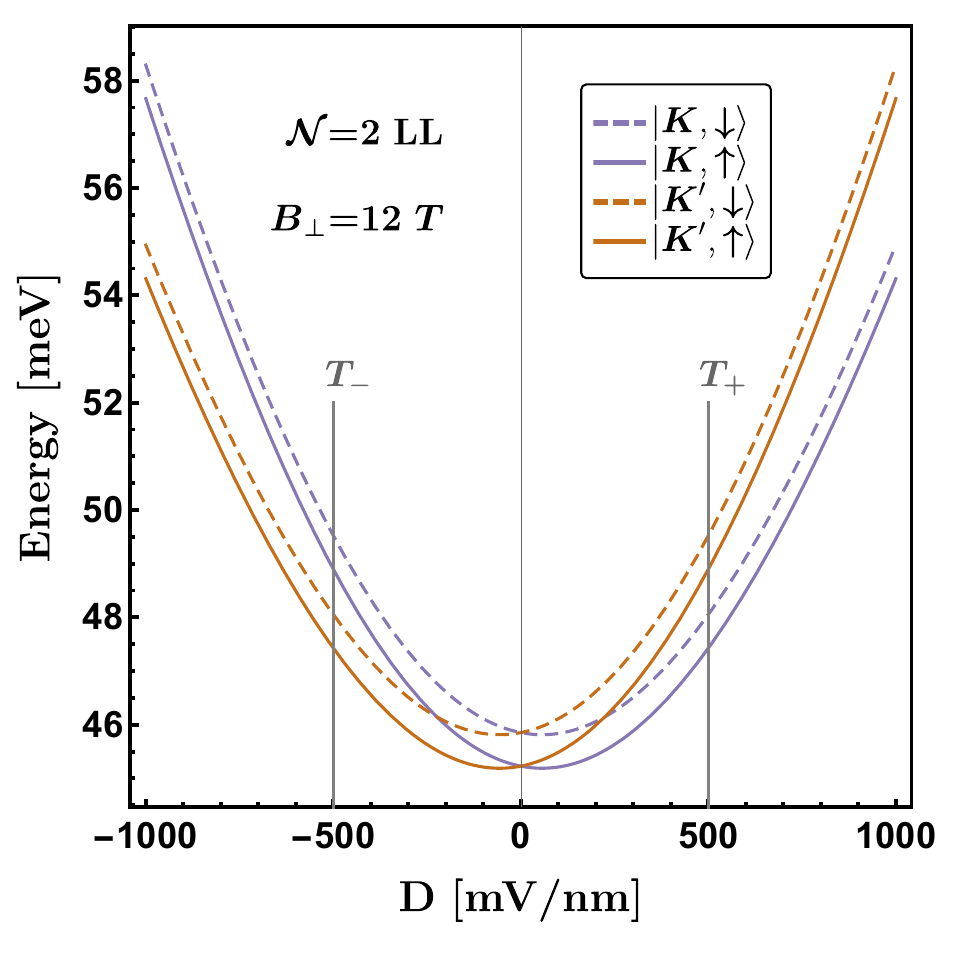}
    \end{tabular}
    \caption{Landau level energies in bilayer graphene. Panel $(a)$ shows the valley-resolved Landau level energies for $\mathcal{N}{=}0,1$, and $\pm2$ as a function of the magnetic field, with the interlayer potential fixed at $U{=}30$ meV. For visual clarity, we have set the Zeeman coupling to zero. Panel $(b)$ shows the energy of each $\mathcal{N}{=}2$ Landau level in bilayer graphene, labeled by spin-valley quantum numbers, as a function of displacement field $D$ at a fixed magnetic field of $B_{\perp}{=}12$ T. We have used a renormalized value of the Land\'e $g$ factor, $g{\approx}0.9$~\cite{Kumar25}. The vertical lines marked by $T_{+}$ and $T_{-}$ in panel $(b)$ highlight the displacement fields $D{=}500$ and $D{=}{-}500$ mV/nm, respectively. }
    \label{fig: BLG2_LLs}
\end{figure}

The LL spectrum of $H_+$ can be analytically solved using the properties of the ladder operators: $a\left|n,k\right\rangle{=}\sqrt{n}\left|n{-}1,k \right\rangle$, and $a^{\dagger}\left|n,k\right\rangle{=}\sqrt{n{+}1}\left|n{+}1,k \right\rangle$. Here, $\left|n,k\right\rangle$, with $n{=}0,1,2,{\cdots}$, are the LL eigenstates of non-relativistic electrons in GaAs, or equivalently, shifted harmonic oscillator eigenstates centered around $x_0{=}k\ell$, where $k$ labels the guiding center. The spectrum of $H_+$ consists of a discrete set of LLs labeled by an integer $\mathcal{N}$. We are interested in the LLs whose energies lie well below the band gap ${\sim}t_1$--- separating the lower and upper conduction bands (equivalently, also the valence bands) of BLG---in the absence of a magnetic field. Within this low-energy window, the BLG LLs consist of ZLL states spanned by the orbital indices $\mathcal{N}{=}0$ and $1$, while the labels $\mathcal{N}{\geq}2$ and $\mathcal{N}{\leq}{-}2$ correspond to positive- and negative-energy states, respectively. The eigenstates of $\mathcal{N}{=}0$, $1$, and $|\mathcal{N}|{\geq}2$ LLs for $\alpha{=}{+}1$ in a given spin sector are given by
\begin{align}
 \label{eq: four_band_model_N_0_two_eigenstates}
  \left|\Phi_{\mathcal{N}{=}0,k,\alpha=+}\right\rangle&=\begin{pmatrix}
      0\\0\\0\\\left|n{=}0,k\right\rangle
  \end{pmatrix},\\[1.5ex]
   \label{eq: four_band_model_N_1_two_eigenstates}
  \left|\Phi_{\mathcal{N}{=}1,k,\alpha=+}\right\rangle&=\begin{pmatrix}
      0\\\sin(\theta_1)\cos(\phi_1)\left|n{=}0,k\right\rangle\\\sin(\theta_1)\sin(\phi_1)\left|n{=}0,k\right\rangle\\\cos(\theta_1)\left|n{=}1,k\right\rangle
  \end{pmatrix}, \\[1.5ex]
  \label{eq: four_band_model_N_geq_two_eigenstates}
  \left|\Phi_{|\mathcal{N}|{\geq}2,k,\alpha=+}\right\rangle&=\begin{pmatrix}
      \cos(\theta_{\mathcal{N}})\left|n-2,k\right\rangle \\ -\sin(\theta_\mathcal{N})\cos(\phi_\mathcal{N})\cos(\tilde{\phi}_{\mathcal{N}})\left|n-1,k\right\rangle\\ \sin(\theta_\mathcal{N})\cos(\phi_\mathcal{N})\sin(\tilde{\phi}_\mathcal{N})\left|n-1,k\right\rangle\\\sin(\theta_\mathcal{N})\sin(\phi_\mathcal{N})\left|n,k\right\rangle
  \end{pmatrix}.
\end{align}
In Eq.~\eqref{eq: four_band_model_N_geq_two_eigenstates} the non-relativistic LL index $n{=}|\mathcal{N}|$. The parameters $\theta_{\mathcal{N}}$, $\phi_{\mathcal{N}}$, and $\tilde{\phi}_{\mathcal{N}}$ generally involve lengthy expressions in terms of the BLG parameters appearing in Eq.~\eqref{eq: single_particle_magnetic_Hmiltonian_K_valley}; nonetheless, they can be evaluated numerically. In particular, the parameters $\theta_1$ and $\phi_1$ in $\left|\Phi_{\mathcal{N}{=}1,k,\alpha=+}\right\rangle$ can be obtained by diagonalizing the Hamiltonian
\begin{align}
    H_{+}^{\rm ZLL}& =
    \hbar\omega_0\begin{pmatrix}
        -\gamma_{U} & 0 & 0 & 0\\
        0 & \gamma_\Delta\gamma_1-\gamma_{U} & \gamma_1 & - \gamma_{4} \\
       0 & \gamma_1 & \gamma_\Delta \gamma_1+\gamma_U & 1\\
        0 & - \gamma_{4} & 1 & \gamma_U
    \end{pmatrix}.
\end{align}
The above Hamiltonian has one eigenvalue ${-}\gamma_U$, which corresponds to $ \left|\Phi_{\mathcal{N}{=}0,k,\alpha=+}\right\rangle$. The eigenstate corresponding to the eigenvalue with the smallest magnitude among the remaining three eigenvalues determines the parameters $\theta_1$ and $\phi_1$ in $\left|\Phi_{\mathcal{N}{=}1,k,\alpha=+}\right\rangle$. Similarly, the parameters $\theta_{\mathcal{N}}$ and $\phi_{\mathcal{N}}$ in $ \left|\Phi_{|\mathcal{N}|{\geq}2,k,\alpha=+}\right\rangle$ can be determined by diagonalizing
\begin{align}
    H_{+}^{|\mathcal{N}|{\geq}2}&= \hbar\omega_0\begin{pmatrix}
        -\gamma_{U} & \sqrt{n-1} & -\gamma_4\sqrt{n-1} & 0\\
         \sqrt{n-1} & \gamma_\Delta\gamma_1-\gamma_{U} & \gamma_1 & - \gamma_{4}\sqrt{n} \\
       -\gamma_4\sqrt{n-1} & \gamma_1 & \gamma_\Delta \gamma_1+\gamma_U & \sqrt{n}\\
        0 & - \gamma_{4}\sqrt{n} & \sqrt{{n}} & \gamma_U
    \end{pmatrix}.
\end{align}
The lowest positive eigenvalue of the above matrix determines the state $\left|\Phi_{\mathcal{N}{\geq}2,k,\alpha=+}\right\rangle$, while the negative eigenvalue with the smallest magnitude corresponds to the state  $\left|\Phi_{\mathcal{N}{\leq}{-}2,k,\alpha=+}\right\rangle$. The rest of the eigenstates lie in the high-energy sector above the band gap ${\sim}t_1$. In Fig.~\ref{fig: BLG2_LLs}$(a)$, we present the energies of a few LLs, including $\mathcal{N}{=}0$, $1$, and $\pm2$, as a function of $B_{\perp}$ for a fixed $U{=}30$ meV. In Fig.~\ref{fig: BLG2_LLs}$(b)$, we show the $\mathcal{N}{=}2$ LL energies by varying the displacement field, while keeping the magnetic field constant at $B_{\perp}{=}12$ T.

\subsubsection{Bilayer graphene Landau levels on the Haldane sphere}
In the previous section, we discussed the LLs of BLG on an infinite plane. For numerical studies, it is often convenient to consider the compact spherical geometry, where $N$ electrons move on the surface of a sphere threaded by a radially outward magnetic field $B_{\perp}$ emanating from a magnetic monopole of strength $2Q$ (gauge invariance mandates that $2Q$ is an integer) placed at the center of the sphere~\cite{Haldane83}. Since $2Qhc/e$ flux quanta thread the sphere, its radius $R$ is related to $Q$ as $R{=}\sqrt{Q}\ell$. The LLs of BLG on the sphere are obtained by translating the planar Hamiltonian in Eq.~\eqref{eq: single_particle_magnetic_Hmiltonian_K_valley} to the sphere, where the ladder operators $a$ and $a^{\dagger}$ transform into the magnetic monopole creation and annihilation operators, respectively~\cite{Arciniaga16}. The resulting LL spinor eigenstates have components expressed in terms of the monopole spherical harmonics~\cite{Wu76, Wu77}: $Y^{Q}_{l{=}Q{+}n,m}{\equiv}Y^{Q}_{n,m}{\equiv}|Q,n,m\rangle$, where $l{=}Q{+}n$ is the shell angular momentum of a LL indexed by $n{=}0,1,2,{\cdots}$, and $m$ is the azimuthal quantum number ranging from $-l{\leq}m{\leq}l$~\cite{Hsiao20, Balram21b, Dora23}. It is useful to note the action of $a$ and $a^{\dagger}$ on $|Q,n,m\rangle$: $a|Q,n,m\rangle {\propto}|Q{+}1,n{-}1,m\rangle$, and $a^{\dagger}|Q,n,m\rangle {\propto}|Q{-}1,n{+}1,m\rangle$~\cite{Arciniaga16}. The positive energy $\mathcal{N}{\geq}2$ LL eigenstates on the sphere are given by
\begin{align}
\label{eq: sphere_four_band_two_band_model_N_geq_2_LLs}
          \left|\Phi_{\mathcal{N}{\geq}2,k,\alpha=+}\right\rangle&=\begin{pmatrix}
      \cos\left(\theta_{\mathcal{N}}\right)~Y^{Q+2}_{\mathcal{N}-2,m} \\ -\sin\left(\theta_{\mathcal{N}}\right)\cos\left(\phi_{\mathcal{N}}\right)\cos\left(\tilde{\phi}_{\mathcal{N}}\right) ~Y^{Q+1}_{\mathcal{N}-1,m}\\ 
\sin\left(\theta_{\mathcal{N}}\right)\cos\left(\phi_{\mathcal{N}}\right)\sin\left(\tilde{\phi}_{\mathcal{N}}\right) ~Y^{Q+1}_{\mathcal{N}-1,m}
     \\
     \sin\left(\theta_{\mathcal{N}}\right)\sin\left(\phi_{\mathcal{N}}\right)~Y^{Q}_{\mathcal{N},m}
  \end{pmatrix}.
\end{align}
The parameters $\theta_{\mathcal{N}},\phi_{\mathcal{N}}$, and $\tilde{\phi}_{\mathcal{N}}$ generally depend on $Q$ as well as on the BLG parameters, such as the hopping strengths, and external parameters, such as the inter-layer displacement field. However, as we intend to compute the energies of many-body states in a given LL in the thermodynamic limit, we take the values of $\theta_{\mathcal{N}},\phi_{\mathcal{N}}$, and $\tilde{\phi}_{\mathcal{N}}$ to be the same as their planar counterparts.

\subsection{Coulomb Hamiltonian}
\label{ssec: interaction_Hamiltonian}
In this section, we present the density-density interaction Hamiltonian governing the dynamics of Coulomb-interacting electrons. For simplicity, we restrict electrons to a single spin-valley-polarized LL and ignore LL mixing. Moreover, we consider only the positive-energy LLs relevant for electron-doped BLG; a similar analysis applies to hole-doped BLG, where the relevant LLs lie at negative energy. Since our primary interest is the interacting phases in the $\mathcal{N}{=}2$ LL, we project the Coulomb interaction onto this LL.

\subsubsection{On the plane}
\label{sssec: planar_interaction_Hamiltonian}
In the planar geometry, we write the Coulomb interaction within the $\mathcal{N}{=}2$ LL in terms of the projected density operators as
\begin{align}
\label{eq: projected_interaction_Hamiltonian_1}
    \bar{H}^{\mathcal{N}{=}2}&=\frac{1}{2}\int\frac{d^2\boldsymbol{q}}{(2\pi)^2} \tilde{v}\left(q\right)\colon \bar{\rho}^{\mathcal{N}=2}\left(-\boldsymbol{q}\right)\bar{\rho}^{\mathcal{N}=2}\left(\boldsymbol{q}\right)\colon,
\end{align}
where $\colon \colon$ denotes the normal-ordering of operators inside it, $\bar{\rho}^{\mathcal{N}{=}2}\left(\boldsymbol{q}\right)$ is the $\mathcal{N}{=}2$ LL projected density operator, and $\tilde{v}\left(q\right){=}\left(2\pi/q\right)[e^2/\epsilon]$ is the Fourier component of the Coulomb interaction $v(r){=}(1/r)[e^2/\epsilon]$, where $\epsilon$ is the dielectric constant of the background host material. Throughout this paper, we quote interaction energy in Coulomb units of $e^2/(\epsilon\ell)$. Since, in the projected density operator, the LL index is fixed, one can encode the nature of the LL by a quantity called the form factor, which allows writing a higher LL projected density operator in terms of the $\mathcal{N}{=}0$ LL projected density operator [which is also the same as that of the LLL of non-relativistic electrons in GaAs, i.e., $n{=}0$ LL]~\cite{Haldane83}. In other words, the $\mathcal{N}{=}2$ LL projected density operator can be written as 
\begin{align}
\bar{\rho}^{\mathcal{N}{=}2}\left(\boldsymbol{q}\right) &= F\left(q;\theta_2,\phi_2\right)~\bar{\rho}^{\rm LLL}\left(\boldsymbol{q}\right),
\end{align}
where $F\left(q;\theta_2,\phi_2\right)$ is the form factor of the $\mathcal{N}{=}2$ LL, expressed in terms of the Laguerre polynomial $L_{n}(x)$ as~\cite{Papic11, Aoki13, Dora23}
\begin{align}
\label{eq: four_band_model_N_2_form_factor}
    F\left(q;\theta_2,\phi_2\right)&= \sin^{2}\left(\theta_2\right)\sin^{2}\left(\phi_2\right) L_{2}\left(-\frac{q^2\ell^2}{2}\right)\nonumber\\
    &+ \sin^{2}\left(\theta_2\right)\cos^{2}\left(\phi_2\right) L_{1}\left(-\frac{q^2\ell^2}{2}\right)\nonumber\\
    &+\cos^{2}\left(\theta_2\right) L_{0}\left(-\frac{q^2\ell^2}{2}\right),
\end{align}
and
\begin{align}
\label{eq: LLL_projected_density_operator}
    \bar{\rho}^{\rm LLL}\left(\boldsymbol{q}\right){=}e^{-q^2\ell^2/4}\bar{\rho}^{\rm LLL}_{g}\left(\boldsymbol{q}\right),
\end{align}
is the LLL projected density operator, where $\bar{\rho}^{\rm LLL}_{g}\left(\boldsymbol{q}\right)$ is the guiding center operator. In the Landau gauge~\cite{Goerbig04}
\begin{align}
\label{eq: Landau_gauge_guding_center_operator}
   \bar{\rho}^{\rm LLL}_{g}\left(\boldsymbol{q}\right) &= \sum_{x_0} e^{i q_x x_0} \chi^{\dagger}_{x_0+q_x\ell^2/2}~\chi_{x_0-q_x\ell^2/2},
\end{align}
where $\chi$ denotes the LLL guiding center annihilation operator labeled by $x_0$, where we have omitted the $n{=}0$ index for brevity. The Landau gauge is particularly useful in computing the energy of the stripe phases, since the LL eigenstates in this gauge are shaped like stripes. On the other hand, to compute the energy of the FQH states, it is useful to work in the symmetric gauge, where $\boldsymbol{A}{=}\left(-B_{\perp}y/2, B_{\perp}x/2,0\right)$. In the symmetric gauge~\cite{Goerbig11}, 
\begin{align}
\label{eq: symmetric_gauge_guding_center_operator}
    \bar{\rho}^{\rm LLL}_{g}\left(\boldsymbol{q}\right) &= \sum_{m,m^{\prime}} \left\langle m^{\prime}\right| e^{i \boldsymbol{q}\cdot\boldsymbol{R}} \left | m\right\rangle \chi^{\dagger}_{m^{\prime}}~\chi_{m},
\end{align}
where $\boldsymbol{R}$ denotes the guiding center coordinate, and the dummy index $m$ corresponds to the states $\left|m\right\rangle{\equiv}\left|0,m\right\rangle$, which denote the symmetric gauge LLL eigenstates. The matrix elements $\left\langle m^{\prime}\right| e^{i \boldsymbol{q}\cdot\boldsymbol{R}} \left | m\right\rangle$ can be readily evaluated following the procedure outlined in Ref.~\cite{Goerbig11}. To this end, the projected Hamiltonian in Eq.~\eqref{eq: projected_interaction_Hamiltonian_1} can be written as 
\begin{align}
    \label{eq: projected_interaction_Hamiltonian_2}
    \bar{H}^{\mathcal{N}{=}2}&=\frac{1}{2}\int\frac{d^2\boldsymbol{q}}{(2\pi)^2} \bar{v}\left(q;\theta_2,\phi_2\right)~\colon \bar{\rho}^{\rm LLL}\left(-\boldsymbol{q}\right)~\bar{\rho}^{\rm LLL}\left(\boldsymbol{q}\right)\colon,
\end{align}
where we have defined $\bar{v}\left(q;\theta_2,\phi_2\right){=}\left[F\left(q;\theta_2,\phi_2 \right)\right]^2 \tilde{v}\left(q\right)$ as the effective interaction potential that simulates the $\mathcal{N}{=}2$ LL Coulomb interaction in the LLL. Next, we simplify the above Hamiltonian by lifting the normal ordering of operators to obtain
\begin{align}
\label{eq: not_normal_ordrerd_projected_interaction_Hamiltonian_2}
    \bar{H}^{\mathcal{N}{=}2}&=\frac{1}{2}\int\frac{d^2\boldsymbol{q}}{(2\pi)^2} \bar{v}\left(q;\theta_2,\phi_2\right)~\bigg [\bar{\rho}^{\rm LLL}\left(-\boldsymbol{q}\right)~\bar{\rho}^{\rm LLL}\left(\boldsymbol{q}\right)\nonumber\\
    &~~~~~~~~~~~~~~~~~~~~~~~~~~~~~~~~~ - N e^{-q^2\ell^2/2}\bigg],
\end{align}
where $N$ is the number of particles.

We note that the effective interaction in the $\mathcal{N}{=}2$ LL of BLG, determined by the form factor in Eq.~\eqref{eq: four_band_model_N_2_form_factor}, can reproduce the effective interactions of various other physical systems for specific values of $\theta_2$ and $\phi_2$. For example, it reduces to the effective interaction in the $n{=}0$ LL of GaAs for $\theta_2{=}0$, while $(\theta_2{=}\pi/2,\phi_2{=}0)$ and $(\theta_2{=}\pi/2,\phi_2{=}\pi/2)$ correspond to the effective interactions in the $n{=}1$ and $n{=}2$ LLs, respectively. Similarly, for $(\theta_{2}{=}\pi/4, \phi_2{=}0)$ and $(\theta_{2}{=}\pi/2, \phi_2{=}\pi/4)$, the effective interaction in the $\mathcal{N}{=}2$ LL of BLG reduces to those in the first and second excited LLs of MLG, which we refer to as the MLG$1$ and MLG$2$ LLs, respectively. The effective interaction in the $\mathcal{N}{=}1$ ZLL of BLG corresponds to $\phi_2{=}0$. For the ideal BLG, where only the intralayer and nearest-neighbor interlayer hoppings are considered, the effective interaction in the first excited LL of BLG, which we refer to as the BLG$2$ LL, is reproduced by setting $(\theta_2{=}\pi/4, \phi_2{=}\pi/2)$ in the form factor of the $\mathcal{N}{=}2$ LL of BLG given in Eq.~\eqref{eq: four_band_model_N_2_form_factor}. Finally, we note that the LLL in MLG and the $\mathcal{N}{=}0$ ZLL in BLG are identical to the LLL ($n{=}0$ LL) of GaAs.  

\subsubsection{On the sphere}
\label{sssec: sphere_interaction_Hamiltonian}
On the sphere, the $\mathcal{N}{=}2$ LL projected Hamiltonian can be written as~\cite{Dora24}
\begin{align}
\label{eq: sphere_N_2_projected_Hamiltonian_1}
    \bar{H}_{Q}^{\mathcal{N}=2}&=\frac{4\pi}{2}\sum_{L=0}^{2(Q+2)}v_{L}\sum_{M{=}-L}^{L}\colon \left[\bar{\rho}^{\mathcal{N}{=}2}_{L,M}\right]^{\dagger}~\bar{\rho}^{\mathcal{N}{=}2}_{L,M} \colon,
\end{align}
where $\bar{\rho}^{\mathcal{N}{=}2}_{L, M}$ is the $\mathcal{N}{=}2$ LL projected angular momentum space density operator on the sphere, and $v_{L}{=}1/[\sqrt{Q} (2L{+}1)] (e^2/\epsilon\ell)$ is the expansion coefficient in the spherical harmonics decomposition of the Coulomb interaction. Similar to the planar case, we write the above Hamiltonian in terms of the LLL projected density operators. To do so, it is useful to appropriately shift the monopole index $Q{\to}Q{-}2$ in each monopole harmonic appearing in the $\mathcal{N}{=}2$ LL spinor [see Eq.~\eqref{eq: sphere_four_band_two_band_model_N_geq_2_LLs}] such that they all have an angular momentum $l{=}Q$, instead of $l{=}Q{+}2$. This shifting of $Q$ fixes the number of orbitals in the $\mathcal{N}{=}2$ LL to be the same as in the LLL with the unshifted $Q$~\cite{Balram15c, Balram21b, Hsiao20, Dora23}. To this end, the $\mathcal{N}{=}2$ LL projected Hamiltonian of Eq.~\eqref{eq: sphere_N_2_projected_Hamiltonian_1} in the LLL is
\begin{align}
\label{eq: sphere_N_2_projected_Hamiltonian_2}
    \bar{H}_{Q}^{\mathcal{N}=2}&=\frac{4\pi}{2}\sum_{L=0}^{2Q}\bar{v}_{L}\left(\theta_2,\phi_2\right)\sum_{M{=}-L}^{L}\colon \left[\bar{\rho}^{\rm LLL}_{L,M}\right]^{\dagger}~\bar{\rho}^{\rm LLL}_{L,M} \colon.
\end{align}
Here, $\bar{v}_{L}\left(\theta_2,\phi_2\right)$ describes the $\mathcal{N}{=}2$ LL effective interaction potential in the LLL, i.e.,  
\begin{align}
\label{eq: effective_Coulomb_interaction_BLG2_LL_sphere}
  \bar{v}_{L}\left(\theta_2,\phi_2\right)&= \left[F\left(L,\theta_2,\phi_2\right)\right]^2v_{L},
\end{align}
where
\begin{align}
  F\left(L,\theta_2,\phi_2\right) &= \frac{\mathcal{W}_2\left(L\right)}{\mathcal{W}_0\left(L\right)}\sin^2\left(\theta_2\right)\sin^2\left(\phi_2\right)\nonumber\\
  &+\frac{\mathcal{W}_1\left(L\right)}{\mathcal{W}_0\left(L\right)}\sin^2\left(\theta_2\right)\cos^2\left(\phi_2\right)+ \cos^{2}\left(\theta_2\right) ,
\end{align}
is the $\mathcal{N}{=}2$ LL form factor on the sphere, with $\mathcal{W}_x$ representing the Wigner $3j$ symbol, defined as
\begin{align}
  \mathcal{W}_r\left(L\right)&= (-1)^{Q+r}(2Q+1)\sqrt{\frac{2L+1}{4\pi}}\begin{pmatrix}
        Q & Q & L\\
        -Q+r & Q-r &0
    \end{pmatrix},  
\end{align}
where $r$ is an integer. We have also defined the LLL projected density operator $\bar{\rho}^{\rm LLL}_{L,M}$ in Eq.~\eqref{eq: sphere_N_2_projected_Hamiltonian_2} as
\begin{align}
    \bar{\rho}^{\rm LLL}_{L,M}&=\mathcal{W}_{0}\sum_{m{=}-Q}^{Q} (-1)^m
    \begin{pmatrix}
        Q & Q & L\\
        -m-M & m & M
    \end{pmatrix} \chi^{\dagger}_{m+M}~ \chi_{m},
\end{align}
where $\chi_{m}$ is the LLL annihilation operator on the sphere; for convenience, here, we have omitted the LL index $Q$. 

Finally, the Hamiltonian in Eq.~\eqref{eq: sphere_N_2_projected_Hamiltonian_2} can be written without the normal-ordering of operators as~\cite{Dora24, Dora25}
\begin{align}
\label{eq: not_normal_ordered_sphere_N_2_projected_Hamiltonian}
    \bar{H}_{Q}^{\mathcal{N}=2}&=\frac{4\pi}{2}\sum_{L=0}^{2Q}\bar{v}_{L}\left(\theta_2,\phi_2\right)\sum_{M{=}-L}^{L}\bigg[\left[\bar{\rho}^{\rm LLL}_{L,M}\right]^{\dagger}~\bar{\rho}^{\rm LLL}_{L,M}\nonumber\\
    &~~~~~~~~~~~~~~~-(-1)^{2Q}\frac{N}{2Q+1}\left(\mathcal{W}_{0}\left(L\right)\right)^2\bigg].
\end{align}

\subsection{Candidate states}
\label{ssec: candidate_states}
In this section, we list the quantum Hall states considered in this study at various filling fractions. As noted in the introduction, multiple candidate FQH states may exist at a given filling, distinguished by their topological properties such as the Wen–Zee shift $\mathcal{S}$~\cite{Wen92}, chiral central charge $c_{-}$~\cite{Kane97}, etc., and these states can compete depending on system parameters. Here, we restrict ourselves to $\nu{\leq}1/2$ since the states at $1{-}\nu$ are related by particle-hole symmetry to those at $\nu$ within a LL and their energies are related to each other for the two-body interactions considered here~\cite{Moller05, Balram15b}. Below, we write the wave functions in the disk geometry, but these can be readily translated to the spherical geometry, which we use for most of our computations. 

\subsubsection{Primary Jain states at $\nu{=}s/(2s{\pm}1)$}
In the LLL, FQH states at $\nu{=}s/(2sp{\pm}1)$ are accurately described by the Jain states~\cite{Jain89, Jain07, Balram13}, in which the CFs carrying $2p$ vortices form an IQH state. Here, we focus on the two-vortex ($p{=}1$) CF states that describe the FQH effect at primary Jain fillings $\nu{=}n/(2s{\pm}1)$ by the Jain wave function~\cite{Jain89}:
\begin{align}
\label{eq: primary_Jain_wf}
    \Psi_{\nu{=}s/(2s{\pm}1)}^{\rm Jain}&= \mathcal{P}_{\rm LLL}~\left(\Phi_1\right)^{2} ~\Phi_{\pm s},
\end{align}
where $\mathcal{P}_{\rm LLL}$ implements LLL-projection that we carry out using the Jain-Kamilla method~\cite{Jain97, Davenport12}. The Jastrow factor $\Phi_1$ is given by
\begin{align}
\label{eq: Jastrow_factor}
    \Phi_1&=\prod_{1{\leq}i<j{\leq}N}(z_i-z_j),
\end{align}
where $z_{i}$ is the two-dimensional coordinate of the $i^{\rm th}$ electron parametrized as a complex number. The ubiquitous Gaussian factor, $\exp\left(-\sum_{1{\leq}i{\leq}N} |z_{i}|^{2}/(4\ell^{2}) \right)$, is suppressed throughout for ease of notation. The square of the Jastrow factor, $\left(\Phi_1\right)^{2}$, attaches two vortices to each electron in the $s$-LL filled IQH state, $\Phi_{+s}$, where the CFs see the magnetic field in the same direction as the electrons (parallel-vortex attachment), or $\Phi_{{-}s}{\equiv}\left[\Phi_{s}\right]^{*}$, where the CFs experience a magnetic field opposite to that the electrons experience (reverse-vortex attachment). Jain CF states are Abelian states, with $\mathcal{S}{=}s{+}2$, and $c_{-}{=}s$ for $\nu{=}s/(2s{+}1)$, while $\mathcal{S}{=}2{-}s$, and $c_{-}{=}2{-}s$ for $\nu{=}s/(2s{-}1)$~\cite{Wen95}.

\subsubsection{Bonderson-Slingerland and anti-Read-Rezayi $3$-cluster states at $\nu{=}2/5$}
Besides the Abelian Jain state at $2/5$ [corresponding to $s{=}2$ and ${+}$ sign in Eq.~\eqref{eq: primary_Jain_wf}], we also consider two topologically distinct non-Abelian states at $2/5$, namely the Bonderson-Slingerland (BS) state and the anti-Read-Rezayi (aRR) $3$-cluster state. The BS state is described by the wave function~\cite{Bonderson08} 
\begin{eqnarray}
\label{eq: BS_2_5}
    \Psi_{\nu{=}2/5}^{\rm BS} &{=}& \mathcal{P}_{\rm LLL}\left[ {\rm Pf} \left( \frac{1}{z_{j}{-}z_{k}} \right) \prod_{j<k}(z_{j}{-}z_{k})^{3} \Phi^{*}_{2} \right] \nonumber \\
    &\sim& \frac{\Psi^{\rm MR}_{\nu{=}1/2}~\Psi^{\rm Jain}_{\nu{=}2/3}}{\Phi_1},
\end{eqnarray}
where the notation ${\rm{Pf}}\left(\left(z_{j}{-}z_{k}\right)^{{-}1}\right)$ denotes the Pfaffian of an antisymmetric matrix with off-diagonal elements $\left(z_j{-}z_{k}\right)^{{-}1}$ and vanishing diagonal entries. The state $\Psi_{\nu{=}1/2}^{\rm MR}$ denotes the MR wave function [see Eq.~\eqref{eq: MR_1_2_state} below for its wave function], and the symbol $\sim$ indicates that the states on either side of the sign are microscopically different but are expected to lie in the same universality class~\cite{Balram16b, Anand22}. Only the wave function written in the second line of Eq.~\eqref{eq: BS_2_5} is amenable to large system evaluations. Owing to the Pfaffian factor, the BS state is non-Abelian and has $c_{-}{=}1/2$ and $\mathcal{S}{=}2$. 

The wave function of the anti-Read-Rezayi (aRR) $3$-cluster state can be obtained by particle-hole conjugating the Read-Rezayi (RR) $3$-cluster state at $3/5$, whose wave function is given by~\cite{Read99}:
\begin{align}
\label{eq: wfn_RR3}
    \Psi^{\rm RR 3}_{\nu{=}3/5}&{=} \Phi_{1}~ \mathbb{S}\bigg[\prod_{1\leq i_1<j_1\leq N/3} \left(z_{i_1,j_1}\right)^{2} \prod_{N/3<i_2<j_2\leq 2N/3}\left(z_{i_2,j_2}\right)^{2} \nonumber\\
    &\times\prod_{2N/3<i_2<j_2\leq N}\left(z_{i_3,j_3}\right)^{2} \bigg]~ e^{-\sum_{i{=}1}^{N}z_{i}^{2}/2}.
\end{align}
In Eq.~\eqref{eq: wfn_RR3}, $N$ is a multiple of three, and $\mathbb{S}$ is the symmetrization operator, which symmetrizes over all possible partitions of the particles into three clusters of equal size, and we have defined $z_{i,j}{=}z_i{-}z_{j}$. The resulting $3$-cluster aRR state, denoted as aRR$3$, has $\mathcal{S}{=}{-}2$ and $c_{-}{=}{-}4/5$~\cite{Bishara08, Balram19}.

The $2/5$ BS state of Eq.~\eqref{eq: BS_2_5} is in close competition with the $2/5$ $3$-cluster aRR state in the vicinity of the second LL Coulomb point~\cite{Bonderson12}, though excitation and entanglement spectrum studies suggest that the latter is favored at $12/5$~\cite{Wojs09, Zhu15, Mong15, Pakrouski16}. Unlike the BS state, the aRR$3$ state is not readily amenable to large-scale Monte Carlo-based numerical evaluation (though some progress in this direction has recently been made in Ref.~\cite{Bose25a}), precluding a reliable estimation of its thermodynamic energies. 

\subsubsection{Parton states at fillings $\nu{=}3/7, 4/9$, and $6/13$}
In contrast to the LLL, the primary Jain states are energetically disfavored relative to topologically distinct parton states at $\nu{=}3/7, 4/9$, and $6/13$ in the SLL of GaAs~\cite{Balram18a, Faugno20a, Balram20b}. In the parton theory~\cite{Jain89b}, candidate FQH states are constructed from the IQH states of $r$ species of partons, which are fictitious particles obtained from dividing electrons into $r$ parts, where $r{>}1$ and is odd. A parton state denoted by $n_{1}n_{2}{\cdots}n_{r}$ corresponds to product of IQH states $\Phi_{n_{\tilde{s}}}$ for each parton species $\tilde{s}$, where $\tilde{s}{=}1,2,{\cdots},r$, i.e., 
\begin{equation}
    \label{eq: parton_wfn}
    \Psi_{\nu}^{n_{1}n_{2}{\cdots}n_{r}}=\mathcal{P}_{\rm LLL}\prod_{\tilde{s}{=}1,2,{\cdots},r}\Phi_{n_{\tilde{s}}},
\end{equation}
where $\mathcal{P}_{\rm LLL}$ implements the projection appropriate for the high magnetic field limit of our interest. The variational wave function of Eq.~\eqref{eq: parton_wfn} for electrons is obtained by identifying the coordinates of all species of partons in the IQH states with electron coordinates, which effectively glues partons back into electrons, and subsequently projecting the resulting product state onto the LLL. Since each parton species has the same density as that of the parent electrons and is exposed to the same external magnetic field, the charge of the $\tilde{s}^{\rm th}$ parton species is $e_{\tilde{s}}{=}-e\nu/n_{\tilde{s}}$, where $-e$ is the charge of an electron. This ensures that the $\tilde{s}^{\rm th}$ parton species fills $n_{\tilde{s}}$ LLs. Moreover, since the charges of the partons should add up to that of the electron, the parton state $n_{1} n_{2}{\cdots}n_{r}$ corresponds to the filling $\nu{=}\left[\sum_{\tilde{s}{=}1}^{r}\left( n_{\tilde{s}}\right)^{-1}\right]^{-1}$. The value of $n_{\tilde{s}}$ can be negative, which we represent as $\bar{n}_{\tilde{s}}$, and in that case the IQH state is $\left[\Phi_{n_{\tilde{s}}}\right]^{*}$. The charge $e_{\tilde{s}}$ is positive for negative $n_{\tilde{s}}$, in which case the corresponding parton species experiences an effective magnetic field in the direction opposite to that of the parent electrons. A parton state is non-Abelian iff it has a repeated factor of $n_{\tilde{s}}$ with $|n_{\tilde{s}}|{\geq}2$~\cite{Wen91}. Interestingly, Abelian single-component parton states exist at even denominator fillings, such as $\bar{6}21$ at $\nu{=}3/4$ (or $10~2~1$ at $\nu{=}5/8$), which has a fundamental quasihole of expected charge $e/8$~\cite{Levin09} arising from the hole in $\bar{6}$.

At $3/7$, the parton state $\bar{3}\bar{3}111$, in short $\bar{3}^{2}1^3$, provides a good representative ground state for the SLL (and also $\mathcal{N}{=}1$ LL of BLG, where signatures of the FQH effect at $3/7$ have been observed experimentally~\cite{Huang21}) with Coulomb interaction~\cite{Faugno20a, Balram21}. Its wave function is given by
\begin{align}
\Psi^{\bar{3}^{2}1^3}_{\nu=3/7}&= \mathcal{P}_{\rm LLL} \left[\Phi_{3}^{2}\right]^{*}\Phi_{1}^{3}\sim \frac{\left[\Psi^{\rm Jain}_{\nu=3/5}\right]^{2}}{\Phi_1}.
\end{align}
The above parton state at $3/7$ is non-Abelian, which follows from the presence of two factors of $\bar{3}$. Moreover, the state has $\mathcal{S}{=}{-}3$, and $c_{-}{=}{-}11/5$~\cite{Faugno20a}. 

Similarly, the parton states that have lower energy than the corresponding Jain CF states for the SLL Coulomb interaction at $4/9$ and $6/13$ are~\cite{Balram18a, Balram20b}:
\begin{align}
    \Psi^{\bar{4}\bar{2}1^3}_{\nu=4/9}&= \left[\Phi_{4}\right]^{*}\left[\Phi_{2}\right]^{*}\Phi_{1}^{3}\sim \frac{\Psi^{\rm Jain}_{\nu=4/7}~\Psi^{\rm Jain}_{\nu=2/3}}{\Phi_1},
\end{align}
and
\begin{align}
    \Psi^{\bar{3}\bar{2}1^3}_{\nu=6/13}&= \left[\Phi_{3}\right]^{*}\left[\Phi_{2}\right]^{*}\Phi_{1}^{3}\sim \frac{\Psi^{\rm Jain}_{\nu=3/5}~\Psi^{\rm Jain}_{\nu=2/3}}{\Phi_1}.
\end{align}
The $\bar{4}\bar{2}1^3$ parton state has $\mathcal{S}{=}{-}3$, $c_{-}{=}{-}3$, while the $\bar{3}\bar{2}1^3$ state has $\mathcal{S}{=}{-}2$, $c_{-}{=}{-}2$~\cite{Balram18a, Balram20b}. We note that the FQH state at $6/13$ has been experimentally observed in the SLL of GaAs~\cite{Kumar10} and in the $\mathcal{N}{=}1$ LL of BLG~\cite{Huang21, Hu24, Kumar24} and in the LLL of wide GaAs quantum wells~\cite{Singh23, Singh25}. No definitive signatures of a FQH state at $4/9$ have been reported in a LL dominated by the $n{=}1$ LL orbital character. 

\begin{table}[h]
	\begin{center}
		\begin{tabular} { | c | c | c | c |}
			\hline
			$\nu$ & state & $\mathcal{S}$ & $c_{-}$\\
			\hline
            \hline
			$2/5$ & $211$, or $2/5$ Jain~\cite{Jain89} & $4$ & $2$\\
			\hline
            $2/5$ & $\bar{2}^3 1^{4}$~\cite{Balram19} $\sim$ aRR$3$~\cite{Read99} & $-2$ & $-4/5$~\cite{Bishara08, Balram19} \\
            \hline
            $2/5$ & Bonderson-Slingerland~\cite{Bonderson08}& $2$ & $1/2$~\cite{Balram18, Faugno20a}\\
			\hline\hline
            $3/7$ & $311$, or $3/7$ Jain~\cite{Jain89} & $5$ & $3$\\
			\hline
            $3/7$ & $\bar{3}^2 1^{3}$~\cite{Faugno20a} & $-11/5$ & $-3$\\
            \hline\hline
            $4/9$ & $411$, or $4/9$ Jain~\cite{Jain89} & $6$ & $4$\\
			\hline
            $4/9$ & $\bar{4} \bar{2} 1^{3}$~\cite{Balram20b} & $-3$ & $-3$\\
            \hline\hline
            $6/13$ & $611$, or $6/13$ Jain~\cite{Jain89} & $8$ & $6$\\
			\hline
            $6/13$ & $\bar{3}\bar{2} 1^{3}$~\cite{Balram18a} & $-2$ & $-2$\\
            \hline\hline
            $1/2$ & $\bar{2}^{2}1^{3}$~\cite{Balram18} $\sim$ anti-Pfaffian~\cite{Lee07, Levin07} & $-1$ & $-1/2$\\
			\hline
            $1/2$ & Moore-Read~\cite{Moore91} & $3$ & $3/2$\\
            \hline
		\end{tabular}
	\end{center}
	\caption{Candidate incompressible fractional quantum Hall states, their fillings $\nu$, shifts $\mathcal{S}$ on the sphere, which is connected to the Hall viscosity~\cite{Read09} $\eta_{H}{=}\hbar(\nu/2\pi\ell^{2})\mathcal{S}/4$, and chiral central charge $c_{-}$, which, assuming full equilibration of edge modes, is related to the thermal Hall conductance $\kappa_{xy}{=}c_{-}[\pi^2 k_{\rm B}^2 /(3h)]T$ at temperature $T$ much below the bulk gap (each filled Landau level makes an additional unit contribution to $c_{-}$).}
    \label{tab: chiral_central_charge_shift_candiadate_states}
\end{table}

\subsubsection{Composite fermion Fermi liquid and Moore-Read states at $\nu{=}1/2$} 
\label{sssec: CFFL_MR_states}
For the CFFL, we use the Rezayi-Read wave function~\cite{Rezayi94}
\begin{equation}
    \label{eq: wfn_CFFL}
    \Psi^{\rm CFFL}_{\nu{=}1/2}=\mathcal{P}_{\rm LLL} \Phi_1^{2} \Phi^{\rm FL},
\end{equation}
where $\Phi^{\rm FL}{=}{\rm Det}[e^{i\vec{k}_{j}{\cdot}\vec{r}_{l}}]$ is the wave function of the Fermi liquid state. On the sphere, we consider filled-shell CF states that are uniform and are known to provide a good description of the CFFL in the half-filled $\mathcal{N}{=}0,1$ LLs of MLG~\cite{Rezayi00, Liu20, Balram21b}. In other words, we model the CFFL on the sphere using the primary Jain states at $\nu{=}s/(2s{+}1)$, constructed with a specific number of particles, $N{=}s^2$, where $s{\geq}2$, which maps to $\nu{=}1/2$ as $s$ becomes large. The relation $N{=}s^2$ results from the fact that in a CFFL, the CFs see a net-zero effective magnetic field, and thus states filling the lowest $s$ LLs on the sphere occur precisely at $N{=}s^2$.  

Another candidate state at $\nu{=}1/2$ is the MR state, whose wave function is given by~\cite{Moore91}:
\begin{align}
\label{eq: MR_1_2_state}
    \Psi^{\rm MR}_{\nu{=}1/2}&=\Phi_{1}^{2} ~\rm{Pf}\left(\frac{1}{z_i-z_{j}}\right).
\end{align}
The above state can be interpreted as the $p$-wave paired superconducting state of CFs. The MR state has $\mathcal{S}{=}3$, and $c_{-}{=}3/2$. The $\bar{2}^{2}1^{3}$ parton state lies in the same universality class as the hole-conjugate of the MR, namely the aPf~\cite{Balram18}. 

A caveat to note is that in the subsequent computations we will use these trial wave functions, so the computation is inherently variational in nature. Moreover, it is not possible to perform exact diagonalization, particularly for fillings with large denominators, where only a few system sizes are accessible to exact diagonalization. For the 1/2 state, one can do a BCS-instability calculation for CFs~\cite{Moller08, Sharma21}, which is more reliable than the use of a particular wave function like MR, especially when energy differences are small~\cite{Sharma22, Sharma23}, but we have not pursued that here. Aside from the MR state, we have also considered the $\bar{2}^{2}1^{3}$ parton state~\cite{Balram18}, which lies in the same universality class as the hole-conjugate MR state, the anti-Pfaffian (aPf)~\cite{Lee07, Levin07}, but presents a better microscopic representation of the exact SLL Coulomb ground state compared to the aPf. Note that since we are only dealing with two-body Hamiltonians here, the Pf and aPf have identical energies for all interactions considered in this work.

In Table~\ref{tab: chiral_central_charge_shift_candiadate_states}, we have summarized the shift $\mathcal{S}$ and chiral central charge $c_-$ of the candidate states considered in this work. In Appendix~\ref{app: chiral_central_charge_parton_BS}, we provide some arguments to infer the chiral central charge of the BS and parton states. In the following Secs.~\ref{ssec: FQH_states_energy_on_plane} and~\ref{ssec: FQH_state_energy_on_the_sphere}, we compute the energies of the FQH states discussed above from their static structure factor, which are obtained numerically using their explicit wave functions. 

\subsubsection{Stripes}
\label{sssec: stripes}
Besides the FQH states, we will also consider the stripe states~\cite{Fogler96, Koulakov96}. These are unidirectional charge density wave states that break translation and rotational symmetry and can compete with the FQH states, particularly near the half-filling of a LL. Generically, stripe phases dominate over the FQH states in higher LLs~\cite{Fogler96, Koulakov96}. In Sec.~\ref{ssec: Energy of the stripe phases}, we compute the energy of the stripe states. 

\section{Energies of candidate states}
\label{sec: Energies_of_candidate_states}
In this section, we lay out the framework to compute the energies of different candidate states, including the FQH, CFFL, and stripe states. We begin by discussing the energy of FQH states in the planar and spherical geometries and then discuss the energy computation for the stripe phase. In Appendix~\ref{app: thermodynamic_energies_candidates_different_LLs}, we list the planar and spherical energies of various candidate states in the $\mathcal{N}{=}2$ LL of BLG for specific $(\theta_2,\phi_2)$ points. 

\subsection{Energy on the plane}
\label{ssec: FQH_states_energy_on_plane}
The energy of a FQH state for a two-body Hamiltonian can be fully determined from the knowledge of its density-density correlation function, i.e., its projected static structure factor $\bar{S}\left(q\right)$. The per-particle energy [see Eq.~\eqref{eq: not_normal_ordrerd_projected_interaction_Hamiltonian_2}] can be written in terms of $\bar{S}\left(q\right)$ as
\begin{align}
\label{eq: energy_S_q_1}
    E\left(\theta_2,\phi_2\right)&\equiv\frac{\left\langle\bar{H}^{\mathcal{N}{=}2}\right\rangle}{N}=\frac{1}{2}\int \frac{d^2\boldsymbol{q}}{(2\pi)^2}~ \bar{v}\left(q;\theta_2,\phi_2\right)\times\nonumber\\
    &~~~~~~~~~~~~~~~~~~~~~~~~~~~~~~~\big[\bar{S}\left(q\right)-e^{-q^2\ell^2/2}\big],
\end{align}
where we have defined $\bar{S}\left(q\right)$ as
\begin{align}
  \bar{S}\left(q\right)&=\frac{\left\langle\bar{\rho}^{\rm LLL}\left(-\boldsymbol{q}\right)~\bar{\rho}^{\rm LLL}\left(\boldsymbol{q}\right)\right\rangle}{N}.
\end{align}
In Eq.~\eqref{eq: energy_S_q_1}, the contribution to the energy from $\bar{S}\left(q\right)$ at $q{=}0$ is exactly canceled by the electron-background interaction. Therefore, to compute the total energy of a FQH state, including the electron-electron and electron-background interactions, we simply set $\bar{S}(0){=}0$.

For FQH states, $\bar{S}(q)$ can be expanded in terms of Laguerre polynomials as~\cite{Girvin84a}
\begin{align}
\label{eq: projected_S(q)_expansion}
    \bar{S}(q)&=(1-\nu)e^{\frac{-q^2\ell^2}{2}} +4\nu\sum_{\mathfrak{m}=0}^{\infty}c_{2\mathfrak{m}+1} L_{2\mathfrak{m}+1}\left(q^2\ell^2\right)e^{-q^2\ell^2},
\end{align}
where $c_{2\mathfrak{m}+1}$ are the expansion coefficients. To obtain the above expansion, one notes the identities: $\bar{S}(q)$ is related to the unprojected structure factor $S(q)$ as $\bar{S}(q){=}S(q)-\left[1-e^{-q^2\ell^2/2}\right]$~\cite{Girvin85, Girvin86}, and $S(q)$ is related to the real-space density-density correlation function, i.e., the pair-correlation function $g(r)$, via a Fourier transform as $S(q){=}1+(\nu n_\ell)\int d^{2}\boldsymbol{r} e^{-i\boldsymbol{q}.\boldsymbol{r}}\left[g(r)-1\right]$. Thus, the $g(r)$ admits the following expansion~\cite{Girvin84a}
\begin{align}
\label{eq: pair_correlation_expansion}
    g(r)&=1-e^{-r^2/2\ell^2}+\sum_{\mathfrak{m}=0}^{\infty}\frac{c_{2\mathfrak{m}+1}~e^{-r^2/4\ell^2}}{(2\mathfrak{m}+1)!}\left(\frac{r^2}{4\ell^2}\right)^{2\mathfrak{m}+1}.
\end{align}
The coefficients $c_{2\mathfrak{m}+1}$ are determined by fitting the numerically computed $g(r)$ to the above expression while ensuring the correct leading long-wavelength behavior of $S(q)$~\cite{Girvin86, Dora24}. A discussion of this can be found in the Appendix~\ref{app: leading_sq_parton}. Note that in Eq.~\eqref{eq: projected_S(q)_expansion}, $\bar{S}(0){=}0$, which follows from the constraint imposed on $c_{2\mathfrak{m}+1}$ arising from the local charge neutrality in FQH ground states. 

To this end, the total energy of a FQH state can be written using Eq.~\eqref{eq: projected_S(q)_expansion} as
\begin{align}
\label{eq: total_energy_FQH_states}
    E\left(\theta_2,\phi_2\right)&=  2\nu\sum_{\mathfrak{m}=0}^{\infty}c_{2\mathfrak{m}+1}  V_{2\mathfrak{m}+1}\left(\theta_2,\phi_2\right) + E^{\rm uc}\left(\theta_2,\phi_2\right).
\end{align}
Here, $ V_{2\mathfrak{m}+1}\left(\theta_2,\phi_2\right)$ is the Haldane pseudopotential (PP)~\cite{Haldane83}, defined as
\begin{align}
     \label{eq: planar_Haldane_PPs}
V_{2\mathfrak{m}+1}\left(\theta_2,\phi_2\right)&=\int \frac{d^2\boldsymbol{q}}{(2\pi)^2} \bar{v}\left(q;\theta_2,\phi_2\right)L_{2\mathfrak{m}+1}\left(q^2\ell^2\right)e^{-q^2\ell^2},
\end{align}
which is the energy of a pair of particles in a state with relative angular momentum $2\mathfrak{m}{+}1$ for the interaction potential $\bar{v}\left(q;\theta_2,\phi_2\right)$, and $E^{\rm uc}$ is an overall uncorrelated energy given by
\begin{align}
\label{eq: planar_uncorrelated_energy}
   E^{\rm uc}\left(\theta_2,\phi_2\right)&= -\frac{\nu}{2}\int \frac{d^2\boldsymbol{q}}{(2\pi)^2}~ \bar{v}\left(q;\theta_2,\phi_2\right) e^{-q^2\ell^2/2}.
\end{align}
We use Eq.~\eqref{eq: total_energy_FQH_states} to compute the total energy of FQH states, and study the competition between candidate states at the same filling fraction. 

To determine the competition between the stripe and FQH states, it is useful to define the cohesive energy for FQH states. Following Ref.~\cite{Goerbig04}, we drop the uncorrelated energy $E^{\rm uc}$ in Eq.~\eqref{eq: total_energy_FQH_states}, and define the cohesive energy of FQH states as
\begin{align}
\label{eq: cohesive_energy_FQH_states}
    E^{\rm coh}\left(\theta_2,\phi_2\right)&= 2\nu\sum_{\mathfrak{m}=0}^{\infty}c_{2\mathfrak{m}+1}  V_{2\mathfrak{m}+1}\left(\theta_2,\phi_2\right).
\end{align}

Unfortunately, the precise behavior of the static structure factor in the long-wavelength limit for the microscopic CFFL remains unclear and, in fact, appears to be in tension with field-theoretic predictions~\cite{Halperin93, Anakru25, Makki26}, which is why we have not attempted to fit its pair-correlation function and determine its planar energies in the thermodynamic limit, as we do for other incompressible FQH states. Its energies are only determined on the sphere via an extrapolation to the thermodynamic limit of finite-size results.

\subsection{Energy on the sphere}
\label{ssec: FQH_state_energy_on_the_sphere}
In a complementary approach to compute the energy of a FQH state, we first compute its energies for finite-size systems on the sphere and extrapolate them to the thermodynamic limit. It follows from Eq.~\eqref{eq: not_normal_ordered_sphere_N_2_projected_Hamiltonian} that the per-particle electron-electron interaction energy of an FQH ground state (that is rotationally invariant on the sphere) for the Hamiltonian $\bar{H}^{\mathcal{N}{=}2}_{Q}$ can be written in terms of the projected static structure factor $\bar{S}\left(L\right)$ as~\cite{Dora24}
\begin{align}
\label{eq: per_particle_energy_on_the_sphere}
E_Q\left(\theta_2,\phi_2\right)&\equiv\frac{\left\langle\bar{H}_{Q}^{\mathcal{N}=2}\right\rangle}{N}
   =\frac{1}{2}\sum_{L=0}^{2Q}\bar{v}_{L}\left(\theta_2,\phi_2\right)\left(2L+1\right)\bar{S}\left(L\right)\nonumber\\
   &~~~~~~~~~~~~~~~~~~~~~~+ E^{\rm uc}_{Q}\left(\theta_2,\phi_2\right).
\end{align}
Here, we have defined the projected static structure factor $\bar{S}\left(L\right)$ of the state on the sphere as~\cite{He94, Dora24}
\begin{align}
\label{eq: projected_structure_factor_on_sphere}
    \bar{S}\left(L\right)&=\frac{4\pi}{N}\left\langle\left[\bar{\rho}^{\rm LLL}_{L,M}\right]^{\dagger}~\bar{\rho}^{\rm LLL}_{L,M}\right\rangle,
\end{align}
where $L$ and $M$ stand for quantum numbers corresponding to the total orbital angular momentum and its $z$-component. The overall uncorrelated energy $E^{\rm uc}_{Q}$ in Eq.~\eqref{eq: per_particle_energy_on_the_sphere} on the sphere is~\cite{Dora24}
\begin{align}
\label{eq: uncorrelated_energy_sphere}
    E^{\rm uc}_{Q}\left(\theta_2,\phi_2\right)&=-\frac{1}{2}\sum_{L=0}^{2Q}\bar{v}_{L}\left(\theta_2,\phi_2\right)\left(2L+1\right)\times\nonumber\\
&~~~~~~~~~~\bigg[\frac{(-1)^{2Q}4\pi\left(\mathcal{W}_{0}\left(L\right)\right)^2}{2Q+1}\bigg].
\end{align}
To account for the contribution from the uniform positively charged background to the total energy, we set $\bar{S}\left(0\right){=}0$~\cite{Kundu26} [instead of $\bar{S}\left(0\right){=}N$], as we did on the plane. Alternatively, to determine the total energy, one can subtract the per-particle electron-background and the background-background contributions to the Coulomb energy, which is $N/(2\sqrt{Q})$~\cite{Jain07}, from $E_Q\left(\theta_2,\phi_2\right)$ in Eq.~\eqref{eq: per_particle_energy_on_the_sphere}. This contribution of $N/(2\sqrt{Q})$ is exactly equal to (and hence, canceled by) the Coulomb contribution to the energy from $\bar{S}\left(L\right)$ at $L{=}0$.

$\bar{S}\left(L\right)$ can be computed exactly in Fock space; however, its evaluation becomes challenging for larger systems, and consequently so does the computation of the energy. Instead, we compute $\bar{S}\left(L\right)$ from its counterpart $S\left(L\right)$, which denotes the unprojected static structure factor, via the relation~\cite{Dora24}
\begin{align}
    \bar{S}\left(L\right){=}S\left(L\right)-\mathbb{O}\left(L\right),
\end{align}
where the offset factor $\mathbb{O}\left(L\right)$ is 
\begin{align}
    \mathbb{O}\left(L\right){=}1-(2Q+1)
    \begin{pmatrix}
        Q & Q & L\\
        -Q & Q &0
    \end{pmatrix}^{2}.
\end{align}
$S\left(L\right)$ can be computed numerically from the real-space wave function of the state using Monte Carlo techniques, and is therefore amenable to large system sizes~\cite{Kamilla97, Balram17}. However, the resulting $\bar{S}\left(L\right)$ obtained from $S\left(L\right)$ is not exact, since each $S\left(L\right)$ is determined only up to a statistical sampling error. As $L$ increases, $\bar{S}\left(L\right)$ approaches zero and eventually becomes comparable to the error bars, rendering the computed $\bar{S}\left(L\right)$ at large $L$ (${\lesssim}2Q$) less reliable. To minimize the resulting error in the energy, we set $\bar{S}\left(L\right){=}0$ for $L{>}L_{\rm cut-off}$ [see Ref.~\cite{Dora24} for a discussion to determine $L_{\rm cut-off}$].

On the sphere, the analog of the planar cohesive energy in Eq.~\eqref{eq: cohesive_energy_FQH_states} can be obtained by adding the quantity $[N/(2Q{+}1)]E^{\rm uc}_{Q}\left(\theta_2,\phi_2\right)$ to the total energy $E_Q\left(\theta_2,\phi_2\right)$ in Eq.~\eqref{eq: per_particle_energy_on_the_sphere}, i.e., the cohesive energy on the sphere is 
\begin{align}
\label{eq: cohesive_energy_on_sphere}
    E^{\rm coh}_{Q}\left(\theta_2,\phi_2\right)&= E_Q\left(\theta_2,\phi_2\right) + \frac{N}{2Q+1} E^{\rm uc}_{Q}\left(\theta_2,\phi_2\right),
\end{align}
where the second term on the right-hand side of the above equation is equivalent to the corresponding sign-reversed second term in the planar expression on the right-hand side of Eq.~\eqref{eq: total_energy_FQH_states}, i.e., ${-}E^{\rm uc}(\theta_2,\phi_2)$.

To this end, we determine thermodynamic limits of the total energy and the cohesive energy by extrapolating $E_Q\left(\theta_2,\phi_2\right)$ and $E_{Q}^{\rm coh}\left(\theta_2,\phi_2\right)$, respectively, as a linear function of the inverse system size $1/N$. Before the extrapolation, to minimize the finite-size effects, we density-correct the energies by multiplying them by a factor of $\sqrt{2Q\nu/N}$~\cite{Morf86b}.

As mentioned in Sec.~\ref{sssec: CFFL_MR_states}, we determine the energy of the CFFL state by extrapolating the energies of filled-shell Jain states at $\nu{=}s/(2s{+})1$ with $N{=}s^2$ particles, considering $s{=}3$ to $6$. 

We note that, like ground state energies, the energy of excited states can also be computed from their structure factors for two-body Hamiltonians. In Appendix~\ref{app: energy_gap_excitations}, we compute the energy of an excited state, which can be used to obtain the excitation gap of the underlying FQH state.

\subsubsection{Energy on the sphere from planar or disk PPs}
\label{sssec: Energy_on_the_sphere_from_planar_PPs}
In Sec.~\ref{ssec: FQH_state_energy_on_the_sphere}, we computed the energy of FQH states for the Coulomb interaction on the sphere. In the following, we discuss the computation of the energy of a given state for rotationally invariant interactions parametrized by a set of spherical Haldane PPs $\{V_L\}$, similar to Eq.~\eqref{eq: total_energy_FQH_states} in the planar geometry. Here, $V_L$ is the weight in the decomposition of the interaction potential in the two-particle basis state with definite total angular orbital momentum $L$. In other words, $V_L$ is the interaction energy of two particles in a definite total angular orbital momentum $L$ state.

The spherical PPs $V_L$ map onto the planar or the disk PPs $V_{\mathfrak{m}}$, via the relation $\mathfrak{m}{=}2Q{-}L$. In what follows, we map the disk PPs of the $\mathcal{N}{=}2$ LL-projected Coulomb interaction in Eq.~\eqref{eq: planar_Haldane_PPs} to the spherical geometry and compute the energies of FQH states for these PPs. Note that, as $L$ ranges from $0$ to $2Q$ in the LLL, $\mathfrak{m}$ also ranges from $0$ to $2Q$; accordingly, we set $V_{\mathfrak{m}}{=}0$ for $\mathfrak{m}{>}2Q$. To determine the energy, it is convenient to first obtain the effective spherical harmonics $\bar{v}_{L}$ corresponding to $\{V_{\mathfrak{m}}\left(\theta_2,\phi_2\right)\}$ using the following expression~\cite{Wooten14}:
\begin{align}
\label{eq: PPs_to_harmonics_relation}
    \bar{v}_{L}\left(\theta_2,\phi_2\right)&=\sum_{\mathfrak{m}=0}^{2 Q}\bigg[\frac{(-1)^{2Q-\mathfrak{m}}}{2L+1}~\left(\mathcal{W}_{0}\left(L\right)\right)^2V_{\mathfrak{m}}\left(\theta_2,\phi_2\right)\nonumber \\
    & \times \left(2 \left(2Q-\mathfrak{m}\right)+1\right)\left\{\begin{array}{lll}
Q & Q & ~~~~L \\
Q & Q & 2Q-\mathfrak{m}
\end{array}\right\} \bigg].
\end{align}
Next, the above-determined $\bar{v}_{L}\left(\theta_2,\phi_2\right)$ can be substituted into either Eqs.~\eqref{eq: per_particle_energy_on_the_sphere} or~\eqref{eq: cohesive_energy_on_sphere} to obtain, respectively, the total energy or the cohesive energy on the sphere. These finite-size energies are then extrapolated to extract their thermodynamic limit values. We expect that the thermodynamic energies computed using the spherical PPs of the Coulomb interaction on the sphere and using the disk PPs of the Coulomb interaction on the sphere would be consistent with each other. Moreover, we expect that the thermodynamic limit of energies computed on the sphere would be consistent with those obtained in the planar geometry, as the sphere becomes equivalent to the plane in the thermodynamic limit.

\subsection{Energy of the stripe phases}
\label{ssec: Energy of the stripe phases}

\begin{figure}
    \centering
    \includegraphics[width=0.99\columnwidth]{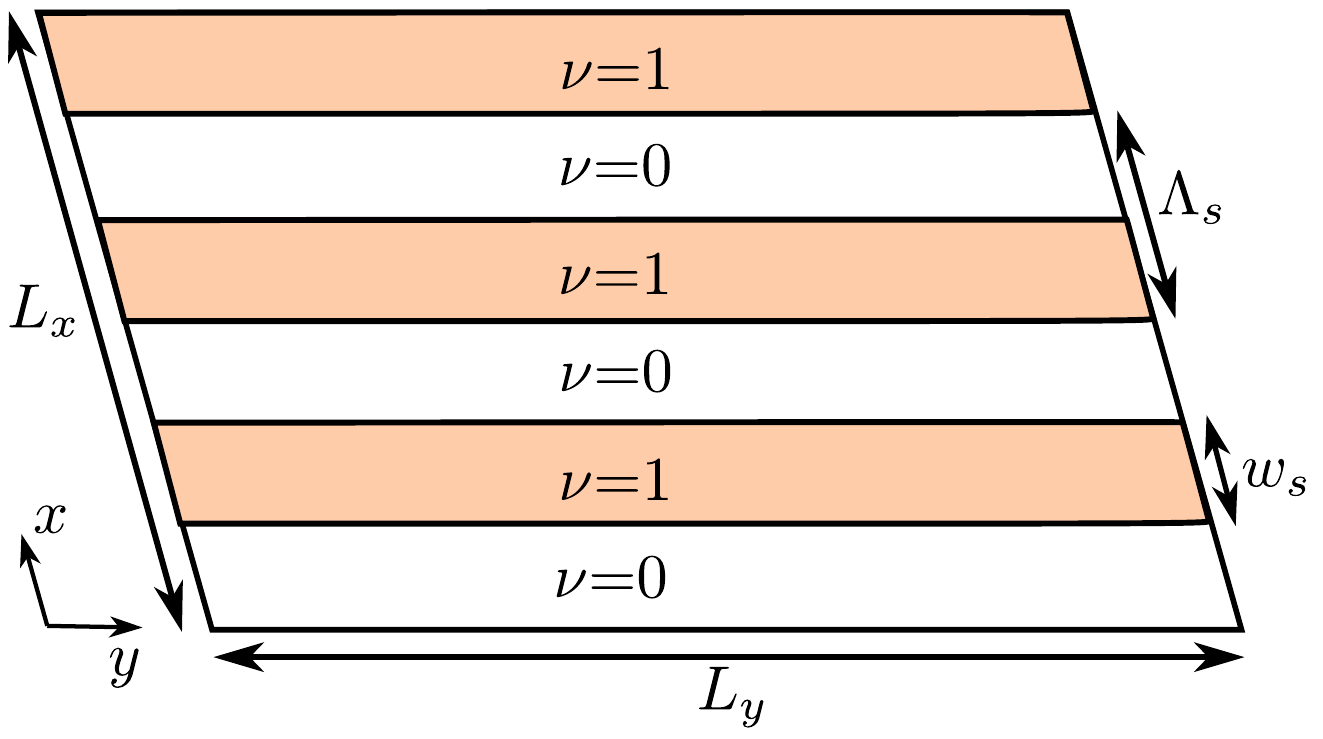}
    \caption{Schematic of the stripe phase that is composed of alternating strips of $\nu{=}0$ and $\nu{=}1$ integer quantum Hall states. The specific case of half-filling is shown, for which $\Lambda_{s}{=}2w_{s}$.}
    \label{fig: schematic_stripe}
\end{figure}

Stripe phases are unidirectional charge density wave states that can compete with FQH states near half-filling of a LL. The stripes are modeled as alternating strips of $\nu{=}0$ and $\nu{=}1$ IQH states, as shown in Fig.~\ref{fig: schematic_stripe}. To compute the energy of the stripe states, we follow the Hartree–Fock (HF) mean-field approach, for which it is convenient to work in the planar geometry. The Hamiltonian in either Eqs.~\eqref{eq: projected_interaction_Hamiltonian_2} or~\eqref{eq: not_normal_ordrerd_projected_interaction_Hamiltonian_2} can be written, within the HF approximation, as~\cite{Goerbig04}
\begin{align}
\label{eq: Hartree_Fock_Hamiltonian}
    \bar{H}^{\mathcal{N}{=}2}_{\rm HF}&=\frac{1}{2}\int\frac{d^2\boldsymbol{q}}{(2\pi)^2} \bar{v}_{\rm HF}\left(q;\theta_{2},\phi_{2}\right) \left\langle\bar{\rho}^{\rm LLL}_{g}\left(-\boldsymbol{q}\right)\right\rangle\bar{\rho}^{\rm LLL}_{g}\left(\boldsymbol{q}\right),
\end{align}
where $\bar{v}_{\rm HF}\left(q;\theta_{2},\phi_{2}\right)$ is the HF interaction potential: $\bar{v}_{\rm HF}\left(q;\theta_{2},\phi_{2}\right){=}\bar{v}_{H}\left(q;\theta_{2},\phi_{2}\right){-}\bar{v}_{F}\left(q;\theta_{2},\phi_{2}\right)$. The Hartree term $\bar{v}_{H}$ and the Fock term $\bar{v}_{F}$, are related to the bare interaction $\bar{v}$ as follows
\begin{align}
    \bar{v}_{H}\left(q;\theta_{2},\phi_{2}\right)&= \bar{v}\left(q;\theta_{2},\phi_{2}\right)e^{-q^2\ell^2/2},\\[1ex]
    \bar{v}_{F}\left(q;\theta_{2},\phi_{2}\right)&=\frac{1}{n_\ell}e^{-q^2\ell^2/2}\int\frac{d^2\boldsymbol{p}}{(2\pi)^2} \bar{v}\left(q;\theta_{2},\phi_{2}\right)e^{i \ell^2\left(\boldsymbol{p}\times \boldsymbol{q}\right)_z},
\end{align}
where we have defined $n_{\ell}{\equiv}\rho_{1}{=}1/(2\pi\ell^2)$ is the density of a uniform $\nu{=}1$ IQH state. The expectation value $\left\langle\bar{\rho}^{\rm LLL}_{g}\left(-\boldsymbol{q}\right)\right\rangle$ in Eq.~\eqref{eq: Hartree_Fock_Hamiltonian} is taken with respect to the specific state under consideration, which in our case is proportional to the stripe order parameter. In the stripe phase, electrons are distributed homogeneously within a stripe of width $w_s$ that extends along, say, the $y{-}$direction, and these stripes repeat periodically with period $\Lambda_s$ along the $x{-}$direction in the two-dimensional plane~\cite{Fogler96, Fogler00} (see Fig.~\ref{fig: schematic_stripe}). Thus, the local filling fraction $\nu\left(\boldsymbol{r}\right)$ serves as the order parameter for the stripe phase, which can be modeled as~\cite{Fogler96, Goerbig04}
\begin{align}
\label{eq: real_space_stripe_order_parameter}
    \nu\left(\boldsymbol{r}\right)&=\sum_{j{=}-\infty}^{\infty}\Theta\left(w_s/2-\left|x-x_j\right|\right),
\end{align}
where $\Theta(x)$ denotes the Heaviside theta function, $x_j{=}j\Lambda_s$ with integer $j$, and we have considered an infinite two-dimensional plane. The above equation describes a uniform local filling of unity within a stripe, and a global total filling $\nu{=}w_{s}/\Lambda_s$. The total filling follows from the number of electrons in a $\nu{=}1$ strip of the stripe phase, $N_s{=}w_sL_yn_{\ell}$, in an area $A_{s}{=}\Lambda_sL_y$---consisting of a stripe and a region void of electrons--- which encloses $A_s n_{\ell}$ flux quanta, where $L_y$ represents the length along the $y{-}$direction of a two-dimensional plane with area $A{=}L_x {\times}L_y$. At $\nu{=}1/2$, the stripe state is particle-hole (PH) symmetric, as the width of a stripe is equal to the width of a void region (see Fig.~\ref{fig: schematic_stripe}). This PH symmetry makes the stripe phase a particularly plausible candidate state in a half-filled LL, particularly when the PH symmetry of the two-body Coulomb interaction is not spontaneously broken. Typically, stripe phases occur at half-filling of the $n{\geq}2$ LLs, wherein the nodes in the single-particle wave function soften the short-range part of the repulsive Coulomb interaction that can favor the formation of stripes over FQH states. Here, we will consider stripe phases at half-filling, as well as at other fillings, and study their competition with FQH states.

The average local guiding center density $\left\langle\rho^{\rm LLL}_{g}\left(\boldsymbol{r}\right)\right\rangle$ is related to $\nu\left(\boldsymbol{r}\right)$ as $\left\langle\rho^{\rm LLL}_{g}\left(\boldsymbol{r}\right)\right\rangle{=}n_{\ell} \nu\left(\boldsymbol{r}\right)$, which upon a Fourier transformation results in
\begin{align}
    \left\langle\bar{\rho}^{\rm LLL}_{g}\left(\boldsymbol{q}\right)\right\rangle&\equiv \int d^{2}\boldsymbol{r} ~e^{i \boldsymbol{q}.\boldsymbol{r}} \left\langle\rho^{\rm LLL}_{g}\left(\boldsymbol{r}\right)\right\rangle\nonumber\\
    &=n_{\ell} \int d^{2}\boldsymbol{r} ~e^{i \boldsymbol{q}.\boldsymbol{r}}\nu\left(\boldsymbol{r}\right). 
\end{align}
Substituting Eq.~\eqref{eq: real_space_stripe_order_parameter} into the above equation, we get
\begin{align}
\label{eq: guiding_center_density_average_stripe}
  \left\langle\bar{\rho}^{\rm LLL}_{g}\left(\boldsymbol{q}\right)\right\rangle &=2n_{\ell}\left(2\pi\delta\left(q_y\right)\right) \frac{1}{q_x}\sin\left(\frac{q_x \Lambda_s\nu}{2}\right)\nonumber\\
  &~~~~~~~~~~\times\frac{2\pi}{\Lambda_s}\sum_{l=-\infty}^{\infty}\delta\left(q_x-\frac{2\pi}{\Lambda_s}l\right).
\end{align}
In obtaining the above equation, we have used the following identity
\begin{align}
\label{eq: summation_over_Delta_functions}
    \sum_{j=-\infty}^{\infty}e^{-ij q_x\Lambda_s}&=\frac{2\pi}{\Lambda_s}\sum_{l=-\infty}^{\infty}\delta\left(q_x-\frac{2\pi}{\Lambda_s}l\right).
\end{align}
Next, we substitute the expression of $\left\langle\bar{\rho}^{\rm LLL}\left(\boldsymbol{q}\right)\right\rangle$ from Eq.~\eqref{eq: guiding_center_density_average_stripe} into Eq.~\eqref{eq: Hartree_Fock_Hamiltonian} and then perform the integration over $\boldsymbol{q}$ to obtain
\begin{align}
\label{eq: HF_Hamiltonian_2}
      \bar{H}^{\mathcal{N}{=}2}_{\rm HF}&=\frac{n_{\ell}}{2\pi}\sum_{l\neq0}\frac{\sin\left(\pi\nu l\right)}{l} \bar{v}_{\rm HF}\left(q=\frac{2\pi}{\Lambda_s}l;\theta_{2},\phi_{2}\right) \nonumber\\
    &~~~~~~~~~~~~~~ \times\bar{\rho}^{\rm LLL}_{g}\left(q_x=\frac{2\pi}{\Lambda_s}l,q_y=0\right).
\end{align}
We exclude the $l{=}0$, or equivalently the $\boldsymbol{q}{=}0$, term, which corresponds to an uncorrelated uniform state, as we are interested in charge density wave states with a non-zero modulation wave vector. Omitting the $l{=}0$ term takes into account the contribution from the uniformly charged positive background, since the direct contribution from the Hartree part $\bar{v}_{H}\left(q{=}0,\theta_{2},\phi_{2}\right)$ cancels the background contribution, and the exchange contribution from the Fock part $\bar{v}_{F}\left(q{=}0,\theta_{2},\phi_{2}\right)$ is chosen as the reference energy from which the total energy is measured~\cite {Lee01}. To highlight this, the per-particle average energy of the stripe phase under the HF Hamiltonian, given in Eq.~\eqref{eq: HF_Hamiltonian_2}, is referred to as the cohesive energy, $E_{\rm coh}$, which is
\begin{align}
 E_{\rm coh}&\equiv\frac{\left\langle \bar{H}^{\mathcal{N}{=}2}_{\rm HF}\right\rangle}{N}=\frac{n_{\ell}}{2\pi N}\sum_{l\neq0}\frac{\sin\left(\pi\nu l\right)}{l} \bar{v}_{\rm HF}\left(q=\frac{2\pi}{\Lambda_s}l;\theta_{2},\phi_{2}\right)\nonumber\\
    &~~~~~~~~~~~~~~~~~~~~~~~~~\times\left\langle\bar{\rho}^{\rm LLL}_{g}\left(q_x=\frac{2\pi}{\Lambda_s}l,q_y=0\right)\right\rangle.
\end{align}
From Eq.~\eqref{eq: summation_over_Delta_functions}, it follows that  
\begin{align}
   \left\langle\bar{\rho}^{\rm LLL}_{g}\left(q_x=\frac{2\pi}{\Lambda_s}l,q_y=0\right)\right\rangle &=2n_{\ell}\left[2\pi\delta\left(q_y=0\right)\right] \frac{\sin\left(\pi\nu l\right)}{l}\nonumber\\
   &\times \sum_{l^{\prime}=-\infty}^{\infty}\delta\left(\frac{2\pi}{\Lambda_s}l-\frac{2\pi}{\Lambda_s}l^{\prime}\right).
\end{align}
To evaluate the right-hand side of the above equation, we convert the Dirac delta function to the Kronecker delta through the relation
\begin{align}
    \delta\left(q_x-a\right)\delta\left(q_y-b\right)&=\frac{A}{(2\pi)^2}\delta_{q_x,a}\delta_{q_y,b},
\end{align}
which results in
\begin{align}
    \left\langle\bar{\rho}^{\rm LLL}_{g}\left(q_x=\frac{2\pi}{\Lambda_s}l,q_y=0\right)\right\rangle &=\frac{A}{2\pi} (2n_\ell) \frac{\sin(\pi\nu l)}{l}.
\end{align}
Consequently, one obtains the following expression for the cohesive energy of the stripe phase
\begin{align}
    E^{\rm stripe}_{\rm coh}&=\frac{n_\ell}{\pi^2\nu}\sum_{l{\geq}1}\frac{\sin^2\left(\pi\nu l\right)}{l^2} \bar{v}_{\rm HF}\left(q=\frac{2\pi}{\Lambda_s}l;\theta_{2},\phi_{2}\right),
\end{align}
where we have used the total number of electrons $N{=}\nu n_{\ell} A$, and restricted the sum to positive integers, as $\bar{v}_{\rm HF}\left(q;\theta_{2},\phi_{2}\right)$ is an even function of $q$. We have evaluated the stripe energy by retaining the first $70$ terms in the summation, which yields a well-converged energy. At a given filling, to determine the energy, one requires the optimal width of the stripe, which we determine by minimizing $E^{\rm stripe}_{\rm coh}$ as a function of $\Lambda_{s}$.

\section{Results}
\label{sec: results}
In the following section, we compare the computed ground state energy of various states at a given filling to obtain the phase diagram as a function of parameters $\cos^2(\theta_2)$ and $\cos^2(\phi_2)$. 

\subsection{Phase diagram at $\nu{=}1/2$}
\label{ssec: competition_1_2}
\begin{figure}[tbh!]
\centering
\begin{tabular}{cc}
        \includegraphics[width=0.499\columnwidth]{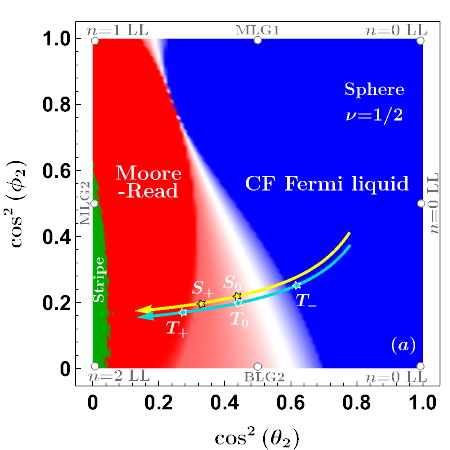}&
     \includegraphics[width=0.499\columnwidth]{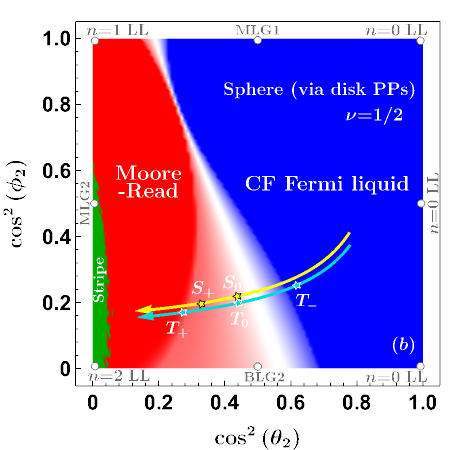}
         \end{tabular}
          \caption{Phase diagram showing the competition between the composite fermion Fermi liquid (blue), Moore-Read (red), and stripe phases (green) in the half-filled $\mathcal{N}{=}2$ LL of bilayer graphene as a function of parameters $\cos^2(\theta_2)$ and $\cos^2(\phi_2)$. The yellow- and cyan colored lines mark the direction of increasing displacement field in the phase diagram at fixed $B_{\perp}{=}13.95$ T and $B_{\perp}{=}12$ T, respectively. The points $S_{0}$ and $S_{+}$ on the yellow line denote the displacement field $D{=}0$ mV/nm, and $D{=}342.9$ mV/nm (relevant for the STM experiment, wherein $U{=}Dd/\epsilon{=}30$ meV with $\epsilon{=}4$, and the layer separation $d{=}0.35$ nm~\cite{Hu24}), respectively. Similarly, the points $T_{\pm}$ and $T_{0}$ denote $D{=}{\pm}500$ mV/nm and $D{=}0$ mV/nm, respectively, relevant for the recent transport experiment of Ref.~\cite{Kumar25}. The panels $(a)$ and $(b)$ represent the phase diagram obtained directly on the sphere and the one obtained by mapping the pseudopotentials (PPs) from the disk geometry onto the sphere, respectively. The open circles at the boundaries of each panel denote the specific LL points. We have used the structure factor of system sizes ranging from $N{\sim}~9$ to $36$ electrons on the sphere to compute the thermodynamic limit energy of the composite fermion Fermi liquid, while for the Moore state, we have considered system sizes ranging from $N{\sim}~12$ to $80$ electrons.}
          \label{fig: CFL_MR_stripe_phase_diagram}
        \end{figure} 

The phase diagram of CFFL and MR states, obtained from the thermodynamic limit extrapolated energies computed entirely on the sphere [see Eq.~\eqref{eq: per_particle_energy_on_the_sphere}], is shown in Fig.~\ref{fig: CFL_MR_stripe_phase_diagram}$(a)$. A similar phase diagram, obtained by mapping the disk PPs onto the sphere [see Sec~\ref{sssec: Energy_on_the_sphere_from_planar_PPs}], is depicted in Fig.~\ref{fig: CFL_MR_stripe_phase_diagram}$(b)$. The close agreement between the two phase diagrams obtained through complementary approaches demonstrates that our thermodynamic limit energies on the sphere are well converged. As noted above, we have not been able to obtain the energy of the CFFL on the plane, so we have not been able to construct the phase diagram at half-filling on the plane.

In the phase diagram, we have marked two contours---a yellow and a cyan curve---with the arrowheads representing the direction of increasing displacement field $D$ at fixed $B_{\perp}{=}13.95$ T and $B_{\perp}{=}12$ T, corresponding to the parameters in two recent experiments based on local STM probes~\cite{Hu24} and macroscopic transport measurements, respectively~\cite{Kumar25}. In the STM measurement, a FQH state was observed at parameters $B_{\perp}{=}13.95$ T and an intrinsic displacement field corresponding to an interlayer potential difference $U{=}30$ meV, equivalent to the $S_{+}$ point marked on the yellow contour~\cite{Hu24}. Unfortunately, in our computation, we could not resolve the nature of the ground state at $S_{+}$, as the energies of both the CFFL and MR states are very close, and their difference is comparable to the error bars. Nevertheless, we identify the trend that, by increasing $D$ further away from the $S_{+}$ point, one can stabilize the MR state in the $\mathcal{N}{=}2$ LL of BLG. Interestingly, this is consistent with the complementary transport measurement, wherein the displacement field between the layers can be explicitly tuned~\cite{Kumar25}.

Ref.~\cite{Kumar25} finds that within the range $4{<}{\nu}{<}6$, at the half filling of each lowest energy spin-valley resolved LLs, the MR state gradually strengthens out of the CFFL, for $D{>}540$ mV/nm in the $\boldsymbol{K}$ valley, and $D{<}{-}540$ mV/nm in the $\boldsymbol{K}^{\prime}$ valley [see Fig.~\ref{fig: BLG2_LLs}($b$) for the ordering of $\mathcal{N}{=}2$ spin-valley resolved LLs as a function of $D$, where we have used same parameters as those in Ref.~\cite{Kumar25}]. This experimental observation is readily captured in our computed phase diagram, shown for the $\boldsymbol{K}$ valley. Specifically, the $D{=}540$ mV/nm point evidently lies above the marked point $T_{+}$ (denoting $D{=}500$ mV/nm) on the cyan contour in Fig.~\ref{fig: CFL_MR_stripe_phase_diagram}, around and above which the MR is the dominant phase. The phase diagram at the $\boldsymbol{K}^{\prime}$ valley is the same as the $\boldsymbol{K}$ valley [see Fig.~\ref{fig: CFL_MR_stripe_phase_diagram}], with the difference that the direction of the contour lines is reversed, including the labeled points, about the corresponding $D{=}$0 mV/nm points. It follows then that $D{=}{-}540$ mV/nm lies below $T_{-}$ (denoting $D{=}{-}500$ mV/nm) on the cyan contour, around and below which the MR state is stabilized in the $\boldsymbol{K}^{\prime}$ valley. Similarly, for $D{>}500$ mV/nm, CFFL dominates in the $\boldsymbol{K}^{\prime}$ valley.
 
Next, we discuss the nature of ground states at the half-filling of higher LLs in the range $6{<}\nu{<}8$, for the same values of $D$ at which FQH states are observed in half-filled LLs in the range $4{<}\nu{<}6$; namely, $D{>}540$ mV/nm in the $\boldsymbol{K}$ valley, and $D{<}{-}540$ mV/nm in the $\boldsymbol{K}^{\prime}$ valley. From Fig.~\ref{fig: BLG2_LLs}($b$), it is evident that for $D{>}500$ mV/nm, the LLs in the range $6{<}\nu{<}8$ are $\boldsymbol{K}^{\prime} $- valley polarized. It follows from our previously envisioned phase diagram at the $\boldsymbol{K}^{\prime}$ valley that, for $D{>}540$ mV/nm, the half-filled states in $6{<}\nu{<}8$ are likely CFFLs. Similarly, for $D{<}{-}540$ mV/nm, the half-filled states within $6{<}\nu{<}8$ in the $\boldsymbol{K}^{\prime}$ valley, are likely CFFLs. These results are consistent with the transport experiment, wherein no incompressible states are observed at the half-fillings in the range $6{<}\nu{<}8$~\cite{Kumar25}. 

At a conceptual level, these results can be understood in light of Ref.~\cite{Kumar25} that the intermediate admixture of the $n{=}2$ and $n{=}1$ LLs in the $\mathcal{N}{=}2$ BLG LL stabilizes the MR state for $D{>}500$ mV/nm in the $\boldsymbol{K}$ valley, and $D{<}{-}500$ mV/nm in the $\boldsymbol{K}^{\prime}$ valley. On the other hand, for $D{<}{-}500$ mV/nm in the $\boldsymbol{K}$ valley and $D{>}500$ mV/nm in the $\boldsymbol{K}^{\prime}$ valley, the $n{=}0$ LL dominates the $\mathcal{N}{=}2$ BLG LL, and no FQH states are observed at half filling.

Finally, we note that as one approaches very close to the $n{=}2$ LL point in the phase diagram [in the vicinity of the bottom left point in Fig.~\ref{fig: CFL_MR_stripe_phase_diagram}], the stripe phase dominates over the MR state. This is consistent with the fact that stripes dominate at the half-filling of the $n{=}2$ LL.

\subsection{Phase diagram at $\nu{=}6/13$}
\label{ssec: competition_6_13}
\begin{figure*}
    \centering
    \includegraphics[width=0.66\columnwidth]{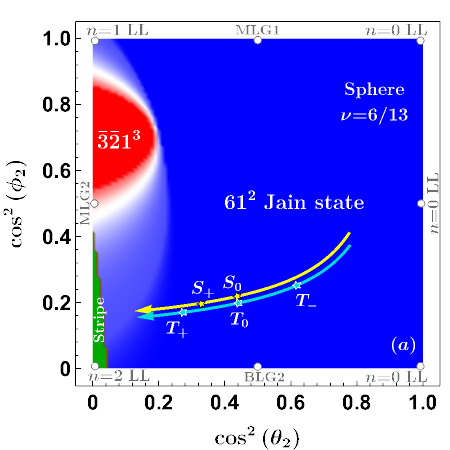}
    \includegraphics[width=0.66\columnwidth]{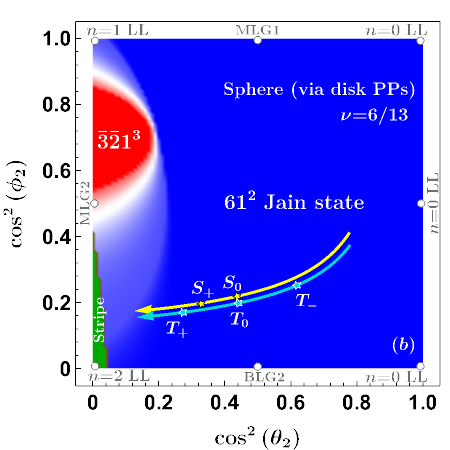}
    \includegraphics[width=0.66\columnwidth]{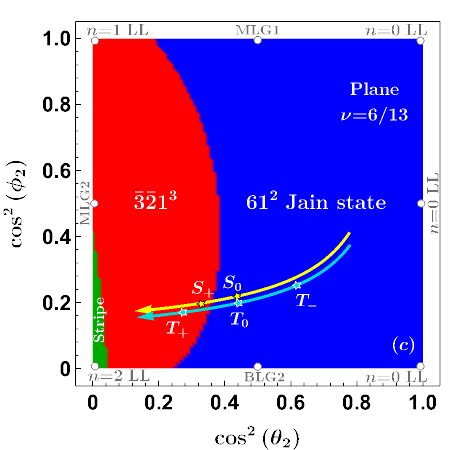}
    \caption{Phase diagram at $\nu{=}6/13$ obtained from the competition between the $6/13$ Jain state, $\bar{3}\bar{2}1^3$ parton state, and stripe phase in the $\mathcal{N}{=}2$ LL of bilayer graphene as a function of parameters $\cos^2(\theta_2)$ and $\cos^2(\phi_2)$. $(a)$ The phase diagram obtained entirely on the spherical geometry. $(b)$ The phase diagram obtained through mapping disk PPs onto the sphere. In both panels $(a)$ and $(b)$, we have considered $N{\sim}36$ to $96$ electrons for the Jain state, and $N{\sim}18$ to $60$ electrons for the $\bar{3}\bar{2}1^3$ parton state, to compute their thermodynamic limit energies. $(c)$ The phase diagram computed in the planar geometry, where we have fitted the pair-correlation data of $N{=}60$ electrons on the sphere to its planar analytic form [see Eq.~\eqref{eq: pair_correlation_expansion}] to obtain the thermodynamic limit energies of the corresponding Jain and parton states. The remaining plot labels are the same as in Fig.~\ref{fig: CFL_MR_stripe_phase_diagram}.}
    \label{fig: 6_13_Jain_parton_states_phase_diagram}
\end{figure*}

In Fig.~\ref{fig: 6_13_Jain_parton_states_phase_diagram}, we present the phase diagram obtained from studying the energetic competition between the $\bar{3}\bar{2}1^3$ parton state, the $61^2$ primary Jain state, and the stripe phase at $\nu{=}6/13$. The phase diagram computed entirely on the sphere [see Fig.~\ref{fig: 6_13_Jain_parton_states_phase_diagram}$(a)$] and that obtained from using the disk PPs on the sphere [see Fig.~\ref{fig: 6_13_Jain_parton_states_phase_diagram}$(b)$] are nearly identical, indicating that the thermodynamic limit energies of FQH states obtained via these two different ways are consistent with each other. In contrast, the phase diagram obtained in the planar geometry see Fig.~\ref{fig: 6_13_Jain_parton_states_phase_diagram}$(c)$, differs slightly from the spherical results. In particular, the $\bar{3}\bar{2}1^3$ state occupies a somewhat larger region, extending slightly beyond the shaded white region in Figs.~\ref{fig: 6_13_Jain_parton_states_phase_diagram}$(a)$ and~\ref{fig: 6_13_Jain_parton_states_phase_diagram}$(b)$. As mentioned before, ideally, we expect the phase diagrams obtained from the thermodynamic limit extrapolated energies on the sphere to agree with those obtained in the planar geometry. However, the energy calculations of the FQH states are inherently numerical and can be plagued by issues stemming from Monte Carlo statistical errors, uncertainties in the fits, and finite-size effects. On the sphere, the energies are evaluated from the numerically computed structure factor, whereas on the plane they are obtained from the numerically computed pair-correlation function. For the pair-correlation function, we have used a large system of $N{=}60$ electrons on the sphere and assumed it to be sufficiently large that it represents the thermodynamic result. Consequently, the energies are accurate only up to a few decimal places. In regions where competing FQH states have very close energies---within the numerical resolution---it becomes difficult to unambiguously identify the true ground state. We believe that this numerical limitation is responsible for the discrepancies between the spherical and planar phase diagrams. More accurate results are required to obtain quantitatively the precise phase boundaries, which is beyond the scope of the current work.

Next, we discuss our phase diagram in light of the experimental observations. The STM experiment, performed at a magnetic field and displacement field (intrinsic to the experimental configuration) corresponding to the marked point $S_{+}$ in our phase diagram [see Fig.~\ref{fig: 6_13_Jain_parton_states_phase_diagram}], observes an incompressible state at $6/13$ in the $\mathcal{N}{=}2$ LL of BLG. This state is observed alongside FQH states in the primary Jain sequence, thereby identifying the $6/13$ state as a Jain state. We find in the phase diagram obtained from the computations on the sphere geometry [see Figs.~\ref{fig: 6_13_Jain_parton_states_phase_diagram}$(a)$ and~\ref{fig: 6_13_Jain_parton_states_phase_diagram}$(b)$ ], the $S_{+}$ point belongs to the Jain state, consistent with the experiment. However, in our planar phase diagram, the $S_{+}$ point is shifted to lie in the region corresponding to the $\bar{3}\bar{2}1^3$ parton state. Nevertheless, our phase diagrams suggest that the strength of the $6/13$ Jain state decreases with increasing the displacement field in the $\boldsymbol{K}$ valley (equivalently, decreasing the displacement field in the $\boldsymbol{K}^{\prime}$ valley).

We note that, in contrast to the STM measurement, the transport experiment at zero interlayer displacement field does not observe the signature of a FQH state at $6/13$~\cite{Diankov16} in the $\mathcal{N}{=}2$ LL. Moreover, the recent transport experiment reported in Ref.~\cite{Kumar25} also does not report the observation of a FQH state at $6/13$ as the interlayer displacement field is tuned. This is likely because the $6/13$ Jain CF state has a low gap.

The $\mathcal{N}{=}2$ LL of BLG for $\cos^{2}\left(\theta_2\right){=}0$ and $\cos^{2}\left(\phi_2\right){=}0$ becomes equivalent to the SLL, i.e., $n{=}1$. Experimentally, the FQH effect has been observed at $2{+}6/13$ in GaAs~\cite{Kumar10}. Moreover, previous studies~\cite{Balram18a, Balram24a} find $\bar{3}\bar{2}1^3$ parton as a better candidate state than the $6/13$ Jain state. This is consistent with our planar phase diagram [see Fig.~\ref{fig: 6_13_Jain_parton_states_phase_diagram}$(c)$], while the regions around the $n{=}1$ LL remain unresolved in the phase diagrams obtained on the spherical geometry [see Fig.~\ref{fig: 6_13_Jain_parton_states_phase_diagram}$(a)$ and~\ref{fig: 6_13_Jain_parton_states_phase_diagram}$(b)$]. In Appendix~\ref{app: 6_13_parton_WQW}, we provide additional results on the $\bar{3}\bar{2}1^3$ state in the LLL of wide quantum well systems, where this state is likely stabilized experimentally~\cite{Singh23, Singh25}. 

The interaction in the $\mathcal{N}{=}1$ LL of BLG appears to stabilize the $1/2$ MR state, and its two daughters at $7/13$ and $8/17$~\cite{Levin09a} at low magnetic fields, i.e., in the vicinity of the SLL Coulomb point (For 1/2, see Fig. 1(b) of Ref.~\cite{Balram21b}, for 6/13 [the results for which apply to 7/13 via hole-conjugation] see Fig. S1 of Ref.~\cite{Balram24a}, and for 8/17 see Fig. 2(b) of Ref.~\cite{Balram24a}, where the ground state has $L{=}0$ for $N{=}20$ at the 8/17 Levin-Halperin (LH) flux away from, but close to the SLL Coulomb point. See also Ref.~\cite{Yutushui25a}.). In our particle-hole symmetric model of two-body interactions, $7/13$ and $6/13$ are considered on the same footing. This is consistent with our planar phase diagram [see Fig.~\ref{fig: 6_13_Jain_parton_states_phase_diagram}$(c)$], but this region remains unresolved in the spherical phase diagrams  [see Figs.~\ref{fig: 6_13_Jain_parton_states_phase_diagram}$(a)$-$(b)$].  

Finally, in the vicinity of the $n{=}2$ LL, the stripe phase becomes the ground state, similar to the case at $\nu{=}1/2$.

\subsection{Phase diagram at $\nu{=}4/9$}

\label{ssec: competition_4_9}
\begin{figure*}
    \centering
    \includegraphics[width=0.66\columnwidth]{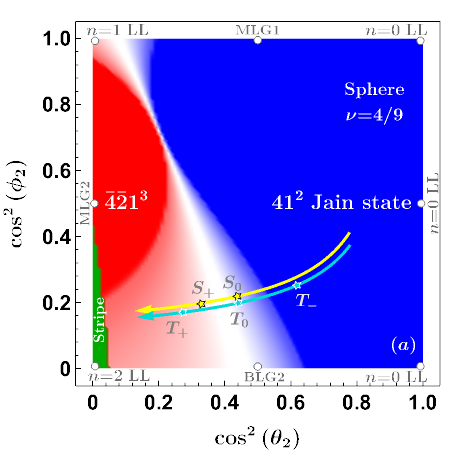}
    \includegraphics[width=0.66\columnwidth]{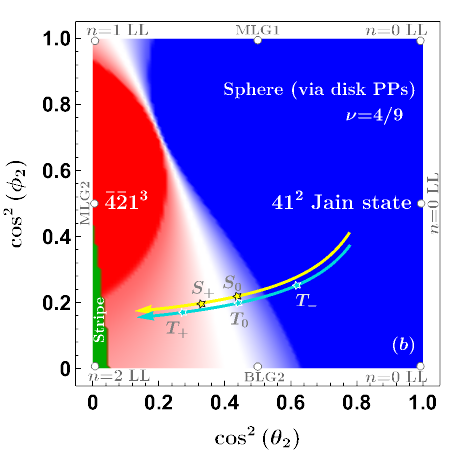}
    \includegraphics[width=0.66\columnwidth]{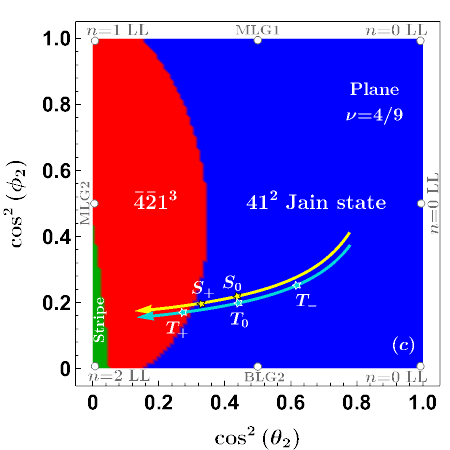}
    \caption{Phase diagram at $\nu{=}4/9$ obtained from the competition between the $4/9$ Jain state, $\bar{4}\bar{2}1^3$ parton state, and stripe phase in the $\mathcal{N}{=}2$ LL of bilayer graphene as a function of parameters $\cos^2\left(\theta_{2}\right)$ and $\cos^2\left(\phi_{2}\right)$. $(a)$ The phase diagram obtained entirely on the spherical geometry, where we have considered $N{\sim}20$ to $84$ electrons for the Jain state, and $N{\sim}20$ to $60$ electrons for the $\bar{4}\bar{2}1^3$ parton state, to extrapolate their thermodynamic limit energies. $(b)$ The phase diagram obtained using the disk PPs in the spherical geometry, with the same system sizes of FQH states as in panel $(a)$. $(c)$ The phase diagram computed in the planar geometry, where the thermodynamic limit energies of FQH states are computed from the analytic pair-correlation function [see Eq.~\eqref{eq: pair_correlation_expansion}] obtained by fitting to its numerically computed data for $N{=}60$ electrons on the sphere. The remaining plot labels are the same as in Fig.~\ref{fig: CFL_MR_stripe_phase_diagram}.}
    \label{fig: 4_9_Jain_parton_states_phase_diagram}
\end{figure*}

The phase diagram at $\nu{=}4/9$ obtained from considering the $\bar{4}\bar{2}1^{3}$ parton, $4/9$ Jain CF, and stripe states is shown in Fig.~\ref{fig: 4_9_Jain_parton_states_phase_diagram}. The phase diagram obtained entirely on the sphere is almost similar to that obtained by mapping the disk PPs onto the sphere. The slight mismatch between the spherical and the planar phase diagrams can be attributed to the reason discussed in Sec.~\ref{ssec: competition_6_13}. We find that the $S_{+}$ point in the planar phase diagram corresponds to the Jain state, consistent with the STM measurement~\cite{Hu24} [The $S_{+}$ point in the spherical phase diagram remains unresolved.]. Moreover, at the points $S_{0}$ and $T_{0}$, corresponding to the zero displacement field, the Jain state is favored.  However, we note that no FQH state at $\nu{=}4/9$ is observed in the transport experiment without an applied displacement field~\cite{Diankov16} or with a tunable displacement field~\cite{Kumar25} in the $\mathcal{N}{=}2$ LL of BLG, presumably because its gap is low.

The expected phases at specific LL points are marked in Fig.~\ref{fig: 4_9_Jain_parton_states_phase_diagram}. Notably, we find that in the SLL ($n{=}1$ LL), the parton state $\bar{4}\bar{2}1^{3}$ wins over the Jain state at $\nu{=}4/9$, consistent with the results discussed in Ref.~\cite{Balram20b}. Similarly, near the third LL ($n{=}2$ LL), the stripe phase prevails over the FQH states.

\subsection{Phase diagram at $\nu{=}3/7$}

\label{ssec: competition_3_7}  

\begin{figure*}
    \centering
    \includegraphics[width=0.66\columnwidth]{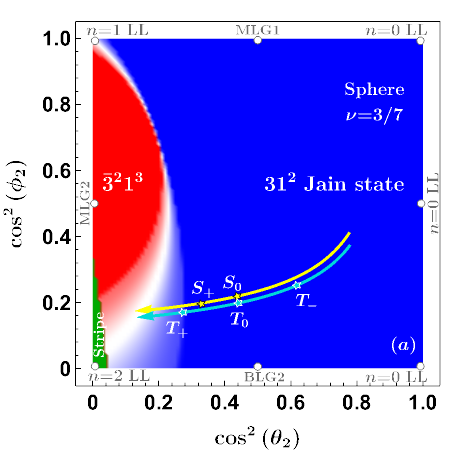}
    \includegraphics[width=0.66\columnwidth]{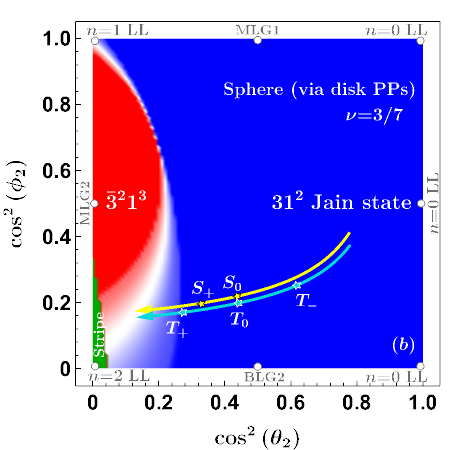}
     \includegraphics[width=0.66\columnwidth]{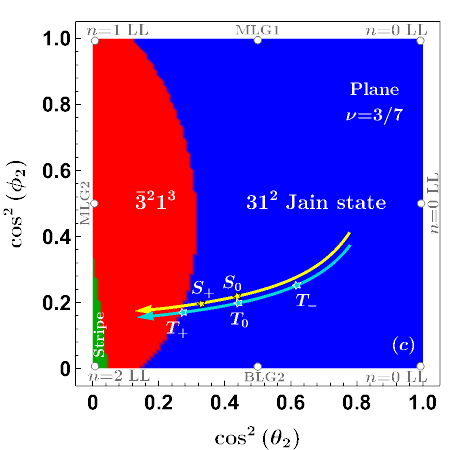}
    \caption{Phase diagram at $\nu{=}3/7$ obtained from the competition between the $3/7$ Jain state, $\bar{3}^21^3$ parton state, and stripe phase in the $\mathcal{N}{=}2$ LL of bilayer graphene as a function of parameters $\cos^2\left(\theta\right)$ and $cos^2\left(\phi\right)$. $(a)$ The phase diagram obtained entirely on the spherical geometry, where we have extrapolated the thermodynamic-limit energies of the $\nu{=}3/7$ Jain state considering system sizes $N{\sim}18$ to 84 electrons, and of the $\bar{3}^21^3$ parton state using $N{\sim}12$ to 60 electrons. $(b)$ The phase diagram obtained through mapping disk PPs onto the sphere, with the same system sizes of FQH states as in panel $(a)$. $(c)$ The phase diagram computed in the planar geometry, where we have used the pair-correlation data for $N{=}60$ electrons on the sphere for each FQH state. The remaining plot labels are the same as in Fig.~\ref{fig: CFL_MR_stripe_phase_diagram}.}
    \label{fig: 3_7_Jain_parton_states_phase_diagram}
\end{figure*}
In this section, we discuss the computed phase diagram at $\nu{=}3/7$ with a focus on the experimental findings. The STM experiment~\cite{Hu24} has observed a FQH state at $\nu{=}3/7$ in the $\mathcal{N}{=}2$ LL of BLG. In our phase diagram, at the equivalent STM experimental point $S_{+}$, the Jain CF state is favored at $\nu{=}3/7$ [see Fig.~\ref{fig: 3_7_Jain_parton_states_phase_diagram}]. We also find that at zero displacement field [see the points $S_0$ and $T_{0}$ in Fig.~\ref{fig: 3_7_Jain_parton_states_phase_diagram}] the Jain CF state is the ground state. However, similar to $\nu{=}6/13$ and $\nu{=}4/9$, no FQH state has so far been observed in transport experiments at $\nu{=}3/7$ for both the zero displacement field~\cite{Diankov16} and the tunable displacement field cases~\cite{Kumar25}.  

Consistent with the previous studies, we find that in the $n{=}1$ LL, the parton state dominates over the Jain state at $\nu{=}3/7$~\cite{Faugno21} in our planar phase diagram, as shown in Fig.~\ref{fig: 3_7_Jain_parton_states_phase_diagram}$(c)$. Similar to the cases at $\nu{=}1/2,6/13$, and $4/9$, the stripe phase continues to dominate around the $n{=}2$ LL at $\nu{=}3/7$.

\subsection{Phase diagram at $\nu{=}2/5$}
\label{ssec: competition_2_5}  

\begin{figure*}
    \centering
    \includegraphics[width=0.66\columnwidth]{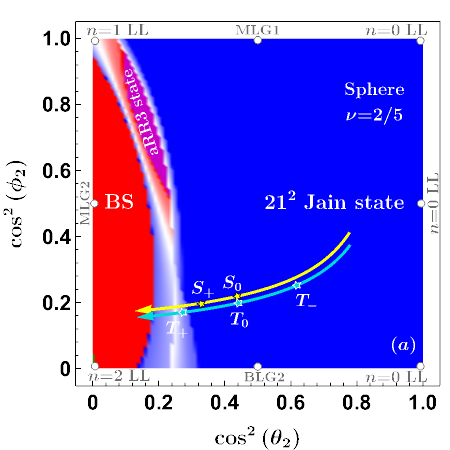}
    \includegraphics[width=0.66\columnwidth]{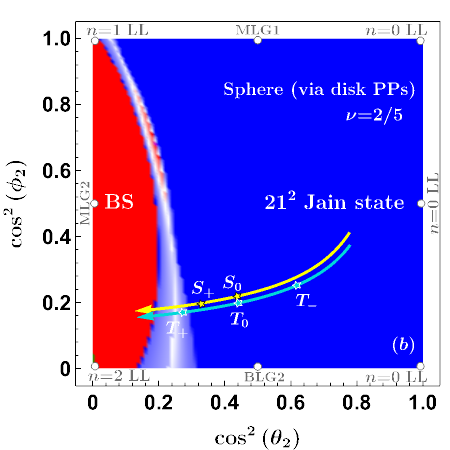}
    \includegraphics[width=0.66 \columnwidth]{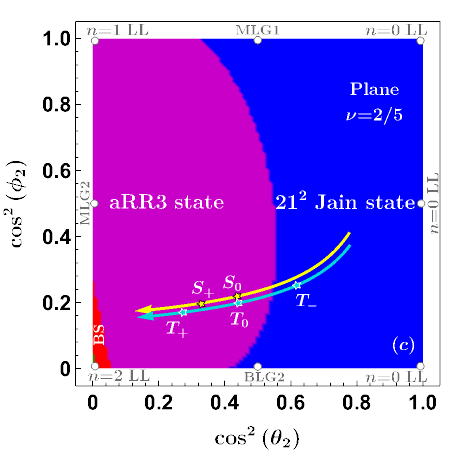}
    \caption{Phase diagram at $\nu{=}2/5$ obtained from the competition between the $2/5$ Jain state, the anti-Read-Rezayi $3-$cluster state, Bonderson-Slingerland state, and stripe phase in the $\mathcal{N}{=}2$ LL of bilayer graphene as a function of parameters $\cos^2\left(\theta\right)$ and $cos^2\left(\phi\right)$. $(a)$ The phase diagram obtained entirely on the spherical geometry. Here, we have considered system sizes spanning from $N{\sim}12$ to $84$ electrons for the $2/5$ Jain state, $N{\sim}10$ to $64$ electrons for the Bonderson-Slingerland state, and $N{\sim}2$ to $14$ electrons for the anti-Read-Rezayi $3-$cluster state. $(b)$ The phase diagram obtained through mapping disk PPs onto the sphere, with the same system sizes as in panel $(a)$. $(c)$ The phase diagram computed in the planar geometry, where we have used the pair-correlation data of $N{=}60$ electrons for both the Jain and Bonderson-Slingerland states, and $N{=}14$ electrons for the anti-Read-Rezayi $3-$cluster state. The remaining plot labels are the same as in Fig.~\ref{fig: CFL_MR_stripe_phase_diagram}.}
    \label{fig: 2_5_Jain_aRR3_BS_states_phase_diagram}
\end{figure*}

The spherical phase diagram [see Fig.~\ref{fig: 2_5_Jain_aRR3_BS_states_phase_diagram}$(a)$] of FQH states at $\nu{=}2/5$ differs significantly from that computed in the planar geometry [see Fig.~\ref{fig: 2_5_Jain_aRR3_BS_states_phase_diagram}$(c)$]. We attribute this discrepancy to finite-size effects, as the planar energy of the aRR$3$ state was computed using pair-correlation data from a relatively small system ($N{=}14$). Consequently, in the following discussion, we primarily rely on the spherical phase diagram for comparison with experimental observations. 

In both the STM experiment~\cite{Hu24} and the transport experiments with tunable~\cite{Kumar25} and zero displacement field~\cite{Diankov16}, a robust FQH state at $\nu{=}2/5$ has been observed. We find that at the STM experimental point $S_{+}$, the Jain $2/5$ CF state is the ground state. Moreover, near the zero displacement field point [see points $S_0$ and $T_0$ in Fig.~\ref{fig: 2_5_Jain_aRR3_BS_states_phase_diagram}$(a)$], the Jain $2/5$ CF state is favored over other candidates. The phase diagrams suggest that the strength of the Jain $2/5$ CF state decreases with increasing displacement field.

The nature of the $\nu{=}2/5$ FQH state is believed to be different than the Jain $2/5$ state in the $n{=}1$ LL~\cite{Bonderson12}. Unfortunately, we could not fully resolve the nature of the ground state at the $n{=}1$ LL point in our phase diagram. Finally, very close to the $n{=}2$ LL point, the stripe phase becomes the ground state, as it does at other fillings.

\section{Conclusion and Discussion}
\label{sec: discussion}

\begin{figure*}
    \centering
    \includegraphics[width=0.66\columnwidth]{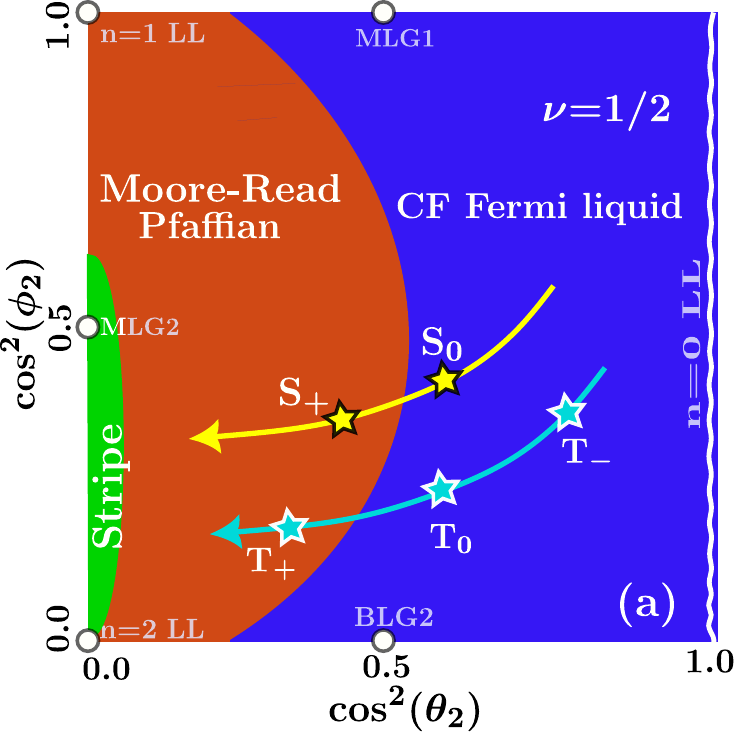}
    \includegraphics[width=0.66\columnwidth]{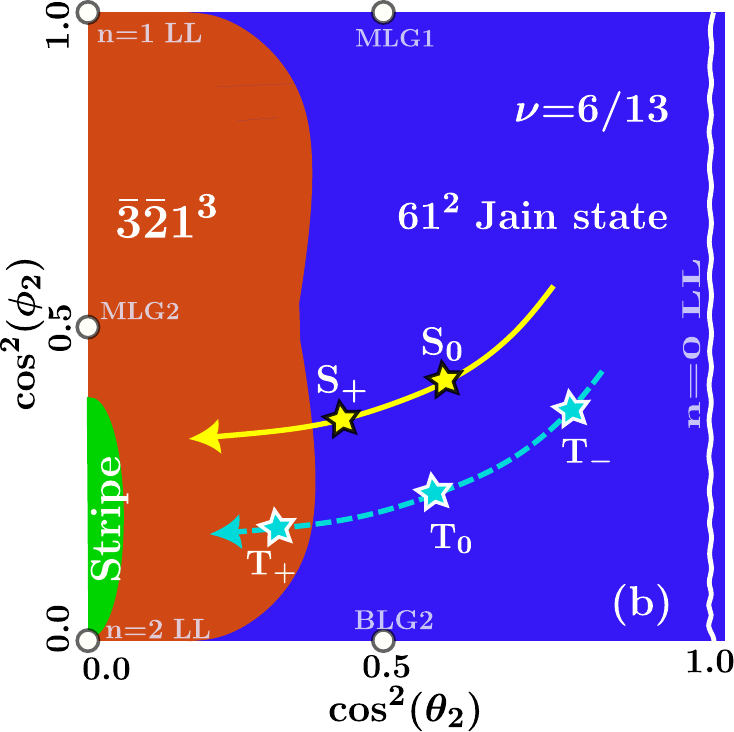}
     \includegraphics[width=0.66\columnwidth]{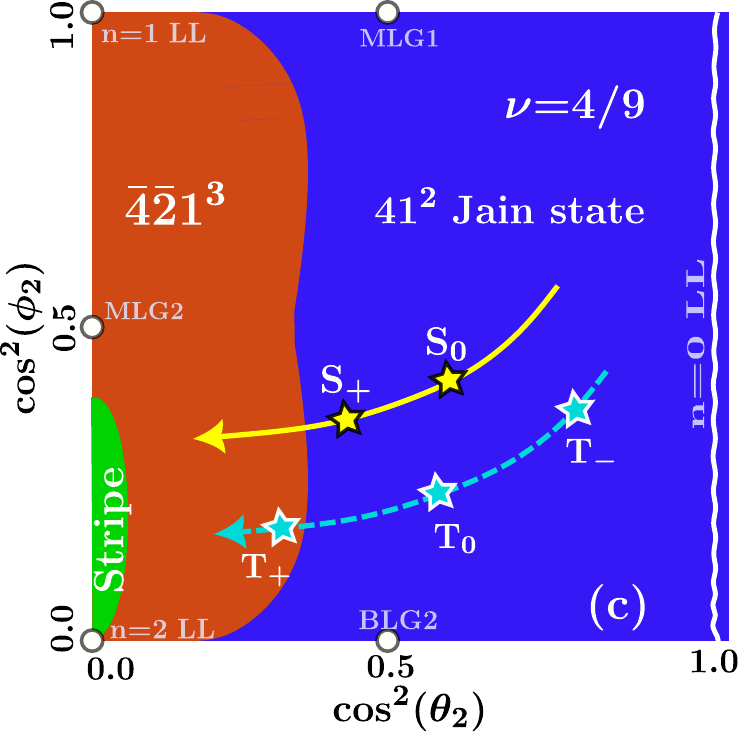}\\
      \includegraphics[width=0.66\columnwidth]{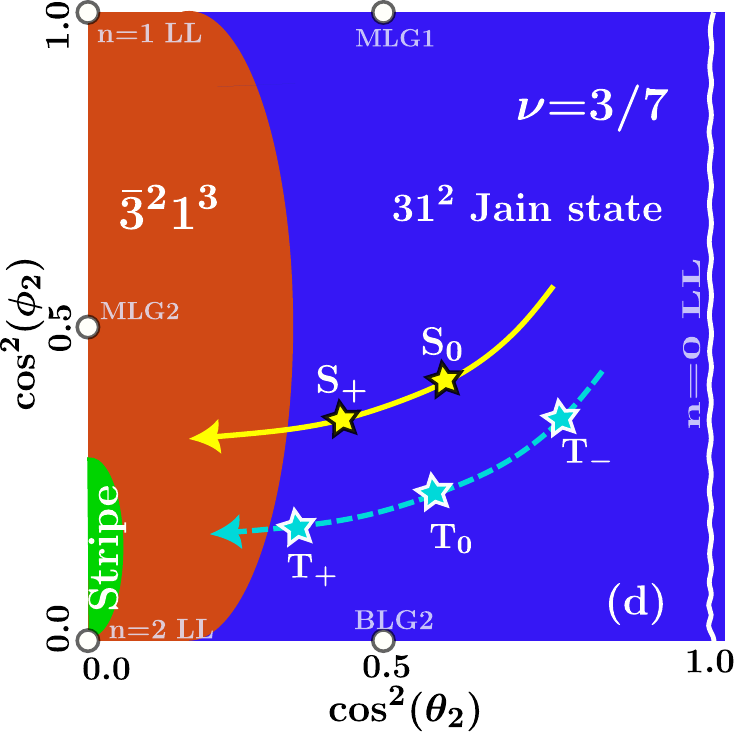}
       \includegraphics[width=0.66\columnwidth]{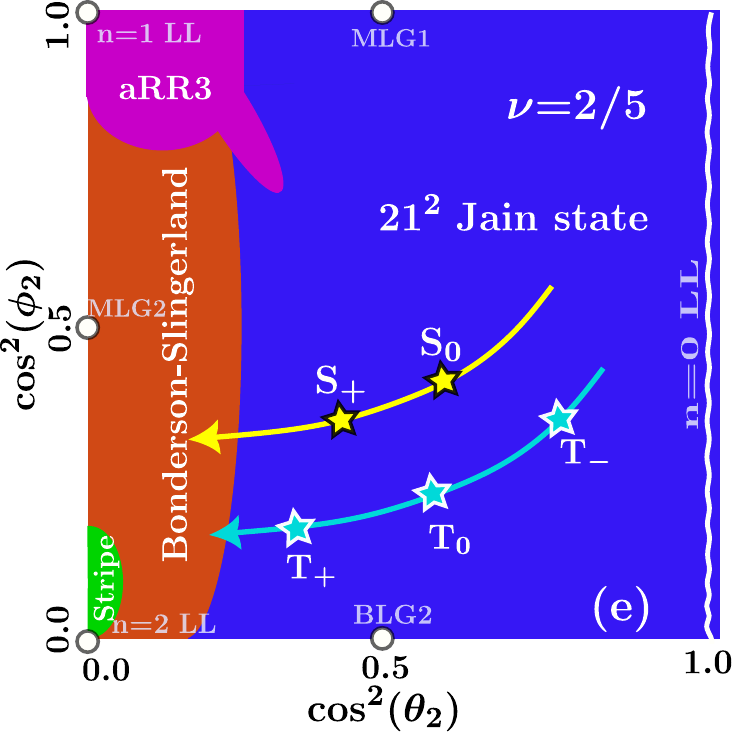}
    \caption{Schematic phase diagrams at various fillings, obtained by combining our theoretical calculations with experimental observations. The cyan colored curves in panels $(b)$-$(d)$ are shown as dashed to highlight that the corresponding FQH states have not been observed in the recent transport experiment reported in Ref.~\cite{Kumar25}.}
    \label{fig: schematic_phase_diagram}
\end{figure*}

In this article, we have investigated the nature of the various quantum Hall states at and around half-filling of a spin-valley-polarized $\mathcal{N}{=}2$ LL in bilayer graphene (BLG) as the interlayer displacement field $D$ is tuned. Specifically, for a given spin, we focus on the $\boldsymbol{K}$ valley, since the corresponding results for the $\boldsymbol{K}^{\prime}$ valley can be obtained by noting that the $\boldsymbol{K}$ valley at positive $D$ is related to the $\boldsymbol{K}^{\prime}$ valley at negative $D$, and vice versa. Furthermore, the phase diagram is identical for spin-up and spin-down states because the single-particle wave functions are independent of the Zeeman coupling.

We find that, as the displacement field increases, the composite Fermi liquid at half-filling gives way to a paired state of composite fermions described by the Moore-Read state [see Fig.~\ref{fig: CFL_MR_stripe_phase_diagram}]. Interestingly, this result is consistent with the recent transport experiment~\cite{Kumar25}, in which such a transition has been observed upon tuning $D$. Similarly, the STM experiment~\cite{Hu24}, in which an interlayer potential difference generates a fixed displacement field, has also observed signatures consistent with the MR state at half-filling of the $\mathcal{N}{=}2$ LL. However, the corresponding STM experimental point, denoted by $S_{+}$, in our phase diagram, lies in the unresolved region, preventing us from unambiguously determining the nature of the state around $S_{+}$. To provide a complete picture of the possible phase diagram over the entire parameter space spanned by $\cos^2\left(\theta_{2}\right)$ and $\cos^2\left(\phi_{2}\right)$, we present a schematic phase diagram at $\nu{=}1/2$ in Fig.~\ref{fig: schematic_phase_diagram}(a), obtained by combining our numerical results with the experimentally established phases at different values of $D$. Similar schematic phase diagrams for $\nu{=}6/13$, $4/9$, $3/7$, and $2/5$ are also presented in Fig.~\ref{fig: schematic_phase_diagram}.

Similar to the case of $\nu{=}1/2$, our results suggest that, at other fillings, an FQH state can transition into states belonging to different topological sectors upon tuning $D$. Specifically, in the $\boldsymbol{K}$ valley, the Jain state at $\nu{=}6/13$ can phase transition into the $\bar{3}\bar{2}1^3$ parton state as $D$ increases. Likewise, the Jain states at $\nu{=}4/9$ and $3/7$ can undergo phase transitions to the $\bar{4}\bar{2}1^3$ and $\bar{3}^2 1^3$ parton states, respectively, with increasing $D$. Finally, the Jain state at $\nu{=}2/5$ can transition to either the anti-Read–Rezayi 3-cluster state or the Bonderson–Slingerland state as $D$ is increased.

Our phase diagram suggests that, for fillings $6{<}\nu{<}8$ in the $\mathcal{N}{=}2$ LL, Jain FQH states at partial fillings $2/5$, $3/7$, $4/9$, and $6/13$ could be observable for $|D|{>}500$ mV/nm. Notably, this is the same range of $D$ in which FQH states at half-fillings are observed for $4{<}\nu{<}6$~\cite{Kumar25}. For $6{<}\nu{<}8$, the $\mathcal{N}{=}2$ LL is $\boldsymbol{K}^{\prime}$ valley polarized for $D{>}500$ mV/nm and $\boldsymbol{K}$-valley polarized for $D{<}-500$ mV/nm [see Fig.~\ref{fig: BLG2_LLs}($b$)]. As shown in Fig.~\ref{fig: schematic_phase_diagram}, the Jain states dominate at partial fillings $2/5$, $3/7$, $4/9$, and $6/13$ for $D{<}-500$ mV/nm. The corresponding phase diagram for $D{>}500$ mV/nm is obtained by reversing the direction of the cyan arrow in the $\boldsymbol{K}$ valley phase diagrams shown in Fig.~\ref{fig: schematic_phase_diagram}. Microscopically, this behavior originates from the fact that the $\mathcal{N}{=}2$ LL for $6{<}\nu{<}8$ is predominantly composed of the $n{=}0$ orbital, which favors Jain states. We note, however, that the transport experiment~\cite{Kumar25} has not yet observed signatures of FQH states in the range $6{<}\nu{<}8$. For $\nu{=}1/3$, there could be a transition from the Laughlin state~\cite{Laughlin83} to a $\mathbb{Z}_{n}$ superconductor of composite bosons, which is described by the $n\bar{n}1^{3}$ parton state~\cite{Balram20, Faugno21, Balram21b}.

The STM experiments also see an FQH state at $\nu{=}5/11$ in the $\mathcal{N}{=}1$ LL of BLG (and also its hole-conjugate at $\nu{=}6/11$). This state could potentially be an FQH state of partons~\cite{Balram24a} described by the wave function $\overline{\left(3{+}1/3\right)}\bar{2}111{\equiv}\overline{\left(10/3\right)}\bar{2}111$. Another candidate state for it is $321$~\cite{Balram18a}, but since its predecessors in the $n21$ parton sequence, namely, the $221$ state at $1/2$ and the $211$ state at $2/5$, are unlikely to be relevant in the $n{=}1$ LL~\cite{Wojs09, Bonderson12, Sharma22}, we believe that it is unlikely that $321$ underlies the FQH at $\nu{=}6/11$ in the $\mathcal{N}{=}1$ LL as the states in its vicinity are also the ones that arise in $n{=}1$ LL. This also suggests that the $\bar{6}11$ Jain state is unlikely to materialize here.

The interaction in the experiment is likely to be much more complex than how we modeled it. In particular, we ignored the effects of screening, for example, by gates, LL mixing, and disorder, which all play a role in precisely determining the phase boundaries. Nevertheless, surprisingly, we find very good qualitative and even semi-quantitative agreement with experiments. We have also not considered the lattice-scale anisotropies, which would have to be incorporated to determine the magnetic and lattice ordering of the FQH phase~\cite{Kharitonov12a, An24a}.

Microscopically slightly different wave functions for many of the phases we considered can be obtained from the successive condensation of vortices or Abelian anyons in Laughlin and Jain states~\cite{Yutushui25} (see App.~\ref{app: anyon_condensation_parton_many_body_wfns}). However, we have not considered these states here since their wave functions are not readily amenable to very large-scale numerical evaluations, unlike those of the parton states.

We close this section by noting that FQH-like states, referred to as fractional quantum anomalous Hall (FQAH) states, have been observed in transition metal dichalcogenides~\cite{FQAH_MoTe2_Xu_2023a, FQAH_MoTe2_Xu_2023b, FQAH_MoTe2_Mak_Shan_2023, FQAHE_MoTe2_Li_2023, Zhao24_MoS2_FQH} and pentalayer graphene~\cite{FQAH_Pentalayer_Graphene_Ju_2024} on hexagonal boron nitride moir\'e heterostructures in the absence of an external magnetic field. In these systems, phase transitions from an FQAH state to other states, such as Fermi liquid or correlated insulator states, have also been observed upon varying the displacement field~\cite{Hadjri26}. Moreover, these systems also realize the zero-field analog of the CFFL state~\cite{FQAH_Pentalayer_Graphene_Ju_2024}, which can also phase transition to an electron Fermi liquid, a correlated insulator, or an extended quantum anomalous Hall state~\cite{Lu25}. The ideas we presented could also be applied to these systems.

\begin{acknowledgments}
We acknowledge useful discussions with Maissam Barkeshli, David Mross, and Dung Xuan Nguyen. Computational portions of this work were undertaken on the Nandadevi and Kamet supercomputers, maintained and supported by the Institute of Mathematical Sciences' High-Performance Computing Center. Some of the numerical calculations were performed using the DiagHam libraries~\cite{DiagHam}. U.K. acknowledges support from the Physical Research Laboratory, Ahmedabad, and the Department of Space (DOS), Government of India. The work was made possible by financial support from the Anusandhan National Research Foundation (ANRF) of the Department of Science and Technology (DST) via the Mathematical Research Impact Centric Support (MATRICS) Grant No. MTR/2023/000002 and the Advanced Research Grant No. ANRF/ARG/2025/000562/PS. 
\end{acknowledgments}

\appendix

\section{Chiral central charge of parton and Bonderson-Slingerland states}
\label{app: chiral_central_charge_parton_BS}
In this section, we discuss a way to infer the chiral central charge of the parton and Bonderson-Slingerland states. The chiral central charge of the parton states is obtained by considering the effective edge theory of their unprojected versions~\cite{Wen91, Wen95, Balram18a, Balram19}. Consider the $\bar{n}^{2}1^{3}$ parton state that occurs at filling factor $\nu{=}\left({-}1/n{-}1/n{+}1^{{-}1}{+}1^{{-}1}{+}1\right)^{{-}1}{=}n/(3n{-}2)$ and shift $\mathcal{S}{=}3{-}2n$. Let us work out the chiral central charge of this state, which is described by the gauge group $SU(2)_{{-}n}{\times}U(1)_{1{-}n}{\times}SU(3)_{1}$~\cite{Wen91, Balram19}. The chiral central charge is given by $c_{{-}}{=}c_{{-},{\rm MF}}{-}c_{{-},{\rm gauge}}$, where $c_{{-},{\rm MF}}$ and $c_{{-},{\rm gauge}}$ are the chiral central charges of the mean-field state and the gauge group, respectively~\cite{Wen91, Balram19}. The chiral central charge of the mean-field state is obtained by simply adding the chiral central charge of the constituent IQH states (keeping the appropriate signs). Let us first evaluate the gauge chiral central charge of the gauge group $SU(2)_{{-}n}$. To do so, we note that the net chiral central charge of the $SU(2)_{{-}n}$ gauge group is $c^{\bar{n}^2}_{-}{=}{-}n(2n{+}1)/(n{+}2)$~\cite{Wen91}. Therefore, the gauge chiral central charge is $c^{\bar{n}^2}_{{-},{\rm gauge}}{=}c^{\bar{n}^2}_{{-},{\rm MF}}{-}c^{\bar{n}^2}_{{-}}{=}-2n{+}n(2n{+}1)/(n{+}2){=}{-}3n/(n{+}2)$. Thus, the gauge chiral central charge of the gauge group $SU(2)_{{-}n}{\times}U(1)_{1{-}n}{\times}SU(3)_{1}$, for $n{\geq}2$, is given by $c^{\bar{n}^{2}1^{3}}_{-,{\rm gauge}}{=}{-}3n/(n{+}2){-}1{+}(3{-}1){=}{-}2(n{-}1)/(n{+}2)$. The gauge contribution of $SU(n)_{1}$ is $(n{-}1)$ since the $1^{n}$ Laughlin state has chiral central charge $c^{1^n}_{-}{=}1$ (as it supports a single chiral boson at its edge), while its mean-field chiral central charge $c^{1^n}_{{-},{\rm MF}}{=}n$, which gives its gauge-chiral central charge to be $c^{\bar{n}^2}_{{-},{\rm gauge}}{=}n{-}1$. Therefore, the chiral central charge of the $\bar{n}^{2}1^{3}$ parton state, for $n{\geq}2$, is $c^{\bar{n}^{2}1^{3}}_{{-}}{=}3{-}2n{-}[{-}2(n{-}1)/(n{+}2)]{=}(4{+}n{-}2n^{2})/(n{+}2)$. For $n{=}2$ and $n{=}3$, corresponding to the $\bar{2}^{2}1^{3}$ and $\bar{3}^{2}1^{3}$ parton states, we get chiral central charges $c^{\bar{2}^{2}1^{3}}_{{-}}{=}{-}1/2$ and $c^{\bar{3}^{2}1^{3}}_{{-}}{=}{-}11/5$, consistent with previous results~\cite{Balram18, Faugno20a}. 

Similarly, for the $\bar{n}\bar{2}1^{3}$ parton states, for $n{>}2$, the gauge group is $U(1)_{{-}2{-}n}{\times}U(1)_{1{-}2}{\times}SU(3)_{1}$~\cite{Wen91, Balram19}, which has $c^{\bar{n}\bar{2}1^{3}}_{-,{\rm gauge}}{=}{-}1{-}1{+}(3{-}1){=}0$. Thus, the chiral central charge of the $\bar{n}\bar{2}1^{3}$ parton states is just given by its mean-field value, i.e., $c^{\bar{n}\bar{2}1^{3}}_{{-}}{=}{-}n{-}{2}{+}3{=}1{-}n$. For $n{=}3$ and $n{=}4$, corresponding to the $\bar{3}\bar{2}1^{3}$ and $\bar{4}\bar{2}1^{3}$ parton states, we get chiral central charges $c^{\bar{3}\bar{2}1^{3}}_{{-}}{=}{-}2$ and $c^{\bar{4}\bar{2}1^{3}}_{{-}}{=}{-}3$, consistent with previous results~\cite{Balram18a, Balram20b}. 

Next, we turn to the chiral central charge of the BS state. Consider the unprojected 2/5 BS state, which is described by the wave function $\Psi^{\rm bosonic-MR}_{1/3}\bar{2}$. The mean-field chiral central charge is $c^{2/5~{\rm BS}}_{{-},{\rm MF}}{=}2{-}2{=}0$, where ${+}2$ is for the bosonic $\nu_{b}{=}1/3$ MR state [can be ascertained by viewing the MR state via a symmetrization of the Halperin-$(4,4,2)$ state~\cite{Halperin83}, which has two chiral bosons at its edge~\cite{Wen95, Wen99}] and ${-}2$ is for $\bar{2}$. The gauge chiral central charge $c^{2/5~{\rm BS}}_{-,{\rm gauge}}{=}1/2{-}1{=}{-}1/2$, where the $1/2$ is for the bosonic $\nu_{b}{=}1/3$ MR state [can be obtained by noting that the MR state has $c^{\rm MR}_{-}{=}3/2$ and $c^{\rm MR}_{{-},{\rm MF}}{=}2$, therefore it has $c^{\rm MR}_{{-}, {\rm gauge}}{=}1/2$, so that $c^{\rm MR}_{{-}}{=}c^{\rm MR}_{{-},{\rm MF}}{-}c^{\rm MR}_{{-},{\rm gauge}}$] and ${-}1$ comes from the $U(1)_{{-}1}$ gauge group of the $U(1)$ gauge field that glues the $1$ [of the MR state] and $\bar{2}$ partons together. Thus, the chiral central charge of the 2/5 BS state is $c^{2/5~{\rm BS}}_{{-}}{=}c^{2/5~{\rm BS}}_{{-},{\rm MF}}{-}c^{2/5~{\rm BS}}_{{-},{\rm gauge}}{=}0{-}({-}1/2){=}1/2$, consistent with the result of Ref.~\cite{Bishara08}. 

Similarly, consider the aPf $({\sim}\bar{2}^{2}1^{3})$ version of the unprojected BS state, that is described by the parton state $(\bar{2}^{2}1^{3})(\bar{2}1){\equiv}(\bar{2}^{2}1^{4})(\bar{2})$ (it is important to not combine the $\bar{2}$ factors to produce $\bar{2}^{3}1^{4}$, which, then results in the topologically distinct anti-RR$3$ state~\cite{Balram19, Bose25}). The mean-field chiral central charge is $c^{2/5~{\rm aPf-BS}}_{{-},{\rm MF}}{=}{-}6{+}4{=}{-}2$. The gauge group is $SU(2)_{-2}{\times}U(1)_{1{-}2}{\times}SU(4)_{1}{\times}U(1)_{1{-}2}$~\cite{Wen91, Balram19}, which has $c^{2/5~{\rm aPf-BS}}_{-,{\rm gauge}}{=}{-}3/2{-}1{+}(4{-}1){-}1{=}{-}1/2$. Thus, the chiral central charge of the aPf version of the BS state is $c^{2/5~{\rm aPf-BS}}_{{-}}{=}c^{2/5~{\rm aPf-BS}}_{{-},{\rm MF}}{-}c^{2/5~{\rm aPf-BS}}_{{-},{\rm gauge}}{=}{-}2{-}({-}1/2){=}{-}3/2$, consistent with the result of Ref.~\cite{Bishara08}.

\section{The leading coefficients in the static structure factor of the parton states}
\label{app: leading_sq_parton}
As stated in the main text, we require the leading coefficients, i.e., the expansion in the long-wavelength $q{\to}0$, of the unprojected static structure factor to fit its pair-correlation function accurately on the plane [see Sec.~\ref{ssec: FQH_states_energy_on_plane}]~\cite{Dora24}. For \emph{chiral} FQH states, Refs.~\cite{Gromov15, Nguyen17} conjectured these to be fixed by topological quantum numbers as
\begin{align}
\label{eq: sq_chiral_FQHE_conjecture}
    S(q) &= \frac{1}{2}q^{2} +  \left(\frac{\mathcal{S}-2}{8}\right)q^{4} -  \left(\frac{b}{8\nu}+\frac{\mathcal{S}-2}{16}\right)q^{6} + \cdots\nonumber\\
    &\equiv s_2 q^2 + s_{4}q^4 + s_{6}q^6+\cdots ,
\end{align}
where the topological quantum number $\nu$ is the filling factor, $\mathcal{S}$ is the Wen-Zee shift, $c_{-}$ is the chiral central charge, ${\rm var}(s)$ is the orbital spin variance and $b{=}\nu \mathcal{S}(2{-}\mathcal{S})/4{+}c_{-}/12{-}\nu {\rm var}(s)$. The parton states of our interest, owing to the $\bar{n}$ factors in them, are not chiral, so we cannot directly employ Eq.~\eqref{eq: sq_chiral_FQHE_conjecture} to read off the leading coefficients in their $S(q)$. However, for states that are the hole conjugates of fully chiral states, one can use the following result of Ref.~\cite{Nguyen17}
\begin{align}
\label{eq: projected_sq_hole_conjugate_fully_chiral_FQHE_conjecture}
   S^{\rm PH}(q) &= \frac{1}{2}q^{2}+\left(\frac{\nu\left(\mathcal{S}-1\right)}{8\left(1-\nu\right)}-\frac{1}{8}\right)q^{4}\nonumber\\
   &~~~+  \left(\frac{-6b+5\nu-3\nu\mathcal{S}}{48\left(1-\nu\right)}+\frac{1}{48}\right)q^{6} + \cdots\nonumber\\
   &\equiv s_{2}q^2 + s_{4}^{\rm PH} q^4 + s_{6}^{\rm PH} q^6 + \cdots,
\end{align}
where the quantum numbers $\nu$, $\mathcal{S}$, $c_{-}$, ${\rm var}(s)$, and $b$, correspond to those of the fully chiral state. Under the operation of particle-hole conjugation, these topological quantum numbers transform as~\cite{Nguyen17}
\begin{eqnarray}
    \label{eq: topological_numbers_PH_transformation}
    \nu^{\rm PH} &=& 1 - \nu, \nonumber \\ 
    \mathcal{S}^{\rm PH} &=& \frac{1 - \nu \mathcal{S}}{1 - \nu}, \nonumber \\ 
    c_{-}^{\rm PH} &=& 1 - c_{-}, \nonumber \\ 
    {\rm var}(s)^{\rm PH} &=& \frac{\nu}{1-\nu}\left(\frac{\left(1-\mathcal{S}\right)^2}{4\left(1-\nu\right)} + {\rm var}(s)\right). 
\end{eqnarray}
We conjecture that the hole-conjugates of all the parton states of our interest are chiral. We will test this conjecture and show numerical evidence in support of it. We then use Eq.~\eqref {eq: topological_numbers_PH_transformation} to find the topological quantum numbers of its conjugate state, which, we reiterate, we conjecture to be chiral, and then use Eq.~\eqref{eq: projected_sq_hole_conjugate_fully_chiral_FQHE_conjecture} to ascertain the leading coefficients of $s(q)$ in the parton states considered in this work. For the Abelian parton states, using their $K$-matrix, we can evaluate the topological quantities and set the values of the topological quantum numbers of their hole-conjugate to be consistent with Eq.~\eqref{eq: topological_numbers_PH_transformation}. For the non-Abelian states, namely $\bar{2}^{2}1^{3}$ and $\bar{3}^{2}1^{3}$, we can compute all topological quantities, except ${\rm var}(s)$. For the hole-conjugate of these two non-Abelian states, we conjecture that ${\rm var}(s){=}0$, and for the other topological quantum numbers, we will use the values that are consistent with Eq.~\eqref{eq: topological_numbers_PH_transformation}. 

As an example, let us first consider the $\bar{2}^{2}1^{3}$ parton state, which is topologically equivalent to the aPf~\cite{Balram18, Levin07, Lee07} and is not fully chiral. However, the aPf's hole conjugate, the MR state, is fully chiral. The leading long-wavelength behavior of the MR state's structure factor is determined solely from its topological quantum numbers, i.e., $\nu{=}1/2$, $\mathcal{S}{=}3$~\cite{Wen92}, $c_{-}{=}3/2$~\cite{Moore91}, ${\rm var}(s){=}0$, $b{=}{-}1/4$, and is given by~\cite{Nguyen17, Dwivedi19}
\begin{equation}
    \label{eq: structure_factor_expansion_Moore_Read}
    S^{\rm MR}_{\nu=1/2}(q)=\frac{1}{2}q^{2}+\frac{1}{8}q^{4}+ 0 q^{6}+ \mathcal{O}(q^{8}).
\end{equation}

To find an expansion for the static structure factor of the aPf or the $\bar{2}^{2}1^{3}$ parton state, we start with the MR and use its aforementioned topological quantum numbers and the fact that its hole-conjugate has the following leading coefficients for its static structure factor [see Eq.~\eqref{eq: projected_sq_hole_conjugate_fully_chiral_FQHE_conjecture}] to obtain
\begin{eqnarray}
    s^{\rm aPf, {\rm via-PH}}_{2}&=& \frac{1}{2},  \nonumber \\ 
   s^{\rm aPf, {\rm via-PH}}_{4}&=& \frac{\nu (\mathcal{S} - 1) }{8 (1 - \nu)}-\frac{1}{8} =\frac{1}{8} , \nonumber \\ 
    s^{\rm aPf, {\rm via-PH}}_{6}&=&\frac{-6 b + 5 \nu - 3 \nu \mathcal{S}}{48 (1 - \nu)} +\frac{1}{48}=0. 
\end{eqnarray}
These coefficients are the same as those of the $\nu{=}1/2$ MR state [see Eq.~\eqref{eq: structure_factor_expansion_Moore_Read}]. This also follows from the relation $\rho_{1{-}\nu}\bar{S}_{1{-}\nu}{=}\rho_{\nu}\bar{S}_{\nu}$~\cite{Balram15b, Nguyen17}, where $\rho_{\nu}{=}\nu/(2\pi\ell^2)$ is the density at $\nu$, for states related by hole-conjugation and by noting the fact that both the MR and aPf states have the same filling and density at $\nu{=}1/2$. Therefore, their projected and unprojected static structure factors are identical. 

To find an expansion for the static structure factor of the $\bar{3}^{2}1^{3}$ parton state at $\nu{=}3/7$, we conjecture that its hole-conjugate state is chiral and has the properties $\nu{=}4/7$, $\mathcal{S}{=}4$, $c_{-}{=}1{-}({-}11/5){=}16/5$, ${\rm var}(s){=}0$, $b{=}{-}92/105$ and use the fact that its hole-conjugate has the following leading coefficients for its static structure factor
\begin{eqnarray}
    s^{\bar{3}^{2}1^{3}, {\rm via-PH}}_{2}&=&\frac{1}{2}, \nonumber \\ 
    s^{\bar{3}^{2}1^{3}, {\rm via-PH}}_{4}&=& \frac{\nu (\mathcal{S} - 1) }{8 (1 - \nu)}-\frac{1}{8} =\frac{3}{8} ,\nonumber \\ 
   s^{\bar{3}^{2}1^{3}, {\rm via-PH}}_{6}&=&\frac{-6 b + 5 \nu - 3 \nu \mathcal{S}}{48 (1 - \nu)}+\frac{1}{48}=\frac{59}{720}. 
\end{eqnarray}
Aside from the conjecture on the hole-conjugate being chiral, there is an additional conjecture here that ${\rm var}(s){=}0$ for the chiral hole-conjugate state. The motivation for this comes from the fact that the hole-conjugate state at $4/7$ likely lends itself to a description in terms of a wave function that is a \emph{sum} of conformal blocks~\cite {Hermanns10, Sreejith11b, Hansson17}. When an FQH state is constructed from a \emph{single} conformal block in a conformal field theory, it is known to have ${\rm var}(s){=}0$~\cite{Gromov15a, Bradlyn15}. Note that our conjecture ${\rm var}(s){=}0$, which affects the $s_6$, does not alter the ground state energies significantly, which we will discuss later. The other topological properties are obtained by hole-conjugating the analogous properties of the $\bar{3}^{2}1^{3}$ state, which can be read off from its microscopic wave function or effective field theory.

The hole-conjugate state of the $\bar{3}\bar{2}1^{3}$ state has $\nu{=}1{-}6/13{=}7/13$, $\mathcal{S}{=}25/7$ [follows from $\mathcal{S}^{\bar{3}\bar{2}1^{3}}{=}{-}2$], $c_{-}{=}1{-}({-}2){=}3$,~${\rm var}(s){=}4/7$ [so that ${\rm var}(s){=}{-}5/2$ for the $\bar{3}\bar{2}1^{3}$, as can be ascertained from its Chern-Simons theory~\cite{Balram18a}]. Using these, we obtain the following leading coefficients for the static structure factor of the $\bar{3}\bar{2}1^{3}$ state
\begin{eqnarray}
    s^{\bar{3}\bar{2}1^{3}, {\rm via-PH}}_{2}&=&\frac{1}{2}, \nonumber \\ 
    s^{\bar{3}\bar{2}1^{3}, {\rm via-PH}}_{4}&=& \frac{\nu (\mathcal{S} - 1) }{8 (1 - \nu)}-\frac{1}{8}=\frac{1}{4}, \nonumber \\ 
   s^{\bar{3}\bar{2}1^{3}, {\rm via-PH}}_{6}&=&\frac{-6 b + 5 \nu - 3 \nu \mathcal{S}}{48 (1 - \nu)}+\frac{1}{48}=\frac{103}{1008}.   
\end{eqnarray}

Similarly, the hole-conjugate state of the $\bar{4}\bar{2}1^{3}$ state has $\nu{=}1{-}4/9{=}5/9$, $\mathcal{S}{=}21/5$ [follows from $\mathcal{S}^{\bar{4}\bar{2}1^{3}}{=}{-}3$], $c_{-}{=}1{-}({-}3){=}4$,~${\rm var}(s){=}23/10$ [so that ${\rm var}(s){=}{-}11/2$ for the $\bar{4}\bar{2}1^{3}$, as can be ascertained from its Chern-Simons theory~\cite{Balram20b}]. Using these, we obtain the following leading coefficients for the static structure factor of the $\bar{4}\bar{2}1^{3}$ state
\begin{eqnarray}
    s^{\rm PH,\bar{4}\bar{2}1^{3}, {\rm via-PH}}_{2}&=&\frac{1}{2} ,\nonumber \\ 
    s^{\rm PH,\bar{4}\bar{2}1^{3}, {\rm via-PH}}_{4}&=& \frac{\nu (\mathcal{S} - 1) }{8 (1 - \nu)}-\frac{1}{8} =\frac{3}{8} , \\ 
   s^{\rm PH,\bar{4}\bar{2}1^{3}, {\rm via-PH}}_{6}&=&\frac{-6 b + 5 \nu - 3 \nu \mathcal{S}}{48 (1 - \nu)}+\frac{1}{48}=\frac{863}{1920}. \nonumber
\end{eqnarray}

For the $\bar{3}\bar{2}1^{3}$~\cite{Balram18a} and $\bar{4}\bar{2}1^{3}$~\cite{Balram20b} states, there exists a null vector~\cite{Haldane95, Balram20, Dora25} $\vec{\Lambda}^{\bar{3}\bar{2}1^{3}}{=}\{0, 1, {-}1, -2\}^{T}$ and $\vec{\Lambda}^{\bar{3}\bar{2}1^{3}}{=}\{0, 0, 1, {-}1, -2\}^{T}$, respectively, such that $\vec{\Lambda}^{T}{\cdot}K{\cdot}\vec{\Lambda}{=}0$, where the $K$ matrices of these states are given in Refs.~\cite{Balram18a} and \cite{Balram20b}, respectively. The existence of null vectors indicates that, under generic perturbations, a pair of counter-propagating modes in these states can become gapped. Gapping out a pair of counter-propagating modes in the $\bar{3}\bar{2}1^{3}$ and $\bar{4}\bar{2}1^{3}$ states leaves the gapless edge modes in these states to be co-propagating, rendering the edge theory fully anti-chiral. Thus, the hole conjugates of these states are expected to be fully chiral, and to those we can apply the aforementioned results on the structure factor.

In the following section, we will validate the above conjecture for the small-$q$ expansion of $S(q)$ for non-chiral parton FQH states. We do so by computing the neutral excitation gaps via Girvin-MacDonald-Platzman (GMP) density wave states~\cite{Girvin85, Girvin86}, as these gaps, particularly in the long-wavelength limit, depend sensitively on the long-wavelength behavior of $S(q)$. 

\subsection{Full structure factor from pair-correlation function and Girvin-MacDonald-Platzman gap}
\label{ssec: full_structure_factor_pair_correlation_function}

\begin{figure}[tbh!]
\centering
\begin{tabular}{cc}
        \includegraphics[width=0.499\columnwidth]{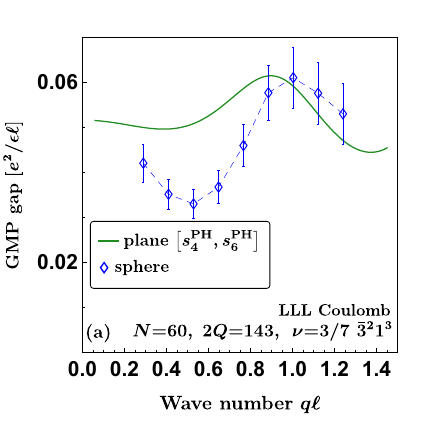}&
     \includegraphics[width=0.499\columnwidth]{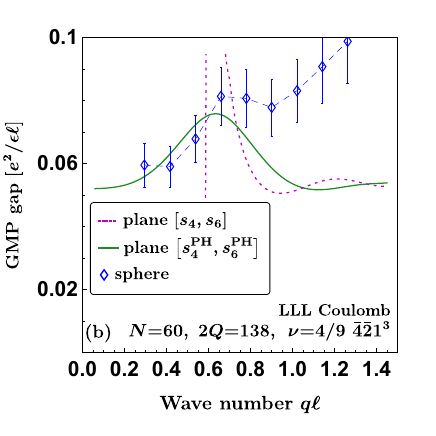}\\
     \includegraphics[width=0.499\columnwidth]{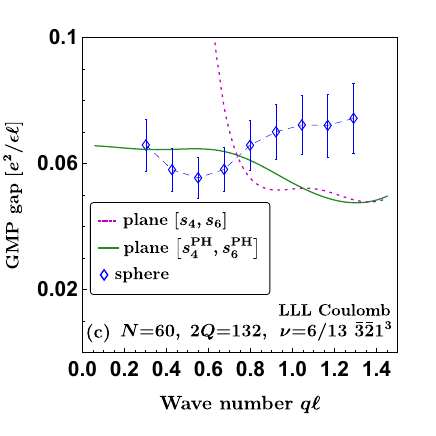}&
     \includegraphics[width=0.499\columnwidth]{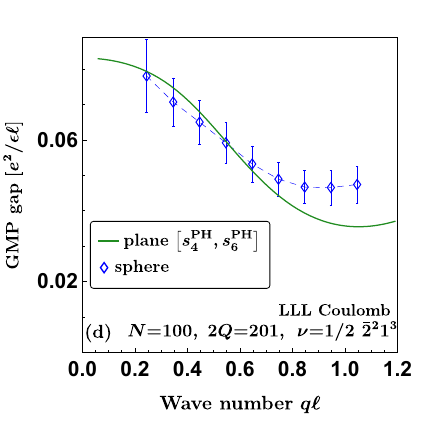}
         \end{tabular}
          \caption{The LLL Coulomb GMP gaps in planar and spherical geometries for the parton states $(a)$ $\bar{3}^{2}1^3$, $(b)$ $\bar{4}\bar{2}1^3$, $(c)$  $\bar{3}\bar{2}1^3$, and $(d)$ $\bar{2}^2 1^3$. The pink dotted line in panels $(b)$ and $(c)$ indicates the planar GMP gaps computed from the fitted pair-correlation functions constrained to reproduce the long-wavelength limit behavior of the structure factor given in Eq.~\eqref{eq: sq_chiral_FQHE_conjecture}. The planar GMP gaps shown by green solid lines are obtained from the fitted pair-correlation functions constrained to reproduce the long-wavelength limit behavior of the structure factor given in Eq.~\eqref{eq: g(r)_expansion_coeffs_constraint_PH_conjugate_states}. No GMP gaps are shown as pink dotted lines for the states in panels $(a)$ and $(d)$, because no analog of Eq.~\eqref{eq: sq_chiral_FQHE_conjecture} is available for these non-Abelian states. The pair-correlation function for $N{=}100$, $\bar{2}^2 1^3$ parton state is taken from Ref.~\cite{Balram18}.}
          \label{fig: GMP_gap_parton_states}
        \end{figure} 

\begin{figure}[tbh!]
\centering
\begin{tabular}{cc}
        \includegraphics[width=0.499\columnwidth]{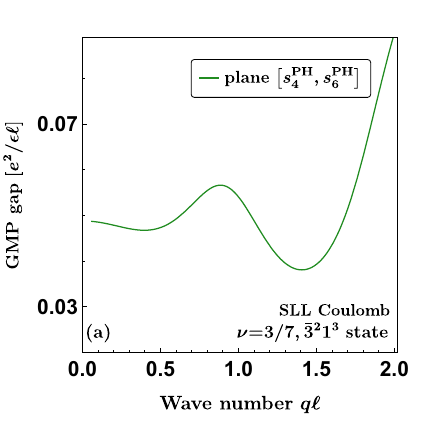}&
     \includegraphics[width=0.499\columnwidth]{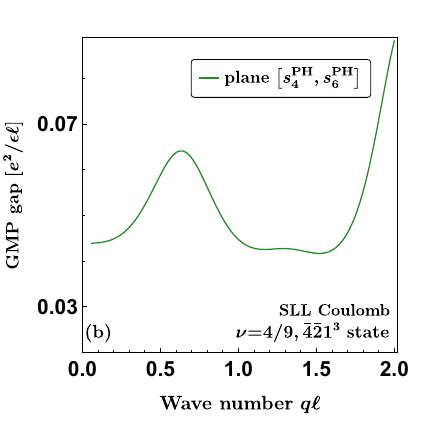}\\
     \includegraphics[width=0.499\columnwidth]{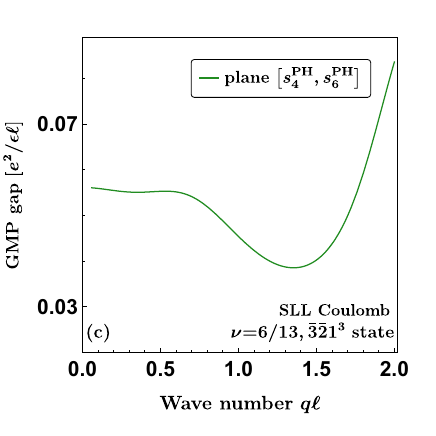}&
     \includegraphics[width=0.499\columnwidth]{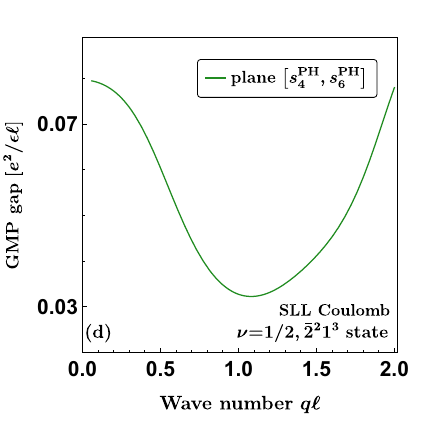}
         \end{tabular}
          \caption{Planar GMP gaps of $(a)$ $\bar{3}^{2}1^3$, $(b)$ $\bar{4}\bar{2}1^3$, $(c)$  $\bar{3}\bar{2}1^3$, and $(d)$ $\bar{2}^2 1^3$ parton states, for the SLL ($n{=}1$ LL) Coulomb interaction. We have used the same system sizes for the corresponding states as in Fig.~\ref{fig: GMP_gap_parton_states}. The pair-correlation function for $N{=}100$ $\bar{2}^2 1^3$ parton state is taken from Ref.~\cite{Balram18}.}
          \label{fig: SLL_Coulomb_GMP_gap_parton_states}
        \end{figure} 

To compute the GMP gap of a FQH state, one requires its full structure factor $S(q)$ at all $q$. As explained in the main text, $S(q)$ can be obtained from the pair-correlation function $g(r)$ [see \ref{ssec: FQH_states_energy_on_plane}]. For convenience, we repeat the relation between $g(r)$ and $S(q)$~\cite{Girvin84a}:
\begin{align}
\label{eq: g(r)_S(q)_relation_app}
S(q)&=1+(\nu n_\ell)\int d^{2}\boldsymbol{r} e^{-i\boldsymbol{q}.\boldsymbol{r}}\left[g(r)-1\right],\\
\label{eq: g(r)_Definition}
    g(r)&=1-e^{-r^2/2\ell^2}+\sum_{\mathfrak{m}=0}^{\infty}\frac{c_{2\mathfrak{m}+1}~e^{-r^2/4\ell^2}}{(2\mathfrak{m}+1)!}\left(\frac{r^2}{4\ell^2}\right)^{2\mathfrak{m}+1}.
\end{align}
This method of obtaining $S(q)$ from $g(r)$ has the advantage that the small-$q$ behavior of $S(q)$ can be controlled by imposing a set of constraints on the expansion coefficients $ c_m$. To ensure the small-$q$ expansion of $S(q)$ is consistent with the results presented in the previous section [see Appendix~\ref{app: leading_sq_parton}], we impose the following constraints on the expansion coefficients $c_{2\mathfrak{m}+1}$ [defined in Eq.~\eqref{eq: pair_correlation_expansion}]:
\begin{align}
 \sum_{\mathfrak{m}{=0}}^{\infty}(2\mathfrak{m}+2)(2\mathfrak{m}+3)(2\mathfrak{m}+4) c_{2\mathfrak{m}+1} &=\frac{9}{\nu}\left(s_{6}^{\rm PH}+\frac{\nu}{48}\right)\nonumber,\\
  \sum_{\mathfrak{m}{=0}}^{\infty}(2\mathfrak{m}+2)(2\mathfrak{m}+3) c_{2\mathfrak{m}+1} &=\frac{1}{\nu}\left(s_{4}^{\rm PH}+\frac{\nu}{8}\right)\nonumber,\\
    \sum_{\mathfrak{m}{=0}}^{\infty}(2\mathfrak{m}+2) c_{2\mathfrak{m}+1} &=  \frac{\nu-1}{4\nu}\nonumber,\\
    \sum_{\mathfrak{m}{=0}}^{\infty} c_{2\mathfrak{m}+1}&= \frac{\nu-1}{4\nu}.
    \label{eq: g(r)_expansion_coeffs_constraint_PH_conjugate_states}
\end{align}
We determine the unknown coefficients, $c_{2\mathfrak{m}+1}$, by fitting the numerically computed $g(r)$ to its analytic form given in Eq.~\eqref{eq: g(r)_Definition}, subject to the constraints in Eq.~\eqref{eq: g(r)_expansion_coeffs_constraint_PH_conjugate_states}. Subsequently, $S(q)$ can be determined from the analytically known expression for $g(r)$ using the relation presented in Eq.~\eqref{eq: g(r)_S(q)_relation_app}. 

Once the $S(q)$ of the non-chiral states is determined, their GMP gaps can be computed~\cite{Girvin85, Girvin86}. In Fig.~\ref{fig: GMP_gap_parton_states}, we present the GMP gaps of various non-chiral parton states for the LLL Coulomb interaction. The long-wavelength GMP gap computed via the above-discussed method using the planar structure factor $S(q)$ depends sensitively on accurate knowledge of the small-$q$ behavior of $S(q)$~\cite{Dora24}. This can be tested by comparing the planar GMP gap with that computed on the sphere~\cite{Dora24}. We find that for $\bar{2}^2 1^3$, $\bar{3}\bar{2}1^3$, $\bar{4}\bar{2}1^3$, the small-$q$ planar GMP gap agrees very accurately with the small-$q$ GMP gap computed on the sphere (using a complementary approach involving the algebra of density operators on the sphere that is detailed in Ref.~\cite{Dora24}). However, for the non-Abelian state, $\bar{3}^2 1^3$, we do not find a very good agreement between the planar and spherical gaps, presumably because either our conjecture on its hole-conjugate having ${\rm var}(s){=}0$ is false, or it has a large correlation length such that larger systems are required to bring the planar and spherical gaps into agreement. 

As a further support for the form of the small-$q$ expansion presented in Eq.~\eqref{eq: projected_sq_hole_conjugate_fully_chiral_FQHE_conjecture} for non-chiral states, we compute the GMP gaps of the non-chiral Abelian states $\bar{3}\bar{2}1^3$ and $\bar{4}\bar{2}1^3$, using the $S(q)$ with the small-$q$ expansion presented in Eq.~\eqref{eq: sq_chiral_FQHE_conjecture}. We find that the long-wavelength planar GMP gaps computed in this way differ significantly from the corresponding spherical GMP gaps [see pink dotted lines in Fig.~\ref{fig: GMP_gap_parton_states}($b$-$c$)]. This is expected because  Eq.~\eqref{eq: sq_chiral_FQHE_conjecture} is valid only for chiral states, whereas $\bar{3}\bar{2}1^3$, $\bar{4}\bar{2}1^3$ states are non-chiral. 

For completeness, in Fig.~\ref{fig: SLL_Coulomb_GMP_gap_parton_states}, we present the GMP gaps of non-chiral parton states for the SLL Coulomb interaction, as these states are good candidate states in the SLL~\cite{Balram18, Balram18a, Balram20b, Faugno20a, Balram21}. We have omitted the spherical GMP gaps because they have large error bars for the SLL Coulomb interaction.

We note that the GMP graviton---long-wavelength limit GMP state---is expected to split into two gravitons in the $\bar{3}\bar{2}1^{3}$ and the $\bar{4}\bar{2}1^{3}$ parton states~\cite{Balram21d, Balram24, Bose25}, while it remains as a single graviton in the $\bar{2}^{2} 1^{3}$ and $\bar{3}^{2} 1^{3}$ states.

Finally, we close this section by discussing the dependence of the ground state energy on the small-$q$ behavior of $S(q)$. We find that the per-particle ground state energy, which is the main quantity of interest in the main text, is not very sensitive to the precise values of these leading coefficients. For example, the fits with and without including the coefficients $s_4$ and $s_6$ yield nearly the same per-particle energies. This is because the small-$q$ terms in $S(q)$ determine the long-distance physics, while the dominant contribution to the energy comes from the short-range part. However, other quantities, especially order-$1$ numbers arising from differences of extensive energies that determine gaps (such as the GMP gap discussed above), are very sensitive to these leading coefficients~\cite{Dora24, Kundu26a}. As a consistency check, the thermodynamic extrapolation of finite-system ground state energy obtained using the sphere and disk Coulomb PPs in the spherical geometry is in close agreement with the Coulomb energy obtained from the fitted planar pair-correlation function, both with or without including the coefficients $s_4$ and $s_6$.

\section{Energies of non-uniform states and excitation gaps on the sphere}
\label{app: energy_gap_excitations}
In this section, we compute the energy of excited states (ESs), $|\Psi^{\rm ES}\rangle$, from their structure factors on the sphere, which subsequently allows us to determine the energy gaps to these excitations. Since the excited states are not uniform, the standard definition of the structure factor, $\langle \Psi^{\rm ES}| \rho_{L, M}^{\dagger} \rho_{L, M}|\Psi^{\rm ES}\rangle$, depends on the $M$ quantum number for a given $L$, where the unprojected density operator $\rho_{L, M}$ is defined as
\begin{align}
    \rho_{L,M}&= \sum_{i{=}1}^{N} Y_{L,M}(\boldsymbol{\Omega}_i).
\end{align}
Here, $Y_{L,M}$ are the spherical harmonics, and $\boldsymbol{\Omega}{=}(\theta,\phi)$ denotes the angular position on a unit sphere. For rotationally invariant interactions, it is useful to define a rotationally invariant version of the structure factor as follows
\begin{align}
    \label{eq: SL_ES}
    S^{\rm ES}(L)&=\frac{4\pi}{N}\frac{1}{2L+1}\left\langle\sum_{M{=}-L}^{L}\bigg|\sum_{i=1}^{N}Y_{L,M}(\boldsymbol{\Omega}_i)\bigg|^2\right\rangle_{\Psi^{\rm ES}}\nonumber\\
    &=\frac{1}{N}\left\langle\sum_{i,j{=}1}^{N}P_{L}\left(\cos\left(|\boldsymbol{\Omega}_i-\boldsymbol{\Omega}_j\right)|\right)\right\rangle_{\Psi^{\rm ES}},
\end{align}
where $P_{L}(x)$ is the Legendre polynomial, and $|\boldsymbol{\Omega}_i-\boldsymbol{\Omega}_j|{=}\cos(\theta_i)\cos(\theta_j)+\sin(\theta_i)\sin(\theta_j)\cos(\phi_i-\phi_j)$ is the arc distance between particles $i$ and $j$ on the unit sphere. The energy of $|\Psi^{\rm ES}\rangle$ for a rotationally invariant interaction Hamiltonian $H$ can be written in terms of $S^{\rm ES}(L)$ as
\begin{align}
    E^{\rm ES}&= \left\langle H \right\rangle_{\Psi^{\rm ES}}=\frac{N}{2}\sum_{L=0}^{2Q}v_{L}~ (2L+1)\left[ S^{\rm ES}(L)-1\right],
\end{align}
where $v_L$ are the expansion coefficients in the spherical harmonics decomposition of the rotationally invariant interaction potential. For the Coulomb interaction in the $\mathcal{N}{=}2$ LL of BLG, $v_L{=}\bar{v}_{L}(\theta_2,\phi_2)$ [see Eq.~\eqref{eq: effective_Coulomb_interaction_BLG2_LL_sphere}]. 

The energy gap $\Delta$ of $|\Psi^{\rm ES}\rangle$ can be obtained by subtracting the ground state (GS) energy $E^{\rm GS}$ from $E^{\rm ES}$. In Eq.~\eqref{eq: per_particle_energy_on_the_sphere} of the main text, we presented the expression of the GS energy in terms of its projected structure factor. The GS energy can alternatively be expressed in terms of the unprojected structure factor $S^{\rm GS}$ as
\begin{align}
     E^{\rm GS}&= \left\langle H \right\rangle_{\Psi^{\rm GS}}=\frac{N}{2}\sum_{L=0}^{2Q}v_{L}~ (2L+1)\left[ S^{\rm GS}(L)-1\right],
\end{align}
where $S^{\rm GS}(L)$, owing to the rotational invariance of the GS, is given by~\cite{Kamilla97, Balram17}
\begin{align}
\label{eq: SL_GS}
    S^{\rm GS}(L)&=\frac{4\pi}{N}\left\langle\bigg|\sum_{i=1}^{N}Y_{L,M}(\boldsymbol{\Omega}_i)\bigg|^2\right\rangle_{\Psi^{\rm GS}}.
\end{align}
For the GS, one can show that Eqs.~\eqref{eq: SL_GS} and~\eqref{eq: SL_ES} produce identical results. However, note that the computation in Eq.~\eqref{eq: SL_GS} involves only a single loop of order-$N$, while that in Eq.~\eqref{eq: SL_ES} involves running two loops, making the computation of order-$N^{2}$. Therefore, for GSs, it is preferable to use the expression given in Eq.~\eqref{eq: SL_GS}.

Next, $\Delta$ can be computed as
\begin{align}
\label{eq: energy_gap}
   \Delta=\frac{N}{2}\sum_{L=0}^{2Q}v_{L}~ (2L+1)\left[ S^{\rm ES}(L)-S^{\rm GS}(L)\right].
\end{align}
We use Eq.~\eqref{eq: energy_gap} to compute the excitation gaps of the primary Jain states at $\nu{=}s/(2s\pm1)$ for the Coulomb interaction. Specifically, we compute the transport gap in the spin-conserving channel by considering the CF exciton (CFE) state at the largest angular momentum, $L{=}L_{\rm max}=N/s{+}s{-}1$~\cite{Balram16d}. At this angular momentum, the constituent CF particle (CFP) and CF hole (CFH) are farthest apart, with the CFP going to the drain and the CFH going to the source~\cite{Jain97b, Zhao22a}. We therefore use the structure factor of the CFE state at $L{=}L_{\rm max}$ and that of the CF ground state as inputs to Eq.~\eqref{eq: energy_gap} to compute the transport gap $\Delta^{\rm CFE}_{L_{\rm max}}$. Here, $M$ in the CFE state can be fixed to $L_{\rm max}$, since the gap is independent of $M$ due to the rotational invariance of $H$. 

We correct for the finite-size effects of the gap on the sphere by adding to  $\Delta^{\rm CFE}_{L_{\rm max}}$, the interaction energy between far-separated CFP and CFH, $V^{\rm CFP-CFH}{=}(e^{*})^{2}/(2\epsilon R){=}\left(1/\left[\left(2s{\pm}1\right)^{2}2\sqrt{Q}\right]\right) e^2/(\epsilon\ell)$, where $2R$ is the diameter of the sphere~\cite{Jain07, Zhao22a}. Thus, we compute $\Delta^{\rm transport}{=}\Delta^{\rm CFE}_{L_{\rm max}}{+V^{\rm CFP-CFH}}$, density-correct~\cite{Morf86b} it, by multiplying it by a factor of $\sqrt{2Q\nu/N}$, and then extrapolate the gap, $\sqrt{2Q\nu/N}\Delta^{\rm transport}$, to the thermodynamic limit as a linear function of $1/N$.

\begin{figure}[tbh!]
\centering
\begin{tabular}{cc}
        \includegraphics[width=0.499\columnwidth]{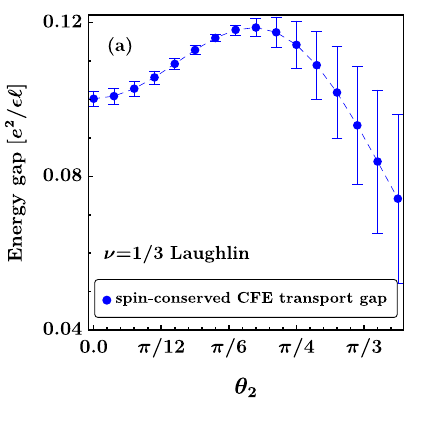}&
     \includegraphics[width=0.499\columnwidth]{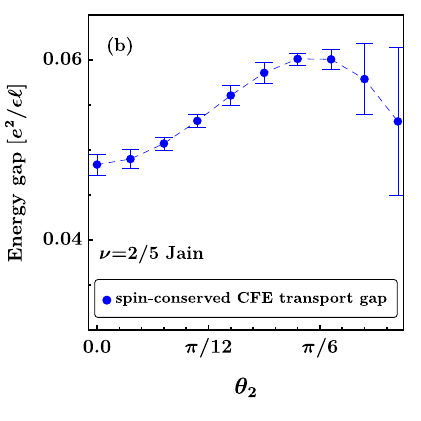}\\
     \includegraphics[width=0.499\columnwidth]{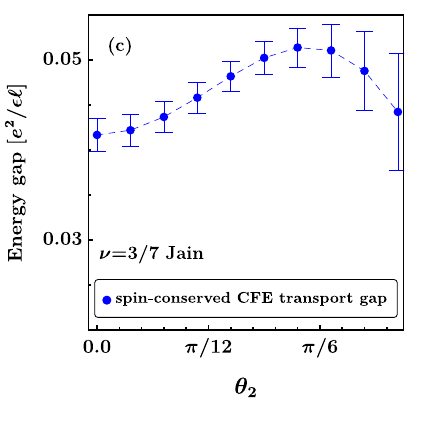}&
     \includegraphics[width=0.499\columnwidth]{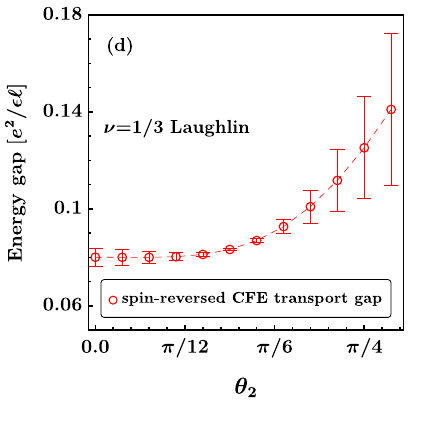}
         \end{tabular}
          \caption{The thermodynamic-limit CF exciton transport gaps for the Coulomb interaction in the $\mathcal{N}{=}1$ zeroth LL of bilayer graphene. Panels $(a)$-$(c)$ show the spin-conserved transport gaps in the $\nu{=}1/3$ Laughlin, $\nu{=}2/5$, and $\nu{=}3/7$ Jain states. The thermodynamic-limit transport gaps are obtained by extrapolating the gaps computed for system sizes ranging from $N{\sim}14$ to $50$ electrons for the $1/3$ Laughlin, from $N{\sim}14$ to $40$ electrons for the $2/5$ Jain, and from $N{\sim}12$ to $42$ electrons for the $3/7$ Jain states on the sphere. Panel $(d)$ shows the spin-reversed CF exciton transport gap in the Laughlin $1/3$ state, obtained by extrapolating the gaps for system sizes ranging from $N{\sim}14$ to $50$ electrons on the sphere.}
          \label{fig: CFE_transport_gaps}
        \end{figure}

In Figs.~\ref{fig: CFE_transport_gaps}$(a)$-$(c)$, we present the spin-conserving transport gaps for the $1/3$ Laughlin, $2/5$ and $3/7$ Jain states in the $\mathcal{N}{=}2$ LL of BLG. We have set $\phi_2{=}0$, resulting in $\mathcal{N}{=}2$ LL containing only the $n{=}0$ and $n{=}1$ LLs, which is equivalent to the $\mathcal{N}{=}1$ ZLL of BLG. In other regions of the parameter space, when a finite component of the $n{=}2$ LL is present, we find that the error bars in the gaps are too high. We note that recently, in Ref.~\cite{Borici26}, the transport gap for the $1/3$ Laughlin state in a cavity was ascertained using this method.

In Fig.~\ref{fig: CFE_transport_gaps}$(d)$, we also present the spin-reversed transport gaps in the $\mathcal{N}{=}1$ ZLL of BLG for the $1/3$ Laughlin state, where we have considered a spin-reversed CFE state at $L_{\rm max}{=}N{-}1$. The analogous computation at $\nu{=}1$ was recently carried out in Ref.~\cite{Kundu26a}.

\section{The $6/13$ parton state in a wide quantum well}
\label{app: 6_13_parton_WQW}
\begin{figure*}[tbh!]
        \includegraphics[width=0.66\columnwidth]{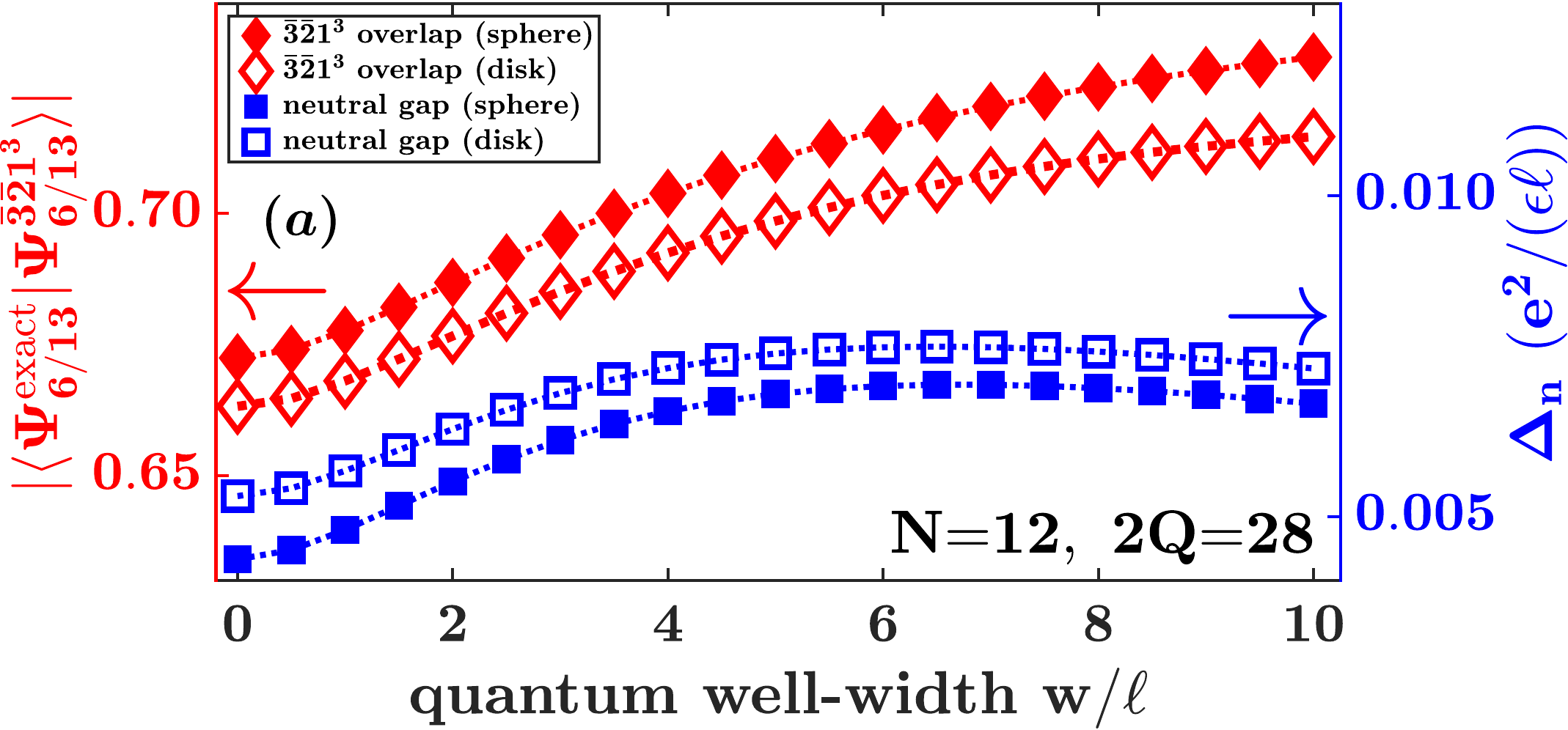}
        \includegraphics[width=0.66\columnwidth]{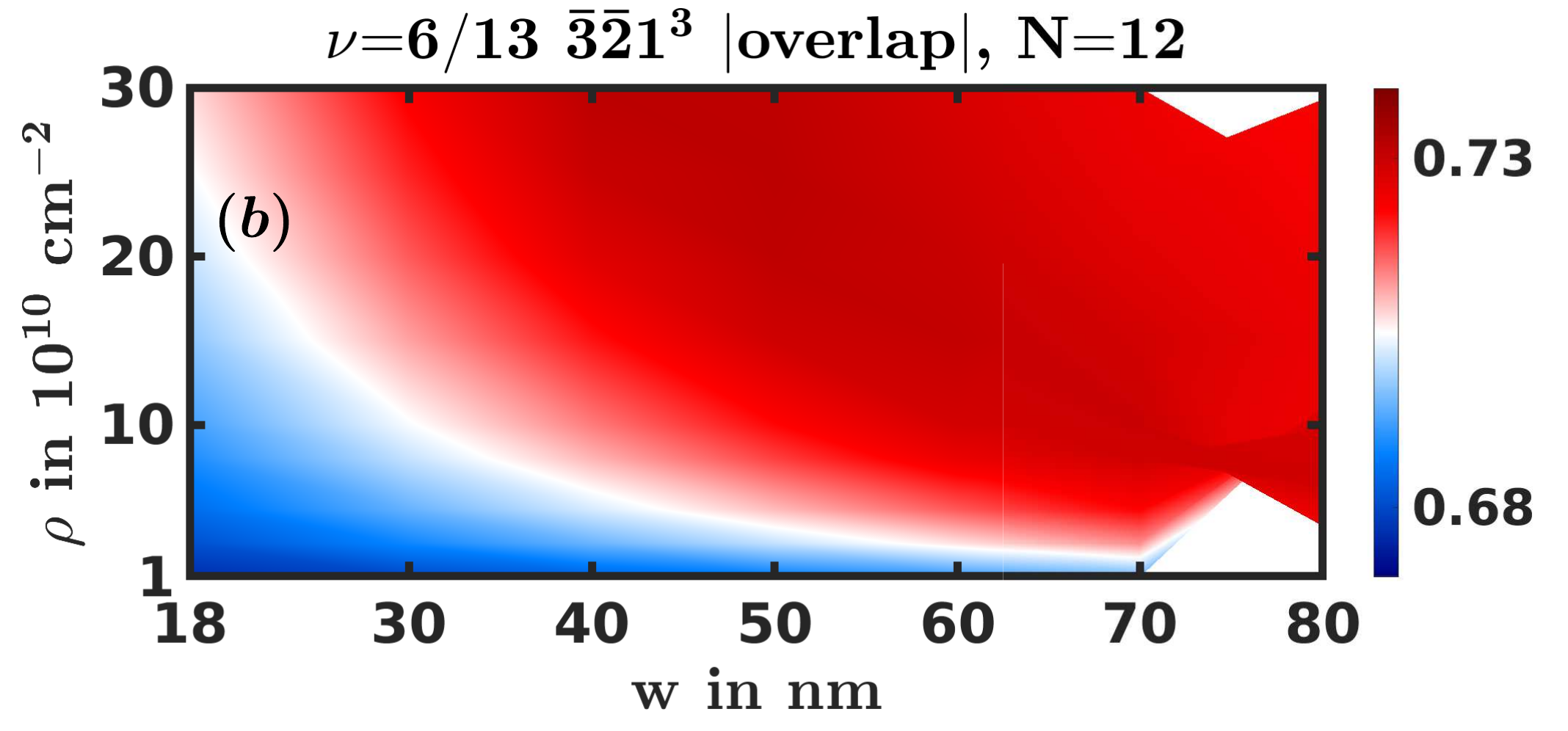}
        \includegraphics[width=0.66\columnwidth]{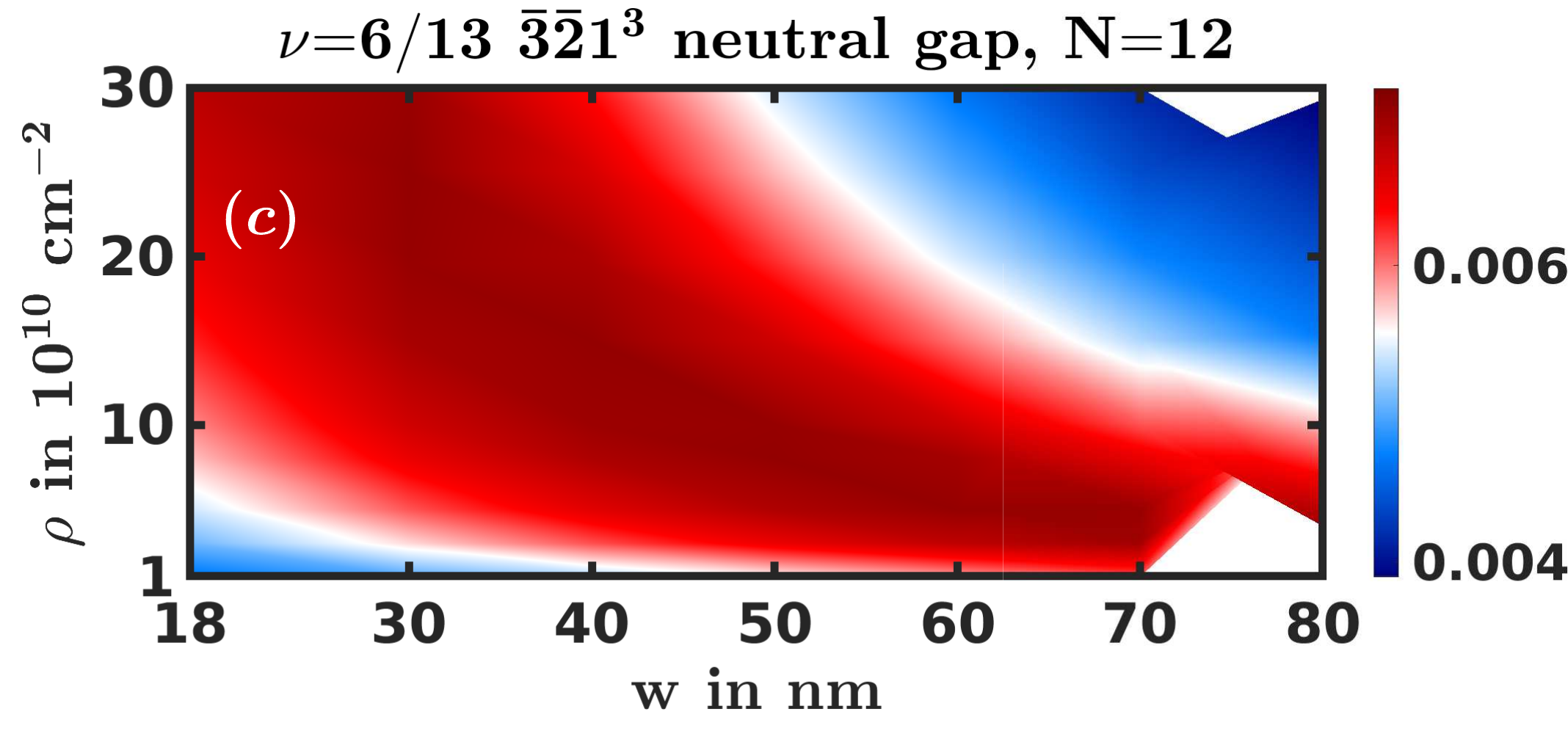}
      \caption{Overlap of the $\bar{3}\bar{2}1^{3}$ parton state for $N{=}12$ electrons on the sphere with the exact Coulomb ground state in the lowest Landau level for a finite-width model, modeled by (a) a cosine square well potential and (b) a local-density approximation which also accounts for density. The neutral gap for the cosine potential is shown in panel (a) itself, and for the local-density approximation, the interaction is shown in panel (c).}
          \label{fig: 6_13_wide_quantum_well}
        \end{figure*}
For a $72.5$ nm wide quantum well with density $1.4{\times}10^{11}$ cm$^{-2}$~\cite{Singh23, Singh25}, FQH at the parent MR fraction of $1/2$ and its two daughters at $7/13$ and $8/17$ in the LLL have been observed. By hole-conjugation, the state at $7/13$ maps to $6/13$, and in models with only two-body interactions that we have worked in throughout, these states are considered on the same level. A simplified model for a finite-width quantum well is to approximate it as a square-well potential. Results for the cosine square well model in the LLL for the $1/2$ MR state [see Fig. S8(b) of Ref.~\cite{Balram20b}] show that as the well-width is increased, the overlap of the exact Coulomb ground state with the MR state increases. The CF superconducting/pairing instability at $1/2$ arises from the softening of the short-range part of the interaction as the well-width is increased~\cite{Sharma23}. Similarly, we find that the overlap of the $\bar{3}\bar{2}1^{3}$ parton state at $\nu{=}6/13$ (and, to a certain extent, its neutral gap) goes up with increasing well-width [see Fig.~\ref{fig: 6_13_wide_quantum_well}].

A more accurate model for the interaction in a finite-width quantum well is a local-density approximation (LDA)~\cite{Ortalano97, Martin20}. For the MR state, it has been shown previously that its overlap with the exact ground state of the LDA interaction increases as the width and density are increased (see Fig. 36 of Ref.~\cite{Zhao21}). Furthermore, the Fermi liquid of CFs undergoes a $p$-wave pairing instability as the width and density are increased~\cite{Sharma23}. We find that the same is true for the $\bar{3}\bar{2}1^3$ state at $6/13$ [see Fig.~\ref{fig: 6_13_wide_quantum_well}]. 

As noted before, the other daughter of the MR state occurs at $8/17$~\cite{Levin09a}. Checking whether the ground state for $N{=}20$ at the $8/17$ daughter flux is uniform in this regime of large widths and densities is beyond the reach of the computational resources we have access to. The smaller system of $N{=}12$ does have an $L{=}0$ ground state, but that is not a true test since this state aliases with the $3/7$ Jain state (results for this are shown in Fig. 35 of Ref.~\cite{Zhao21}).

\section{Viewing anyon condensation from partonic many-body wave functions}
\label{app: anyon_condensation_parton_many_body_wfns}
In this appendix, we show how the complementary approach of obtaining certain Abelian and non-Abelian phases from the condensation of Abelian anyons~\cite{Yutushui25}, or vortex condensation, can be viewed through the lens of many-body wave functions that have been previously constructed for these states from the parton theory. 

One can go from the Abelian $2/3$ Jain state, i.e., $\bar{2}11$ in the parton language, to a parton state in the same topological phase as the non-Abelian aPf by multiplying by $\bar{2}1$, i.e., aPf${\sim}\bar{2}^{2} 1^3$~\cite{Balram18}, where ${\sim}$ denotes topological (and not microscopic) equivalence. Using the knowledge of the parton theory, this process can be viewed as the analog of going from $1/3$ described by $111$, which is the same as the Laughlin state~\cite{Laughlin83}, to $2/7$ at the first stage of the Haldane-Halperin hierarchy~\cite{Haldane83, Halperin83}, which in the Jain CF theory is $\bar{2}1111$, i.e., we take $111$ and multiply it by $\bar{2}1$ to arrive at $\bar{2}1111$. Therefore, in the parton framework, we can reinterpret this operation as replacing a ``$1$" by ``$\bar{2}11$." This process can now be iterated to go from a state in the aPf universality class to one in the non-Abelian aRR3 phase by another multiplication by $\bar{2}1$, i.e., aRR$3{\sim}\bar{2}^{3} 1^4$~\cite{Balram19}.
 
Similarly, one can obtain an Abelian state at 6/13 from the Abelian $3/5$ Jain state [$\bar{3}11$] by multiplying by $\bar{2}1$, i.e., the $\bar{3}\bar{2}111$ state at $6/13$~\cite{Balram18a}. This state is topologically equivalent to the Levin-Halperin daughter state of the aPf~\cite{Levin09a}. From here, one can further go to $3/8$ by another multiplication by $\bar{2}1$, i.e., the non-Abelian $3/8$ state described by the $\bar{3}\bar{2}^{2} 1^4$ wave function~\cite{Balram21}.

Thus, $\bar{n}\bar{2}1^{3}$ and $\bar{n}\bar{2}^{2}1^{4}$ are the two parton sequences that produce the above set of states, and except for $3/8$, states in the same universality class were recently obtained via an anyon-condensation approach~\cite{Yutushui25}. The $n{=}1,2,3$ states of these two parton sequences are known to describe the most prominent plateaus in the second Landau level of GaAs~\cite{Bose23}, and the states corresponding to $n{=}4$ are what one expects to observe when the sample quality improves~\cite{Balram20b, Bose23}.

One can extend this parton construction to obtain other states. For example, starting from the $\bar{3}\bar{2}111$ state at $6/13$ we can get the following three states:
\begin{itemize}
    \item the Abelian $\bar{4}\bar{2}111$ state~\cite{Balram20b} at $4/9$: obtained by replacing the $\bar{3}$ by $\bar{4}$ in $\bar{3}\bar{2}111$ which is reinterpreted as condensing the Abelian $2e/13$ charged quasiholes of $\bar{3}$ in $\bar{3}\bar{2}111$, 
    \item the non-Abelian $\bar{3}\bar{3}111$ state at $3/7$~\cite{Faugno20a}: obtained by replacing the $\bar{2}$ by $\bar{3}$ in $\bar{3}\bar{2}111$ which is reinterpreted as condensing the Abelian $3e/13$ charged quasiholes of $\bar{2}$ in $\bar{3}\bar{2}111$,
    \item the non-Abelian $\bar{3}\bar{2}^{2} 1^{4}$ state at $3/8$: obtained by replacing one of the ``$1$"s by $\bar{2}11$ in $\bar{3}\bar{2}111$ which is reinterpreted as condensing the Abelian $6e/13$ charged quasiholes of $1$ in $\bar{3}\bar{2}111$.
\end{itemize}
Recently, similar constructions have also been carried out using a Chern-Simons-Ginzburg-Landau approach to FQH hierarchies~\cite{Lee26}.

\section{Thermodynamic energies of the candidate states in various Landau levels}
\label{app: thermodynamic_energies_candidates_different_LLs}
In this appendix, we present the thermodynamic limit energies of candidate states at various fillings for certain specific $(\theta_{2},\phi_{2})$ points of the $\mathcal{N}{=}2$ LL of BLG [see Eq.~\eqref{eq: four_band_model_N_geq_two_eigenstates}] that correspond to certain LLs in GaAs and multilayer graphene. In Tables~\ref{tab: thermodynamic_limit_energies_n_0_LL}-\ref{tab: thermodynamic_limit_energies_n_2_LL}, we provide the energies of candidate states in the $n{=}0$, $n{=}1$, and $n{=}2$ LLs. Similarly, in Tables~\ref{tab: thermodynamic_limit_energies_MLG1_LL} and~\ref{tab: thermodynamic_limit_energies_MLG2_LL}, we present the energies of the candidate states in the first and second excited LLs of MLG, referred to as the MLG$1$ and MLG$2$ LLs, respectively. Finally, in Table~\ref{tab: thermodynamic_limit_energies_BLG2_LL}, we present the thermodynamic limit energies of candidate states in the $\mathcal{N}{=}2$ LL of ideal BLG, referred to as the BLG$2$ LL. 

The Coulomb energies of the $1/2$ MR and all of the $2/5,~3/7,~4/9$, and $6/13$ Jain states in the $n{=}0$ LL and $\mathcal{N}{=}1$ LL of MLG were computed previously via a direct Monte Carlo evaluation~\cite{Jain97b, Park98b, Balram15a, Balram15c, Balram17} and are consistent with the numbers given here.

\begin{table*}[t]
	\begin{center}
		\begin{tabular}{| c | c | c | c | c | c |}
			\hline
			$\nu$ & state &~ energy (sphere)~ & ~energy (disk PPs on sphere)~ & ~energy (plane)~ & ~cohesive energy (plane)~\\
			\hline\noalign{\vskip 1.5mm}\hline
\multirow{4}{*}{$2/5$} 
    & $2/5$ Jain & $-0.4327(1)$  & $-0.4329(1)$  & $-0.4321$ & $-0.1814$ \\
    
     & aRR3 & $-0.418(3)$ & $-0.417(1)$ & $-$ & $-$ \\
      
    &  BS   & $-0.4104(1)$  & $-0.4106(1)$  & $-0.4107$ & $-0.1601$ \\
  
    & stripe     & $-$  & $-$  & $-$ & $-0.1413$ \\
\hline\noalign{\vskip 1.1mm}\hline

\multirow{3}{*}{$3/7$} 
   & $3/7$ Jain  & $-0.4424(1)$  & $-0.4426(1)$ & $-0.4417$ & $-0.1732$ \\
   
   & $\bar{3}^2 1^{3}$ & $-0.4254(2)$ & $-0.4250(1)$ & $-0.4303$ &$-0.1617$  \\
    
   & stripe            & $-$ & $-$ & $-$ & $-0.1357$ \\
 \hline\noalign{\vskip 1.1mm}\hline
 
\multirow{3}{*}{$4/9$} 
   & $4/9$ Jain & $-0.4474(1)$ & $-0.4476(1)$ & $-0.4464$ & $-0.1679$ \\
   
   & $\bar{4}\bar{2} 1^{3}$ & $-0.4320(4)$ & $-0.4318(4)$ & $-0.4373$ & $-0.1588$ \\
   
  & stripe & $-$ & $-$ & $-$ & $-0.1323$ \\
  \hline\noalign{\vskip 1.1mm}\hline

\multirow{3}{*}{$6/13$} 
  & $6/13$ Jain & $-0.4529(3)$ & $-0.4530(3)$ & $-0.4500$ & $-0.1607$  \\
  
  & $\bar{3}\bar{2} 1^{3}$ & $-0.4369(5)$ & $-0.4366(5)$ & $-0.4421$ & $-0.1528$ \\
   
  & stripe & $-$ & $-$ & $-$ & $-0.1286$ \\
 \hline\noalign{\vskip 1.1mm}\hline
 
\multirow{3}{*}{$1/2$} 
  & CF Fermi liquid & $-0.4655(1)$ & $-0.4655(3)$ & $-$ & $-$ \\
   
  & MR & $-0.4572(1)$ & $-0.4575(1)$ & $-0.4566$     & $-0.1433$   \\
 
  & stripe & $-$ & $-$ & $-$ & $-0.1197$\\
\hline
		\end{tabular}
	\end{center}
	\caption{Thermodynamic limit energies of candidate states in the $n{=}0$ LL. The third, fourth, and fifth columns list the total energies computed on the sphere, using disk PPs on the sphere, and on the plane, respectively. The sixth column gives the cohesive energy in the planar geometry. The fitting error from the linear thermodynamic extrapolation of the sphere energies as a function of $1/N$ is indicated in the parentheses. The symbol ``$-$" indicates cases where the energy is unavailable for a given computational scheme, i.e., (i) stripe energies on the sphere, since stripe states are defined here only in the planar geometry, and (ii) CFFL and aRR3 energies on the plane, due to the difficulty in computing their pair-correlation functions for large system sizes.}
    \label{tab: thermodynamic_limit_energies_n_0_LL}
\end{table*}


\begin{table*}[t]
	\begin{center}
		\begin{tabular}{| c | c | c | c | c | c |}
			\hline
			$\nu$ & state &~ energy (sphere)~ & ~energy (disk PPs on sphere)~ & ~energy (plane)~ & ~cohesive energy (plane)~\\
			\hline\noalign{\vskip 1.5mm}\hline
\multirow{4}{*}{$2/5$} 
  & $2/5$ Jain & $-0.3361(8)$ & $-0.3363(8)$ & $-0.3366$ & $-0.1486$\\
  & $2/5$ aRR3 & $-0.3407(3)$ & $-0.3387(4)$ & $-$ & $-$ \\
  & $2/5$ BS & $-0.341(2)$ & $-0.341(2)$ & $-0.3406$ & $-0.1526$ \\
  & stripe & $-$ & $-$ & $-$ & $-0.1353$ \\
\hline\noalign{\vskip 1.1mm}\hline

\multirow{3}{*}{$3/7$} 
    & $3/7$ Jain & $-0.343(1)$ & $-0.343(1)$ & $-0.3423$ & $-0.1408$ \\
     
    & $\bar{3}^2 1^{3}$ & $-0.343(2)$ & $-0.342(2)$ & $-0.3493$ & $-0.1479$  \\
     
    & stripe & $-$ & $-$ & $-$ & $-0.1296$ \\
 \hline\noalign{\vskip 1.1mm}\hline
 
\multirow{3}{*}{$4/9$} 
   & $4/9$ Jain & $-0.344(2)$ & $-0.345(2)$ & $-0.3448$ & $-0.1360$ \\
    
   & $\bar{4}\bar{2} 1^{3}$ & $-0.35(1)$ & $-0.35(1)$ & $-0.3525$ & $-0.1436$  \\
   
   & stripe & $-$ & $-$ & $-$ & $-0.1263$ \\
  \hline\noalign{\vskip 1.1mm}\hline

\multirow{3}{*}{$6/13$} 
   & $6/13$ Jain & $-0.350(5)$ & $-0.350(4)$ & $-0.3469$ & $-0.1300$\\
    
   & $\bar{3}\bar{2} 1^{3}$ & $-0.33(2)$ & $-0.33(2)$ & $-0.3560$ & $-0.1390$ \\
    
   & stripe & $-$ & $-$ & $-$ & $-0.1226$ \\
 \hline\noalign{\vskip 1.1mm}\hline
 
\multirow{3}{*}{$1/2$} 
   & CF Fermi liquid & $-0.354(2)$ & $-0.354(2)$ & $-$ & $-$ \\
    
   & MR & $-0.3628(4)$ & $-0.3631(4)$ & $-0.3623$ &  $-0.1273$\\
    
   & stripe &$-$ & $-$ & $-$ & $-0.1140$ \\
\hline
		\end{tabular}
	\end{center}
	\caption{Same as Table~\ref{tab: thermodynamic_limit_energies_n_0_LL}, but for the $n{=}1$ LL.}
    \label{tab: thermodynamic_limit_energies_n_1_LL}
\end{table*}


\begin{table*}[t]
	\begin{center}
		\begin{tabular}{| c | c | c | c | c | c |}
			\hline
			$\nu$ & state &~ energy (sphere)~ & ~energy (disk PPs on sphere)~ & ~energy (plane)~ & ~cohesive energy (plane)~\\
			\hline\noalign{\vskip 1.5mm}\hline
\multirow{4}{*}{$2/5$} 
 & $2/5$ Jain & $-0.275(5)$ & $-0.273(5)$ & $-0.2747$ & $-0.1141$  \\
 & $2/5$ aRR3 & $-0.283(2)$ & $-0.2809(8)$ & $-$ & $-$ \\
 & $2/5$ BS & $-0.296(9)$ & $-0.293(8)$ & $-0.2859$ & $-0.1253$ \\
 & stripe & $-$ & $-$ & $-$ & $-0.1263$ \\
\hline\noalign{\vskip 1.1mm}\hline

\multirow{3}{*}{$3/7$} 
  & $3/7$ Jain & $-0.281(7)$ & $-0.279(7)$ & $-0.2802$ & $-0.1082$  \\
  & $\bar{3}^2 1^{3}$ & $-0.28(2)$ & $-0.28(1)$ & $-0.2857$ & $-0.1137$ \\
 & stripe & $-$ & $-$ & $-$ & $-0.1211$ \\
 \hline\noalign{\vskip 1.1mm}\hline
 
\multirow{3}{*}{$4/9$} 
 & $4/9$ Jain & $-0.273(9)$ & $-0.272(9)$ & $-0.2832$ & $-0.1048$ \\
 & $\bar{4}\bar{2} 1^{3}$ & $-0.32(6)$ & $-0.31(6)$ & $-0.2881$ & $-0.1097$ \\
 & stripe & $-$ & $-$ & $-$ & $-0.1181$ \\
  \hline\noalign{\vskip 1.1mm}\hline

\multirow{3}{*}{$6/13$} 
 & $6/13$ Jain & $-0.32(3)$ & $-0.31(3)$ & $-0.2847$ & $-0.0994$ \\
 & $\bar{3}\bar{2} 1^{3}$ & $-0.2(1)$ & $-0.2(1)$ & $-0.2928$ & $-0.1075$ \\
& stripe & $-$ & $-$ & $-$ & $-0.1147$ \\
 \hline\noalign{\vskip 1.1mm}\hline
 
\multirow{3}{*}{$1/2$} 
  & CF Fermi liquid & $-0.25(4)$ & $-0.26(3)$ & $-$ & $-$ \\
 & MR & $-0.300(5)$ & $-0.299(5)$ & $-0.3008$ & $-0.1001$  \\
& stripe & $-$ & $-$ & $-$ & $-0.1067$ \\
\hline
		\end{tabular}
	\end{center}
	\caption{Same as Table~\ref{tab: thermodynamic_limit_energies_n_0_LL}, but for the $n{=}2$ LL.}
    \label{tab: thermodynamic_limit_energies_n_2_LL}
\end{table*}


\begin{table*}[t]
	\begin{center}
		\begin{tabular}{| c | c | c | c | c | c |}
			\hline
			$\nu$ & state &~ energy (sphere)~ & ~energy (disk PPs on sphere)~ & ~energy (plane)~ & ~cohesive energy (plane)~\\
			\hline\noalign{\vskip 1.5mm}\hline
\multirow{4}{*}{$2/5$} 
& $2/5$ Jain & $-0.3786(1)$ & $-0.3788(2)$ & $-0.3779$ & $-0.2056$ \\
& $2/5$ aRR3 & $-0.361(3)$ & $-0.361(2)$ & $-$ & $-$ \\
& $2/5$ BS & $-0.3517(3)$ & $-0.3518(3)$ & $-0.3521$ & $-0.1797$ \\
& stripe & $-$ & $-$ & $-$ & $-0.1587$ \\
\hline\noalign{\vskip 1.1mm}\hline

\multirow{3}{*}{$3/7$} 
  & $3/7$ Jain & $-0.3817(2)$ & $-0.3818(2)$ & $-0.3809$ & $-0.1963$  \\
  & $\bar{3}^2 1^{3}$ & $-0.3614(4)$ & $-0.3612(4)$ & $-0.3674$ & $-0.1828$  \\
 & stripe & $-$ & $-$ & $-$ & $-0.1525$ \\
 \hline\noalign{\vskip 1.1mm}\hline
 
\multirow{3}{*}{$4/9$} 
 & $4/9$ Jain & $-0.3830(3)$ & $-0.3831(3)$ & $-0.3817$ & $-0.1902$ \\
 & $\bar{4}\bar{2} 1^{3}$ & $-0.366(2)$ & $-0.365(2)$ & $-0.3710$ & $-0.1795$  \\
 & stripe & $-$ & $-$ & $-$ & $-0.1488$ \\
  \hline\noalign{\vskip 1.1mm}\hline

\multirow{3}{*}{$6/13$} 
& $6/13$ Jain & $-0.3851(8)$ & $-0.3852(8)$ & $-0.3809$ & $-0.1821$ \\
& $\bar{3}\bar{2} 1^{3}$ & $-0.363(3)$ & $-0.363(3)$ & $-0.3715$ & $-0.1727$ \\
 & stripe & $-$ & $-$ & $-$ & $-0.1447$ \\
 \hline\noalign{\vskip 1.1mm}\hline
 
\multirow{3}{*}{$1/2$} 
 & CF Fermi liquid & $-0.3872(5)$ & $-0.3873(6)$ & $-$ & $-$ \\
 & MR & $-0.3780(1)$ & $-0.3782(1)$ & $-0.3773$ & $-0.1619$ \\
 & stripe & $-$ & $-$ & $-$ & $-0.1347$ \\
\hline
		\end{tabular}
	\end{center}
	\caption{Same as Table~\ref{tab: thermodynamic_limit_energies_n_0_LL}, but for the first excited LL of monolayer graphene.}
    \label{tab: thermodynamic_limit_energies_MLG1_LL}
\end{table*}


\begin{table*}[t]
	\begin{center}
		\begin{tabular}{| c | c | c | c | c | c |}
			\hline
			$\nu$ & state &~ energy (sphere)~ & ~energy (disk PPs on sphere)~ & ~energy (plane)~ & ~cohesive energy (plane)~\\
			\hline\noalign{\vskip 1.5mm}\hline
\multirow{4}{*}{$2/5$} 
 & $2/5$ Jain & $-0.2689(6)$ & $-0.2686(6)$ & $-0.2681$ & $-0.1261$ \\
 & $2/5$ aRR3 & $-0.286(2)$ & $-0.2843(9)$ & $-$ & $-$ \\
 & $2/5$ BS & $-0.2926(9)$ & $-0.2924(8)$ & $-0.2913$ & $-0.1493$ \\
& stripe & $-$ & $-$ & $-$ & $-0.1408$ \\
\hline\noalign{\vskip 1.1mm}\hline

\multirow{3}{*}{$3/7$} 
 & $3/7$ Jain & $-0.2711(7)$ & $-0.2708(7)$ & $-0.2708$ & $-0.1187$ \\
 & $\bar{3}^2 1^{3}$ & $-0.289(2)$ & $-0.288(2)$ & $-0.2916$ & $-0.1394$ \\
& stripe & $-$ & $-$ & $-$ & $-0.1355$ \\
 \hline\noalign{\vskip 1.1mm}\hline
 
\multirow{3}{*}{$4/9$} 
 & $4/9$ Jain & $-0.2712(9)$ & $-0.2711(9)$ & $-0.2718$ & $-0.1141$ \\
 & $\bar{4}\bar{2} 1^{3}$ & $-0.294(6)$ & $-0.293(6)$ & $-0.2918$ & $-0.1340$ \\
 & stripe & $-$ & $-$ & $-$ & $-0.1324$ \\
  \hline\noalign{\vskip 1.1mm}\hline

\multirow{3}{*}{$6/13$} 
 & $6/13$ Jain & $-0.278(4)$ & $-0.277(3)$ & $-0.2726$ & $-0.1088$ \\
& $\bar{3}\bar{2} 1^{3}$ & $-0.28(1)$ & $-0.28(1)$ & $-0.2944$ & $-0.1305$ \\
& stripe & $-$ & $-$ & $-$ & $-0.1288$ \\
 \hline\noalign{\vskip 1.1mm}\hline
 
\multirow{3}{*}{$1/2$} 
 & CF Fermi liquid & $-0.274(5)$ & $-0.275(5)$ & $-$ & $-$ \\
 & MR & $-0.2950(7)$ & $-0.2948(7)$ & $-0.2946$ & $-0.1171$ \\
& stripe & $-$ & $-$ & $-$ & $-0.1200$ \\
\hline
		\end{tabular}
	\end{center}
	\caption{Same as Table~\ref{tab: thermodynamic_limit_energies_n_0_LL}, but for the second excited LL  of monolayer graphene.}
     \label{tab: thermodynamic_limit_energies_MLG2_LL}
\end{table*}


\begin{table*}[t]
	\begin{center}
		\begin{tabular}{| c | c | c | c | c | c |}
			\hline
			$\nu$ & state &~ energy (sphere)~ & ~energy (disk PPs on sphere)~ & ~energy (plane)~ & ~cohesive energy (plane)~\\
			\hline\noalign{\vskip 1.5mm}\hline
\multirow{4}{*}{$2/5$} 
 & $2/5$ Jain & $-0.339(2)$ & $-0.339(1)$ & $-0.3386$ & $-0.1888$ \\
 & $2/5$ aRR3 & $-0.327(2)$ & $-0.327(1)$ & $-$ & $-$ \\
 & $2/5$ BS & $-0.323(2)$ & $-0.323(2)$ & $-0.3209$ & $-0.1711$ \\
 & stripe & $-$ & $-$ & $-$ & $-0.1490$ \\
\hline\noalign{\vskip 1.1mm}\hline

\multirow{3}{*}{$3/7$} 
 & $3/7$ Jain & $-0.341(2)$ & $-0.341(2)$ & $-0.3405$ & $-0.1800$ \\
 & $\bar{3}^2 1^{3}$ & $-0.325(4)$ & $-0.324(4)$ & $-0.3313$ & $-0.1707$ \\
 & stripe & $-$ & $-$ & $-$ & $-0.1428$ \\
 \hline\noalign{\vskip 1.1mm}\hline
 
\multirow{3}{*}{$4/9$} 
 & $4/9$ Jain & $-0.339(2)$ & $-0.339(2)$ & $-0.3409$ & $-0.1745$ \\
 & $\bar{4}\bar{2} 1^{3}$ & $-0.34(2)$ & $-0.34(2)$ & $-0.3337$ & $-0.1673$ \\
& stripe & $-$ & $-$ & $-$ & $-0.1393$ \\
  \hline\noalign{\vskip 1.1mm}\hline

\multirow{3}{*}{$6/13$} 
 & $6/13$ Jain & $-0.351(9)$ & $-0.350(8)$ & $-0.3396$ & $-0.1667$ \\
 & $\bar{3}\bar{2} 1^{3}$ & $-0.30(3)$ & $-0.30(3)$ & $-0.3342$ & $-0.1613$ \\
 & stripe & $-$ & $-$ & $-$ & $-0.1353$ \\
 \hline\noalign{\vskip 1.1mm}\hline
 
\multirow{3}{*}{$1/2$} 
& CF Fermi liquid & $-0.33(1)$ & $-0.336(9)$ & $-$ & $-$ \\
 & MR & $-0.339(1)$ & $-0.339(1)$ & $-0.3383$ & $-0.1511$ \\
 & stripe & $-$ & $-$ & $-$ & $-0.1259$ \\
\hline
		\end{tabular}
	\end{center}
	\caption{Same as Table~\ref{tab: thermodynamic_limit_energies_n_0_LL}, but for the first excited LL of ideal bilayer graphene.}
     \label{tab: thermodynamic_limit_energies_BLG2_LL}
\end{table*}

\bibliography{biblio_fqhe}

\begin{thebibliography}{167}%
\makeatletter
\providecommand \@ifxundefined [1]{%
 \@ifx{#1\undefined}
}%
\providecommand \@ifnum [1]{%
 \ifnum #1\expandafter \@firstoftwo
 \else \expandafter \@secondoftwo
 \fi
}%
\providecommand \@ifx [1]{%
 \ifx #1\expandafter \@firstoftwo
 \else \expandafter \@secondoftwo
 \fi
}%
\providecommand \natexlab [1]{#1}%
\providecommand \enquote  [1]{``#1''}%
\providecommand \bibnamefont  [1]{#1}%
\providecommand \bibfnamefont [1]{#1}%
\providecommand \citenamefont [1]{#1}%
\providecommand \href@noop [0]{\@secondoftwo}%
\providecommand \href [0]{\begingroup \@sanitize@url \@href}%
\providecommand \@href[1]{\@@startlink{#1}\@@href}%
\providecommand \@@href[1]{\endgroup#1\@@endlink}%
\providecommand \@sanitize@url [0]{\catcode `\\12\catcode `\$12\catcode `\&12\catcode `\#12\catcode `\^12\catcode `\_12\catcode `\%12\relax}%
\providecommand \@@startlink[1]{}%
\providecommand \@@endlink[0]{}%
\providecommand \url  [0]{\begingroup\@sanitize@url \@url }%
\providecommand \@url [1]{\endgroup\@href {#1}{\urlprefix }}%
\providecommand \urlprefix  [0]{URL }%
\providecommand \Eprint [0]{\href }%
\providecommand \doibase [0]{https://doi.org/}%
\providecommand \selectlanguage [0]{\@gobble}%
\providecommand \bibinfo  [0]{\@secondoftwo}%
\providecommand \bibfield  [0]{\@secondoftwo}%
\providecommand \translation [1]{[#1]}%
\providecommand \BibitemOpen [0]{}%
\providecommand \bibitemStop [0]{}%
\providecommand \bibitemNoStop [0]{.\EOS\space}%
\providecommand \EOS [0]{\spacefactor3000\relax}%
\providecommand \BibitemShut  [1]{\csname bibitem#1\endcsname}%
\let\auto@bib@innerbib\@empty
\bibitem [{\citenamefont {Tsui}\ \emph {et~al.}(1982)\citenamefont {Tsui}, \citenamefont {Stormer},\ and\ \citenamefont {Gossard}}]{Tsui82}%
  \BibitemOpen
  \bibfield  {author} {\bibinfo {author} {\bibfnamefont {D.~C.}\ \bibnamefont {Tsui}}, \bibinfo {author} {\bibfnamefont {H.~L.}\ \bibnamefont {Stormer}},\ and\ \bibinfo {author} {\bibfnamefont {A.~C.}\ \bibnamefont {Gossard}},\ }\bibfield  {title} {\bibinfo {title} {Two-dimensional magnetotransport in the extreme quantum limit},\ }\href {https://doi.org/10.1103/PhysRevLett.48.1559} {\bibfield  {journal} {\bibinfo  {journal} {Phys. Rev. Lett.}\ }\textbf {\bibinfo {volume} {48}},\ \bibinfo {pages} {1559} (\bibinfo {year} {1982})}\BibitemShut {NoStop}%
\bibitem [{\citenamefont {Novoselov}\ \emph {et~al.}(2004)\citenamefont {Novoselov}, \citenamefont {Geim}, \citenamefont {Morozov}, \citenamefont {Jiang}, \citenamefont {Zhang}, \citenamefont {Dubonos}, \citenamefont {Grigorieva},\ and\ \citenamefont {Firsov}}]{Novoselov04}%
  \BibitemOpen
  \bibfield  {author} {\bibinfo {author} {\bibfnamefont {K.~S.}\ \bibnamefont {Novoselov}}, \bibinfo {author} {\bibfnamefont {A.~K.}\ \bibnamefont {Geim}}, \bibinfo {author} {\bibfnamefont {S.~V.}\ \bibnamefont {Morozov}}, \bibinfo {author} {\bibfnamefont {D.}~\bibnamefont {Jiang}}, \bibinfo {author} {\bibfnamefont {Y.}~\bibnamefont {Zhang}}, \bibinfo {author} {\bibfnamefont {S.~V.}\ \bibnamefont {Dubonos}}, \bibinfo {author} {\bibfnamefont {I.~V.}\ \bibnamefont {Grigorieva}},\ and\ \bibinfo {author} {\bibfnamefont {A.~A.}\ \bibnamefont {Firsov}},\ }\bibfield  {title} {\bibinfo {title} {Electric field effect in atomically thin carbon films},\ }\href {https://doi.org/10.1126/science.1102896} {\bibfield  {journal} {\bibinfo  {journal} {Science}\ }\textbf {\bibinfo {volume} {306}},\ \bibinfo {pages} {666} (\bibinfo {year} {2004})}\BibitemShut {NoStop}%
\bibitem [{\citenamefont {Zhang}\ \emph {et~al.}(2005)\citenamefont {Zhang}, \citenamefont {Tan}, \citenamefont {Stormer},\ and\ \citenamefont {Kim}}]{Zhang05}%
  \BibitemOpen
  \bibfield  {author} {\bibinfo {author} {\bibfnamefont {Y.}~\bibnamefont {Zhang}}, \bibinfo {author} {\bibfnamefont {Y.-W.}\ \bibnamefont {Tan}}, \bibinfo {author} {\bibfnamefont {H.~L.}\ \bibnamefont {Stormer}},\ and\ \bibinfo {author} {\bibfnamefont {P.}~\bibnamefont {Kim}},\ }\bibfield  {title} {\bibinfo {title} {Experimental observation of the quantum {Hall} effect and {Berry}'s phase in graphene},\ }\href {https://doi.org/10.1038/nature04235} {\bibfield  {journal} {\bibinfo  {journal} {Nature}\ }\textbf {\bibinfo {volume} {438}},\ \bibinfo {pages} {201} (\bibinfo {year} {2005})}\BibitemShut {NoStop}%
\bibitem [{\citenamefont {Castro~Neto}\ \emph {et~al.}(2009)\citenamefont {Castro~Neto}, \citenamefont {Guinea}, \citenamefont {Peres}, \citenamefont {Novoselov},\ and\ \citenamefont {Geim}}]{Neto09}%
  \BibitemOpen
  \bibfield  {author} {\bibinfo {author} {\bibfnamefont {A.~H.}\ \bibnamefont {Castro~Neto}}, \bibinfo {author} {\bibfnamefont {F.}~\bibnamefont {Guinea}}, \bibinfo {author} {\bibfnamefont {N.~M.~R.}\ \bibnamefont {Peres}}, \bibinfo {author} {\bibfnamefont {K.~S.}\ \bibnamefont {Novoselov}},\ and\ \bibinfo {author} {\bibfnamefont {A.~K.}\ \bibnamefont {Geim}},\ }\bibfield  {title} {\bibinfo {title} {The electronic properties of graphene},\ }\href {https://doi.org/10.1103/RevModPhys.81.109} {\bibfield  {journal} {\bibinfo  {journal} {Rev. Mod. Phys.}\ }\textbf {\bibinfo {volume} {81}},\ \bibinfo {pages} {109} (\bibinfo {year} {2009})}\BibitemShut {NoStop}%
\bibitem [{\citenamefont {Aoki}\ and\ \citenamefont {Dresselhaus}(2013)}]{Aoki13}%
  \BibitemOpen
  \bibfield  {author} {\bibinfo {author} {\bibfnamefont {H.}~\bibnamefont {Aoki}}\ and\ \bibinfo {author} {\bibfnamefont {M.~S.}\ \bibnamefont {Dresselhaus}},\ }\href {https://doi.org/10.1007/978-3-319-02633-6} {\emph {\bibinfo {title} {Physics of Graphene}}}\ (\bibinfo  {publisher} {Springer International Publishing},\ \bibinfo {year} {2013})\BibitemShut {NoStop}%
\bibitem [{\citenamefont {Das~Sarma}\ and\ \citenamefont {Pinczuk}(2007)}]{DasSarma07}%
  \BibitemOpen
  \bibinfo {editor} {\bibfnamefont {S.}~\bibnamefont {Das~Sarma}}\ and\ \bibinfo {editor} {\bibfnamefont {A.}~\bibnamefont {Pinczuk}},\ eds.,\ \href@noop {} {\emph {\bibinfo {title} {Perspectives in Quantum {Hall} Effects}}}\ (\bibinfo  {publisher} {Wiley-VCH Verlag GmbH},\ \bibinfo {year} {2007})\BibitemShut {NoStop}%
\bibitem [{\citenamefont {Halperin}\ and\ \citenamefont {Jain}(2020)}]{Halperin20}%
  \BibitemOpen
  \bibinfo {editor} {\bibfnamefont {B.~I.}\ \bibnamefont {Halperin}}\ and\ \bibinfo {editor} {\bibfnamefont {J.~K.}\ \bibnamefont {Jain}},\ eds.,\ \href {https://doi.org/10.1142/11751} {\emph {\bibinfo {title} {{Fractional} {Quantum} {Hall} {Effects} {New} {Developments}}}}\ (\bibinfo  {publisher} {World Scientific},\ \bibinfo {year} {2020})\ \Eprint {https://arxiv.org/abs/https://worldscientific.com/doi/pdf/10.1142/11751} {https://worldscientific.com/doi/pdf/10.1142/11751} \BibitemShut {NoStop}%
\bibitem [{\citenamefont {Jain}(1989{\natexlab{a}})}]{Jain89}%
  \BibitemOpen
  \bibfield  {author} {\bibinfo {author} {\bibfnamefont {J.~K.}\ \bibnamefont {Jain}},\ }\bibfield  {title} {\bibinfo {title} {Composite-fermion approach for the fractional quantum {Hall} effect},\ }\href {https://doi.org/10.1103/PhysRevLett.63.199} {\bibfield  {journal} {\bibinfo  {journal} {Phys. Rev. Lett.}\ }\textbf {\bibinfo {volume} {63}},\ \bibinfo {pages} {199} (\bibinfo {year} {1989}{\natexlab{a}})}\BibitemShut {NoStop}%
\bibitem [{\citenamefont {Jain}(2007)}]{Jain07}%
  \BibitemOpen
  \bibfield  {author} {\bibinfo {author} {\bibfnamefont {J.~K.}\ \bibnamefont {Jain}},\ }\href@noop {} {\emph {\bibinfo {title} {Composite Fermions}}}\ (\bibinfo  {publisher} {Cambridge University Press, New York, US},\ \bibinfo {year} {2007})\BibitemShut {NoStop}%
\bibitem [{\citenamefont {Halperin}\ \emph {et~al.}(1993)\citenamefont {Halperin}, \citenamefont {Lee},\ and\ \citenamefont {Read}}]{Halperin93}%
  \BibitemOpen
  \bibfield  {author} {\bibinfo {author} {\bibfnamefont {B.~I.}\ \bibnamefont {Halperin}}, \bibinfo {author} {\bibfnamefont {P.~A.}\ \bibnamefont {Lee}},\ and\ \bibinfo {author} {\bibfnamefont {N.}~\bibnamefont {Read}},\ }\bibfield  {title} {\bibinfo {title} {Theory of the half-filled {Landau} level},\ }\href {https://doi.org/10.1103/PhysRevB.47.7312} {\bibfield  {journal} {\bibinfo  {journal} {Phys. Rev. B}\ }\textbf {\bibinfo {volume} {47}},\ \bibinfo {pages} {7312} (\bibinfo {year} {1993})}\BibitemShut {NoStop}%
\bibitem [{\citenamefont {Willett}\ \emph {et~al.}(1987)\citenamefont {Willett}, \citenamefont {Eisenstein}, \citenamefont {St\"ormer}, \citenamefont {Tsui}, \citenamefont {Gossard},\ and\ \citenamefont {English}}]{Willett87}%
  \BibitemOpen
  \bibfield  {author} {\bibinfo {author} {\bibfnamefont {R.}~\bibnamefont {Willett}}, \bibinfo {author} {\bibfnamefont {J.~P.}\ \bibnamefont {Eisenstein}}, \bibinfo {author} {\bibfnamefont {H.~L.}\ \bibnamefont {St\"ormer}}, \bibinfo {author} {\bibfnamefont {D.~C.}\ \bibnamefont {Tsui}}, \bibinfo {author} {\bibfnamefont {A.~C.}\ \bibnamefont {Gossard}},\ and\ \bibinfo {author} {\bibfnamefont {J.~H.}\ \bibnamefont {English}},\ }\bibfield  {title} {\bibinfo {title} {Observation of an even-denominator quantum number in the fractional quantum {Hall} effect},\ }\href {https://doi.org/10.1103/PhysRevLett.59.1776} {\bibfield  {journal} {\bibinfo  {journal} {Phys. Rev. Lett.}\ }\textbf {\bibinfo {volume} {59}},\ \bibinfo {pages} {1776} (\bibinfo {year} {1987})}\BibitemShut {NoStop}%
\bibitem [{\citenamefont {Pan}\ \emph {et~al.}(2008)\citenamefont {Pan}, \citenamefont {Xia}, \citenamefont {Stormer}, \citenamefont {Tsui}, \citenamefont {Vicente}, \citenamefont {Adams}, \citenamefont {Sullivan}, \citenamefont {Pfeiffer}, \citenamefont {Baldwin},\ and\ \citenamefont {West}}]{Pan08}%
  \BibitemOpen
  \bibfield  {author} {\bibinfo {author} {\bibfnamefont {W.}~\bibnamefont {Pan}}, \bibinfo {author} {\bibfnamefont {J.~S.}\ \bibnamefont {Xia}}, \bibinfo {author} {\bibfnamefont {H.~L.}\ \bibnamefont {Stormer}}, \bibinfo {author} {\bibfnamefont {D.~C.}\ \bibnamefont {Tsui}}, \bibinfo {author} {\bibfnamefont {C.}~\bibnamefont {Vicente}}, \bibinfo {author} {\bibfnamefont {E.~D.}\ \bibnamefont {Adams}}, \bibinfo {author} {\bibfnamefont {N.~S.}\ \bibnamefont {Sullivan}}, \bibinfo {author} {\bibfnamefont {L.~N.}\ \bibnamefont {Pfeiffer}}, \bibinfo {author} {\bibfnamefont {K.~W.}\ \bibnamefont {Baldwin}},\ and\ \bibinfo {author} {\bibfnamefont {K.~W.}\ \bibnamefont {West}},\ }\bibfield  {title} {\bibinfo {title} {Experimental studies of the fractional quantum {Hall} effect in the first excited {Landau} level},\ }\href {https://doi.org/10.1103/PhysRevB.77.075307} {\bibfield  {journal} {\bibinfo  {journal} {Phys. Rev. B}\ }\textbf {\bibinfo {volume} {77}},\ \bibinfo {pages} {075307} (\bibinfo {year}
  {2008})}\BibitemShut {NoStop}%
\bibitem [{\citenamefont {Choi}\ \emph {et~al.}(2008)\citenamefont {Choi}, \citenamefont {Kang}, \citenamefont {Das~Sarma}, \citenamefont {Pfeiffer},\ and\ \citenamefont {West}}]{Choi08}%
  \BibitemOpen
  \bibfield  {author} {\bibinfo {author} {\bibfnamefont {H.~C.}\ \bibnamefont {Choi}}, \bibinfo {author} {\bibfnamefont {W.}~\bibnamefont {Kang}}, \bibinfo {author} {\bibfnamefont {S.}~\bibnamefont {Das~Sarma}}, \bibinfo {author} {\bibfnamefont {L.~N.}\ \bibnamefont {Pfeiffer}},\ and\ \bibinfo {author} {\bibfnamefont {K.~W.}\ \bibnamefont {West}},\ }\bibfield  {title} {\bibinfo {title} {Activation gaps of fractional quantum {Hall} effect in the second {Landau} level},\ }\href {https://doi.org/10.1103/PhysRevB.77.081301} {\bibfield  {journal} {\bibinfo  {journal} {Phys. Rev. B}\ }\textbf {\bibinfo {volume} {77}},\ \bibinfo {pages} {081301} (\bibinfo {year} {2008})}\BibitemShut {NoStop}%
\bibitem [{\citenamefont {Kumar}\ \emph {et~al.}(2010)\citenamefont {Kumar}, \citenamefont {Cs\'athy}, \citenamefont {Manfra}, \citenamefont {Pfeiffer},\ and\ \citenamefont {West}}]{Kumar10}%
  \BibitemOpen
  \bibfield  {author} {\bibinfo {author} {\bibfnamefont {A.}~\bibnamefont {Kumar}}, \bibinfo {author} {\bibfnamefont {G.~A.}\ \bibnamefont {Cs\'athy}}, \bibinfo {author} {\bibfnamefont {M.~J.}\ \bibnamefont {Manfra}}, \bibinfo {author} {\bibfnamefont {L.~N.}\ \bibnamefont {Pfeiffer}},\ and\ \bibinfo {author} {\bibfnamefont {K.~W.}\ \bibnamefont {West}},\ }\bibfield  {title} {\bibinfo {title} {Nonconventional odd-denominator fractional quantum {Hall} states in the second {Landau} level},\ }\href {https://doi.org/10.1103/PhysRevLett.105.246808} {\bibfield  {journal} {\bibinfo  {journal} {Phys. Rev. Lett.}\ }\textbf {\bibinfo {volume} {105}},\ \bibinfo {pages} {246808} (\bibinfo {year} {2010})}\BibitemShut {NoStop}%
\bibitem [{\citenamefont {Zhang}\ \emph {et~al.}(2012)\citenamefont {Zhang}, \citenamefont {Huan}, \citenamefont {Xia}, \citenamefont {Sullivan}, \citenamefont {Pan}, \citenamefont {Baldwin}, \citenamefont {West}, \citenamefont {Pfeiffer},\ and\ \citenamefont {Tsui}}]{Zhang12}%
  \BibitemOpen
  \bibfield  {author} {\bibinfo {author} {\bibfnamefont {C.}~\bibnamefont {Zhang}}, \bibinfo {author} {\bibfnamefont {C.}~\bibnamefont {Huan}}, \bibinfo {author} {\bibfnamefont {J.~S.}\ \bibnamefont {Xia}}, \bibinfo {author} {\bibfnamefont {N.~S.}\ \bibnamefont {Sullivan}}, \bibinfo {author} {\bibfnamefont {W.}~\bibnamefont {Pan}}, \bibinfo {author} {\bibfnamefont {K.~W.}\ \bibnamefont {Baldwin}}, \bibinfo {author} {\bibfnamefont {K.~W.}\ \bibnamefont {West}}, \bibinfo {author} {\bibfnamefont {L.~N.}\ \bibnamefont {Pfeiffer}},\ and\ \bibinfo {author} {\bibfnamefont {D.~C.}\ \bibnamefont {Tsui}},\ }\bibfield  {title} {\bibinfo {title} {Spin polarization of the $\nu=12/5$ fractional quantum {Hall} state},\ }\href {https://doi.org/10.1103/PhysRevB.85.241302} {\bibfield  {journal} {\bibinfo  {journal} {Phys. Rev. B}\ }\textbf {\bibinfo {volume} {85}},\ \bibinfo {pages} {241302} (\bibinfo {year} {2012})}\BibitemShut {NoStop}%
\bibitem [{\citenamefont {Greiter}\ \emph {et~al.}(1991)\citenamefont {Greiter}, \citenamefont {Wen},\ and\ \citenamefont {Wilczek}}]{Greiter91}%
  \BibitemOpen
  \bibfield  {author} {\bibinfo {author} {\bibfnamefont {M.}~\bibnamefont {Greiter}}, \bibinfo {author} {\bibfnamefont {X.-G.}\ \bibnamefont {Wen}},\ and\ \bibinfo {author} {\bibfnamefont {F.}~\bibnamefont {Wilczek}},\ }\bibfield  {title} {\bibinfo {title} {Paired {Hall} state at half filling},\ }\href {https://doi.org/10.1103/PhysRevLett.66.3205} {\bibfield  {journal} {\bibinfo  {journal} {Phys. Rev. Lett.}\ }\textbf {\bibinfo {volume} {66}},\ \bibinfo {pages} {3205} (\bibinfo {year} {1991})}\BibitemShut {NoStop}%
\bibitem [{\citenamefont {Greiter}\ \emph {et~al.}(1992)\citenamefont {Greiter}, \citenamefont {Wen},\ and\ \citenamefont {Wilczek}}]{Greiter92a}%
  \BibitemOpen
  \bibfield  {author} {\bibinfo {author} {\bibfnamefont {M.}~\bibnamefont {Greiter}}, \bibinfo {author} {\bibfnamefont {X.}~\bibnamefont {Wen}},\ and\ \bibinfo {author} {\bibfnamefont {F.}~\bibnamefont {Wilczek}},\ }\bibfield  {title} {\bibinfo {title} {Paired {Hall} states},\ }\href {https://doi.org/http://dx.doi.org/10.1016/0550-3213(92)90401-V} {\bibfield  {journal} {\bibinfo  {journal} {Nucl. Phys. B}\ }\textbf {\bibinfo {volume} {374}},\ \bibinfo {pages} {567 } (\bibinfo {year} {1992})}\BibitemShut {NoStop}%
\bibitem [{\citenamefont {Read}\ and\ \citenamefont {Green}(2000)}]{Read00}%
  \BibitemOpen
  \bibfield  {author} {\bibinfo {author} {\bibfnamefont {N.}~\bibnamefont {Read}}\ and\ \bibinfo {author} {\bibfnamefont {D.}~\bibnamefont {Green}},\ }\bibfield  {title} {\bibinfo {title} {Paired states of fermions in two dimensions with breaking of parity and time-reversal symmetries and the fractional quantum {Hall} effect},\ }\href {https://doi.org/10.1103/PhysRevB.61.10267} {\bibfield  {journal} {\bibinfo  {journal} {Phys. Rev. B}\ }\textbf {\bibinfo {volume} {61}},\ \bibinfo {pages} {10267} (\bibinfo {year} {2000})}\BibitemShut {NoStop}%
\bibitem [{\citenamefont {Morf}(1998)}]{Morf98}%
  \BibitemOpen
  \bibfield  {author} {\bibinfo {author} {\bibfnamefont {R.~H.}\ \bibnamefont {Morf}},\ }\bibfield  {title} {\bibinfo {title} {Transition from quantum {Hall} to compressible states in the second {Landau} level: New light on the $\nu=5/2$ enigma},\ }\href {https://doi.org/10.1103/PhysRevLett.80.1505} {\bibfield  {journal} {\bibinfo  {journal} {Phys. Rev. Lett.}\ }\textbf {\bibinfo {volume} {80}},\ \bibinfo {pages} {1505} (\bibinfo {year} {1998})}\BibitemShut {NoStop}%
\bibitem [{\citenamefont {Scarola}\ \emph {et~al.}(2002)\citenamefont {Scarola}, \citenamefont {Jain},\ and\ \citenamefont {Rezayi}}]{Scarola02b}%
  \BibitemOpen
  \bibfield  {author} {\bibinfo {author} {\bibfnamefont {V.~W.}\ \bibnamefont {Scarola}}, \bibinfo {author} {\bibfnamefont {J.~K.}\ \bibnamefont {Jain}},\ and\ \bibinfo {author} {\bibfnamefont {E.~H.}\ \bibnamefont {Rezayi}},\ }\bibfield  {title} {\bibinfo {title} {Possible pairing-induced even-denominator fractional quantum {Hall} effect in the lowest {Landau} level},\ }\href {https://doi.org/10.1103/PhysRevLett.88.216804} {\bibfield  {journal} {\bibinfo  {journal} {Phys. Rev. Lett.}\ }\textbf {\bibinfo {volume} {88}},\ \bibinfo {pages} {216804} (\bibinfo {year} {2002})}\BibitemShut {NoStop}%
\bibitem [{\citenamefont {Sharma}\ \emph {et~al.}(2024)\citenamefont {Sharma}, \citenamefont {Balram},\ and\ \citenamefont {Jain}}]{Sharma23}%
  \BibitemOpen
  \bibfield  {author} {\bibinfo {author} {\bibfnamefont {A.}~\bibnamefont {Sharma}}, \bibinfo {author} {\bibfnamefont {A.~C.}\ \bibnamefont {Balram}},\ and\ \bibinfo {author} {\bibfnamefont {J.~K.}\ \bibnamefont {Jain}},\ }\bibfield  {title} {\bibinfo {title} {Composite-fermion pairing at half-filled and quarter-filled lowest {Landau} level},\ }\href {https://doi.org/10.1103/PhysRevB.109.035306} {\bibfield  {journal} {\bibinfo  {journal} {Phys. Rev. B}\ }\textbf {\bibinfo {volume} {109}},\ \bibinfo {pages} {035306} (\bibinfo {year} {2024})}\BibitemShut {NoStop}%
\bibitem [{\citenamefont {Suen}\ \emph {et~al.}(1992)\citenamefont {Suen}, \citenamefont {Engel}, \citenamefont {Santos}, \citenamefont {Shayegan},\ and\ \citenamefont {Tsui}}]{Suen92}%
  \BibitemOpen
  \bibfield  {author} {\bibinfo {author} {\bibfnamefont {Y.~W.}\ \bibnamefont {Suen}}, \bibinfo {author} {\bibfnamefont {L.~W.}\ \bibnamefont {Engel}}, \bibinfo {author} {\bibfnamefont {M.~B.}\ \bibnamefont {Santos}}, \bibinfo {author} {\bibfnamefont {M.}~\bibnamefont {Shayegan}},\ and\ \bibinfo {author} {\bibfnamefont {D.~C.}\ \bibnamefont {Tsui}},\ }\bibfield  {title} {\bibinfo {title} {Observation of a $\nu=1/2$ fractional quantum {Hall} state in a double-layer electron system},\ }\href {https://doi.org/10.1103/PhysRevLett.68.1379} {\bibfield  {journal} {\bibinfo  {journal} {Phys. Rev. Lett.}\ }\textbf {\bibinfo {volume} {68}},\ \bibinfo {pages} {1379} (\bibinfo {year} {1992})}\BibitemShut {NoStop}%
\bibitem [{\citenamefont {Suen}\ \emph {et~al.}(1994)\citenamefont {Suen}, \citenamefont {Manoharan}, \citenamefont {Ying}, \citenamefont {Santos},\ and\ \citenamefont {Shayegan}}]{Suen94}%
  \BibitemOpen
  \bibfield  {author} {\bibinfo {author} {\bibfnamefont {Y.}~\bibnamefont {Suen}}, \bibinfo {author} {\bibfnamefont {H.}~\bibnamefont {Manoharan}}, \bibinfo {author} {\bibfnamefont {X.}~\bibnamefont {Ying}}, \bibinfo {author} {\bibfnamefont {M.}~\bibnamefont {Santos}},\ and\ \bibinfo {author} {\bibfnamefont {M.}~\bibnamefont {Shayegan}},\ }\bibfield  {title} {\bibinfo {title} {One-component to two-component transitions of fractional quantum {Hall} states in a wide quantum well},\ }\href {https://doi.org/https://doi.org/10.1016/0039-6028(94)90852-4} {\bibfield  {journal} {\bibinfo  {journal} {Surface Science}\ }\textbf {\bibinfo {volume} {305}},\ \bibinfo {pages} {13 } (\bibinfo {year} {1994})}\BibitemShut {NoStop}%
\bibitem [{\citenamefont {Luhman}\ \emph {et~al.}(2008)\citenamefont {Luhman}, \citenamefont {Pan}, \citenamefont {Tsui}, \citenamefont {Pfeiffer}, \citenamefont {Baldwin},\ and\ \citenamefont {West}}]{Luhman08}%
  \BibitemOpen
  \bibfield  {author} {\bibinfo {author} {\bibfnamefont {D.~R.}\ \bibnamefont {Luhman}}, \bibinfo {author} {\bibfnamefont {W.}~\bibnamefont {Pan}}, \bibinfo {author} {\bibfnamefont {D.~C.}\ \bibnamefont {Tsui}}, \bibinfo {author} {\bibfnamefont {L.~N.}\ \bibnamefont {Pfeiffer}}, \bibinfo {author} {\bibfnamefont {K.~W.}\ \bibnamefont {Baldwin}},\ and\ \bibinfo {author} {\bibfnamefont {K.~W.}\ \bibnamefont {West}},\ }\bibfield  {title} {\bibinfo {title} {Observation of a fractional quantum {Hall} state at $\ensuremath{\nu}=1/4$ in a wide {Ga}{As} quantum well},\ }\href {https://doi.org/10.1103/PhysRevLett.101.266804} {\bibfield  {journal} {\bibinfo  {journal} {Phys. Rev. Lett.}\ }\textbf {\bibinfo {volume} {101}},\ \bibinfo {pages} {266804} (\bibinfo {year} {2008})}\BibitemShut {NoStop}%
\bibitem [{\citenamefont {Shabani}\ \emph {et~al.}(2009{\natexlab{a}})\citenamefont {Shabani}, \citenamefont {Gokmen},\ and\ \citenamefont {Shayegan}}]{Shabani09a}%
  \BibitemOpen
  \bibfield  {author} {\bibinfo {author} {\bibfnamefont {J.}~\bibnamefont {Shabani}}, \bibinfo {author} {\bibfnamefont {T.}~\bibnamefont {Gokmen}},\ and\ \bibinfo {author} {\bibfnamefont {M.}~\bibnamefont {Shayegan}},\ }\bibfield  {title} {\bibinfo {title} {Correlated states of electrons in wide quantum wells at low fillings: The role of charge distribution symmetry},\ }\href {https://doi.org/10.1103/PhysRevLett.103.046805} {\bibfield  {journal} {\bibinfo  {journal} {Phys. Rev. Lett.}\ }\textbf {\bibinfo {volume} {103}},\ \bibinfo {pages} {046805} (\bibinfo {year} {2009}{\natexlab{a}})}\BibitemShut {NoStop}%
\bibitem [{\citenamefont {Shabani}\ \emph {et~al.}(2009{\natexlab{b}})\citenamefont {Shabani}, \citenamefont {Gokmen}, \citenamefont {Chiu},\ and\ \citenamefont {Shayegan}}]{Shabani09b}%
  \BibitemOpen
  \bibfield  {author} {\bibinfo {author} {\bibfnamefont {J.}~\bibnamefont {Shabani}}, \bibinfo {author} {\bibfnamefont {T.}~\bibnamefont {Gokmen}}, \bibinfo {author} {\bibfnamefont {Y.~T.}\ \bibnamefont {Chiu}},\ and\ \bibinfo {author} {\bibfnamefont {M.}~\bibnamefont {Shayegan}},\ }\bibfield  {title} {\bibinfo {title} {Evidence for developing fractional quantum {Hall} states at even denominator $1/2$ and $1/4$ fillings in asymmetric wide quantum wells},\ }\href {https://doi.org/10.1103/PhysRevLett.103.256802} {\bibfield  {journal} {\bibinfo  {journal} {Phys. Rev. Lett.}\ }\textbf {\bibinfo {volume} {103}},\ \bibinfo {pages} {256802} (\bibinfo {year} {2009}{\natexlab{b}})}\BibitemShut {NoStop}%
\bibitem [{\citenamefont {Singh}\ \emph {et~al.}(2024)\citenamefont {Singh}, \citenamefont {Wang}, \citenamefont {Tai}, \citenamefont {Calhoun}, \citenamefont {Villegas~Rosales}, \citenamefont {Madathil}, \citenamefont {Gupta}, \citenamefont {Baldwin}, \citenamefont {Pfeiffer},\ and\ \citenamefont {Shayegan}}]{Singh23}%
  \BibitemOpen
  \bibfield  {author} {\bibinfo {author} {\bibfnamefont {S.~K.}\ \bibnamefont {Singh}}, \bibinfo {author} {\bibfnamefont {C.}~\bibnamefont {Wang}}, \bibinfo {author} {\bibfnamefont {C.~T.}\ \bibnamefont {Tai}}, \bibinfo {author} {\bibfnamefont {C.~S.}\ \bibnamefont {Calhoun}}, \bibinfo {author} {\bibfnamefont {K.~A.}\ \bibnamefont {Villegas~Rosales}}, \bibinfo {author} {\bibfnamefont {P.~T.}\ \bibnamefont {Madathil}}, \bibinfo {author} {\bibfnamefont {A.}~\bibnamefont {Gupta}}, \bibinfo {author} {\bibfnamefont {K.~W.}\ \bibnamefont {Baldwin}}, \bibinfo {author} {\bibfnamefont {L.~N.}\ \bibnamefont {Pfeiffer}},\ and\ \bibinfo {author} {\bibfnamefont {M.}~\bibnamefont {Shayegan}},\ }\bibfield  {title} {\bibinfo {title} {Topological phase transition between {Jain} states and daughter states of the $\nu${\thinspace}={\thinspace}1/2 fractional quantum {Hall} state},\ }\bibfield  {journal} {\bibinfo  {journal} {Nature Physics}\ }\href {https://doi.org/10.1038/s41567-024-02517-w} {10.1038/s41567-024-02517-w}
  (\bibinfo {year} {2024})\BibitemShut {NoStop}%
\bibitem [{\citenamefont {Singh}\ \emph {et~al.}(2025)\citenamefont {Singh}, \citenamefont {Wang}, \citenamefont {Gupta}, \citenamefont {Baldwin}, \citenamefont {Pfeiffer},\ and\ \citenamefont {Shayegan}}]{Singh25}%
  \BibitemOpen
  \bibfield  {author} {\bibinfo {author} {\bibfnamefont {S.~K.}\ \bibnamefont {Singh}}, \bibinfo {author} {\bibfnamefont {C.}~\bibnamefont {Wang}}, \bibinfo {author} {\bibfnamefont {A.}~\bibnamefont {Gupta}}, \bibinfo {author} {\bibfnamefont {K.~W.}\ \bibnamefont {Baldwin}}, \bibinfo {author} {\bibfnamefont {L.~N.}\ \bibnamefont {Pfeiffer}},\ and\ \bibinfo {author} {\bibfnamefont {M.}~\bibnamefont {Shayegan}},\ }\bibfield  {title} {\bibinfo {title} {Fractional quantum {Hall} state at $\ensuremath{\nu}\text{}=\text{}1/2$ with energy gap up to 6 {K} and possible transition from the one- to two-component state},\ }\href {https://doi.org/10.1103/ywpx-qm7d} {\bibfield  {journal} {\bibinfo  {journal} {Phys. Rev. Lett.}\ }\textbf {\bibinfo {volume} {135}},\ \bibinfo {pages} {246603} (\bibinfo {year} {2025})}\BibitemShut {NoStop}%
\bibitem [{\citenamefont {Shabani}\ \emph {et~al.}(2013)\citenamefont {Shabani}, \citenamefont {Liu}, \citenamefont {Shayegan}, \citenamefont {Pfeiffer}, \citenamefont {West},\ and\ \citenamefont {Baldwin}}]{Shabani13}%
  \BibitemOpen
  \bibfield  {author} {\bibinfo {author} {\bibfnamefont {J.}~\bibnamefont {Shabani}}, \bibinfo {author} {\bibfnamefont {Y.}~\bibnamefont {Liu}}, \bibinfo {author} {\bibfnamefont {M.}~\bibnamefont {Shayegan}}, \bibinfo {author} {\bibfnamefont {L.~N.}\ \bibnamefont {Pfeiffer}}, \bibinfo {author} {\bibfnamefont {K.~W.}\ \bibnamefont {West}},\ and\ \bibinfo {author} {\bibfnamefont {K.~W.}\ \bibnamefont {Baldwin}},\ }\bibfield  {title} {\bibinfo {title} {Phase diagrams for the stability of the $\ensuremath{\nu}=\frac{1}{2}$ fractional quantum {Hall} effect in electron systems confined to symmetric, wide {Ga}{As} quantum wells},\ }\href {https://doi.org/10.1103/PhysRevB.88.245413} {\bibfield  {journal} {\bibinfo  {journal} {Phys. Rev. B}\ }\textbf {\bibinfo {volume} {88}},\ \bibinfo {pages} {245413} (\bibinfo {year} {2013})}\BibitemShut {NoStop}%
\bibitem [{\citenamefont {Moore}\ and\ \citenamefont {Read}(1991)}]{Moore91}%
  \BibitemOpen
  \bibfield  {author} {\bibinfo {author} {\bibfnamefont {G.}~\bibnamefont {Moore}}\ and\ \bibinfo {author} {\bibfnamefont {N.}~\bibnamefont {Read}},\ }\bibfield  {title} {\bibinfo {title} {Nonabelions in the fractional quantum {Hall} effect},\ }\href {https://doi.org/10.1016/0550-3213(91)90407-O} {\bibfield  {journal} {\bibinfo  {journal} {Nucl. Phys. B}\ }\textbf {\bibinfo {volume} {360}},\ \bibinfo {pages} {362 } (\bibinfo {year} {1991})}\BibitemShut {NoStop}%
\bibitem [{\citenamefont {Min}\ and\ \citenamefont {MacDonald}(2008)}]{Min08}%
  \BibitemOpen
  \bibfield  {author} {\bibinfo {author} {\bibfnamefont {H.}~\bibnamefont {Min}}\ and\ \bibinfo {author} {\bibfnamefont {A.~H.}\ \bibnamefont {MacDonald}},\ }\bibfield  {title} {\bibinfo {title} {Chiral decomposition in the electronic structure of graphene multilayers},\ }\href {https://doi.org/10.1103/PhysRevB.77.155416} {\bibfield  {journal} {\bibinfo  {journal} {Phys. Rev. B}\ }\textbf {\bibinfo {volume} {77}},\ \bibinfo {pages} {155416} (\bibinfo {year} {2008})}\BibitemShut {NoStop}%
\bibitem [{\citenamefont {Barlas}\ \emph {et~al.}(2012)\citenamefont {Barlas}, \citenamefont {Yang},\ and\ \citenamefont {MacDonald}}]{Barlas12}%
  \BibitemOpen
  \bibfield  {author} {\bibinfo {author} {\bibfnamefont {Y.}~\bibnamefont {Barlas}}, \bibinfo {author} {\bibfnamefont {K.}~\bibnamefont {Yang}},\ and\ \bibinfo {author} {\bibfnamefont {A.~H.}\ \bibnamefont {MacDonald}},\ }\bibfield  {title} {\bibinfo {title} {Quantum {Hall} effects in graphene-based two-dimensional electron systems},\ }\href {https://doi.org/10.1088/0957-4484/23/5/052001} {\bibfield  {journal} {\bibinfo  {journal} {Nanotechnology}\ }\textbf {\bibinfo {volume} {23}},\ \bibinfo {pages} {052001} (\bibinfo {year} {2012})}\BibitemShut {NoStop}%
\bibitem [{\citenamefont {Du}\ \emph {et~al.}(2009)\citenamefont {Du}, \citenamefont {Skachko}, \citenamefont {Duerr}, \citenamefont {Luican},\ and\ \citenamefont {Andrei}}]{Du09}%
  \BibitemOpen
  \bibfield  {author} {\bibinfo {author} {\bibfnamefont {X.}~\bibnamefont {Du}}, \bibinfo {author} {\bibfnamefont {I.}~\bibnamefont {Skachko}}, \bibinfo {author} {\bibfnamefont {F.}~\bibnamefont {Duerr}}, \bibinfo {author} {\bibfnamefont {A.}~\bibnamefont {Luican}},\ and\ \bibinfo {author} {\bibfnamefont {E.~Y.}\ \bibnamefont {Andrei}},\ }\bibfield  {title} {\bibinfo {title} {Fractional quantum {Hall} effect and insulating phase of {Dirac} electrons in graphene},\ }\href {http://dx.doi.org/10.1038/nature08522} {\bibfield  {journal} {\bibinfo  {journal} {Nature}\ }\textbf {\bibinfo {volume} {462}},\ \bibinfo {pages} {192 EP } (\bibinfo {year} {2009})}\BibitemShut {NoStop}%
\bibitem [{\citenamefont {Dean}\ \emph {et~al.}(2011)\citenamefont {Dean}, \citenamefont {Young}, \citenamefont {Cadden-Zimansky}, \citenamefont {Wang}, \citenamefont {Ren}, \citenamefont {Watanabe}, \citenamefont {Taniguchi}, \citenamefont {Kim}, \citenamefont {Hone},\ and\ \citenamefont {Shepard}}]{Dean11}%
  \BibitemOpen
  \bibfield  {author} {\bibinfo {author} {\bibfnamefont {C.~R.}\ \bibnamefont {Dean}}, \bibinfo {author} {\bibfnamefont {A.~F.}\ \bibnamefont {Young}}, \bibinfo {author} {\bibfnamefont {P.}~\bibnamefont {Cadden-Zimansky}}, \bibinfo {author} {\bibfnamefont {L.}~\bibnamefont {Wang}}, \bibinfo {author} {\bibfnamefont {H.}~\bibnamefont {Ren}}, \bibinfo {author} {\bibfnamefont {K.}~\bibnamefont {Watanabe}}, \bibinfo {author} {\bibfnamefont {T.}~\bibnamefont {Taniguchi}}, \bibinfo {author} {\bibfnamefont {P.}~\bibnamefont {Kim}}, \bibinfo {author} {\bibfnamefont {J.}~\bibnamefont {Hone}},\ and\ \bibinfo {author} {\bibfnamefont {K.~L.}\ \bibnamefont {Shepard}},\ }\bibfield  {title} {\bibinfo {title} {Multicomponent fractional quantum {Hall} effect in graphene},\ }\href@noop {} {\bibfield  {journal} {\bibinfo  {journal} {Nature Physics}\ }\textbf {\bibinfo {volume} {7}},\ \bibinfo {pages} {693} (\bibinfo {year} {2011})}\BibitemShut {NoStop}%
\bibitem [{\citenamefont {Feldman}\ \emph {et~al.}(2012)\citenamefont {Feldman}, \citenamefont {Krauss}, \citenamefont {Smet},\ and\ \citenamefont {Yacoby}}]{Feldman12}%
  \BibitemOpen
  \bibfield  {author} {\bibinfo {author} {\bibfnamefont {B.~E.}\ \bibnamefont {Feldman}}, \bibinfo {author} {\bibfnamefont {B.}~\bibnamefont {Krauss}}, \bibinfo {author} {\bibfnamefont {J.~H.}\ \bibnamefont {Smet}},\ and\ \bibinfo {author} {\bibfnamefont {A.}~\bibnamefont {Yacoby}},\ }\bibfield  {title} {\bibinfo {title} {Unconventional sequence of fractional quantum {Hall} states in suspended graphene},\ }\href {https://doi.org/10.1126/science.1224784} {\bibfield  {journal} {\bibinfo  {journal} {Science}\ }\textbf {\bibinfo {volume} {337}},\ \bibinfo {pages} {1196} (\bibinfo {year} {2012})},\ \Eprint {https://arxiv.org/abs/http://www.sciencemag.org/content/337/6099/1196.full.pdf} {http://www.sciencemag.org/content/337/6099/1196.full.pdf} \BibitemShut {NoStop}%
\bibitem [{\citenamefont {Feldman}\ \emph {et~al.}(2013)\citenamefont {Feldman}, \citenamefont {Levin}, \citenamefont {Krauss}, \citenamefont {Abanin}, \citenamefont {Halperin}, \citenamefont {Smet},\ and\ \citenamefont {Yacoby}}]{Feldman13}%
  \BibitemOpen
  \bibfield  {author} {\bibinfo {author} {\bibfnamefont {B.~E.}\ \bibnamefont {Feldman}}, \bibinfo {author} {\bibfnamefont {A.~J.}\ \bibnamefont {Levin}}, \bibinfo {author} {\bibfnamefont {B.}~\bibnamefont {Krauss}}, \bibinfo {author} {\bibfnamefont {D.~A.}\ \bibnamefont {Abanin}}, \bibinfo {author} {\bibfnamefont {B.~I.}\ \bibnamefont {Halperin}}, \bibinfo {author} {\bibfnamefont {J.~H.}\ \bibnamefont {Smet}},\ and\ \bibinfo {author} {\bibfnamefont {A.}~\bibnamefont {Yacoby}},\ }\bibfield  {title} {\bibinfo {title} {Fractional quantum {Hall} phase transitions and four-flux states in graphene},\ }\href {https://doi.org/10.1103/PhysRevLett.111.076802} {\bibfield  {journal} {\bibinfo  {journal} {Phys. Rev. Lett.}\ }\textbf {\bibinfo {volume} {111}},\ \bibinfo {pages} {076802} (\bibinfo {year} {2013})}\BibitemShut {NoStop}%
\bibitem [{\citenamefont {Amet}\ \emph {et~al.}(2015)\citenamefont {Amet}, \citenamefont {Bestwick}, \citenamefont {Williams}, \citenamefont {Balicas}, \citenamefont {Watanabe}, \citenamefont {Taniguchi},\ and\ \citenamefont {Goldhaber-Gordon}}]{Amet15}%
  \BibitemOpen
  \bibfield  {author} {\bibinfo {author} {\bibfnamefont {F.}~\bibnamefont {Amet}}, \bibinfo {author} {\bibfnamefont {A.~J.}\ \bibnamefont {Bestwick}}, \bibinfo {author} {\bibfnamefont {J.~R.}\ \bibnamefont {Williams}}, \bibinfo {author} {\bibfnamefont {L.}~\bibnamefont {Balicas}}, \bibinfo {author} {\bibfnamefont {K.}~\bibnamefont {Watanabe}}, \bibinfo {author} {\bibfnamefont {T.}~\bibnamefont {Taniguchi}},\ and\ \bibinfo {author} {\bibfnamefont {D.}~\bibnamefont {Goldhaber-Gordon}},\ }\bibfield  {title} {\bibinfo {title} {Composite fermions and broken symmetries in graphene},\ }\href {https://doi.org/10.1038/ncomms6838} {\bibfield  {journal} {\bibinfo  {journal} {Nat. Commun.}\ }\textbf {\bibinfo {volume} {6}},\ \bibinfo {pages} {5838} (\bibinfo {year} {2015})}\BibitemShut {NoStop}%
\bibitem [{\citenamefont {Zibrov}\ \emph {et~al.}(2018)\citenamefont {Zibrov}, \citenamefont {Spanton}, \citenamefont {Zhou}, \citenamefont {Kometter}, \citenamefont {Taniguchi}, \citenamefont {Watanabe},\ and\ \citenamefont {Young}}]{Zibrov17}%
  \BibitemOpen
  \bibfield  {author} {\bibinfo {author} {\bibfnamefont {A.~A.}\ \bibnamefont {Zibrov}}, \bibinfo {author} {\bibfnamefont {E.~M.}\ \bibnamefont {Spanton}}, \bibinfo {author} {\bibfnamefont {H.}~\bibnamefont {Zhou}}, \bibinfo {author} {\bibfnamefont {C.}~\bibnamefont {Kometter}}, \bibinfo {author} {\bibfnamefont {T.}~\bibnamefont {Taniguchi}}, \bibinfo {author} {\bibfnamefont {K.}~\bibnamefont {Watanabe}},\ and\ \bibinfo {author} {\bibfnamefont {A.~F.}\ \bibnamefont {Young}},\ }\bibfield  {title} {\bibinfo {title} {Even-denominator fractional quantum {Hall} states at an isospin transition in monolayer graphene},\ }\href {https://doi.org/10.1038/s41567-018-0190-0} {\bibfield  {journal} {\bibinfo  {journal} {Nature Physics}\ }\textbf {\bibinfo {volume} {14}},\ \bibinfo {pages} {930} (\bibinfo {year} {2018})}\BibitemShut {NoStop}%
\bibitem [{\citenamefont {Chanda}\ \emph {et~al.}(2026)\citenamefont {Chanda}, \citenamefont {Kaur}, \citenamefont {Singh}, \citenamefont {Watanabe}, \citenamefont {Taniguchi}, \citenamefont {Jain}, \citenamefont {Khanna}, \citenamefont {Balram},\ and\ \citenamefont {Bid}}]{Chanda25}%
  \BibitemOpen
  \bibfield  {author} {\bibinfo {author} {\bibfnamefont {T.}~\bibnamefont {Chanda}}, \bibinfo {author} {\bibfnamefont {S.}~\bibnamefont {Kaur}}, \bibinfo {author} {\bibfnamefont {H.}~\bibnamefont {Singh}}, \bibinfo {author} {\bibfnamefont {K.}~\bibnamefont {Watanabe}}, \bibinfo {author} {\bibfnamefont {T.}~\bibnamefont {Taniguchi}}, \bibinfo {author} {\bibfnamefont {M.}~\bibnamefont {Jain}}, \bibinfo {author} {\bibfnamefont {U.}~\bibnamefont {Khanna}}, \bibinfo {author} {\bibfnamefont {A.~C.}\ \bibnamefont {Balram}},\ and\ \bibinfo {author} {\bibfnamefont {A.}~\bibnamefont {Bid}},\ }\bibfield  {title} {\bibinfo {title} {Even denominator fractional quantum {Hall} states in the zeroth {Landau} level of {A}{B}{A} trilayer graphene},\ }\href {https://doi.org/10.1103/tsnc-4jjl} {\bibfield  {journal} {\bibinfo  {journal} {Phys. Rev. Lett.}\ }\textbf {\bibinfo {volume} {137}},\ \bibinfo {pages} {036601} (\bibinfo {year} {2026})}\BibitemShut {NoStop}%
\bibitem [{\citenamefont {Kim}\ \emph {et~al.}(2019)\citenamefont {Kim}, \citenamefont {Balram}, \citenamefont {Taniguchi}, \citenamefont {Watanabe}, \citenamefont {Jain},\ and\ \citenamefont {Smet}}]{Kim19}%
  \BibitemOpen
  \bibfield  {author} {\bibinfo {author} {\bibfnamefont {Y.}~\bibnamefont {Kim}}, \bibinfo {author} {\bibfnamefont {A.~C.}\ \bibnamefont {Balram}}, \bibinfo {author} {\bibfnamefont {T.}~\bibnamefont {Taniguchi}}, \bibinfo {author} {\bibfnamefont {K.}~\bibnamefont {Watanabe}}, \bibinfo {author} {\bibfnamefont {J.~K.}\ \bibnamefont {Jain}},\ and\ \bibinfo {author} {\bibfnamefont {J.~H.}\ \bibnamefont {Smet}},\ }\bibfield  {title} {\bibinfo {title} {Even denominator fractional quantum {Hall} states in higher {Landau} levels of graphene},\ }\href {https://doi.org/10.1038/s41567-018-0355-x} {\bibfield  {journal} {\bibinfo  {journal} {Nature Physics}\ }\textbf {\bibinfo {volume} {15}},\ \bibinfo {pages} {154} (\bibinfo {year} {2019})}\BibitemShut {NoStop}%
\bibitem [{\citenamefont {Sharma}\ \emph {et~al.}(2023)\citenamefont {Sharma}, \citenamefont {Pu}, \citenamefont {Balram},\ and\ \citenamefont {Jain}}]{Sharma22}%
  \BibitemOpen
  \bibfield  {author} {\bibinfo {author} {\bibfnamefont {A.}~\bibnamefont {Sharma}}, \bibinfo {author} {\bibfnamefont {S.}~\bibnamefont {Pu}}, \bibinfo {author} {\bibfnamefont {A.~C.}\ \bibnamefont {Balram}},\ and\ \bibinfo {author} {\bibfnamefont {J.~K.}\ \bibnamefont {Jain}},\ }\bibfield  {title} {\bibinfo {title} {Fractional quantum {Hall} effect with unconventional pairing in monolayer graphene},\ }\href {https://doi.org/10.1103/PhysRevLett.130.126201} {\bibfield  {journal} {\bibinfo  {journal} {Phys. Rev. Lett.}\ }\textbf {\bibinfo {volume} {130}},\ \bibinfo {pages} {126201} (\bibinfo {year} {2023})}\BibitemShut {NoStop}%
\bibitem [{\citenamefont {{Zibrov}}\ \emph {et~al.}(2017)\citenamefont {{Zibrov}}, \citenamefont {{Kometter}}, \citenamefont {{Zhou}}, \citenamefont {{Spanton}}, \citenamefont {{Taniguchi}}, \citenamefont {{Watanabe}}, \citenamefont {{Zaletel}},\ and\ \citenamefont {{Young}}}]{Zibrov16}%
  \BibitemOpen
  \bibfield  {author} {\bibinfo {author} {\bibfnamefont {A.~A.}\ \bibnamefont {{Zibrov}}}, \bibinfo {author} {\bibfnamefont {C.~R.}\ \bibnamefont {{Kometter}}}, \bibinfo {author} {\bibfnamefont {H.}~\bibnamefont {{Zhou}}}, \bibinfo {author} {\bibfnamefont {E.~M.}\ \bibnamefont {{Spanton}}}, \bibinfo {author} {\bibfnamefont {T.}~\bibnamefont {{Taniguchi}}}, \bibinfo {author} {\bibfnamefont {K.}~\bibnamefont {{Watanabe}}}, \bibinfo {author} {\bibfnamefont {M.~P.}\ \bibnamefont {{Zaletel}}},\ and\ \bibinfo {author} {\bibfnamefont {A.~F.}\ \bibnamefont {{Young}}},\ }\bibfield  {title} {\bibinfo {title} {Tunable interacting composite fermion phases in a half-filled bilayer-graphene {Landau} level},\ }\href {https://doi.org/10.1038/nature23893} {\bibfield  {journal} {\bibinfo  {journal} {Nature}\ }\textbf {\bibinfo {volume} {549}},\ \bibinfo {pages} {360} (\bibinfo {year} {2017})}\BibitemShut {NoStop}%
\bibitem [{\citenamefont {Huang}\ \emph {et~al.}(2022)\citenamefont {Huang}, \citenamefont {Fu}, \citenamefont {Hickey}, \citenamefont {Alem}, \citenamefont {Lin}, \citenamefont {Watanabe}, \citenamefont {Taniguchi},\ and\ \citenamefont {Zhu}}]{Huang21}%
  \BibitemOpen
  \bibfield  {author} {\bibinfo {author} {\bibfnamefont {K.}~\bibnamefont {Huang}}, \bibinfo {author} {\bibfnamefont {H.}~\bibnamefont {Fu}}, \bibinfo {author} {\bibfnamefont {D.~R.}\ \bibnamefont {Hickey}}, \bibinfo {author} {\bibfnamefont {N.}~\bibnamefont {Alem}}, \bibinfo {author} {\bibfnamefont {X.}~\bibnamefont {Lin}}, \bibinfo {author} {\bibfnamefont {K.}~\bibnamefont {Watanabe}}, \bibinfo {author} {\bibfnamefont {T.}~\bibnamefont {Taniguchi}},\ and\ \bibinfo {author} {\bibfnamefont {J.}~\bibnamefont {Zhu}},\ }\bibfield  {title} {\bibinfo {title} {Valley isospin controlled fractional quantum {Hall} states in bilayer graphene},\ }\href {https://doi.org/10.1103/PhysRevX.12.031019} {\bibfield  {journal} {\bibinfo  {journal} {Phys. Rev. X}\ }\textbf {\bibinfo {volume} {12}},\ \bibinfo {pages} {031019} (\bibinfo {year} {2022})}\BibitemShut {NoStop}%
\bibitem [{\citenamefont {Kumar}\ \emph {et~al.}(2025{\natexlab{a}})\citenamefont {Kumar}, \citenamefont {Haug}, \citenamefont {Kim}, \citenamefont {Yutushui}, \citenamefont {Khudiakov}, \citenamefont {Bhardwaj}, \citenamefont {Ilin}, \citenamefont {Watanabe}, \citenamefont {Taniguchi}, \citenamefont {Mross},\ and\ \citenamefont {Ronen}}]{Kumar24}%
  \BibitemOpen
  \bibfield  {author} {\bibinfo {author} {\bibfnamefont {R.}~\bibnamefont {Kumar}}, \bibinfo {author} {\bibfnamefont {A.}~\bibnamefont {Haug}}, \bibinfo {author} {\bibfnamefont {J.}~\bibnamefont {Kim}}, \bibinfo {author} {\bibfnamefont {M.}~\bibnamefont {Yutushui}}, \bibinfo {author} {\bibfnamefont {K.}~\bibnamefont {Khudiakov}}, \bibinfo {author} {\bibfnamefont {V.}~\bibnamefont {Bhardwaj}}, \bibinfo {author} {\bibfnamefont {A.}~\bibnamefont {Ilin}}, \bibinfo {author} {\bibfnamefont {K.}~\bibnamefont {Watanabe}}, \bibinfo {author} {\bibfnamefont {T.}~\bibnamefont {Taniguchi}}, \bibinfo {author} {\bibfnamefont {D.~F.}\ \bibnamefont {Mross}},\ and\ \bibinfo {author} {\bibfnamefont {Y.}~\bibnamefont {Ronen}},\ }\bibfield  {title} {\bibinfo {title} {Quarter- and half-filled quantum {Hall} states and their topological orders revealed by daughter states in bilayer graphene},\ }\href {https://doi.org/10.1038/s41467-025-62650-9} {\bibfield  {journal} {\bibinfo  {journal} {Nature Communications}\ }\textbf {\bibinfo
  {volume} {16}},\ \bibinfo {pages} {7255} (\bibinfo {year} {2025}{\natexlab{a}})}\BibitemShut {NoStop}%
\bibitem [{\citenamefont {Huang}\ \emph {et~al.}(2025)\citenamefont {Huang}, \citenamefont {Balram}, \citenamefont {Fu}, \citenamefont {Guo}, \citenamefont {Watanabe}, \citenamefont {Taniguchi}, \citenamefont {Jain},\ and\ \citenamefont {Zhu}}]{Huang25}%
  \BibitemOpen
  \bibfield  {author} {\bibinfo {author} {\bibfnamefont {K.}~\bibnamefont {Huang}}, \bibinfo {author} {\bibfnamefont {A.~C.}\ \bibnamefont {Balram}}, \bibinfo {author} {\bibfnamefont {H.}~\bibnamefont {Fu}}, \bibinfo {author} {\bibfnamefont {C.}~\bibnamefont {Guo}}, \bibinfo {author} {\bibfnamefont {K.}~\bibnamefont {Watanabe}}, \bibinfo {author} {\bibfnamefont {T.}~\bibnamefont {Taniguchi}}, \bibinfo {author} {\bibfnamefont {J.~K.}\ \bibnamefont {Jain}},\ and\ \bibinfo {author} {\bibfnamefont {J.}~\bibnamefont {Zhu}},\ }\bibfield  {title} {\bibinfo {title} {Hetero-orbital two-component fractional quantum {Hall} states in bilayer graphene},\ }\href {https://doi.org/10.1103/kfn2-qggs} {\bibfield  {journal} {\bibinfo  {journal} {Phys. Rev. X}\ }\textbf {\bibinfo {volume} {15}},\ \bibinfo {pages} {031023} (\bibinfo {year} {2025})}\BibitemShut {NoStop}%
\bibitem [{\citenamefont {Hu}\ \emph {et~al.}(2025)\citenamefont {Hu}, \citenamefont {Tsui}, \citenamefont {He}, \citenamefont {Kamber}, \citenamefont {Wang}, \citenamefont {Mohammadi}, \citenamefont {Watanabe}, \citenamefont {Taniguchi}, \citenamefont {Papi{\'c}}, \citenamefont {Zaletel},\ and\ \citenamefont {Yazdani}}]{Hu24}%
  \BibitemOpen
  \bibfield  {author} {\bibinfo {author} {\bibfnamefont {Y.}~\bibnamefont {Hu}}, \bibinfo {author} {\bibfnamefont {Y.-C.}\ \bibnamefont {Tsui}}, \bibinfo {author} {\bibfnamefont {M.}~\bibnamefont {He}}, \bibinfo {author} {\bibfnamefont {U.}~\bibnamefont {Kamber}}, \bibinfo {author} {\bibfnamefont {T.}~\bibnamefont {Wang}}, \bibinfo {author} {\bibfnamefont {A.~S.}\ \bibnamefont {Mohammadi}}, \bibinfo {author} {\bibfnamefont {K.}~\bibnamefont {Watanabe}}, \bibinfo {author} {\bibfnamefont {T.}~\bibnamefont {Taniguchi}}, \bibinfo {author} {\bibfnamefont {Z.}~\bibnamefont {Papi{\'c}}}, \bibinfo {author} {\bibfnamefont {M.~P.}\ \bibnamefont {Zaletel}},\ and\ \bibinfo {author} {\bibfnamefont {A.}~\bibnamefont {Yazdani}},\ }\bibfield  {title} {\bibinfo {title} {High-resolution tunnelling spectroscopy of fractional quantum {Hall} states},\ }\href {https://doi.org/10.1038/s41567-025-02830-y} {\bibfield  {journal} {\bibinfo  {journal} {Nature Physics}\ }\textbf {\bibinfo {volume} {21}},\ \bibinfo {pages} {716} (\bibinfo
  {year} {2025})}\BibitemShut {NoStop}%
\bibitem [{\citenamefont {Apalkov}\ and\ \citenamefont {Chakraborty}(2011)}]{Apalkov11}%
  \BibitemOpen
  \bibfield  {author} {\bibinfo {author} {\bibfnamefont {V.~M.}\ \bibnamefont {Apalkov}}\ and\ \bibinfo {author} {\bibfnamefont {T.}~\bibnamefont {Chakraborty}},\ }\bibfield  {title} {\bibinfo {title} {Stable {Pfaffian} state in bilayer graphene},\ }\href {https://doi.org/10.1103/PhysRevLett.107.186803} {\bibfield  {journal} {\bibinfo  {journal} {Phys. Rev. Lett.}\ }\textbf {\bibinfo {volume} {107}},\ \bibinfo {pages} {186803} (\bibinfo {year} {2011})}\BibitemShut {NoStop}%
\bibitem [{\citenamefont {Zhu}\ \emph {et~al.}(2020)\citenamefont {Zhu}, \citenamefont {Sheng},\ and\ \citenamefont {Sodemann}}]{Zhu20a}%
  \BibitemOpen
  \bibfield  {author} {\bibinfo {author} {\bibfnamefont {Z.}~\bibnamefont {Zhu}}, \bibinfo {author} {\bibfnamefont {D.~N.}\ \bibnamefont {Sheng}},\ and\ \bibinfo {author} {\bibfnamefont {I.}~\bibnamefont {Sodemann}},\ }\bibfield  {title} {\bibinfo {title} {Widely tunable quantum phase transition from {Moore}-{Read} to composite {Fermi} liquid in bilayer graphene},\ }\href {https://doi.org/10.1103/PhysRevLett.124.097604} {\bibfield  {journal} {\bibinfo  {journal} {Phys. Rev. Lett.}\ }\textbf {\bibinfo {volume} {124}},\ \bibinfo {pages} {097604} (\bibinfo {year} {2020})}\BibitemShut {NoStop}%
\bibitem [{\citenamefont {Balram}(2022)}]{Balram21b}%
  \BibitemOpen
  \bibfield  {author} {\bibinfo {author} {\bibfnamefont {A.~C.}\ \bibnamefont {Balram}},\ }\bibfield  {title} {\bibinfo {title} {Transitions from {Abelian} composite fermion to non-{Abelian} parton fractional quantum {Hall} states in the zeroth {Landau} level of bilayer graphene},\ }\href {https://doi.org/10.1103/PhysRevB.105.L121406} {\bibfield  {journal} {\bibinfo  {journal} {Phys. Rev. B}\ }\textbf {\bibinfo {volume} {105}},\ \bibinfo {pages} {L121406} (\bibinfo {year} {2022})}\BibitemShut {NoStop}%
\bibitem [{\citenamefont {Kumar}\ \emph {et~al.}(2025{\natexlab{b}})\citenamefont {Kumar}, \citenamefont {Firon}, \citenamefont {Haug}, \citenamefont {Yutushui}, \citenamefont {Gaon}, \citenamefont {Watanabe}, \citenamefont {Taniguchi}, \citenamefont {Mross},\ and\ \citenamefont {Ronen}}]{Kumar25}%
  \BibitemOpen
  \bibfield  {author} {\bibinfo {author} {\bibfnamefont {R.}~\bibnamefont {Kumar}}, \bibinfo {author} {\bibfnamefont {T.}~\bibnamefont {Firon}}, \bibinfo {author} {\bibfnamefont {A.}~\bibnamefont {Haug}}, \bibinfo {author} {\bibfnamefont {M.}~\bibnamefont {Yutushui}}, \bibinfo {author} {\bibfnamefont {A.~N.}\ \bibnamefont {Gaon}}, \bibinfo {author} {\bibfnamefont {K.}~\bibnamefont {Watanabe}}, \bibinfo {author} {\bibfnamefont {T.}~\bibnamefont {Taniguchi}}, \bibinfo {author} {\bibfnamefont {D.~F.}\ \bibnamefont {Mross}},\ and\ \bibinfo {author} {\bibfnamefont {Y.}~\bibnamefont {Ronen}},\ }\href {https://arxiv.org/abs/2512.21383} {\bibinfo {title} {Orbitally tuned composite-fermion metal-to-superfluid transitions}} (\bibinfo {year} {2025}{\natexlab{b}}),\ \Eprint {https://arxiv.org/abs/2512.21383} {arXiv:2512.21383 [cond-mat.mes-hall]} \BibitemShut {NoStop}%
\bibitem [{\citenamefont {Diankov}\ \emph {et~al.}(2016)\citenamefont {Diankov}, \citenamefont {Liang}, \citenamefont {Amet}, \citenamefont {Gallagher}, \citenamefont {Lee}, \citenamefont {Bestwick}, \citenamefont {Tharratt}, \citenamefont {Coniglio}, \citenamefont {Jaroszynski}, \citenamefont {Watanabe}, \citenamefont {Taniguchi},\ and\ \citenamefont {Goldhaber-Gordon}}]{Diankov16}%
  \BibitemOpen
  \bibfield  {author} {\bibinfo {author} {\bibfnamefont {G.}~\bibnamefont {Diankov}}, \bibinfo {author} {\bibfnamefont {C.-T.}\ \bibnamefont {Liang}}, \bibinfo {author} {\bibfnamefont {F.}~\bibnamefont {Amet}}, \bibinfo {author} {\bibfnamefont {P.}~\bibnamefont {Gallagher}}, \bibinfo {author} {\bibfnamefont {M.}~\bibnamefont {Lee}}, \bibinfo {author} {\bibfnamefont {A.~J.}\ \bibnamefont {Bestwick}}, \bibinfo {author} {\bibfnamefont {K.}~\bibnamefont {Tharratt}}, \bibinfo {author} {\bibfnamefont {W.}~\bibnamefont {Coniglio}}, \bibinfo {author} {\bibfnamefont {J.}~\bibnamefont {Jaroszynski}}, \bibinfo {author} {\bibfnamefont {K.}~\bibnamefont {Watanabe}}, \bibinfo {author} {\bibfnamefont {T.}~\bibnamefont {Taniguchi}},\ and\ \bibinfo {author} {\bibfnamefont {D.}~\bibnamefont {Goldhaber-Gordon}},\ }\bibfield  {title} {\bibinfo {title} {Robust fractional quantum {Hall} effect in the {N}=2 {Landau} level in bilayer graphene},\ }\href {http://dx.doi.org/10.1038/ncomms13908} {\bibfield  {journal} {\bibinfo
  {journal} {Nature Communications}\ }\textbf {\bibinfo {volume} {7}},\ \bibinfo {pages} {13908 EP } (\bibinfo {year} {2016})},\ \bibinfo {note} {article}\BibitemShut {NoStop}%
\bibitem [{\citenamefont {Levin}\ and\ \citenamefont {Halperin}(2009)}]{Levin09a}%
  \BibitemOpen
  \bibfield  {author} {\bibinfo {author} {\bibfnamefont {M.}~\bibnamefont {Levin}}\ and\ \bibinfo {author} {\bibfnamefont {B.~I.}\ \bibnamefont {Halperin}},\ }\bibfield  {title} {\bibinfo {title} {Collective states of non-abelian quasiparticles in a magnetic field},\ }\href {https://doi.org/10.1103/PhysRevB.79.205301} {\bibfield  {journal} {\bibinfo  {journal} {Phys. Rev. B}\ }\textbf {\bibinfo {volume} {79}},\ \bibinfo {pages} {205301} (\bibinfo {year} {2009})}\BibitemShut {NoStop}%
\bibitem [{\citenamefont {Papi\ifmmode~\acute{c}\else \'{c}\fi{}}\ \emph {et~al.}(2011)\citenamefont {Papi\ifmmode~\acute{c}\else \'{c}\fi{}}, \citenamefont {Abanin}, \citenamefont {Barlas},\ and\ \citenamefont {Bhatt}}]{Papic11}%
  \BibitemOpen
  \bibfield  {author} {\bibinfo {author} {\bibfnamefont {Z.}~\bibnamefont {Papi\ifmmode~\acute{c}\else \'{c}\fi{}}}, \bibinfo {author} {\bibfnamefont {D.~A.}\ \bibnamefont {Abanin}}, \bibinfo {author} {\bibfnamefont {Y.}~\bibnamefont {Barlas}},\ and\ \bibinfo {author} {\bibfnamefont {R.~N.}\ \bibnamefont {Bhatt}},\ }\bibfield  {title} {\bibinfo {title} {Tunable interactions and phase transitions in {Dirac} materials in a magnetic field},\ }\href {https://doi.org/10.1103/PhysRevB.84.241306} {\bibfield  {journal} {\bibinfo  {journal} {Phys. Rev. B}\ }\textbf {\bibinfo {volume} {84}},\ \bibinfo {pages} {241306} (\bibinfo {year} {2011})}\BibitemShut {NoStop}%
\bibitem [{\citenamefont {Balram}(2021)}]{Balram21}%
  \BibitemOpen
  \bibfield  {author} {\bibinfo {author} {\bibfnamefont {A.~C.}\ \bibnamefont {Balram}},\ }\bibfield  {title} {\bibinfo {title} {{A non-Abelian parton state for the $\ensuremath{\nu}=2+3/8$ fractional quantum Hall effect}},\ }\href {https://doi.org/10.21468/SciPostPhys.10.4.083} {\bibfield  {journal} {\bibinfo  {journal} {SciPost Phys.}\ }\textbf {\bibinfo {volume} {10}},\ \bibinfo {pages} {83} (\bibinfo {year} {2021})}\BibitemShut {NoStop}%
\bibitem [{\citenamefont {Jain}(1989{\natexlab{b}})}]{Jain89b}%
  \BibitemOpen
  \bibfield  {author} {\bibinfo {author} {\bibfnamefont {J.~K.}\ \bibnamefont {Jain}},\ }\bibfield  {title} {\bibinfo {title} {Incompressible quantum {Hall} states},\ }\href {https://doi.org/10.1103/PhysRevB.40.8079} {\bibfield  {journal} {\bibinfo  {journal} {Phys. Rev. B}\ }\textbf {\bibinfo {volume} {40}},\ \bibinfo {pages} {8079} (\bibinfo {year} {1989}{\natexlab{b}})}\BibitemShut {NoStop}%
\bibitem [{\citenamefont {Balram}\ \emph {et~al.}(2018{\natexlab{a}})\citenamefont {Balram}, \citenamefont {Barkeshli},\ and\ \citenamefont {Rudner}}]{Balram18}%
  \BibitemOpen
  \bibfield  {author} {\bibinfo {author} {\bibfnamefont {A.~C.}\ \bibnamefont {Balram}}, \bibinfo {author} {\bibfnamefont {M.}~\bibnamefont {Barkeshli}},\ and\ \bibinfo {author} {\bibfnamefont {M.~S.}\ \bibnamefont {Rudner}},\ }\bibfield  {title} {\bibinfo {title} {Parton construction of a wave function in the anti-{Pfaffian} phase},\ }\href {https://doi.org/10.1103/PhysRevB.98.035127} {\bibfield  {journal} {\bibinfo  {journal} {Phys. Rev. B}\ }\textbf {\bibinfo {volume} {98}},\ \bibinfo {pages} {035127} (\bibinfo {year} {2018}{\natexlab{a}})}\BibitemShut {NoStop}%
\bibitem [{\citenamefont {Balram}\ \emph {et~al.}(2018{\natexlab{b}})\citenamefont {Balram}, \citenamefont {Mukherjee}, \citenamefont {Park}, \citenamefont {Barkeshli}, \citenamefont {Rudner},\ and\ \citenamefont {Jain}}]{Balram18a}%
  \BibitemOpen
  \bibfield  {author} {\bibinfo {author} {\bibfnamefont {A.~C.}\ \bibnamefont {Balram}}, \bibinfo {author} {\bibfnamefont {S.}~\bibnamefont {Mukherjee}}, \bibinfo {author} {\bibfnamefont {K.}~\bibnamefont {Park}}, \bibinfo {author} {\bibfnamefont {M.}~\bibnamefont {Barkeshli}}, \bibinfo {author} {\bibfnamefont {M.~S.}\ \bibnamefont {Rudner}},\ and\ \bibinfo {author} {\bibfnamefont {J.~K.}\ \bibnamefont {Jain}},\ }\bibfield  {title} {\bibinfo {title} {Fractional quantum {Hall} effect at $\ensuremath{\nu}=2+6/13$: The parton paradigm for the second {Landau} level},\ }\href {https://doi.org/10.1103/PhysRevLett.121.186601} {\bibfield  {journal} {\bibinfo  {journal} {Phys. Rev. Lett.}\ }\textbf {\bibinfo {volume} {121}},\ \bibinfo {pages} {186601} (\bibinfo {year} {2018}{\natexlab{b}})}\BibitemShut {NoStop}%
\bibitem [{\citenamefont {Balram}\ \emph {et~al.}(2019)\citenamefont {Balram}, \citenamefont {Barkeshli},\ and\ \citenamefont {Rudner}}]{Balram19}%
  \BibitemOpen
  \bibfield  {author} {\bibinfo {author} {\bibfnamefont {A.~C.}\ \bibnamefont {Balram}}, \bibinfo {author} {\bibfnamefont {M.}~\bibnamefont {Barkeshli}},\ and\ \bibinfo {author} {\bibfnamefont {M.~S.}\ \bibnamefont {Rudner}},\ }\bibfield  {title} {\bibinfo {title} {Parton construction of particle-hole-conjugate {Read}-{Rezayi} parafermion fractional quantum {Hall} states and beyond},\ }\href {https://doi.org/10.1103/PhysRevB.99.241108} {\bibfield  {journal} {\bibinfo  {journal} {Phys. Rev. B}\ }\textbf {\bibinfo {volume} {99}},\ \bibinfo {pages} {241108} (\bibinfo {year} {2019})}\BibitemShut {NoStop}%
\bibitem [{\citenamefont {Bose}\ and\ \citenamefont {Balram}(2023)}]{Bose23}%
  \BibitemOpen
  \bibfield  {author} {\bibinfo {author} {\bibfnamefont {K.}~\bibnamefont {Bose}}\ and\ \bibinfo {author} {\bibfnamefont {A.~C.}\ \bibnamefont {Balram}},\ }\bibfield  {title} {\bibinfo {title} {Prediction of non-abelian fractional quantum {Hall} effect at $\ensuremath{\nu}=2+\frac{4}{11}$},\ }\href {https://doi.org/10.1103/PhysRevB.107.235111} {\bibfield  {journal} {\bibinfo  {journal} {Phys. Rev. B}\ }\textbf {\bibinfo {volume} {107}},\ \bibinfo {pages} {235111} (\bibinfo {year} {2023})}\BibitemShut {NoStop}%
\bibitem [{\citenamefont {Fogler}\ \emph {et~al.}(1996)\citenamefont {Fogler}, \citenamefont {Koulakov},\ and\ \citenamefont {Shklovskii}}]{Fogler96}%
  \BibitemOpen
  \bibfield  {author} {\bibinfo {author} {\bibfnamefont {M.~M.}\ \bibnamefont {Fogler}}, \bibinfo {author} {\bibfnamefont {A.~A.}\ \bibnamefont {Koulakov}},\ and\ \bibinfo {author} {\bibfnamefont {B.~I.}\ \bibnamefont {Shklovskii}},\ }\bibfield  {title} {\bibinfo {title} {Ground state of a two-dimensional electron liquid in a weak magnetic field},\ }\href {https://doi.org/10.1103/PhysRevB.54.1853} {\bibfield  {journal} {\bibinfo  {journal} {Phys. Rev. B}\ }\textbf {\bibinfo {volume} {54}},\ \bibinfo {pages} {1853} (\bibinfo {year} {1996})}\BibitemShut {NoStop}%
\bibitem [{\citenamefont {Koulakov}\ \emph {et~al.}(1996)\citenamefont {Koulakov}, \citenamefont {Fogler},\ and\ \citenamefont {Shklovskii}}]{Koulakov96}%
  \BibitemOpen
  \bibfield  {author} {\bibinfo {author} {\bibfnamefont {A.~A.}\ \bibnamefont {Koulakov}}, \bibinfo {author} {\bibfnamefont {M.~M.}\ \bibnamefont {Fogler}},\ and\ \bibinfo {author} {\bibfnamefont {B.~I.}\ \bibnamefont {Shklovskii}},\ }\bibfield  {title} {\bibinfo {title} {Charge density wave in two-dimensional electron liquid in weak magnetic field},\ }\href {https://doi.org/10.1103/PhysRevLett.76.499} {\bibfield  {journal} {\bibinfo  {journal} {Phys. Rev. Lett.}\ }\textbf {\bibinfo {volume} {76}},\ \bibinfo {pages} {499} (\bibinfo {year} {1996})}\BibitemShut {NoStop}%
\bibitem [{\citenamefont {Lam}\ and\ \citenamefont {Girvin}(1984)}]{Lam84}%
  \BibitemOpen
  \bibfield  {author} {\bibinfo {author} {\bibfnamefont {P.~K.}\ \bibnamefont {Lam}}\ and\ \bibinfo {author} {\bibfnamefont {S.~M.}\ \bibnamefont {Girvin}},\ }\bibfield  {title} {\bibinfo {title} {Liquid-solid transition and the fractional quantum-{Hall} effect},\ }\href {https://doi.org/10.1103/PhysRevB.30.473} {\bibfield  {journal} {\bibinfo  {journal} {Phys. Rev. B}\ }\textbf {\bibinfo {volume} {30}},\ \bibinfo {pages} {473} (\bibinfo {year} {1984})}\BibitemShut {NoStop}%
\bibitem [{\citenamefont {Goerbig}\ \emph {et~al.}(2004{\natexlab{a}})\citenamefont {Goerbig}, \citenamefont {Lederer},\ and\ \citenamefont {Smith}}]{Goerbig04a}%
  \BibitemOpen
  \bibfield  {author} {\bibinfo {author} {\bibfnamefont {M.~O.}\ \bibnamefont {Goerbig}}, \bibinfo {author} {\bibfnamefont {P.}~\bibnamefont {Lederer}},\ and\ \bibinfo {author} {\bibfnamefont {C.~M.}\ \bibnamefont {Smith}},\ }\bibfield  {title} {\bibinfo {title} {Competition between quantum-liquid and electron-solid phases in intermediate {Landau} levels},\ }\href {https://doi.org/10.1103/PhysRevB.69.115327} {\bibfield  {journal} {\bibinfo  {journal} {Phys. Rev. B}\ }\textbf {\bibinfo {volume} {69}},\ \bibinfo {pages} {115327} (\bibinfo {year} {2004}{\natexlab{a}})}\BibitemShut {NoStop}%
\bibitem [{\citenamefont {Archer}\ \emph {et~al.}(2013)\citenamefont {Archer}, \citenamefont {Park},\ and\ \citenamefont {Jain}}]{Archer13}%
  \BibitemOpen
  \bibfield  {author} {\bibinfo {author} {\bibfnamefont {A.~C.}\ \bibnamefont {Archer}}, \bibinfo {author} {\bibfnamefont {K.}~\bibnamefont {Park}},\ and\ \bibinfo {author} {\bibfnamefont {J.~K.}\ \bibnamefont {Jain}},\ }\bibfield  {title} {\bibinfo {title} {Competing crystal phases in the lowest {Landau} level},\ }\href {https://doi.org/10.1103/PhysRevLett.111.146804} {\bibfield  {journal} {\bibinfo  {journal} {Phys. Rev. Lett.}\ }\textbf {\bibinfo {volume} {111}},\ \bibinfo {pages} {146804} (\bibinfo {year} {2013})}\BibitemShut {NoStop}%
\bibitem [{\citenamefont {Zuo}\ \emph {et~al.}(2020)\citenamefont {Zuo}, \citenamefont {Balram}, \citenamefont {Pu}, \citenamefont {Zhao}, \citenamefont {Jolicoeur}, \citenamefont {W\'ojs},\ and\ \citenamefont {Jain}}]{Zuo20}%
  \BibitemOpen
  \bibfield  {author} {\bibinfo {author} {\bibfnamefont {Z.-W.}\ \bibnamefont {Zuo}}, \bibinfo {author} {\bibfnamefont {A.~C.}\ \bibnamefont {Balram}}, \bibinfo {author} {\bibfnamefont {S.}~\bibnamefont {Pu}}, \bibinfo {author} {\bibfnamefont {J.}~\bibnamefont {Zhao}}, \bibinfo {author} {\bibfnamefont {T.}~\bibnamefont {Jolicoeur}}, \bibinfo {author} {\bibfnamefont {A.}~\bibnamefont {W\'ojs}},\ and\ \bibinfo {author} {\bibfnamefont {J.~K.}\ \bibnamefont {Jain}},\ }\bibfield  {title} {\bibinfo {title} {Interplay between fractional quantum {Hall} liquid and crystal phases at low filling},\ }\href {https://doi.org/10.1103/PhysRevB.102.075307} {\bibfield  {journal} {\bibinfo  {journal} {Phys. Rev. B}\ }\textbf {\bibinfo {volume} {102}},\ \bibinfo {pages} {075307} (\bibinfo {year} {2020})}\BibitemShut {NoStop}%
\bibitem [{\citenamefont {Yutushui}\ and\ \citenamefont {Mross}(2026)}]{Yutushui26}%
  \BibitemOpen
  \bibfield  {author} {\bibinfo {author} {\bibfnamefont {M.}~\bibnamefont {Yutushui}}\ and\ \bibinfo {author} {\bibfnamefont {D.~F.}\ \bibnamefont {Mross}},\ }\href {https://arxiv.org/abs/2601.15386} {\bibinfo {title} {Theory of next-generation even-denominator states}} (\bibinfo {year} {2026}),\ \Eprint {https://arxiv.org/abs/2601.15386} {arXiv:2601.15386 [cond-mat.str-el]} \BibitemShut {NoStop}%
\bibitem [{\citenamefont {McClure}(1957)}]{McClure57}%
  \BibitemOpen
  \bibfield  {author} {\bibinfo {author} {\bibfnamefont {J.~W.}\ \bibnamefont {McClure}},\ }\bibfield  {title} {\bibinfo {title} {Band structure of graphite and de {Haas}-van {Alphen} effect},\ }\href {https://doi.org/10.1103/PhysRev.108.612} {\bibfield  {journal} {\bibinfo  {journal} {Phys. Rev.}\ }\textbf {\bibinfo {volume} {108}},\ \bibinfo {pages} {612} (\bibinfo {year} {1957})}\BibitemShut {NoStop}%
\bibitem [{\citenamefont {Hunt}\ \emph {et~al.}(2017)\citenamefont {Hunt}, \citenamefont {Li}, \citenamefont {Zibrov}, \citenamefont {Wang}, \citenamefont {Taniguchi}, \citenamefont {Watanabe}, \citenamefont {Hone}, \citenamefont {Dean}, \citenamefont {Zaletel}, \citenamefont {Ashoori},\ and\ \citenamefont {Young}}]{Hunt17}%
  \BibitemOpen
  \bibfield  {author} {\bibinfo {author} {\bibfnamefont {B.~M.}\ \bibnamefont {Hunt}}, \bibinfo {author} {\bibfnamefont {J.~I.~A.}\ \bibnamefont {Li}}, \bibinfo {author} {\bibfnamefont {A.~A.}\ \bibnamefont {Zibrov}}, \bibinfo {author} {\bibfnamefont {L.}~\bibnamefont {Wang}}, \bibinfo {author} {\bibfnamefont {T.}~\bibnamefont {Taniguchi}}, \bibinfo {author} {\bibfnamefont {K.}~\bibnamefont {Watanabe}}, \bibinfo {author} {\bibfnamefont {J.}~\bibnamefont {Hone}}, \bibinfo {author} {\bibfnamefont {C.~R.}\ \bibnamefont {Dean}}, \bibinfo {author} {\bibfnamefont {M.}~\bibnamefont {Zaletel}}, \bibinfo {author} {\bibfnamefont {R.~C.}\ \bibnamefont {Ashoori}},\ and\ \bibinfo {author} {\bibfnamefont {A.~F.}\ \bibnamefont {Young}},\ }\bibfield  {title} {\bibinfo {title} {Direct measurement of discrete valley and orbital quantum numbers in bilayer graphene},\ }\href {https://doi.org/10.1038/s41467-017-00824-w} {\bibfield  {journal} {\bibinfo  {journal} {Nature Communications}\ }\textbf {\bibinfo {volume} {8}},\ \bibinfo
  {pages} {948} (\bibinfo {year} {2017})}\BibitemShut {NoStop}%
\bibitem [{\citenamefont {Khanna}\ \emph {et~al.}(2023)\citenamefont {Khanna}, \citenamefont {Huang}, \citenamefont {Murthy}, \citenamefont {Fertig}, \citenamefont {Watanabe}, \citenamefont {Taniguchi}, \citenamefont {Zhu},\ and\ \citenamefont {Shimshoni}}]{Khanna23}%
  \BibitemOpen
  \bibfield  {author} {\bibinfo {author} {\bibfnamefont {U.}~\bibnamefont {Khanna}}, \bibinfo {author} {\bibfnamefont {K.}~\bibnamefont {Huang}}, \bibinfo {author} {\bibfnamefont {G.}~\bibnamefont {Murthy}}, \bibinfo {author} {\bibfnamefont {H.~A.}\ \bibnamefont {Fertig}}, \bibinfo {author} {\bibfnamefont {K.}~\bibnamefont {Watanabe}}, \bibinfo {author} {\bibfnamefont {T.}~\bibnamefont {Taniguchi}}, \bibinfo {author} {\bibfnamefont {J.}~\bibnamefont {Zhu}},\ and\ \bibinfo {author} {\bibfnamefont {E.}~\bibnamefont {Shimshoni}},\ }\bibfield  {title} {\bibinfo {title} {Phase diagram of the $\ensuremath{\nu}=2$ quantum {Hall} state in bilayer graphene},\ }\href {https://doi.org/10.1103/PhysRevB.108.L041107} {\bibfield  {journal} {\bibinfo  {journal} {Phys. Rev. B}\ }\textbf {\bibinfo {volume} {108}},\ \bibinfo {pages} {L041107} (\bibinfo {year} {2023})}\BibitemShut {NoStop}%
\bibitem [{\citenamefont {McCann}\ and\ \citenamefont {Fal'ko}(2006)}]{McCann06}%
  \BibitemOpen
  \bibfield  {author} {\bibinfo {author} {\bibfnamefont {E.}~\bibnamefont {McCann}}\ and\ \bibinfo {author} {\bibfnamefont {V.~I.}\ \bibnamefont {Fal'ko}},\ }\bibfield  {title} {\bibinfo {title} {{Landau}-level degeneracy and quantum {Hall} effect in a graphite bilayer},\ }\href {https://doi.org/10.1103/PhysRevLett.96.086805} {\bibfield  {journal} {\bibinfo  {journal} {Phys. Rev. Lett.}\ }\textbf {\bibinfo {volume} {96}},\ \bibinfo {pages} {086805} (\bibinfo {year} {2006})}\BibitemShut {NoStop}%
\bibitem [{\citenamefont {Haldane}(1983)}]{Haldane83}%
  \BibitemOpen
  \bibfield  {author} {\bibinfo {author} {\bibfnamefont {F.~D.~M.}\ \bibnamefont {Haldane}},\ }\bibfield  {title} {\bibinfo {title} {Fractional quantization of the {Hall} effect: A hierarchy of incompressible quantum fluid states},\ }\href {https://doi.org/10.1103/PhysRevLett.51.605} {\bibfield  {journal} {\bibinfo  {journal} {Phys. Rev. Lett.}\ }\textbf {\bibinfo {volume} {51}},\ \bibinfo {pages} {605} (\bibinfo {year} {1983})}\BibitemShut {NoStop}%
\bibitem [{\citenamefont {Arciniaga}\ and\ \citenamefont {Peterson}(2016)}]{Arciniaga16}%
  \BibitemOpen
  \bibfield  {author} {\bibinfo {author} {\bibfnamefont {M.}~\bibnamefont {Arciniaga}}\ and\ \bibinfo {author} {\bibfnamefont {M.~R.}\ \bibnamefont {Peterson}},\ }\bibfield  {title} {\bibinfo {title} {Landau level quantization for massless {Dirac} fermions in the spherical geometry: Graphene fractional quantum {Hall} effect on the {Haldane} sphere},\ }\href {https://doi.org/10.1103/PhysRevB.94.035105} {\bibfield  {journal} {\bibinfo  {journal} {Phys. Rev. B}\ }\textbf {\bibinfo {volume} {94}},\ \bibinfo {pages} {035105} (\bibinfo {year} {2016})}\BibitemShut {NoStop}%
\bibitem [{\citenamefont {Wu}\ and\ \citenamefont {Yang}(1976)}]{Wu76}%
  \BibitemOpen
  \bibfield  {author} {\bibinfo {author} {\bibfnamefont {T.~T.}\ \bibnamefont {Wu}}\ and\ \bibinfo {author} {\bibfnamefont {C.~N.}\ \bibnamefont {Yang}},\ }\bibfield  {title} {\bibinfo {title} {Dirac monopole without strings: Monopole harmonics},\ }\href {https://doi.org/10.1016/0550-3213(76)90143-7} {\bibfield  {journal} {\bibinfo  {journal} {Nucl. Phys. B}\ }\textbf {\bibinfo {volume} {107}},\ \bibinfo {pages} {365} (\bibinfo {year} {1976})}\BibitemShut {NoStop}%
\bibitem [{\citenamefont {Wu}\ and\ \citenamefont {Yang}(1977)}]{Wu77}%
  \BibitemOpen
  \bibfield  {author} {\bibinfo {author} {\bibfnamefont {T.~T.}\ \bibnamefont {Wu}}\ and\ \bibinfo {author} {\bibfnamefont {C.~N.}\ \bibnamefont {Yang}},\ }\bibfield  {title} {\bibinfo {title} {Some properties of monopole harmonics},\ }\href {https://doi.org/10.1103/PhysRevD.16.1018} {\bibfield  {journal} {\bibinfo  {journal} {Phys. Rev. D}\ }\textbf {\bibinfo {volume} {16}},\ \bibinfo {pages} {1018} (\bibinfo {year} {1977})}\BibitemShut {NoStop}%
\bibitem [{\citenamefont {Hsiao}(2020)}]{Hsiao20}%
  \BibitemOpen
  \bibfield  {author} {\bibinfo {author} {\bibfnamefont {W.-H.}\ \bibnamefont {Hsiao}},\ }\bibfield  {title} {\bibinfo {title} {{Landau} quantization of multilayer graphene on a {Haldane} sphere},\ }\href {https://doi.org/10.1103/PhysRevB.101.155310} {\bibfield  {journal} {\bibinfo  {journal} {Phys. Rev. B}\ }\textbf {\bibinfo {volume} {101}},\ \bibinfo {pages} {155310} (\bibinfo {year} {2020})}\BibitemShut {NoStop}%
\bibitem [{\citenamefont {Dora}\ and\ \citenamefont {Balram}(2023)}]{Dora23}%
  \BibitemOpen
  \bibfield  {author} {\bibinfo {author} {\bibfnamefont {R.~K.}\ \bibnamefont {Dora}}\ and\ \bibinfo {author} {\bibfnamefont {A.~C.}\ \bibnamefont {Balram}},\ }\bibfield  {title} {\bibinfo {title} {Competition between fractional quantum {Hall} liquid and electron solid phases in the {Landau} levels of multilayer graphene},\ }\href {https://doi.org/10.1103/PhysRevB.108.235153} {\bibfield  {journal} {\bibinfo  {journal} {Phys. Rev. B}\ }\textbf {\bibinfo {volume} {108}},\ \bibinfo {pages} {235153} (\bibinfo {year} {2023})}\BibitemShut {NoStop}%
\bibitem [{\citenamefont {Goerbig}\ \emph {et~al.}(2004{\natexlab{b}})\citenamefont {Goerbig}, \citenamefont {Lederer},\ and\ \citenamefont {Smith}}]{Goerbig04}%
  \BibitemOpen
  \bibfield  {author} {\bibinfo {author} {\bibfnamefont {M.~O.}\ \bibnamefont {Goerbig}}, \bibinfo {author} {\bibfnamefont {P.}~\bibnamefont {Lederer}},\ and\ \bibinfo {author} {\bibfnamefont {C.~M.}\ \bibnamefont {Smith}},\ }\bibfield  {title} {\bibinfo {title} {Second generation of composite fermions in the {Hamiltonian} theory},\ }\href {https://doi.org/10.1103/PhysRevB.69.155324} {\bibfield  {journal} {\bibinfo  {journal} {Phys. Rev. B}\ }\textbf {\bibinfo {volume} {69}},\ \bibinfo {pages} {155324} (\bibinfo {year} {2004}{\natexlab{b}})}\BibitemShut {NoStop}%
\bibitem [{\citenamefont {Goerbig}(2011)}]{Goerbig11}%
  \BibitemOpen
  \bibfield  {author} {\bibinfo {author} {\bibfnamefont {M.~O.}\ \bibnamefont {Goerbig}},\ }\bibfield  {title} {\bibinfo {title} {Electronic properties of graphene in a strong magnetic field},\ }\href {https://doi.org/10.1103/RevModPhys.83.1193} {\bibfield  {journal} {\bibinfo  {journal} {Rev. Mod. Phys.}\ }\textbf {\bibinfo {volume} {83}},\ \bibinfo {pages} {1193} (\bibinfo {year} {2011})}\BibitemShut {NoStop}%
\bibitem [{\citenamefont {Dora}\ and\ \citenamefont {Balram}(2025)}]{Dora24}%
  \BibitemOpen
  \bibfield  {author} {\bibinfo {author} {\bibfnamefont {R.~K.}\ \bibnamefont {Dora}}\ and\ \bibinfo {author} {\bibfnamefont {A.~C.}\ \bibnamefont {Balram}},\ }\bibfield  {title} {\bibinfo {title} {Static structure factor and the dispersion of the {Girvin}-{MacDonald}-{Platzman} density mode for fractional quantum {Hall} fluids on the {Haldane} sphere},\ }\href {https://doi.org/10.1103/PhysRevB.111.115132} {\bibfield  {journal} {\bibinfo  {journal} {Phys. Rev. B}\ }\textbf {\bibinfo {volume} {111}},\ \bibinfo {pages} {115132} (\bibinfo {year} {2025})}\BibitemShut {NoStop}%
\bibitem [{\citenamefont {Balram}\ \emph {et~al.}(2015{\natexlab{a}})\citenamefont {Balram}, \citenamefont {T\ifmmode~\mbox{\H{o}}\else \H{o}\fi{}ke}, \citenamefont {W\'ojs},\ and\ \citenamefont {Jain}}]{Balram15c}%
  \BibitemOpen
  \bibfield  {author} {\bibinfo {author} {\bibfnamefont {A.~C.}\ \bibnamefont {Balram}}, \bibinfo {author} {\bibfnamefont {C.}~\bibnamefont {T\ifmmode~\mbox{\H{o}}\else \H{o}\fi{}ke}}, \bibinfo {author} {\bibfnamefont {A.}~\bibnamefont {W\'ojs}},\ and\ \bibinfo {author} {\bibfnamefont {J.~K.}\ \bibnamefont {Jain}},\ }\bibfield  {title} {\bibinfo {title} {Spontaneous polarization of composite fermions in the $n=1$ {Landau} level of graphene},\ }\href {https://doi.org/10.1103/PhysRevB.92.205120} {\bibfield  {journal} {\bibinfo  {journal} {Phys. Rev. B}\ }\textbf {\bibinfo {volume} {92}},\ \bibinfo {pages} {205120} (\bibinfo {year} {2015}{\natexlab{a}})}\BibitemShut {NoStop}%
\bibitem [{\citenamefont {Dora}\ and\ \citenamefont {Balram}(2026)}]{Dora25}%
  \BibitemOpen
  \bibfield  {author} {\bibinfo {author} {\bibfnamefont {R.~K.}\ \bibnamefont {Dora}}\ and\ \bibinfo {author} {\bibfnamefont {A.~C.}\ \bibnamefont {Balram}},\ }\bibfield  {title} {\bibinfo {title} {Dispersion of collective modes in spinful fractional quantum {Hall} states on the sphere},\ }\href {https://doi.org/10.1103/17p5-jkym} {\bibfield  {journal} {\bibinfo  {journal} {Phys. Rev. B}\ }\textbf {\bibinfo {volume} {113}},\ \bibinfo {pages} {115420} (\bibinfo {year} {2026})}\BibitemShut {NoStop}%
\bibitem [{\citenamefont {Wen}\ and\ \citenamefont {Zee}(1992)}]{Wen92}%
  \BibitemOpen
  \bibfield  {author} {\bibinfo {author} {\bibfnamefont {X.~G.}\ \bibnamefont {Wen}}\ and\ \bibinfo {author} {\bibfnamefont {A.}~\bibnamefont {Zee}},\ }\bibfield  {title} {\bibinfo {title} {Shift and spin vector: New topological quantum numbers for the {Hall} fluids},\ }\href {https://doi.org/10.1103/PhysRevLett.69.953} {\bibfield  {journal} {\bibinfo  {journal} {Phys. Rev. Lett.}\ }\textbf {\bibinfo {volume} {69}},\ \bibinfo {pages} {953} (\bibinfo {year} {1992})}\BibitemShut {NoStop}%
\bibitem [{\citenamefont {Kane}\ and\ \citenamefont {Fisher}(1997)}]{Kane97}%
  \BibitemOpen
  \bibfield  {author} {\bibinfo {author} {\bibfnamefont {C.~L.}\ \bibnamefont {Kane}}\ and\ \bibinfo {author} {\bibfnamefont {M.~P.~A.}\ \bibnamefont {Fisher}},\ }\bibfield  {title} {\bibinfo {title} {Quantized thermal transport in the fractional quantum {Hall} effect},\ }\href {https://doi.org/10.1103/PhysRevB.55.15832} {\bibfield  {journal} {\bibinfo  {journal} {Phys. Rev. B}\ }\textbf {\bibinfo {volume} {55}},\ \bibinfo {pages} {15832} (\bibinfo {year} {1997})}\BibitemShut {NoStop}%
\bibitem [{\citenamefont {M\"oller}\ and\ \citenamefont {Simon}(2005)}]{Moller05}%
  \BibitemOpen
  \bibfield  {author} {\bibinfo {author} {\bibfnamefont {G.}~\bibnamefont {M\"oller}}\ and\ \bibinfo {author} {\bibfnamefont {S.~H.}\ \bibnamefont {Simon}},\ }\bibfield  {title} {\bibinfo {title} {Composite fermions in a negative effective magnetic field: A {Monte} {Carlo} study},\ }\href {https://doi.org/10.1103/PhysRevB.72.045344} {\bibfield  {journal} {\bibinfo  {journal} {Phys. Rev. B}\ }\textbf {\bibinfo {volume} {72}},\ \bibinfo {pages} {045344} (\bibinfo {year} {2005})}\BibitemShut {NoStop}%
\bibitem [{\citenamefont {Balram}\ \emph {et~al.}(2015{\natexlab{b}})\citenamefont {Balram}, \citenamefont {T\ifmmode~\mbox{\H{o}}\else \H{o}\fi{}ke},\ and\ \citenamefont {Jain}}]{Balram15b}%
  \BibitemOpen
  \bibfield  {author} {\bibinfo {author} {\bibfnamefont {A.~C.}\ \bibnamefont {Balram}}, \bibinfo {author} {\bibfnamefont {C.}~\bibnamefont {T\ifmmode~\mbox{\H{o}}\else \H{o}\fi{}ke}},\ and\ \bibinfo {author} {\bibfnamefont {J.~K.}\ \bibnamefont {Jain}},\ }\bibfield  {title} {\bibinfo {title} {Luttinger theorem for the strongly correlated {Fermi} liquid of composite fermions},\ }\href {https://doi.org/10.1103/PhysRevLett.115.186805} {\bibfield  {journal} {\bibinfo  {journal} {Phys. Rev. Lett.}\ }\textbf {\bibinfo {volume} {115}},\ \bibinfo {pages} {186805} (\bibinfo {year} {2015}{\natexlab{b}})}\BibitemShut {NoStop}%
\bibitem [{\citenamefont {Balram}\ \emph {et~al.}(2013)\citenamefont {Balram}, \citenamefont {W\'ojs},\ and\ \citenamefont {Jain}}]{Balram13}%
  \BibitemOpen
  \bibfield  {author} {\bibinfo {author} {\bibfnamefont {A.~C.}\ \bibnamefont {Balram}}, \bibinfo {author} {\bibfnamefont {A.}~\bibnamefont {W\'ojs}},\ and\ \bibinfo {author} {\bibfnamefont {J.~K.}\ \bibnamefont {Jain}},\ }\bibfield  {title} {\bibinfo {title} {State counting for excited bands of the fractional quantum {Hall} effect: Exclusion rules for bound excitons},\ }\href {https://doi.org/10.1103/PhysRevB.88.205312} {\bibfield  {journal} {\bibinfo  {journal} {Phys. Rev. B}\ }\textbf {\bibinfo {volume} {88}},\ \bibinfo {pages} {205312} (\bibinfo {year} {2013})}\BibitemShut {NoStop}%
\bibitem [{\citenamefont {Jain}\ and\ \citenamefont {Kamilla}(1997{\natexlab{a}})}]{Jain97}%
  \BibitemOpen
  \bibfield  {author} {\bibinfo {author} {\bibfnamefont {J.~K.}\ \bibnamefont {Jain}}\ and\ \bibinfo {author} {\bibfnamefont {R.~K.}\ \bibnamefont {Kamilla}},\ }\bibfield  {title} {\bibinfo {title} {Composite fermions in the {Hilbert} space of the lowest electronic {Landau} level},\ }\href {https://doi.org/10.1142/S0217979297001301} {\bibfield  {journal} {\bibinfo  {journal} {Int. J. Mod. Phys. B}\ }\textbf {\bibinfo {volume} {11}},\ \bibinfo {pages} {2621} (\bibinfo {year} {1997}{\natexlab{a}})}\BibitemShut {NoStop}%
\bibitem [{\citenamefont {Davenport}\ and\ \citenamefont {Simon}(2012)}]{Davenport12}%
  \BibitemOpen
  \bibfield  {author} {\bibinfo {author} {\bibfnamefont {S.~C.}\ \bibnamefont {Davenport}}\ and\ \bibinfo {author} {\bibfnamefont {S.~H.}\ \bibnamefont {Simon}},\ }\bibfield  {title} {\bibinfo {title} {Spinful composite fermions in a negative effective field},\ }\href {https://doi.org/10.1103/PhysRevB.85.245303} {\bibfield  {journal} {\bibinfo  {journal} {Phys. Rev. B}\ }\textbf {\bibinfo {volume} {85}},\ \bibinfo {pages} {245303} (\bibinfo {year} {2012})}\BibitemShut {NoStop}%
\bibitem [{\citenamefont {Wen}(1995)}]{Wen95}%
  \BibitemOpen
  \bibfield  {author} {\bibinfo {author} {\bibfnamefont {X.-G.}\ \bibnamefont {Wen}},\ }\bibfield  {title} {\bibinfo {title} {Topological orders and edge excitations in fractional quantum {Hall} states},\ }\href {https://doi.org/10.1080/00018739500101566} {\bibfield  {journal} {\bibinfo  {journal} {Advances in Physics}\ }\textbf {\bibinfo {volume} {44}},\ \bibinfo {pages} {405} (\bibinfo {year} {1995})},\ \Eprint {https://arxiv.org/abs/http://www.tandfonline.com/doi/pdf/10.1080/00018739500101566} {http://www.tandfonline.com/doi/pdf/10.1080/00018739500101566} \BibitemShut {NoStop}%
\bibitem [{\citenamefont {Bonderson}\ and\ \citenamefont {Slingerland}(2008)}]{Bonderson08}%
  \BibitemOpen
  \bibfield  {author} {\bibinfo {author} {\bibfnamefont {P.}~\bibnamefont {Bonderson}}\ and\ \bibinfo {author} {\bibfnamefont {J.~K.}\ \bibnamefont {Slingerland}},\ }\bibfield  {title} {\bibinfo {title} {Fractional quantum {Hall} hierarchy and the second {Landau} level},\ }\href {https://doi.org/10.1103/PhysRevB.78.125323} {\bibfield  {journal} {\bibinfo  {journal} {Phys. Rev. B}\ }\textbf {\bibinfo {volume} {78}},\ \bibinfo {pages} {125323} (\bibinfo {year} {2008})}\BibitemShut {NoStop}%
\bibitem [{\citenamefont {Balram}\ and\ \citenamefont {Jain}(2016)}]{Balram16b}%
  \BibitemOpen
  \bibfield  {author} {\bibinfo {author} {\bibfnamefont {A.~C.}\ \bibnamefont {Balram}}\ and\ \bibinfo {author} {\bibfnamefont {J.~K.}\ \bibnamefont {Jain}},\ }\bibfield  {title} {\bibinfo {title} {Nature of composite fermions and the role of particle-hole symmetry: A microscopic account},\ }\href {https://doi.org/10.1103/PhysRevB.93.235152} {\bibfield  {journal} {\bibinfo  {journal} {Phys. Rev. B}\ }\textbf {\bibinfo {volume} {93}},\ \bibinfo {pages} {235152} (\bibinfo {year} {2016})}\BibitemShut {NoStop}%
\bibitem [{\citenamefont {Anand}\ \emph {et~al.}(2022)\citenamefont {Anand}, \citenamefont {Patil}, \citenamefont {Balram},\ and\ \citenamefont {Sreejith}}]{Anand22}%
  \BibitemOpen
  \bibfield  {author} {\bibinfo {author} {\bibfnamefont {A.}~\bibnamefont {Anand}}, \bibinfo {author} {\bibfnamefont {R.~A.}\ \bibnamefont {Patil}}, \bibinfo {author} {\bibfnamefont {A.~C.}\ \bibnamefont {Balram}},\ and\ \bibinfo {author} {\bibfnamefont {G.~J.}\ \bibnamefont {Sreejith}},\ }\bibfield  {title} {\bibinfo {title} {Real-space entanglement spectra of parton states in fractional quantum {Hall} systems},\ }\href {https://doi.org/10.1103/PhysRevB.106.085136} {\bibfield  {journal} {\bibinfo  {journal} {Phys. Rev. B}\ }\textbf {\bibinfo {volume} {106}},\ \bibinfo {pages} {085136} (\bibinfo {year} {2022})}\BibitemShut {NoStop}%
\bibitem [{\citenamefont {Read}\ and\ \citenamefont {Rezayi}(1999)}]{Read99}%
  \BibitemOpen
  \bibfield  {author} {\bibinfo {author} {\bibfnamefont {N.}~\bibnamefont {Read}}\ and\ \bibinfo {author} {\bibfnamefont {E.}~\bibnamefont {Rezayi}},\ }\bibfield  {title} {\bibinfo {title} {Beyond paired quantum {Hall} states: Parafermions and incompressible states in the first excited {Landau} level},\ }\href {https://doi.org/10.1103/PhysRevB.59.8084} {\bibfield  {journal} {\bibinfo  {journal} {Phys. Rev. B}\ }\textbf {\bibinfo {volume} {59}},\ \bibinfo {pages} {8084} (\bibinfo {year} {1999})}\BibitemShut {NoStop}%
\bibitem [{\citenamefont {Bishara}\ \emph {et~al.}(2008)\citenamefont {Bishara}, \citenamefont {Fiete},\ and\ \citenamefont {Nayak}}]{Bishara08}%
  \BibitemOpen
  \bibfield  {author} {\bibinfo {author} {\bibfnamefont {W.}~\bibnamefont {Bishara}}, \bibinfo {author} {\bibfnamefont {G.~A.}\ \bibnamefont {Fiete}},\ and\ \bibinfo {author} {\bibfnamefont {C.}~\bibnamefont {Nayak}},\ }\bibfield  {title} {\bibinfo {title} {Quantum {Hall} states at $\ensuremath{\nu}=\frac{2}{k+2}$: Analysis of the particle-hole conjugates of the general level-$k$ {Read}-{Rezayi} states},\ }\href {https://doi.org/10.1103/PhysRevB.77.241306} {\bibfield  {journal} {\bibinfo  {journal} {Phys. Rev. B}\ }\textbf {\bibinfo {volume} {77}},\ \bibinfo {pages} {241306} (\bibinfo {year} {2008})}\BibitemShut {NoStop}%
\bibitem [{\citenamefont {Bonderson}\ \emph {et~al.}(2012)\citenamefont {Bonderson}, \citenamefont {Feiguin}, \citenamefont {M\"oller},\ and\ \citenamefont {Slingerland}}]{Bonderson12}%
  \BibitemOpen
  \bibfield  {author} {\bibinfo {author} {\bibfnamefont {P.}~\bibnamefont {Bonderson}}, \bibinfo {author} {\bibfnamefont {A.~E.}\ \bibnamefont {Feiguin}}, \bibinfo {author} {\bibfnamefont {G.}~\bibnamefont {M\"oller}},\ and\ \bibinfo {author} {\bibfnamefont {J.~K.}\ \bibnamefont {Slingerland}},\ }\bibfield  {title} {\bibinfo {title} {Competing topological orders in the $\ensuremath{\nu}=12/5$ quantum {Hall} state},\ }\href {https://doi.org/10.1103/PhysRevLett.108.036806} {\bibfield  {journal} {\bibinfo  {journal} {Phys. Rev. Lett.}\ }\textbf {\bibinfo {volume} {108}},\ \bibinfo {pages} {036806} (\bibinfo {year} {2012})}\BibitemShut {NoStop}%
\bibitem [{\citenamefont {W\'ojs}(2009)}]{Wojs09}%
  \BibitemOpen
  \bibfield  {author} {\bibinfo {author} {\bibfnamefont {A.}~\bibnamefont {W\'ojs}},\ }\bibfield  {title} {\bibinfo {title} {Transition from abelian to non-abelian quantum liquids in the second {Landau} level},\ }\href {https://doi.org/10.1103/PhysRevB.80.041104} {\bibfield  {journal} {\bibinfo  {journal} {Phys. Rev. B}\ }\textbf {\bibinfo {volume} {80}},\ \bibinfo {pages} {041104} (\bibinfo {year} {2009})}\BibitemShut {NoStop}%
\bibitem [{\citenamefont {Zhu}\ \emph {et~al.}(2015)\citenamefont {Zhu}, \citenamefont {Gong}, \citenamefont {Haldane},\ and\ \citenamefont {Sheng}}]{Zhu15}%
  \BibitemOpen
  \bibfield  {author} {\bibinfo {author} {\bibfnamefont {W.}~\bibnamefont {Zhu}}, \bibinfo {author} {\bibfnamefont {S.~S.}\ \bibnamefont {Gong}}, \bibinfo {author} {\bibfnamefont {F.~D.~M.}\ \bibnamefont {Haldane}},\ and\ \bibinfo {author} {\bibfnamefont {D.~N.}\ \bibnamefont {Sheng}},\ }\bibfield  {title} {\bibinfo {title} {Fractional quantum {Hall} states at $\ensuremath{\nu}=13/5$ and $12/5$ and their non-abelian nature},\ }\href {https://doi.org/10.1103/PhysRevLett.115.126805} {\bibfield  {journal} {\bibinfo  {journal} {Phys. Rev. Lett.}\ }\textbf {\bibinfo {volume} {115}},\ \bibinfo {pages} {126805} (\bibinfo {year} {2015})}\BibitemShut {NoStop}%
\bibitem [{\citenamefont {Mong}\ \emph {et~al.}(2017)\citenamefont {Mong}, \citenamefont {Zaletel}, \citenamefont {Pollmann},\ and\ \citenamefont {Papi\ifmmode~\acute{c}\else \'{c}\fi{}}}]{Mong15}%
  \BibitemOpen
  \bibfield  {author} {\bibinfo {author} {\bibfnamefont {R.~S.~K.}\ \bibnamefont {Mong}}, \bibinfo {author} {\bibfnamefont {M.~P.}\ \bibnamefont {Zaletel}}, \bibinfo {author} {\bibfnamefont {F.}~\bibnamefont {Pollmann}},\ and\ \bibinfo {author} {\bibfnamefont {Z.}~\bibnamefont {Papi\ifmmode~\acute{c}\else \'{c}\fi{}}},\ }\bibfield  {title} {\bibinfo {title} {Fibonacci anyons and charge density order in the 12/5 and 13/5 quantum {Hall} plateaus},\ }\href {https://doi.org/10.1103/PhysRevB.95.115136} {\bibfield  {journal} {\bibinfo  {journal} {Phys. Rev. B}\ }\textbf {\bibinfo {volume} {95}},\ \bibinfo {pages} {115136} (\bibinfo {year} {2017})}\BibitemShut {NoStop}%
\bibitem [{\citenamefont {Pakrouski}\ \emph {et~al.}(2016)\citenamefont {Pakrouski}, \citenamefont {Troyer}, \citenamefont {Wu}, \citenamefont {Das~Sarma},\ and\ \citenamefont {Peterson}}]{Pakrouski16}%
  \BibitemOpen
  \bibfield  {author} {\bibinfo {author} {\bibfnamefont {K.}~\bibnamefont {Pakrouski}}, \bibinfo {author} {\bibfnamefont {M.}~\bibnamefont {Troyer}}, \bibinfo {author} {\bibfnamefont {Y.-L.}\ \bibnamefont {Wu}}, \bibinfo {author} {\bibfnamefont {S.}~\bibnamefont {Das~Sarma}},\ and\ \bibinfo {author} {\bibfnamefont {M.~R.}\ \bibnamefont {Peterson}},\ }\bibfield  {title} {\bibinfo {title} {Enigmatic 12/5 fractional quantum {Hall} effect},\ }\href {https://doi.org/10.1103/PhysRevB.94.075108} {\bibfield  {journal} {\bibinfo  {journal} {Phys. Rev. B}\ }\textbf {\bibinfo {volume} {94}},\ \bibinfo {pages} {075108} (\bibinfo {year} {2016})}\BibitemShut {NoStop}%
\bibitem [{\citenamefont {Bose}\ \emph {et~al.}(2026)\citenamefont {Bose}, \citenamefont {Simon},\ and\ \citenamefont {Balram}}]{Bose25a}%
  \BibitemOpen
  \bibfield  {author} {\bibinfo {author} {\bibfnamefont {K.}~\bibnamefont {Bose}}, \bibinfo {author} {\bibfnamefont {S.~H.}\ \bibnamefont {Simon}},\ and\ \bibinfo {author} {\bibfnamefont {A.~C.}\ \bibnamefont {Balram}},\ }\bibfield  {title} {\bibinfo {title} {Monte {Carlo} sampling for wave functions requiring symmetrization or antisymmetrization},\ }\href {https://doi.org/10.1103/d9xb-4vq9} {\bibfield  {journal} {\bibinfo  {journal} {Phys. Rev. Lett.}\ }\textbf {\bibinfo {volume} {137}},\ \bibinfo {pages} {056502} (\bibinfo {year} {2026})}\BibitemShut {NoStop}%
\bibitem [{\citenamefont {Faugno}\ \emph {et~al.}(2020)\citenamefont {Faugno}, \citenamefont {Jain},\ and\ \citenamefont {Balram}}]{Faugno20a}%
  \BibitemOpen
  \bibfield  {author} {\bibinfo {author} {\bibfnamefont {W.~N.}\ \bibnamefont {Faugno}}, \bibinfo {author} {\bibfnamefont {J.~K.}\ \bibnamefont {Jain}},\ and\ \bibinfo {author} {\bibfnamefont {A.~C.}\ \bibnamefont {Balram}},\ }\bibfield  {title} {\bibinfo {title} {Non-abelian fractional quantum {Hall} state at $3/7$-filled {Landau} level},\ }\href {https://doi.org/10.1103/PhysRevResearch.2.033223} {\bibfield  {journal} {\bibinfo  {journal} {Phys. Rev. Research}\ }\textbf {\bibinfo {volume} {2}},\ \bibinfo {pages} {033223} (\bibinfo {year} {2020})}\BibitemShut {NoStop}%
\bibitem [{\citenamefont {Balram}\ and\ \citenamefont {W\'ojs}(2020)}]{Balram20b}%
  \BibitemOpen
  \bibfield  {author} {\bibinfo {author} {\bibfnamefont {A.~C.}\ \bibnamefont {Balram}}\ and\ \bibinfo {author} {\bibfnamefont {A.}~\bibnamefont {W\'ojs}},\ }\bibfield  {title} {\bibinfo {title} {Fractional quantum {Hall} effect at $\ensuremath{\nu}=2+4/9$},\ }\href {https://doi.org/10.1103/PhysRevResearch.2.032035} {\bibfield  {journal} {\bibinfo  {journal} {Phys. Rev. Research}\ }\textbf {\bibinfo {volume} {2}},\ \bibinfo {pages} {032035} (\bibinfo {year} {2020})}\BibitemShut {NoStop}%
\bibitem [{\citenamefont {Wen}(1991)}]{Wen91}%
  \BibitemOpen
  \bibfield  {author} {\bibinfo {author} {\bibfnamefont {X.~G.}\ \bibnamefont {Wen}},\ }\bibfield  {title} {\bibinfo {title} {Non-abelian statistics in the fractional quantum {Hall} states},\ }\href {https://doi.org/10.1103/PhysRevLett.66.802} {\bibfield  {journal} {\bibinfo  {journal} {Phys. Rev. Lett.}\ }\textbf {\bibinfo {volume} {66}},\ \bibinfo {pages} {802} (\bibinfo {year} {1991})}\BibitemShut {NoStop}%
\bibitem [{\citenamefont {Levin}\ and\ \citenamefont {Stern}(2009)}]{Levin09}%
  \BibitemOpen
  \bibfield  {author} {\bibinfo {author} {\bibfnamefont {M.}~\bibnamefont {Levin}}\ and\ \bibinfo {author} {\bibfnamefont {A.}~\bibnamefont {Stern}},\ }\bibfield  {title} {\bibinfo {title} {Fractional topological insulators},\ }\href@noop {} {\bibfield  {journal} {\bibinfo  {journal} {Phys. Rev. Lett.}\ }\textbf {\bibinfo {volume} {103}},\ \bibinfo {pages} {1} (\bibinfo {year} {2009})}\BibitemShut {NoStop}%
\bibitem [{\citenamefont {Lee}\ \emph {et~al.}(2007)\citenamefont {Lee}, \citenamefont {Ryu}, \citenamefont {Nayak},\ and\ \citenamefont {Fisher}}]{Lee07}%
  \BibitemOpen
  \bibfield  {author} {\bibinfo {author} {\bibfnamefont {S.-S.}\ \bibnamefont {Lee}}, \bibinfo {author} {\bibfnamefont {S.}~\bibnamefont {Ryu}}, \bibinfo {author} {\bibfnamefont {C.}~\bibnamefont {Nayak}},\ and\ \bibinfo {author} {\bibfnamefont {M.~P.~A.}\ \bibnamefont {Fisher}},\ }\bibfield  {title} {\bibinfo {title} {Particle-hole symmetry and the $\nu=5/2$ quantum {Hall} state},\ }\href {https://doi.org/10.1103/PhysRevLett.99.236807} {\bibfield  {journal} {\bibinfo  {journal} {Phys. Rev. Lett.}\ }\textbf {\bibinfo {volume} {99}},\ \bibinfo {pages} {236807} (\bibinfo {year} {2007})}\BibitemShut {NoStop}%
\bibitem [{\citenamefont {Levin}\ \emph {et~al.}(2007)\citenamefont {Levin}, \citenamefont {Halperin},\ and\ \citenamefont {Rosenow}}]{Levin07}%
  \BibitemOpen
  \bibfield  {author} {\bibinfo {author} {\bibfnamefont {M.}~\bibnamefont {Levin}}, \bibinfo {author} {\bibfnamefont {B.~I.}\ \bibnamefont {Halperin}},\ and\ \bibinfo {author} {\bibfnamefont {B.}~\bibnamefont {Rosenow}},\ }\bibfield  {title} {\bibinfo {title} {Particle-hole symmetry and the {Pfaffian} state},\ }\href {https://doi.org/10.1103/PhysRevLett.99.236806} {\bibfield  {journal} {\bibinfo  {journal} {Phys. Rev. Lett.}\ }\textbf {\bibinfo {volume} {99}},\ \bibinfo {pages} {236806} (\bibinfo {year} {2007})}\BibitemShut {NoStop}%
\bibitem [{\citenamefont {Read}(2009)}]{Read09}%
  \BibitemOpen
  \bibfield  {author} {\bibinfo {author} {\bibfnamefont {N.}~\bibnamefont {Read}},\ }\bibfield  {title} {\bibinfo {title} {Non-abelian adiabatic statistics and {Hall} viscosity in quantum {Hall} states and ${p}_{x}+i{p}_{y}$ paired superfluids},\ }\href {https://doi.org/10.1103/PhysRevB.79.045308} {\bibfield  {journal} {\bibinfo  {journal} {Phys. Rev. B}\ }\textbf {\bibinfo {volume} {79}},\ \bibinfo {pages} {045308} (\bibinfo {year} {2009})}\BibitemShut {NoStop}%
\bibitem [{\citenamefont {Rezayi}\ and\ \citenamefont {Read}(1994)}]{Rezayi94}%
  \BibitemOpen
  \bibfield  {author} {\bibinfo {author} {\bibfnamefont {E.}~\bibnamefont {Rezayi}}\ and\ \bibinfo {author} {\bibfnamefont {N.}~\bibnamefont {Read}},\ }\bibfield  {title} {\bibinfo {title} {Fermi-liquid-like state in a half-filled {Landau} level},\ }\href {https://doi.org/10.1103/PhysRevLett.72.900} {\bibfield  {journal} {\bibinfo  {journal} {Phys. Rev. Lett.}\ }\textbf {\bibinfo {volume} {72}},\ \bibinfo {pages} {900} (\bibinfo {year} {1994})}\BibitemShut {NoStop}%
\bibitem [{\citenamefont {Rezayi}\ and\ \citenamefont {Haldane}(2000)}]{Rezayi00}%
  \BibitemOpen
  \bibfield  {author} {\bibinfo {author} {\bibfnamefont {E.~H.}\ \bibnamefont {Rezayi}}\ and\ \bibinfo {author} {\bibfnamefont {F.~D.~M.}\ \bibnamefont {Haldane}},\ }\bibfield  {title} {\bibinfo {title} {Incompressible paired {Hall} state, stripe order, and the composite fermion liquid phase in half-filled {Landau} levels},\ }\href {https://doi.org/10.1103/PhysRevLett.84.4685} {\bibfield  {journal} {\bibinfo  {journal} {Phys. Rev. Lett.}\ }\textbf {\bibinfo {volume} {84}},\ \bibinfo {pages} {4685} (\bibinfo {year} {2000})}\BibitemShut {NoStop}%
\bibitem [{\citenamefont {Liu}\ \emph {et~al.}(2021)\citenamefont {Liu}, \citenamefont {Balram}, \citenamefont {Papi\ifmmode~\acute{c}\else \'{c}\fi{}},\ and\ \citenamefont {Gromov}}]{Liu20}%
  \BibitemOpen
  \bibfield  {author} {\bibinfo {author} {\bibfnamefont {Z.}~\bibnamefont {Liu}}, \bibinfo {author} {\bibfnamefont {A.~C.}\ \bibnamefont {Balram}}, \bibinfo {author} {\bibfnamefont {Z.}~\bibnamefont {Papi\ifmmode~\acute{c}\else \'{c}\fi{}}},\ and\ \bibinfo {author} {\bibfnamefont {A.}~\bibnamefont {Gromov}},\ }\bibfield  {title} {\bibinfo {title} {Quench dynamics of collective modes in fractional quantum {Hall} bilayers},\ }\href {https://doi.org/10.1103/PhysRevLett.126.076604} {\bibfield  {journal} {\bibinfo  {journal} {Phys. Rev. Lett.}\ }\textbf {\bibinfo {volume} {126}},\ \bibinfo {pages} {076604} (\bibinfo {year} {2021})}\BibitemShut {NoStop}%
\bibitem [{\citenamefont {M\"oller}\ and\ \citenamefont {Simon}(2008)}]{Moller08}%
  \BibitemOpen
  \bibfield  {author} {\bibinfo {author} {\bibfnamefont {G.}~\bibnamefont {M\"oller}}\ and\ \bibinfo {author} {\bibfnamefont {S.~H.}\ \bibnamefont {Simon}},\ }\bibfield  {title} {\bibinfo {title} {Paired composite-fermion wave functions},\ }\href {https://doi.org/10.1103/PhysRevB.77.075319} {\bibfield  {journal} {\bibinfo  {journal} {Phys. Rev. B}\ }\textbf {\bibinfo {volume} {77}},\ \bibinfo {pages} {075319} (\bibinfo {year} {2008})}\BibitemShut {NoStop}%
\bibitem [{\citenamefont {Sharma}\ \emph {et~al.}(2021)\citenamefont {Sharma}, \citenamefont {Pu},\ and\ \citenamefont {Jain}}]{Sharma21}%
  \BibitemOpen
  \bibfield  {author} {\bibinfo {author} {\bibfnamefont {A.}~\bibnamefont {Sharma}}, \bibinfo {author} {\bibfnamefont {S.}~\bibnamefont {Pu}},\ and\ \bibinfo {author} {\bibfnamefont {J.~K.}\ \bibnamefont {Jain}},\ }\bibfield  {title} {\bibinfo {title} {Bardeen-{Cooper}-{Schrieffer} pairing of composite fermions},\ }\href {https://doi.org/10.1103/PhysRevB.104.205303} {\bibfield  {journal} {\bibinfo  {journal} {Phys. Rev. B}\ }\textbf {\bibinfo {volume} {104}},\ \bibinfo {pages} {205303} (\bibinfo {year} {2021})}\BibitemShut {NoStop}%
\bibitem [{\citenamefont {Girvin}(1984)}]{Girvin84a}%
  \BibitemOpen
  \bibfield  {author} {\bibinfo {author} {\bibfnamefont {S.~M.}\ \bibnamefont {Girvin}},\ }\bibfield  {title} {\bibinfo {title} {Anomalous quantum {Hall} effect and two-dimensional classical plasmas: Analytic approximations for correlation functions and ground-state energies},\ }\href {https://doi.org/10.1103/PhysRevB.30.558} {\bibfield  {journal} {\bibinfo  {journal} {Phys. Rev. B}\ }\textbf {\bibinfo {volume} {30}},\ \bibinfo {pages} {558} (\bibinfo {year} {1984})}\BibitemShut {NoStop}%
\bibitem [{\citenamefont {Girvin}\ \emph {et~al.}(1985)\citenamefont {Girvin}, \citenamefont {MacDonald},\ and\ \citenamefont {Platzman}}]{Girvin85}%
  \BibitemOpen
  \bibfield  {author} {\bibinfo {author} {\bibfnamefont {S.~M.}\ \bibnamefont {Girvin}}, \bibinfo {author} {\bibfnamefont {A.~H.}\ \bibnamefont {MacDonald}},\ and\ \bibinfo {author} {\bibfnamefont {P.~M.}\ \bibnamefont {Platzman}},\ }\bibfield  {title} {\bibinfo {title} {Collective-excitation gap in the fractional quantum {Hall} effect},\ }\href {https://doi.org/10.1103/PhysRevLett.54.581} {\bibfield  {journal} {\bibinfo  {journal} {Phys. Rev. Lett.}\ }\textbf {\bibinfo {volume} {54}},\ \bibinfo {pages} {581} (\bibinfo {year} {1985})}\BibitemShut {NoStop}%
\bibitem [{\citenamefont {Girvin}\ \emph {et~al.}(1986)\citenamefont {Girvin}, \citenamefont {MacDonald},\ and\ \citenamefont {Platzman}}]{Girvin86}%
  \BibitemOpen
  \bibfield  {author} {\bibinfo {author} {\bibfnamefont {S.~M.}\ \bibnamefont {Girvin}}, \bibinfo {author} {\bibfnamefont {A.~H.}\ \bibnamefont {MacDonald}},\ and\ \bibinfo {author} {\bibfnamefont {P.~M.}\ \bibnamefont {Platzman}},\ }\bibfield  {title} {\bibinfo {title} {Magneto-roton theory of collective excitations in the fractional quantum {Hall} effect},\ }\href {https://doi.org/10.1103/PhysRevB.33.2481} {\bibfield  {journal} {\bibinfo  {journal} {Phys. Rev. B}\ }\textbf {\bibinfo {volume} {33}},\ \bibinfo {pages} {2481} (\bibinfo {year} {1986})}\BibitemShut {NoStop}%
\bibitem [{\citenamefont {Anakru}\ \emph {et~al.}(2025)\citenamefont {Anakru}, \citenamefont {Gattu}, \citenamefont {Balram}, \citenamefont {Wu}, \citenamefont {Kumar}, \citenamefont {Bi},\ and\ \citenamefont {Jain}}]{Anakru25}%
  \BibitemOpen
  \bibfield  {author} {\bibinfo {author} {\bibfnamefont {A.}~\bibnamefont {Anakru}}, \bibinfo {author} {\bibfnamefont {M.}~\bibnamefont {Gattu}}, \bibinfo {author} {\bibfnamefont {A.~C.}\ \bibnamefont {Balram}}, \bibinfo {author} {\bibfnamefont {X.-C.}\ \bibnamefont {Wu}}, \bibinfo {author} {\bibfnamefont {P.}~\bibnamefont {Kumar}}, \bibinfo {author} {\bibfnamefont {Z.}~\bibnamefont {Bi}},\ and\ \bibinfo {author} {\bibfnamefont {J.~K.}\ \bibnamefont {Jain}},\ }\bibfield  {title} {\bibinfo {title} {Emergent gauge field in composite-fermion metals: A large-scale microscopic study},\ }\href {https://doi.org/10.1103/cjys-fx46} {\bibfield  {journal} {\bibinfo  {journal} {Phys. Rev. Lett.}\ }\textbf {\bibinfo {volume} {135}},\ \bibinfo {pages} {246503} (\bibinfo {year} {2025})}\BibitemShut {NoStop}%
\bibitem [{\citenamefont {Makki}\ \emph {et~al.}(2026)\citenamefont {Makki}, \citenamefont {Gattu},\ and\ \citenamefont {Jain}}]{Makki26}%
  \BibitemOpen
  \bibfield  {author} {\bibinfo {author} {\bibfnamefont {A.~A.}\ \bibnamefont {Makki}}, \bibinfo {author} {\bibfnamefont {M.}~\bibnamefont {Gattu}},\ and\ \bibinfo {author} {\bibfnamefont {J.~K.}\ \bibnamefont {Jain}},\ }\href {https://arxiv.org/abs/2608.19604} {\bibinfo {title} {Universality of long-wavelength behavior of composite-fermion {Fermi} liquid}} (\bibinfo {year} {2026}),\ \Eprint {https://arxiv.org/abs/2608.19604} {arXiv:2608.19604 [cond-mat.str-el]} \BibitemShut {NoStop}%
\bibitem [{\citenamefont {He}\ \emph {et~al.}(1994)\citenamefont {He}, \citenamefont {Simon},\ and\ \citenamefont {Halperin}}]{He94}%
  \BibitemOpen
  \bibfield  {author} {\bibinfo {author} {\bibfnamefont {S.}~\bibnamefont {He}}, \bibinfo {author} {\bibfnamefont {S.~H.}\ \bibnamefont {Simon}},\ and\ \bibinfo {author} {\bibfnamefont {B.~I.}\ \bibnamefont {Halperin}},\ }\bibfield  {title} {\bibinfo {title} {Response function of the fractional quantized {Hall} state on a sphere. ii. exact diagonalization},\ }\href {https://doi.org/10.1103/PhysRevB.50.1823} {\bibfield  {journal} {\bibinfo  {journal} {Phys. Rev. B}\ }\textbf {\bibinfo {volume} {50}},\ \bibinfo {pages} {1823} (\bibinfo {year} {1994})}\BibitemShut {NoStop}%
\bibitem [{\citenamefont {Kundu}\ and\ \citenamefont {Balram}(2026)}]{Kundu26}%
  \BibitemOpen
  \bibfield  {author} {\bibinfo {author} {\bibfnamefont {R.}~\bibnamefont {Kundu}}\ and\ \bibinfo {author} {\bibfnamefont {A.~C.}\ \bibnamefont {Balram}},\ }\href {https://arxiv.org/abs/2604.22628} {\bibinfo {title} {Relations between density-density correlators of states in the maximal spin multiplet}} (\bibinfo {year} {2026}),\ \Eprint {https://arxiv.org/abs/2604.22628} {arXiv:2604.22628 [cond-mat.str-el]} \BibitemShut {NoStop}%
\bibitem [{\citenamefont {Kamilla}\ \emph {et~al.}(1997)\citenamefont {Kamilla}, \citenamefont {Jain},\ and\ \citenamefont {Girvin}}]{Kamilla97}%
  \BibitemOpen
  \bibfield  {author} {\bibinfo {author} {\bibfnamefont {R.~K.}\ \bibnamefont {Kamilla}}, \bibinfo {author} {\bibfnamefont {J.~K.}\ \bibnamefont {Jain}},\ and\ \bibinfo {author} {\bibfnamefont {S.~M.}\ \bibnamefont {Girvin}},\ }\bibfield  {title} {\bibinfo {title} {Fermi-sea-like correlations in a partially filled {Landau} level},\ }\href {https://doi.org/10.1103/PhysRevB.56.12411} {\bibfield  {journal} {\bibinfo  {journal} {Phys. Rev. B}\ }\textbf {\bibinfo {volume} {56}},\ \bibinfo {pages} {12411} (\bibinfo {year} {1997})}\BibitemShut {NoStop}%
\bibitem [{\citenamefont {Balram}\ and\ \citenamefont {Jain}(2017)}]{Balram17}%
  \BibitemOpen
  \bibfield  {author} {\bibinfo {author} {\bibfnamefont {A.~C.}\ \bibnamefont {Balram}}\ and\ \bibinfo {author} {\bibfnamefont {J.~K.}\ \bibnamefont {Jain}},\ }\bibfield  {title} {\bibinfo {title} {Fermi wave vector for the partially spin-polarized composite-fermion {Fermi} sea},\ }\href {https://doi.org/10.1103/PhysRevB.96.235102} {\bibfield  {journal} {\bibinfo  {journal} {Phys. Rev. B}\ }\textbf {\bibinfo {volume} {96}},\ \bibinfo {pages} {235102} (\bibinfo {year} {2017})}\BibitemShut {NoStop}%
\bibitem [{\citenamefont {Morf}\ and\ \citenamefont {Halperin}(1986)}]{Morf86b}%
  \BibitemOpen
  \bibfield  {author} {\bibinfo {author} {\bibfnamefont {R.}~\bibnamefont {Morf}}\ and\ \bibinfo {author} {\bibfnamefont {B.~I.}\ \bibnamefont {Halperin}},\ }\bibfield  {title} {\bibinfo {title} {{Monte} {Carlo} evaluation of trial wave functions for the fractional quantized {Hall} effect: Disk geometry},\ }\href {https://doi.org/10.1103/PhysRevB.33.2221} {\bibfield  {journal} {\bibinfo  {journal} {Phys. Rev. B}\ }\textbf {\bibinfo {volume} {33}},\ \bibinfo {pages} {2221} (\bibinfo {year} {1986})}\BibitemShut {NoStop}%
\bibitem [{\citenamefont {Wooten}\ and\ \citenamefont {Macek}(2014)}]{Wooten14}%
  \BibitemOpen
  \bibfield  {author} {\bibinfo {author} {\bibfnamefont {R.}~\bibnamefont {Wooten}}\ and\ \bibinfo {author} {\bibfnamefont {J.}~\bibnamefont {Macek}},\ }\bibfield  {title} {\bibinfo {title} {Configuration interaction matrix elements for the quantum {Hall} effect},\ }\href@noop {} {\bibfield  {journal} {\bibinfo  {journal} {ArXiv e-prints}\ ,\ \bibinfo {pages} {13}} (\bibinfo {year} {2014})},\ \Eprint {https://arxiv.org/abs/1408.5379} {arXiv:1408.5379 [cond-mat.str-el]} \BibitemShut {NoStop}%
\bibitem [{\citenamefont {Fogler}\ and\ \citenamefont {Huse}(2000)}]{Fogler00}%
  \BibitemOpen
  \bibfield  {author} {\bibinfo {author} {\bibfnamefont {M.~M.}\ \bibnamefont {Fogler}}\ and\ \bibinfo {author} {\bibfnamefont {D.~A.}\ \bibnamefont {Huse}},\ }\bibfield  {title} {\bibinfo {title} {Dynamical response of a pinned two-dimensional {Wigner} crystal},\ }\href {https://doi.org/10.1103/PhysRevB.62.7553} {\bibfield  {journal} {\bibinfo  {journal} {Phys. Rev. B}\ }\textbf {\bibinfo {volume} {62}},\ \bibinfo {pages} {7553} (\bibinfo {year} {2000})}\BibitemShut {NoStop}%
\bibitem [{\citenamefont {Lee}\ \emph {et~al.}(2001)\citenamefont {Lee}, \citenamefont {Scarola},\ and\ \citenamefont {Jain}}]{Lee01}%
  \BibitemOpen
  \bibfield  {author} {\bibinfo {author} {\bibfnamefont {S.-Y.}\ \bibnamefont {Lee}}, \bibinfo {author} {\bibfnamefont {V.~W.}\ \bibnamefont {Scarola}},\ and\ \bibinfo {author} {\bibfnamefont {J.~K.}\ \bibnamefont {Jain}},\ }\bibfield  {title} {\bibinfo {title} {Stripe formation in the fractional quantum {Hall} regime},\ }\href {https://doi.org/10.1103/PhysRevLett.87.256803} {\bibfield  {journal} {\bibinfo  {journal} {Phys. Rev. Lett.}\ }\textbf {\bibinfo {volume} {87}},\ \bibinfo {pages} {256803} (\bibinfo {year} {2001})}\BibitemShut {NoStop}%
\bibitem [{\citenamefont {Balram}\ and\ \citenamefont {Regnault}(2024)}]{Balram24a}%
  \BibitemOpen
  \bibfield  {author} {\bibinfo {author} {\bibfnamefont {A.~C.}\ \bibnamefont {Balram}}\ and\ \bibinfo {author} {\bibfnamefont {N.}~\bibnamefont {Regnault}},\ }\bibfield  {title} {\bibinfo {title} {Fractional quantum {Hall} effect of partons and the nature of the 8/17 state in the zeroth {Landau} level of bilayer graphene},\ }\href {https://doi.org/10.1103/PhysRevB.110.L081114} {\bibfield  {journal} {\bibinfo  {journal} {Phys. Rev. B}\ }\textbf {\bibinfo {volume} {110}},\ \bibinfo {pages} {L081114} (\bibinfo {year} {2024})}\BibitemShut {NoStop}%
\bibitem [{\citenamefont {Yutushui}\ \emph {et~al.}(2025{\natexlab{a}})\citenamefont {Yutushui}, \citenamefont {Dey},\ and\ \citenamefont {Mross}}]{Yutushui25a}%
  \BibitemOpen
  \bibfield  {author} {\bibinfo {author} {\bibfnamefont {M.}~\bibnamefont {Yutushui}}, \bibinfo {author} {\bibfnamefont {A.}~\bibnamefont {Dey}},\ and\ \bibinfo {author} {\bibfnamefont {D.~F.}\ \bibnamefont {Mross}},\ }\href {https://arxiv.org/abs/2508.14162} {\bibinfo {title} {The numerical case for identifying paired quantum {Hall} phases by their daughters}} (\bibinfo {year} {2025}{\natexlab{a}}),\ \Eprint {https://arxiv.org/abs/2508.14162} {arXiv:2508.14162 [cond-mat.str-el]} \BibitemShut {NoStop}%
\bibitem [{\citenamefont {Faugno}\ \emph {et~al.}(2021)\citenamefont {Faugno}, \citenamefont {Zhao}, \citenamefont {Balram}, \citenamefont {Jolicoeur},\ and\ \citenamefont {Jain}}]{Faugno21}%
  \BibitemOpen
  \bibfield  {author} {\bibinfo {author} {\bibfnamefont {W.~N.}\ \bibnamefont {Faugno}}, \bibinfo {author} {\bibfnamefont {T.}~\bibnamefont {Zhao}}, \bibinfo {author} {\bibfnamefont {A.~C.}\ \bibnamefont {Balram}}, \bibinfo {author} {\bibfnamefont {T.}~\bibnamefont {Jolicoeur}},\ and\ \bibinfo {author} {\bibfnamefont {J.~K.}\ \bibnamefont {Jain}},\ }\bibfield  {title} {\bibinfo {title} {Unconventional ${\mathbb{z}}_{n}$ parton states at $\ensuremath{\nu}=7/3$: {Role} of finite width},\ }\href {https://doi.org/10.1103/PhysRevB.103.085303} {\bibfield  {journal} {\bibinfo  {journal} {Phys. Rev. B}\ }\textbf {\bibinfo {volume} {103}},\ \bibinfo {pages} {085303} (\bibinfo {year} {2021})}\BibitemShut {NoStop}%
\bibitem [{\citenamefont {Laughlin}(1983)}]{Laughlin83}%
  \BibitemOpen
  \bibfield  {author} {\bibinfo {author} {\bibfnamefont {R.~B.}\ \bibnamefont {Laughlin}},\ }\bibfield  {title} {\bibinfo {title} {Anomalous quantum {Hall} effect: An incompressible quantum fluid with fractionally charged excitations},\ }\href {https://doi.org/10.1103/PhysRevLett.50.1395} {\bibfield  {journal} {\bibinfo  {journal} {Phys. Rev. Lett.}\ }\textbf {\bibinfo {volume} {50}},\ \bibinfo {pages} {1395} (\bibinfo {year} {1983})}\BibitemShut {NoStop}%
\bibitem [{\citenamefont {Balram}\ \emph {et~al.}(2020)\citenamefont {Balram}, \citenamefont {Jain},\ and\ \citenamefont {Barkeshli}}]{Balram20}%
  \BibitemOpen
  \bibfield  {author} {\bibinfo {author} {\bibfnamefont {A.~C.}\ \bibnamefont {Balram}}, \bibinfo {author} {\bibfnamefont {J.~K.}\ \bibnamefont {Jain}},\ and\ \bibinfo {author} {\bibfnamefont {M.}~\bibnamefont {Barkeshli}},\ }\bibfield  {title} {\bibinfo {title} {${\mathbb{z}}_{n}$ superconductivity of composite bosons and the $7/3$ fractional quantum {Hall} effect},\ }\href {https://doi.org/10.1103/PhysRevResearch.2.013349} {\bibfield  {journal} {\bibinfo  {journal} {Phys. Rev. Research}\ }\textbf {\bibinfo {volume} {2}},\ \bibinfo {pages} {013349} (\bibinfo {year} {2020})}\BibitemShut {NoStop}%
\bibitem [{\citenamefont {Kharitonov}(2012)}]{Kharitonov12a}%
  \BibitemOpen
  \bibfield  {author} {\bibinfo {author} {\bibfnamefont {M.}~\bibnamefont {Kharitonov}},\ }\bibfield  {title} {\bibinfo {title} {Phase diagram for the $\ensuremath{\nu}=0$ quantum {Hall} state in monolayer graphene},\ }\href {https://doi.org/10.1103/PhysRevB.85.155439} {\bibfield  {journal} {\bibinfo  {journal} {Phys. Rev. B}\ }\textbf {\bibinfo {volume} {85}},\ \bibinfo {pages} {155439} (\bibinfo {year} {2012})}\BibitemShut {NoStop}%
\bibitem [{\citenamefont {An}\ \emph {et~al.}(2025)\citenamefont {An}, \citenamefont {Balram}, \citenamefont {Khanna},\ and\ \citenamefont {Murthy}}]{An24a}%
  \BibitemOpen
  \bibfield  {author} {\bibinfo {author} {\bibfnamefont {J.}~\bibnamefont {An}}, \bibinfo {author} {\bibfnamefont {A.~C.}\ \bibnamefont {Balram}}, \bibinfo {author} {\bibfnamefont {U.}~\bibnamefont {Khanna}},\ and\ \bibinfo {author} {\bibfnamefont {G.}~\bibnamefont {Murthy}},\ }\bibfield  {title} {\bibinfo {title} {Fractional quantum {Hall} coexistence phases in higher {Landau} levels of graphene},\ }\href {https://doi.org/10.1103/PhysRevB.111.045110} {\bibfield  {journal} {\bibinfo  {journal} {Phys. Rev. B}\ }\textbf {\bibinfo {volume} {111}},\ \bibinfo {pages} {045110} (\bibinfo {year} {2025})}\BibitemShut {NoStop}%
\bibitem [{\citenamefont {Yutushui}\ \emph {et~al.}(2025{\natexlab{b}})\citenamefont {Yutushui}, \citenamefont {Hermanns},\ and\ \citenamefont {Mross}}]{Yutushui25}%
  \BibitemOpen
  \bibfield  {author} {\bibinfo {author} {\bibfnamefont {M.}~\bibnamefont {Yutushui}}, \bibinfo {author} {\bibfnamefont {M.}~\bibnamefont {Hermanns}},\ and\ \bibinfo {author} {\bibfnamefont {D.~F.}\ \bibnamefont {Mross}},\ }\bibfield  {title} {\bibinfo {title} {Non-{Abelian} phases from the condensation of {Abelian} anyons},\ }\href {https://doi.org/10.1103/3yvl-4hws} {\bibfield  {journal} {\bibinfo  {journal} {Phys. Rev. Lett.}\ }\textbf {\bibinfo {volume} {135}},\ \bibinfo {pages} {056501} (\bibinfo {year} {2025}{\natexlab{b}})}\BibitemShut {NoStop}%
\bibitem [{\citenamefont {Cai}\ \emph {et~al.}(2023)\citenamefont {Cai}, \citenamefont {Anderson}, \citenamefont {Wang}, \citenamefont {Zhang}, \citenamefont {Liu}, \citenamefont {Holtzmann}, \citenamefont {Zhang}, \citenamefont {Fan}, \citenamefont {Taniguchi}, \citenamefont {Watanabe}, \citenamefont {Ran}, \citenamefont {Cao}, \citenamefont {Fu}, \citenamefont {Xiao}, \citenamefont {Yao},\ and\ \citenamefont {Xu}}]{FQAH_MoTe2_Xu_2023a}%
  \BibitemOpen
  \bibfield  {author} {\bibinfo {author} {\bibfnamefont {J.}~\bibnamefont {Cai}}, \bibinfo {author} {\bibfnamefont {E.}~\bibnamefont {Anderson}}, \bibinfo {author} {\bibfnamefont {C.}~\bibnamefont {Wang}}, \bibinfo {author} {\bibfnamefont {X.}~\bibnamefont {Zhang}}, \bibinfo {author} {\bibfnamefont {X.}~\bibnamefont {Liu}}, \bibinfo {author} {\bibfnamefont {W.}~\bibnamefont {Holtzmann}}, \bibinfo {author} {\bibfnamefont {Y.}~\bibnamefont {Zhang}}, \bibinfo {author} {\bibfnamefont {F.}~\bibnamefont {Fan}}, \bibinfo {author} {\bibfnamefont {T.}~\bibnamefont {Taniguchi}}, \bibinfo {author} {\bibfnamefont {K.}~\bibnamefont {Watanabe}}, \bibinfo {author} {\bibfnamefont {Y.}~\bibnamefont {Ran}}, \bibinfo {author} {\bibfnamefont {T.}~\bibnamefont {Cao}}, \bibinfo {author} {\bibfnamefont {L.}~\bibnamefont {Fu}}, \bibinfo {author} {\bibfnamefont {D.}~\bibnamefont {Xiao}}, \bibinfo {author} {\bibfnamefont {W.}~\bibnamefont {Yao}},\ and\ \bibinfo {author} {\bibfnamefont {X.}~\bibnamefont {Xu}},\ }\bibfield  {title}
  {\bibinfo {title} {Signatures of fractional quantum anomalous {Hall} states in twisted {M}o{T}e2},\ }\href {https://doi.org/10.1038/s41586-023-06289-w} {\bibfield  {journal} {\bibinfo  {journal} {Nature}\ }\textbf {\bibinfo {volume} {622}},\ \bibinfo {pages} {63} (\bibinfo {year} {2023})}\BibitemShut {NoStop}%
\bibitem [{\citenamefont {Park}\ \emph {et~al.}(2023)\citenamefont {Park}, \citenamefont {Cai}, \citenamefont {Anderson}, \citenamefont {Zhang}, \citenamefont {Zhu}, \citenamefont {Liu}, \citenamefont {Wang}, \citenamefont {Holtzmann}, \citenamefont {Hu}, \citenamefont {Liu}, \citenamefont {Taniguchi}, \citenamefont {Watanabe}, \citenamefont {Chu}, \citenamefont {Cao}, \citenamefont {Fu}, \citenamefont {Yao}, \citenamefont {Chang}, \citenamefont {Cobden}, \citenamefont {Xiao},\ and\ \citenamefont {Xu}}]{FQAH_MoTe2_Xu_2023b}%
  \BibitemOpen
  \bibfield  {author} {\bibinfo {author} {\bibfnamefont {H.}~\bibnamefont {Park}}, \bibinfo {author} {\bibfnamefont {J.}~\bibnamefont {Cai}}, \bibinfo {author} {\bibfnamefont {E.}~\bibnamefont {Anderson}}, \bibinfo {author} {\bibfnamefont {Y.}~\bibnamefont {Zhang}}, \bibinfo {author} {\bibfnamefont {J.}~\bibnamefont {Zhu}}, \bibinfo {author} {\bibfnamefont {X.}~\bibnamefont {Liu}}, \bibinfo {author} {\bibfnamefont {C.}~\bibnamefont {Wang}}, \bibinfo {author} {\bibfnamefont {W.}~\bibnamefont {Holtzmann}}, \bibinfo {author} {\bibfnamefont {C.}~\bibnamefont {Hu}}, \bibinfo {author} {\bibfnamefont {Z.}~\bibnamefont {Liu}}, \bibinfo {author} {\bibfnamefont {T.}~\bibnamefont {Taniguchi}}, \bibinfo {author} {\bibfnamefont {K.}~\bibnamefont {Watanabe}}, \bibinfo {author} {\bibfnamefont {J.-H.}\ \bibnamefont {Chu}}, \bibinfo {author} {\bibfnamefont {T.}~\bibnamefont {Cao}}, \bibinfo {author} {\bibfnamefont {L.}~\bibnamefont {Fu}}, \bibinfo {author} {\bibfnamefont {W.}~\bibnamefont {Yao}}, \bibinfo {author}
  {\bibfnamefont {C.-Z.}\ \bibnamefont {Chang}}, \bibinfo {author} {\bibfnamefont {D.}~\bibnamefont {Cobden}}, \bibinfo {author} {\bibfnamefont {D.}~\bibnamefont {Xiao}},\ and\ \bibinfo {author} {\bibfnamefont {X.}~\bibnamefont {Xu}},\ }\bibfield  {title} {\bibinfo {title} {Observation of fractionally quantized anomalous {Hall} effect},\ }\href {https://doi.org/10.1038/s41586-023-06536-0} {\bibfield  {journal} {\bibinfo  {journal} {Nature}\ }\textbf {\bibinfo {volume} {622}},\ \bibinfo {pages} {74} (\bibinfo {year} {2023})}\BibitemShut {NoStop}%
\bibitem [{\citenamefont {Zeng}\ \emph {et~al.}(2023)\citenamefont {Zeng}, \citenamefont {Xia}, \citenamefont {Kang}, \citenamefont {Zhu}, \citenamefont {Kn{\"u}ppel}, \citenamefont {Vaswani}, \citenamefont {Watanabe}, \citenamefont {Taniguchi}, \citenamefont {Mak},\ and\ \citenamefont {Shan}}]{FQAH_MoTe2_Mak_Shan_2023}%
  \BibitemOpen
  \bibfield  {author} {\bibinfo {author} {\bibfnamefont {Y.}~\bibnamefont {Zeng}}, \bibinfo {author} {\bibfnamefont {Z.}~\bibnamefont {Xia}}, \bibinfo {author} {\bibfnamefont {K.}~\bibnamefont {Kang}}, \bibinfo {author} {\bibfnamefont {J.}~\bibnamefont {Zhu}}, \bibinfo {author} {\bibfnamefont {P.}~\bibnamefont {Kn{\"u}ppel}}, \bibinfo {author} {\bibfnamefont {C.}~\bibnamefont {Vaswani}}, \bibinfo {author} {\bibfnamefont {K.}~\bibnamefont {Watanabe}}, \bibinfo {author} {\bibfnamefont {T.}~\bibnamefont {Taniguchi}}, \bibinfo {author} {\bibfnamefont {K.~F.}\ \bibnamefont {Mak}},\ and\ \bibinfo {author} {\bibfnamefont {J.}~\bibnamefont {Shan}},\ }\bibfield  {title} {\bibinfo {title} {Thermodynamic evidence of fractional {Chern} insulator in moir{\'e} {M}o{T}e2},\ }\href {https://doi.org/10.1038/s41586-023-06452-3} {\bibfield  {journal} {\bibinfo  {journal} {Nature}\ }\textbf {\bibinfo {volume} {622}},\ \bibinfo {pages} {69} (\bibinfo {year} {2023})}\BibitemShut {NoStop}%
\bibitem [{\citenamefont {Xu}\ \emph {et~al.}(2023)\citenamefont {Xu}, \citenamefont {Sun}, \citenamefont {Jia}, \citenamefont {Liu}, \citenamefont {Xu}, \citenamefont {Li}, \citenamefont {Gu}, \citenamefont {Watanabe}, \citenamefont {Taniguchi}, \citenamefont {Tong}, \citenamefont {Jia}, \citenamefont {Shi}, \citenamefont {Jiang}, \citenamefont {Zhang}, \citenamefont {Liu},\ and\ \citenamefont {Li}}]{FQAHE_MoTe2_Li_2023}%
  \BibitemOpen
  \bibfield  {author} {\bibinfo {author} {\bibfnamefont {F.}~\bibnamefont {Xu}}, \bibinfo {author} {\bibfnamefont {Z.}~\bibnamefont {Sun}}, \bibinfo {author} {\bibfnamefont {T.}~\bibnamefont {Jia}}, \bibinfo {author} {\bibfnamefont {C.}~\bibnamefont {Liu}}, \bibinfo {author} {\bibfnamefont {C.}~\bibnamefont {Xu}}, \bibinfo {author} {\bibfnamefont {C.}~\bibnamefont {Li}}, \bibinfo {author} {\bibfnamefont {Y.}~\bibnamefont {Gu}}, \bibinfo {author} {\bibfnamefont {K.}~\bibnamefont {Watanabe}}, \bibinfo {author} {\bibfnamefont {T.}~\bibnamefont {Taniguchi}}, \bibinfo {author} {\bibfnamefont {B.}~\bibnamefont {Tong}}, \bibinfo {author} {\bibfnamefont {J.}~\bibnamefont {Jia}}, \bibinfo {author} {\bibfnamefont {Z.}~\bibnamefont {Shi}}, \bibinfo {author} {\bibfnamefont {S.}~\bibnamefont {Jiang}}, \bibinfo {author} {\bibfnamefont {Y.}~\bibnamefont {Zhang}}, \bibinfo {author} {\bibfnamefont {X.}~\bibnamefont {Liu}},\ and\ \bibinfo {author} {\bibfnamefont {T.}~\bibnamefont {Li}},\ }\bibfield  {title} {\bibinfo {title}
  {Observation of integer and fractional quantum anomalous {Hall} effects in twisted bilayer ${\mathrm{{m}o{t}e}}_{2}$},\ }\href {https://doi.org/10.1103/PhysRevX.13.031037} {\bibfield  {journal} {\bibinfo  {journal} {Phys. Rev. X}\ }\textbf {\bibinfo {volume} {13}},\ \bibinfo {pages} {031037} (\bibinfo {year} {2023})}\BibitemShut {NoStop}%
\bibitem [{\citenamefont {Zhao}\ \emph {et~al.}(2024)\citenamefont {Zhao}, \citenamefont {Huang}, \citenamefont {Cr{\'e}pel}, \citenamefont {Xiong}, \citenamefont {Wu}, \citenamefont {Zhang}, \citenamefont {Wang}, \citenamefont {Han}, \citenamefont {Li}, \citenamefont {Xi}, \citenamefont {Pan}, \citenamefont {Wang}, \citenamefont {Kuang}, \citenamefont {Luo}, \citenamefont {Shen}, \citenamefont {Yang}, \citenamefont {Zhou}, \citenamefont {Watanabe}, \citenamefont {Taniguchi}, \citenamefont {Sac{\'e}p{\'e}}, \citenamefont {Zhang}, \citenamefont {Wang}, \citenamefont {Lu}, \citenamefont {Regnault},\ and\ \citenamefont {Han}}]{Zhao24_MoS2_FQH}%
  \BibitemOpen
  \bibfield  {author} {\bibinfo {author} {\bibfnamefont {S.}~\bibnamefont {Zhao}}, \bibinfo {author} {\bibfnamefont {J.}~\bibnamefont {Huang}}, \bibinfo {author} {\bibfnamefont {V.}~\bibnamefont {Cr{\'e}pel}}, \bibinfo {author} {\bibfnamefont {Z.}~\bibnamefont {Xiong}}, \bibinfo {author} {\bibfnamefont {X.}~\bibnamefont {Wu}}, \bibinfo {author} {\bibfnamefont {T.}~\bibnamefont {Zhang}}, \bibinfo {author} {\bibfnamefont {H.}~\bibnamefont {Wang}}, \bibinfo {author} {\bibfnamefont {X.}~\bibnamefont {Han}}, \bibinfo {author} {\bibfnamefont {Z.}~\bibnamefont {Li}}, \bibinfo {author} {\bibfnamefont {C.}~\bibnamefont {Xi}}, \bibinfo {author} {\bibfnamefont {S.}~\bibnamefont {Pan}}, \bibinfo {author} {\bibfnamefont {Z.}~\bibnamefont {Wang}}, \bibinfo {author} {\bibfnamefont {G.}~\bibnamefont {Kuang}}, \bibinfo {author} {\bibfnamefont {J.}~\bibnamefont {Luo}}, \bibinfo {author} {\bibfnamefont {Q.}~\bibnamefont {Shen}}, \bibinfo {author} {\bibfnamefont {J.}~\bibnamefont {Yang}}, \bibinfo {author} {\bibfnamefont
  {R.}~\bibnamefont {Zhou}}, \bibinfo {author} {\bibfnamefont {K.}~\bibnamefont {Watanabe}}, \bibinfo {author} {\bibfnamefont {T.}~\bibnamefont {Taniguchi}}, \bibinfo {author} {\bibfnamefont {B.}~\bibnamefont {Sac{\'e}p{\'e}}}, \bibinfo {author} {\bibfnamefont {J.}~\bibnamefont {Zhang}}, \bibinfo {author} {\bibfnamefont {N.}~\bibnamefont {Wang}}, \bibinfo {author} {\bibfnamefont {J.}~\bibnamefont {Lu}}, \bibinfo {author} {\bibfnamefont {N.}~\bibnamefont {Regnault}},\ and\ \bibinfo {author} {\bibfnamefont {Z.~V.}\ \bibnamefont {Han}},\ }\bibfield  {title} {\bibinfo {title} {Fractional quantum {Hall} phases in high-mobility n-type molybdenum disulfide transistors},\ }\href {https://doi.org/10.1038/s41928-024-01274-1} {\bibfield  {journal} {\bibinfo  {journal} {Nature Electronics}\ }\textbf {\bibinfo {volume} {7}},\ \bibinfo {pages} {1117} (\bibinfo {year} {2024})}\BibitemShut {NoStop}%
\bibitem [{\citenamefont {Lu}\ \emph {et~al.}(2024)\citenamefont {Lu}, \citenamefont {Han}, \citenamefont {Yao}, \citenamefont {Reddy}, \citenamefont {Yang}, \citenamefont {Seo}, \citenamefont {Watanabe}, \citenamefont {Taniguchi}, \citenamefont {Fu},\ and\ \citenamefont {Ju}}]{FQAH_Pentalayer_Graphene_Ju_2024}%
  \BibitemOpen
  \bibfield  {author} {\bibinfo {author} {\bibfnamefont {Z.}~\bibnamefont {Lu}}, \bibinfo {author} {\bibfnamefont {T.}~\bibnamefont {Han}}, \bibinfo {author} {\bibfnamefont {Y.}~\bibnamefont {Yao}}, \bibinfo {author} {\bibfnamefont {A.~P.}\ \bibnamefont {Reddy}}, \bibinfo {author} {\bibfnamefont {J.}~\bibnamefont {Yang}}, \bibinfo {author} {\bibfnamefont {J.}~\bibnamefont {Seo}}, \bibinfo {author} {\bibfnamefont {K.}~\bibnamefont {Watanabe}}, \bibinfo {author} {\bibfnamefont {T.}~\bibnamefont {Taniguchi}}, \bibinfo {author} {\bibfnamefont {L.}~\bibnamefont {Fu}},\ and\ \bibinfo {author} {\bibfnamefont {L.}~\bibnamefont {Ju}},\ }\bibfield  {title} {\bibinfo {title} {Fractional quantum anomalous {Hall} effect in multilayer graphene},\ }\href {https://doi.org/10.1038/s41586-023-07010-7} {\bibfield  {journal} {\bibinfo  {journal} {Nature}\ }\textbf {\bibinfo {volume} {626}},\ \bibinfo {pages} {759} (\bibinfo {year} {2024})}\BibitemShut {NoStop}%
\bibitem [{\citenamefont {Hadjri}\ \emph {et~al.}(2026)\citenamefont {Hadjri}, \citenamefont {Yue}, \citenamefont {Han}, \citenamefont {Yao}, \citenamefont {Lu}, \citenamefont {Ye}, \citenamefont {Seo}, \citenamefont {Yang}, \citenamefont {Watanabe}, \citenamefont {Taniguchi}, \citenamefont {Fu}, \citenamefont {Stern},\ and\ \citenamefont {Ju}}]{Hadjri26}%
  \BibitemOpen
  \bibfield  {author} {\bibinfo {author} {\bibfnamefont {Z.}~\bibnamefont {Hadjri}}, \bibinfo {author} {\bibfnamefont {X.}~\bibnamefont {Yue}}, \bibinfo {author} {\bibfnamefont {T.}~\bibnamefont {Han}}, \bibinfo {author} {\bibfnamefont {Y.}~\bibnamefont {Yao}}, \bibinfo {author} {\bibfnamefont {Z.}~\bibnamefont {Lu}}, \bibinfo {author} {\bibfnamefont {S.}~\bibnamefont {Ye}}, \bibinfo {author} {\bibfnamefont {J.}~\bibnamefont {Seo}}, \bibinfo {author} {\bibfnamefont {J.}~\bibnamefont {Yang}}, \bibinfo {author} {\bibfnamefont {K.}~\bibnamefont {Watanabe}}, \bibinfo {author} {\bibfnamefont {T.}~\bibnamefont {Taniguchi}}, \bibinfo {author} {\bibfnamefont {L.}~\bibnamefont {Fu}}, \bibinfo {author} {\bibfnamefont {A.}~\bibnamefont {Stern}},\ and\ \bibinfo {author} {\bibfnamefont {L.}~\bibnamefont {Ju}},\ }\href {https://arxiv.org/abs/2609.09422} {\bibinfo {title} {Quantum phase transitions and fractional quantized anomalous {Hall} insulators in rhombohedral graphene}} (\bibinfo {year} {2026}),\ \Eprint
  {https://arxiv.org/abs/2609.09422} {arXiv:2609.09422 [cond-mat.mes-hall]} \BibitemShut {NoStop}%
\bibitem [{\citenamefont {Lu}\ \emph {et~al.}(2025)\citenamefont {Lu}, \citenamefont {Han}, \citenamefont {Yao}, \citenamefont {Hadjri}, \citenamefont {Yang}, \citenamefont {Seo}, \citenamefont {Shi}, \citenamefont {Ye}, \citenamefont {Watanabe}, \citenamefont {Taniguchi},\ and\ \citenamefont {Ju}}]{Lu25}%
  \BibitemOpen
  \bibfield  {author} {\bibinfo {author} {\bibfnamefont {Z.}~\bibnamefont {Lu}}, \bibinfo {author} {\bibfnamefont {T.}~\bibnamefont {Han}}, \bibinfo {author} {\bibfnamefont {Y.}~\bibnamefont {Yao}}, \bibinfo {author} {\bibfnamefont {Z.}~\bibnamefont {Hadjri}}, \bibinfo {author} {\bibfnamefont {J.}~\bibnamefont {Yang}}, \bibinfo {author} {\bibfnamefont {J.}~\bibnamefont {Seo}}, \bibinfo {author} {\bibfnamefont {L.}~\bibnamefont {Shi}}, \bibinfo {author} {\bibfnamefont {S.}~\bibnamefont {Ye}}, \bibinfo {author} {\bibfnamefont {K.}~\bibnamefont {Watanabe}}, \bibinfo {author} {\bibfnamefont {T.}~\bibnamefont {Taniguchi}},\ and\ \bibinfo {author} {\bibfnamefont {L.}~\bibnamefont {Ju}},\ }\bibfield  {title} {\bibinfo {title} {Extended quantum anomalous hall states in graphene/hbn moir{\'e} superlattices},\ }\href {https://doi.org/10.1038/s41586-024-08470-1} {\bibfield  {journal} {\bibinfo  {journal} {Nature}\ }\textbf {\bibinfo {volume} {637}},\ \bibinfo {pages} {1090} (\bibinfo {year} {2025})}\BibitemShut
  {NoStop}%
\bibitem [{Dia()}]{DiagHam}%
  \BibitemOpen
  \href@noop {} {}\bibinfo {note} {Diag{H}am, \url{https://www.nick-ux.org/diagham}}\BibitemShut {NoStop}%
\bibitem [{\citenamefont {Halperin}(1983)}]{Halperin83}%
  \BibitemOpen
  \bibfield  {author} {\bibinfo {author} {\bibfnamefont {B.~I.}\ \bibnamefont {Halperin}},\ }\bibfield  {title} {\bibinfo {title} {Theory of the quantized {Hall} conductance},\ }\href@noop {} {\bibfield  {journal} {\bibinfo  {journal} {Helvetica Physica Acta}\ }\textbf {\bibinfo {volume} {56}},\ \bibinfo {pages} {75} (\bibinfo {year} {1983})}\BibitemShut {NoStop}%
\bibitem [{\citenamefont {Wen}(1999)}]{Wen99}%
  \BibitemOpen
  \bibfield  {author} {\bibinfo {author} {\bibfnamefont {X.-G.}\ \bibnamefont {Wen}},\ }\bibfield  {title} {\bibinfo {title} {Projective construction of non-abelian quantum {Hall} liquids},\ }\href {https://doi.org/10.1103/PhysRevB.60.8827} {\bibfield  {journal} {\bibinfo  {journal} {Phys. Rev. B}\ }\textbf {\bibinfo {volume} {60}},\ \bibinfo {pages} {8827} (\bibinfo {year} {1999})}\BibitemShut {NoStop}%
\bibitem [{\citenamefont {Bose}\ and\ \citenamefont {Balram}(2025)}]{Bose25}%
  \BibitemOpen
  \bibfield  {author} {\bibinfo {author} {\bibfnamefont {K.}~\bibnamefont {Bose}}\ and\ \bibinfo {author} {\bibfnamefont {A.~C.}\ \bibnamefont {Balram}},\ }\bibfield  {title} {\bibinfo {title} {Dispersion of neutral collective modes in partonic fractional quantum {Hall} states and its applications to paired states of composite fermions},\ }\href {https://doi.org/10.1103/p5sq-kczs} {\bibfield  {journal} {\bibinfo  {journal} {Phys. Rev. B}\ }\textbf {\bibinfo {volume} {112}},\ \bibinfo {pages} {035136} (\bibinfo {year} {2025})}\BibitemShut {NoStop}%
\bibitem [{\citenamefont {Gromov}\ \emph {et~al.}(2015{\natexlab{a}})\citenamefont {Gromov}, \citenamefont {Cho}, \citenamefont {You}, \citenamefont {Abanov},\ and\ \citenamefont {Fradkin}}]{Gromov15}%
  \BibitemOpen
  \bibfield  {author} {\bibinfo {author} {\bibfnamefont {A.}~\bibnamefont {Gromov}}, \bibinfo {author} {\bibfnamefont {G.~Y.}\ \bibnamefont {Cho}}, \bibinfo {author} {\bibfnamefont {Y.}~\bibnamefont {You}}, \bibinfo {author} {\bibfnamefont {A.~G.}\ \bibnamefont {Abanov}},\ and\ \bibinfo {author} {\bibfnamefont {E.}~\bibnamefont {Fradkin}},\ }\bibfield  {title} {\bibinfo {title} {Framing anomaly in the effective theory of the fractional quantum {Hall} effect},\ }\href {https://doi.org/10.1103/PhysRevLett.114.016805} {\bibfield  {journal} {\bibinfo  {journal} {Phys. Rev. Lett.}\ }\textbf {\bibinfo {volume} {114}},\ \bibinfo {pages} {016805} (\bibinfo {year} {2015}{\natexlab{a}})}\BibitemShut {NoStop}%
\bibitem [{\citenamefont {Nguyen}\ \emph {et~al.}(2017)\citenamefont {Nguyen}, \citenamefont {Can},\ and\ \citenamefont {Gromov}}]{Nguyen17}%
  \BibitemOpen
  \bibfield  {author} {\bibinfo {author} {\bibfnamefont {D.~X.}\ \bibnamefont {Nguyen}}, \bibinfo {author} {\bibfnamefont {T.}~\bibnamefont {Can}},\ and\ \bibinfo {author} {\bibfnamefont {A.}~\bibnamefont {Gromov}},\ }\bibfield  {title} {\bibinfo {title} {Particle-hole duality in the lowest {Landau} level},\ }\href {https://doi.org/10.1103/PhysRevLett.118.206602} {\bibfield  {journal} {\bibinfo  {journal} {Phys. Rev. Lett.}\ }\textbf {\bibinfo {volume} {118}},\ \bibinfo {pages} {206602} (\bibinfo {year} {2017})}\BibitemShut {NoStop}%
\bibitem [{\citenamefont {Dwivedi}\ and\ \citenamefont {Klevtsov}(2019)}]{Dwivedi19}%
  \BibitemOpen
  \bibfield  {author} {\bibinfo {author} {\bibfnamefont {V.}~\bibnamefont {Dwivedi}}\ and\ \bibinfo {author} {\bibfnamefont {S.}~\bibnamefont {Klevtsov}},\ }\bibfield  {title} {\bibinfo {title} {Geometric responses of the {Pfaffian} state},\ }\href {https://doi.org/10.1103/PhysRevB.99.205158} {\bibfield  {journal} {\bibinfo  {journal} {Phys. Rev. B}\ }\textbf {\bibinfo {volume} {99}},\ \bibinfo {pages} {205158} (\bibinfo {year} {2019})}\BibitemShut {NoStop}%
\bibitem [{\citenamefont {Hermanns}(2010)}]{Hermanns10}%
  \BibitemOpen
  \bibfield  {author} {\bibinfo {author} {\bibfnamefont {M.}~\bibnamefont {Hermanns}},\ }\bibfield  {title} {\bibinfo {title} {Condensing non-abelian quasiparticles},\ }\href {https://doi.org/10.1103/PhysRevLett.104.056803} {\bibfield  {journal} {\bibinfo  {journal} {Phys. Rev. Lett.}\ }\textbf {\bibinfo {volume} {104}},\ \bibinfo {pages} {056803} (\bibinfo {year} {2010})}\BibitemShut {NoStop}%
\bibitem [{\citenamefont {Sreejith}\ \emph {et~al.}(2011)\citenamefont {Sreejith}, \citenamefont {T\H{o}ke}, \citenamefont {W\'ojs},\ and\ \citenamefont {Jain}}]{Sreejith11b}%
  \BibitemOpen
  \bibfield  {author} {\bibinfo {author} {\bibfnamefont {G.~J.}\ \bibnamefont {Sreejith}}, \bibinfo {author} {\bibfnamefont {C.}~\bibnamefont {T\H{o}ke}}, \bibinfo {author} {\bibfnamefont {A.}~\bibnamefont {W\'ojs}},\ and\ \bibinfo {author} {\bibfnamefont {J.~K.}\ \bibnamefont {Jain}},\ }\bibfield  {title} {\bibinfo {title} {Bipartite composite fermion states},\ }\href {https://doi.org/10.1103/PhysRevLett.107.086806} {\bibfield  {journal} {\bibinfo  {journal} {Phys. Rev. Lett.}\ }\textbf {\bibinfo {volume} {107}},\ \bibinfo {pages} {086806} (\bibinfo {year} {2011})}\BibitemShut {NoStop}%
\bibitem [{\citenamefont {Hansson}\ \emph {et~al.}(2017)\citenamefont {Hansson}, \citenamefont {Hermanns}, \citenamefont {Simon},\ and\ \citenamefont {Viefers}}]{Hansson17}%
  \BibitemOpen
  \bibfield  {author} {\bibinfo {author} {\bibfnamefont {T.~H.}\ \bibnamefont {Hansson}}, \bibinfo {author} {\bibfnamefont {M.}~\bibnamefont {Hermanns}}, \bibinfo {author} {\bibfnamefont {S.~H.}\ \bibnamefont {Simon}},\ and\ \bibinfo {author} {\bibfnamefont {S.~F.}\ \bibnamefont {Viefers}},\ }\bibfield  {title} {\bibinfo {title} {Quantum {Hall} physics: Hierarchies and conformal field theory techniques},\ }\href {https://doi.org/10.1103/RevModPhys.89.025005} {\bibfield  {journal} {\bibinfo  {journal} {Rev. Mod. Phys.}\ }\textbf {\bibinfo {volume} {89}},\ \bibinfo {pages} {025005} (\bibinfo {year} {2017})}\BibitemShut {NoStop}%
\bibitem [{\citenamefont {Gromov}\ \emph {et~al.}(2015{\natexlab{b}})\citenamefont {Gromov}, \citenamefont {Cho}, \citenamefont {You}, \citenamefont {Abanov},\ and\ \citenamefont {Fradkin}}]{Gromov15a}%
  \BibitemOpen
  \bibfield  {author} {\bibinfo {author} {\bibfnamefont {A.}~\bibnamefont {Gromov}}, \bibinfo {author} {\bibfnamefont {G.~Y.}\ \bibnamefont {Cho}}, \bibinfo {author} {\bibfnamefont {Y.}~\bibnamefont {You}}, \bibinfo {author} {\bibfnamefont {A.~G.}\ \bibnamefont {Abanov}},\ and\ \bibinfo {author} {\bibfnamefont {E.}~\bibnamefont {Fradkin}},\ }\bibfield  {title} {\bibinfo {title} {Erratum: Framing anomaly in the effective theory of the fractional quantum {Hall} effect [{Phys}. {Rev}. {Lett}. 114, 016805 (2015)]},\ }\href {https://doi.org/10.1103/PhysRevLett.114.149902} {\bibfield  {journal} {\bibinfo  {journal} {Phys. Rev. Lett.}\ }\textbf {\bibinfo {volume} {114}},\ \bibinfo {pages} {149902} (\bibinfo {year} {2015}{\natexlab{b}})}\BibitemShut {NoStop}%
\bibitem [{\citenamefont {Bradlyn}\ and\ \citenamefont {Read}(2015)}]{Bradlyn15}%
  \BibitemOpen
  \bibfield  {author} {\bibinfo {author} {\bibfnamefont {B.}~\bibnamefont {Bradlyn}}\ and\ \bibinfo {author} {\bibfnamefont {N.}~\bibnamefont {Read}},\ }\bibfield  {title} {\bibinfo {title} {Topological central charge from {Berry} curvature: Gravitational anomalies in trial wave functions for topological phases},\ }\href {https://doi.org/10.1103/PhysRevB.91.165306} {\bibfield  {journal} {\bibinfo  {journal} {Phys. Rev. B}\ }\textbf {\bibinfo {volume} {91}},\ \bibinfo {pages} {165306} (\bibinfo {year} {2015})}\BibitemShut {NoStop}%
\bibitem [{\citenamefont {Haldane}(1995)}]{Haldane95}%
  \BibitemOpen
  \bibfield  {author} {\bibinfo {author} {\bibfnamefont {F.~D.~M.}\ \bibnamefont {Haldane}},\ }\bibfield  {title} {\bibinfo {title} {Stability of chiral {Luttinger} liquids and abelian quantum {Hall} states},\ }\href {https://doi.org/10.1103/PhysRevLett.74.2090} {\bibfield  {journal} {\bibinfo  {journal} {Phys. Rev. Lett.}\ }\textbf {\bibinfo {volume} {74}},\ \bibinfo {pages} {2090} (\bibinfo {year} {1995})}\BibitemShut {NoStop}%
\bibitem [{\citenamefont {Balram}\ \emph {et~al.}(2022)\citenamefont {Balram}, \citenamefont {Liu}, \citenamefont {Gromov},\ and\ \citenamefont {Papi\ifmmode~\acute{c}\else \'{c}\fi{}}}]{Balram21d}%
  \BibitemOpen
  \bibfield  {author} {\bibinfo {author} {\bibfnamefont {A.~C.}\ \bibnamefont {Balram}}, \bibinfo {author} {\bibfnamefont {Z.}~\bibnamefont {Liu}}, \bibinfo {author} {\bibfnamefont {A.}~\bibnamefont {Gromov}},\ and\ \bibinfo {author} {\bibfnamefont {Z.}~\bibnamefont {Papi\ifmmode~\acute{c}\else \'{c}\fi{}}},\ }\bibfield  {title} {\bibinfo {title} {Very-high-energy collective states of partons in fractional quantum {Hall} liquids},\ }\href {https://doi.org/10.1103/PhysRevX.12.021008} {\bibfield  {journal} {\bibinfo  {journal} {Phys. Rev. X}\ }\textbf {\bibinfo {volume} {12}},\ \bibinfo {pages} {021008} (\bibinfo {year} {2022})}\BibitemShut {NoStop}%
\bibitem [{\citenamefont {Balram}\ \emph {et~al.}(2024)\citenamefont {Balram}, \citenamefont {Sreejith},\ and\ \citenamefont {Jain}}]{Balram24}%
  \BibitemOpen
  \bibfield  {author} {\bibinfo {author} {\bibfnamefont {A.~C.}\ \bibnamefont {Balram}}, \bibinfo {author} {\bibfnamefont {G.~J.}\ \bibnamefont {Sreejith}},\ and\ \bibinfo {author} {\bibfnamefont {J.~K.}\ \bibnamefont {Jain}},\ }\bibfield  {title} {\bibinfo {title} {Splitting of the {Girvin}-{MacDonald}-{Platzman} density wave and the nature of chiral gravitons in the fractional quantum {Hall} effect},\ }\href {https://doi.org/10.1103/PhysRevLett.133.246605} {\bibfield  {journal} {\bibinfo  {journal} {Phys. Rev. Lett.}\ }\textbf {\bibinfo {volume} {133}},\ \bibinfo {pages} {246605} (\bibinfo {year} {2024})}\BibitemShut {NoStop}%
\bibitem [{\citenamefont {Kundu}\ \emph {et~al.}(2026)\citenamefont {Kundu}, \citenamefont {Dora}, \citenamefont {Nguyen},\ and\ \citenamefont {Balram}}]{Kundu26a}%
  \BibitemOpen
  \bibfield  {author} {\bibinfo {author} {\bibfnamefont {R.}~\bibnamefont {Kundu}}, \bibinfo {author} {\bibfnamefont {R.~K.}\ \bibnamefont {Dora}}, \bibinfo {author} {\bibfnamefont {D.~X.}\ \bibnamefont {Nguyen}},\ and\ \bibinfo {author} {\bibfnamefont {A.~C.}\ \bibnamefont {Balram}},\ }\href {https://arxiv.org/abs/2608.11133} {\bibinfo {title} {Sum rules and density-wave modes in spin-singlet fractional quantum {Hall} fluids}} (\bibinfo {year} {2026}),\ \Eprint {https://arxiv.org/abs/2608.11133} {arXiv:2608.11133 [cond-mat.str-el]} \BibitemShut {NoStop}%
\bibitem [{\citenamefont {Balram}\ and\ \citenamefont {Pu}(2017)}]{Balram16d}%
  \BibitemOpen
  \bibfield  {author} {\bibinfo {author} {\bibfnamefont {A.~C.}\ \bibnamefont {Balram}}\ and\ \bibinfo {author} {\bibfnamefont {S.}~\bibnamefont {Pu}},\ }\bibfield  {title} {\bibinfo {title} {Positions of the magnetoroton minima in the fractional quantum {Hall} effect},\ }\href {https://doi.org/10.1140/epjb/e2017-80177-5} {\bibfield  {journal} {\bibinfo  {journal} {The European Physical Journal B}\ }\textbf {\bibinfo {volume} {90}},\ \bibinfo {pages} {124} (\bibinfo {year} {2017})}\BibitemShut {NoStop}%
\bibitem [{\citenamefont {Jain}\ and\ \citenamefont {Kamilla}(1997{\natexlab{b}})}]{Jain97b}%
  \BibitemOpen
  \bibfield  {author} {\bibinfo {author} {\bibfnamefont {J.~K.}\ \bibnamefont {Jain}}\ and\ \bibinfo {author} {\bibfnamefont {R.~K.}\ \bibnamefont {Kamilla}},\ }\bibfield  {title} {\bibinfo {title} {Quantitative study of large composite-fermion systems},\ }\href {https://doi.org/10.1103/PhysRevB.55.R4895} {\bibfield  {journal} {\bibinfo  {journal} {Phys. Rev. B}\ }\textbf {\bibinfo {volume} {55}},\ \bibinfo {pages} {R4895} (\bibinfo {year} {1997}{\natexlab{b}})}\BibitemShut {NoStop}%
\bibitem [{\citenamefont {Zhao}\ \emph {et~al.}(2022)\citenamefont {Zhao}, \citenamefont {Kudo}, \citenamefont {Faugno}, \citenamefont {Balram},\ and\ \citenamefont {Jain}}]{Zhao22a}%
  \BibitemOpen
  \bibfield  {author} {\bibinfo {author} {\bibfnamefont {T.}~\bibnamefont {Zhao}}, \bibinfo {author} {\bibfnamefont {K.}~\bibnamefont {Kudo}}, \bibinfo {author} {\bibfnamefont {W.~N.}\ \bibnamefont {Faugno}}, \bibinfo {author} {\bibfnamefont {A.~C.}\ \bibnamefont {Balram}},\ and\ \bibinfo {author} {\bibfnamefont {J.~K.}\ \bibnamefont {Jain}},\ }\bibfield  {title} {\bibinfo {title} {Revisiting excitation gaps in the fractional quantum hall effect},\ }\href {https://doi.org/10.1103/PhysRevB.105.205147} {\bibfield  {journal} {\bibinfo  {journal} {Phys. Rev. B}\ }\textbf {\bibinfo {volume} {105}},\ \bibinfo {pages} {205147} (\bibinfo {year} {2022})}\BibitemShut {NoStop}%
\bibitem [{\citenamefont {Boriçi}\ \emph {et~al.}(2026)\citenamefont {Boriçi}, \citenamefont {Regnault},\ and\ \citenamefont {Ciuti}}]{Borici26}%
  \BibitemOpen
  \bibfield  {author} {\bibinfo {author} {\bibfnamefont {D.}~\bibnamefont {Boriçi}}, \bibinfo {author} {\bibfnamefont {N.}~\bibnamefont {Regnault}},\ and\ \bibinfo {author} {\bibfnamefont {C.}~\bibnamefont {Ciuti}},\ }\href {https://arxiv.org/abs/2607.06298} {\bibinfo {title} {Composite-fermion study of cavity-modified fractional quantum {Hall} excitation gaps}} (\bibinfo {year} {2026}),\ \Eprint {https://arxiv.org/abs/2607.06298} {arXiv:2607.06298 [cond-mat.mes-hall]} \BibitemShut {NoStop}%
\bibitem [{\citenamefont {Ortalano}\ \emph {et~al.}(1997)\citenamefont {Ortalano}, \citenamefont {He},\ and\ \citenamefont {Das~Sarma}}]{Ortalano97}%
  \BibitemOpen
  \bibfield  {author} {\bibinfo {author} {\bibfnamefont {M.~W.}\ \bibnamefont {Ortalano}}, \bibinfo {author} {\bibfnamefont {S.}~\bibnamefont {He}},\ and\ \bibinfo {author} {\bibfnamefont {S.}~\bibnamefont {Das~Sarma}},\ }\bibfield  {title} {\bibinfo {title} {Realistic calculations of correlated incompressible electronic states in {Ga}{As}-${\mathrm{al}}_{\mathrm{x}}$${\mathrm{ga}}_{1\mathrm{-}\mathrm{x}}$as heterostructures and quantum wells},\ }\href {https://doi.org/10.1103/PhysRevB.55.7702} {\bibfield  {journal} {\bibinfo  {journal} {Phys. Rev. B}\ }\textbf {\bibinfo {volume} {55}},\ \bibinfo {pages} {7702} (\bibinfo {year} {1997})}\BibitemShut {NoStop}%
\bibitem [{\citenamefont {Rother}(2020)}]{Martin20}%
  \BibitemOpen
  \bibfield  {author} {\bibinfo {author} {\bibfnamefont {M.}~\bibnamefont {Rother}},\ }\href@noop {} {\bibinfo {title} {2{D} {Schroedinger} {Poisson} solver {A}{Q}{I}{L}{A}}},\ \bibinfo {howpublished} {\url{https://www.mathworks.com/matlabcentral/fileexchange/3344-2d-schroedinger-poisson-solver-aquila}} (\bibinfo {year} {2009--2020})\BibitemShut {NoStop}%
\bibitem [{\citenamefont {Zhao}\ \emph {et~al.}(2021)\citenamefont {Zhao}, \citenamefont {Faugno}, \citenamefont {Pu}, \citenamefont {Balram},\ and\ \citenamefont {Jain}}]{Zhao21}%
  \BibitemOpen
  \bibfield  {author} {\bibinfo {author} {\bibfnamefont {T.}~\bibnamefont {Zhao}}, \bibinfo {author} {\bibfnamefont {W.~N.}\ \bibnamefont {Faugno}}, \bibinfo {author} {\bibfnamefont {S.}~\bibnamefont {Pu}}, \bibinfo {author} {\bibfnamefont {A.~C.}\ \bibnamefont {Balram}},\ and\ \bibinfo {author} {\bibfnamefont {J.~K.}\ \bibnamefont {Jain}},\ }\bibfield  {title} {\bibinfo {title} {Origin of the $\ensuremath{\nu}=1/2$ fractional quantum {Hall} effect in wide quantum wells},\ }\href {https://doi.org/10.1103/PhysRevB.103.155306} {\bibfield  {journal} {\bibinfo  {journal} {Phys. Rev. B}\ }\textbf {\bibinfo {volume} {103}},\ \bibinfo {pages} {155306} (\bibinfo {year} {2021})}\BibitemShut {NoStop}%
\bibitem [{\citenamefont {Lee}\ \emph {et~al.}(2026)\citenamefont {Lee}, \citenamefont {Cho},\ and\ \citenamefont {Seo}}]{Lee26}%
  \BibitemOpen
  \bibfield  {author} {\bibinfo {author} {\bibfnamefont {T.}~\bibnamefont {Lee}}, \bibinfo {author} {\bibfnamefont {G.~Y.}\ \bibnamefont {Cho}},\ and\ \bibinfo {author} {\bibfnamefont {D.}~\bibnamefont {Seo}},\ }\href {https://arxiv.org/abs/2604.09542} {\bibinfo {title} {$\mathrm{U}(2)$ {Chern}-{Simons}-{Ginzburg}-{Landau} theory of fractional quantum {Hall} hierarchies}} (\bibinfo {year} {2026}),\ \Eprint {https://arxiv.org/abs/2604.09542} {arXiv:2604.09542 [cond-mat.str-el]} \BibitemShut {NoStop}%
\bibitem [{\citenamefont {Park}\ \emph {et~al.}(1998)\citenamefont {Park}, \citenamefont {Melik-Alaverdian}, \citenamefont {Bonesteel},\ and\ \citenamefont {Jain}}]{Park98b}%
  \BibitemOpen
  \bibfield  {author} {\bibinfo {author} {\bibfnamefont {K.}~\bibnamefont {Park}}, \bibinfo {author} {\bibfnamefont {V.}~\bibnamefont {Melik-Alaverdian}}, \bibinfo {author} {\bibfnamefont {N.~E.}\ \bibnamefont {Bonesteel}},\ and\ \bibinfo {author} {\bibfnamefont {J.~K.}\ \bibnamefont {Jain}},\ }\bibfield  {title} {\bibinfo {title} {Possibility of $p$-wave pairing of composite fermions at $\nu=1/2$},\ }\href {https://doi.org/10.1103/PhysRevB.58.R10167} {\bibfield  {journal} {\bibinfo  {journal} {Phys. Rev. B}\ }\textbf {\bibinfo {volume} {58}},\ \bibinfo {pages} {R10167} (\bibinfo {year} {1998})}\BibitemShut {NoStop}%
\bibitem [{\citenamefont {Balram}\ \emph {et~al.}(2015{\natexlab{c}})\citenamefont {Balram}, \citenamefont {T\"oke}, \citenamefont {W\'ojs},\ and\ \citenamefont {Jain}}]{Balram15a}%
  \BibitemOpen
  \bibfield  {author} {\bibinfo {author} {\bibfnamefont {A.~C.}\ \bibnamefont {Balram}}, \bibinfo {author} {\bibfnamefont {C.}~\bibnamefont {T\"oke}}, \bibinfo {author} {\bibfnamefont {A.}~\bibnamefont {W\'ojs}},\ and\ \bibinfo {author} {\bibfnamefont {J.~K.}\ \bibnamefont {Jain}},\ }\bibfield  {title} {\bibinfo {title} {Fractional quantum {Hall} effect in graphene: Quantitative comparison between theory and experiment},\ }\href {https://doi.org/10.1103/PhysRevB.92.075410} {\bibfield  {journal} {\bibinfo  {journal} {Phys. Rev. B}\ }\textbf {\bibinfo {volume} {92}},\ \bibinfo {pages} {075410} (\bibinfo {year} {2015}{\natexlab{c}})}\BibitemShut {NoStop}%
\end{thebibliography}%

\end{document}